\documentclass[useAMS,usenatbib]{mn2e}
\usepackage{longtable}
\usepackage{rotating}
\usepackage{sidecap}
\usepackage{array}
\usepackage{subfigure}
\usepackage{epsfig}
\usepackage{natbib}
\usepackage{threeparttable}
\usepackage{amsmath,amsfonts,amssymb}
\usepackage{txfonts}

\title[SZ study of 19 LoCuSS galaxy clusters with AMI out to the virial radius]
{Detailed SZ study with AMI of 19 LoCuSS galaxy clusters: masses and temperatures out to the virial radius
\thanks{We
   request that any reference to this paper cites `AMI Consortium:
   Rodr\'{i}guez-Gonz\'{a}lvez et~al.\ 2012'}}
\label{firstpage}
\author[AMI Consortium: Rodr{\'i}guez-Gonz{\'a}lvez et~al.]
{AMI Consortium:
Carmen Rodr{\'i}guez-Gonz{\'a}lvez$^{1,2}$,
Timothy W. Shimwell$^{1,5}$,
\newauthor
 Matthew L. Davies$^1$,
 Farhan Feroz$^1$,
 Thomas M. O. Franzen$^{1,3}$,
 Keith J. B. Grainge$^{1,4}$,
\newauthor
 Michael P. Hobson$^1$,
 Natasha Hurley-Walker$^{1,5}$,
 Anthony N. Lasenby$^{1,4}$,
 Malak Olamaie$^1$,
\newauthor
 Guy Pooley$^1$,
 Richard D. E. Saunders$^{1,4}$,
 Anna M. M. Scaife$^{7}$,
 Michel P. Schammel$^1$,
\newauthor
 Paul F. Scott$^1$,
 David J. Titterington$^1$,
 Elizabeth M. Waldram$^1$
 \vspace{0.03in}\\
$^1$ Astrophysics Group, Cavendish Laboratory, J J Thomson Avenue, Cambridge
CB3 0HE\\
$^2$ Spitzer Science Center, MS 220-6, California Institute of Technology,
Pasadena, CA 91125, USA \\
$^3$ CSIRO Astronomy \& Space Science, Australia Telescope National Facility,
PO Box 76, Epping, NSW 1710, Australia \\
$^4$ Kavli Institute for Cosmology Cambridge, Madingley Road, Cambridge, CB3
0HA\\
$^5$ International Centre for Radio Astronomy Research, Curtin Institute of
Radio Astronomy, 1 Turner Avenue, Technology Park, Bentley, WA 6845,
Australia\\
$^7$ School of Physics \& Astronomy, University of Southampton, Highfield,
Southampton, SO17 1BJ\\
}

\newcommand{\vect}[1]{\mathbf{#1}}

\date{Accepted ---; received ---; in original form \today}

\begin{document}
\maketitle

\begin{abstract}
We present detailed $16$-GHz interferometric observations using the
Arcminute Microkelvin Imager (AMI) of 19 clusters with $L_X > 7\times10^{37}$\,W ($h_{50}=1$)
selected from the Local Cluster
Substructure Survey (LoCuSS; $0.142 \le z \le
0.295$) and of Abell~1758b, which is in the field of view of Abell~1758a. 
We detect and resolve Sunyaev-Zel'dovich (SZ) signals towards 17 clusters, with 
 peak surface brightnesses between 5 and 23$\sigma$.
We use a fast, Bayesian cluster analysis to
 obtain cluster parameter
estimates in the presence of radio point sources, receiver
noise and primordial CMB anisotropy. We fit isothermal $\beta$-models to our data and assume the clusters are virialized
(with all the kinetic energy in gas internal energy). Our gas temperature, $T_{\rm{AMI}}$, 
is derived from AMI SZ data and \emph{not} from X-ray spectroscopy. Cluster parameters internal to
$r_{500}$ are derived under the assumption of hydrostatic equilibrium.   
We find the following.
(i) Different gNFW parameterizations yield significantly different parameter degeneracies.
(ii) For $h_{70}=1$, we find the classical virial
radius, $r_{200}$, to be typically 1.6$\pm$0.1\,Mpc and the total mass 
$M_{\rm{T}}(r_{200})$ typically to be 2.0-2.5$\times$ $M_{\rm{T}}(r_{500})$.
(iii) Where we have found $M_{\rm{T}}(r_{500})$
and $M_{\rm{T}}(r_{200})$ X-ray and weak-lensing values in the literature, there 
is good agreement between weak-lensing and AMI estimates (with $M_{\rm{T},\rm{AMI}}/M_{\rm{T},WL}=1.2^{+0.2}_{-0.3}$ and $=1.0\pm 0.1$ for
 $r_{500}$ and $r_{200}$, respectively). In comparison, most {\sc{Suzaku}}/{\sc{Chandra}}
  estimates are \emph{higher} than for AMI (with  $M_{\rm{T}, X}/M_{\rm{T},\rm{AMI}}=1.7 \pm {0.2}$ within $r_{500}$), particularly for the stronger mergers.
(iv) Comparison of $T_{\rm{AMI}}$ to $T_X$ sheds light on high X-ray masses: even at
large radius, $T_X$ can substantially exceed $T_{\rm{AMI}}$ in mergers.
The use of these higher $T_X$ values will give higher X-ray masses. We stress
that large-radius $T_{\rm{AMI}}$ and $T_X$ data are scarce and must be increased.
(v) Despite the paucity of data, there is an indication of a relation between merger activity and SZ ellipticity.
(vi) At small radius (but away from any cooling flow) the SZ signal (and $T_{\rm{AMI}}$) is less sensitive to ICM disturbance 
than the X-ray signal (and $T_X$) and, even at high radius, mergers affect $n^2$-weighted X-ray data more than
 $n$-weighted SZ, implying that significant shocking or clumping or both occur in even the outer parts of mergers.

\end{abstract}

\begin{keywords}
 cosmology: observations -- cosmic microwave background --
 galaxies: clusters -- Sunyaev--Zel'dovich X-ray -- galaxies: clusters:
  individual (Abell~586, Abell~611, Abell~ 621, Abell~773, Abell~781, Abell~990, Abell~1413, Abell~1423,
 Abell~1704, Abell~1758a, Abell~1758b, Abell~2009, Abell~2111, Abell~2146, Abell~2218, Abell~2409, RXJ0142+2131, RXJ1720.1+2638,
 Zw0857.9+2107, Zw1454.8+2233) 
\end{keywords}

\section{Introduction}\label{Introduction}

The virtues of galaxy clusters are often extolled 
as, for example, being the largest gravitationally bound systems in the
Universe, or being excellent samplers of the matter field on large scales, or simply
as being of fundamental importance to astrophysics and cosmology (see e.g., \citealt{white1993}, \citealt{eke1996} and \citealt{joy2001}).
 To make full use of these virtues one needs observations that, amongst others things, reach large distances
away from cluster centres. It would often be very useful to reach the classical
virial radius $\approx$ $r_{200}$ of a cluster, internal to which the average
density is 200 times the closure density. 
Studying clusters on these scales is important for many reasons. First, these measurements
 are needed to characterize the entire cluster volume. Second, they can be key for any attempt at precision cosmology, including calibrating scaling relations  \citep{kaiser1986},
as they are believed to be less susceptible to the complicated physics of the core region from
 e.g., star formation, energy feedback from active galactic nuclei and gas cooling. 
 Third, the virial radius marks the transition between the accreting matter and the gravitationally-bound, virialized gas of 
 clusters and thus contains information on the current processes responsible for large-scale structure formation.
However, there are few such observations due to the difficulties of obtaining a signal far away from the cluster centre.
We now comment on four methods of estimating cluster masses (see \citealt{allen2011} for a recent, overall review):
\begin{itemize}
\item Spectroscopic measurements of the velocity dispersion of cluster members
require very high sensitivity at moderate to high redshift, and confusion
becomes worse as redshift increases and as distance on the sky from the cluster
centre increases. Cluster masses have recently been obtained this way in e.g., \cite{rines2010} and \cite{sifon2012}.

\item X-ray observations of the Bremsstrahlung (free-free radiation) from the intracluster plasma
(by convention referred to as `gas') have delivered a great deal of information
on cluster physics on a large number of clusters (see e.g., \citealt{ebeling1998}, \citealt{bona_chandra} and \citealt{kotov2005}).
Observations are, of course, difficult at high redshift due to cosmic dimming, and because the X-ray signal is 
$\propto \int n^{2} f(T) dl$, where $n$ is the 
electron density, $T$ is electron temperature, $f(T)$ is a
weak function of $T$, and $l$ is the line of sight through the cluster, there is
significant bias to gas concentration, which makes reaching a high radius
difficult -- however, at low to intermediate redshift there is a small but
growing number of observations that approach or reach $r_{200}$ mainly with the {\sc{Suzaku}} satellite, though the sky
background subtraction is challenging (e.g., \citealt{george2009}, \citealt{hoshino2010}).  

\item Gravitational lensing of background galaxies
 gives the distribution of all the matter in the cluster directly, 
without relying on assumptions obout the dynamical state of the cluster. Any mass concentrations along the line-of-sight not
 associated with the cluster will lead to an overestimate of the weak lensing cluster mass. 
 But the `shear'
signal is proportional to the rate of change with radius of the 
gravitational potential, which changes increasingly slowly with radius at large radius,
so reaching large radius is difficult.
Confusion also bears strongly on this difficulty, and measurement is of course harder
 as redshift increases. Example weak-lensing cluster studies include \cite{okabe2008} and \cite{corless2009}, for analyses of individual high-mass clusters,
 and \cite{mandelbaum2007} and \cite{rozo2011} for analyses of stacked lensing profiles for many low-mass  clusters.

\item The Sunyaev Zel'dovich (SZ; \citealt{sunyaev1970}; see e.g \citealt{birkinshaw1999} and \citealt{carlstrom2002} for
reviews) signal from inverse Compton scattering of the CMB by the cluster gas
has relatively little bias to gas concentration since it is $\propto \int n T dl$, and has remarkably little sensitivity to redshift
over moderate to high redshift; both of these properties make the SZ
effect extremely attractive. The problem with SZ is that it is intrinsically very faint. The
first generation of SZ telescopes, including the OVRO 40-m (see e.g.,
\citealt{birkinshaw1984}), the OVRO 5-m (see e.g., \citealt{herbig1995}), the
OVRO/BIMA arrays \citep[e.g.,][]{carlstrom1996} and the Ryle Telescope (see e.g.,
\citealt{1993MNRAS.265L..57G}) had to integrate for a very long time to get a
significant SZ detection of a single known cluster. The new generation,
including ACT (see e.g., \citealt{Hincks_2010} and \citealt{Marriage_2011}), AMI
(see e.g., \citealt{zwart2008} and \citealt{SHIMWELL}), AMiBA (see e.g., \citealt{Lo_2001} and \citealt{Wu_2008}), MUSTANG (see e.g., \citealt{Korngut_2011} and \citealt{Mroczkowski_2011}), OCRA (see e.g.,
\citealt{Browne_2000} and \citealt{Lancaster_2011}), Planck (see e.g  \citealt{tauber2010} and \citealt{planck2011}), SPT (see e.g \citealt{carlstrom2011} and \citealt{Williamson_2011}) and SZA (see e.g., \citealt{carlstrom1996} and \citealt{Muchovej_2011}) are all much more sensitive. 

The new generation of SZ facilities include two types of instrument:
 ACT, Planck and SPT are instruments with
wide fields of view (FoV) optimized for detecting CMB imprints in large sky
areas in a short amount of time -- this is a very important ability but, for the
imaging of a particular cluster, a wide FoV is of no benefit; in contrast, AMI,
AMiBA, MUSTANG, OCRA and SZA are designed to go deep and to measure 
the masses of the majority of clusters.
\end{itemize}

 In \cite{zwart2010}, we reported initial SZ observations of seven X-ray clusters
 (selected to have low radio flux-densities to limit confusion) that approach or reach 
$r_{200}$. In this paper we report on resolved, interferometric SZ observations with arcminute resolution
 that approach or reach $r_{200}$ in a substantial sample of X-ray clusters selected
above an X-ray flux-density limit (plus a radio flux-density limit) and over a
limited redshift-range (which limits the effects of cosmic evolution); as far as we
are aware, this is the first time such SZ observations of a large cluster sample have been undertaken. These
 measurements are timely since complementary large-$r$ X-ray data have recently been
 obtained with {\sc{Suzaku}} (e.g., \citealt{bautz2009}, \citealt{hoshino2010} and \citealt{kawaharada2010}).
 These early {\sc{Suzaku}} measurements, despite the large
 model uncertainties, are already showing that ICM profiles on these scales appear to disagree with 
 predictions from hydrodynamical cluster simulations (e.g., \citealt{george2009}) and have drawn attention 
 to possible causes such as ICM clumping \citep{nagai2011} and the breakdown of assumptions such as 
 hydrostatic equilibrium (e.g., \citealt{evrard1997}), which can bias the X-ray masses (e.g., \citealt{rasia2004}, \citealt{meneghetti2010} and \citealt{fabjan2011}).
 We stress that SZ  observations, like those in the optical/IR and X-ray, also have their contaminants and 
systematics, and \emph{all} four methods are also hampered by projection effects.

Studying large samples of clusters using 
 multiple techniques is important for building a thorough understanding of cluster physics.
 Well-calibrated mass-observable relations are
crucial for current and future cosmological studies -- see e.g., \cite{allen2011}. 
 To our knowledge, this is the largest cluster-by-cluster study
 for which masses have been derived from SZ targeted observations out to the virial radius. The results
 from this work will be very valuable for detailed comparisons of cluster mass estimates.
This paper is organized as follows. 
In Sec. 2 
we describe the sample selection. The data and instrument are introduced in Sec. 3, while Sec. 4 focuses on the methods applied 
 for mapping the data, identifying radio source foregrounds and removing them from the maps. The analysis of the cluster + radio sources
 environment is outlined in Sec. 5.  Given the difficulty of comparing cluster mass measurements from different data, 
 we provide considerable detail in our results section, Sec. 6. In particular, we present: maps; details on the radio source environment towards
 the clusters; full cluster parameter
posterior distributions internal to two overdensities, $r_{500}$ and $r_{200}$; an investigation of contaminating radio
sources (our main source of systematic error); and we compare our $\beta$-model parameterization
with several generalized Navarro-Frenk-White (gNFW) parameterizations. In Sec. 7 we illustrate the ability of our methodology 
 to recover the cluster mass even for a cluster with a challenging source environment. In Sec. 7
 we discuss our results, in particular, the morphology and dynamical state of the
  clusters and the comparison of SZ-, weak lensing- and X-ray-derived cluster masses and large $r$ X-ray and SZ temperatures. The conclusions of our 
 study are summarized in Sec. 9.

Throughout, we assume a concordance $\Lambda$CDM cosmology
with $\Omega_{\rm{m},0}=0.3$, $\Omega_{\Lambda,0}=0.7$, $\Omega_k=0$,
$\Omega_{b,0}=0.041$, $w_0=-1$, $w_a=0$ and $\sigma_8=0.8$. For the probability
distribution plots and the tables, we take $h=H_0/100$ km\,s$^{-1}$\,Mpc$^{-1}$;
elsewhere we take $H_0=70$ km\,s$^{-1}$\,Mpc$^{-1}$ as the default value and also 
refer when necessary to $h_{X}=H_0/X$\,kms$^{-1}$\,Mpc$^{-1}$. All
coordinates are at epoch J2000.

\section{The LoCuSS catalogue and our sub-sample}\label{sec:sample}

LoCuSS \citep{smith2003, smith2004} is a multi-wavelength survey of 164 X-ray luminous
($L_{\rm{X}}$ $\geq 2\times10^{37}$\,W over the 0.1-2.4\,keV band in the
cluster rest frame (\citealt{ebeling1998} and \citealt{ebeling2000}, $h_{50}=1$)
 galaxy clusters. The narrow range of redshifts $z$ ($0.142 \le z \le
0.295$ minimises cosmic evolution. The clusters have been selected from the \emph{ROSAT} All-Sky Survey \citep{ebeling1998,
ebeling2000, bohringer2004} without taking
into account their structures or dynamical states. Relevant LoCuSS
papers include \cite{marrone2009} and \cite{zhang2010}.

In this work, we study a sub-sample of 19 clusters from the LoCuSS catalogue and Abell~1758b\footnote{Abell~1758b was serendipitously 
 observed in the field of view of Abell~1758a, a LoCuSS cluster.}
(Tab. \ref{tab:clusdetails}) using 16-GHz interferometric AMI data with arcminute resolution.
Our sub-sample includes only those clusters with $\delta >20^{\circ}$. AMI can observe down to
 lower declinations but suffers from poorer $uv$-coverage and satellite interference at  $\delta <20^{\circ}$.
 We also applied an X-ray luminosity cut, $L_X > 7\times10^{37}$\,W (0.1-2.4 keV restframe, $h_{50}=1$),
 lower-luminosity clusters  tend to be fainter in SZ. Contamination from radio sources at 16\,GHz can significantly affect our
SZ detections. For this reason, we have chosen to exclude
clusters with sources brighter than 10\,mJy\,$\rm{beam}^{-1}$ within 10$\arcmin$
of the cluster X-ray centre.

Several studies of the LoCuSS sample of clusters are ongoing. These include both ground-based (Gemini, Keck, MMT, NOAO, Palomar, Subaru, SZA, UKIRT and VLT) 
 as well as space-based ({\sc{Chandra}}, HST, GALEX,  {{\sc{XMM-Newton}} and Spitzer) facilities.
Our AMI SZ data are complementary to other data taken towards these clusters as they probe the large-scale gas structure, are sensitive
 to gas from destroyed density peaks and are particularly beneficial for obtaining robust cluster masses since the SZ signal
 has long been recognised as a good mass proxy (see e.g., \citealt{motl2005}).

\begin{table*}
\caption{Cluster details. It should be noted that Abell~1758b is not part of LoCuSS.}
\label{tab:clusdetails}
\begin{tabular}{lccccccc}
\hline
Cluster         & Right ascension  & Declination     & Redshift      & X-ray
luminosity          & Alternative cluster names \\
                &  (J2000)         & (J2000)         &               & /$
10^{37}$\,W            & \\
                &                  &                 &               &
($h_{50}=1$; see text)    & \\ \hline
Abell 586       & 07 32 12         & +31 37 30       & 0.171         & 11.1
& \\
Abell 611       & 08 00 56         & +36 03 40       & 0.288         & 13.6
& \\
Abell 621       & 08 11 09         & +70 02 45       & 0.223         & 7.8
& \\
Abell 773       & 09 17 54         & +51 42 58       & 0.217         & 13.1
& RXJ0917.8+5143  \\
Abell 781       & 09 20 25         & +30 31 32       & 0.298         & 17.2
& \\
Abell 990       & 10 23 39         & +49 08 13       & 0.144         & 7.7
& \\
Abell 1413      & 11 55 18         & +23 24 29       & 0.143         & 13.3
& \\
Abell 1423      & 11 57 18         & +33 36 47       & 0.213         & 10.0
& RXJ1157.3+3336  \\
Abell 1704      & 13 14 18         & +64 33 27       & 0.216         & 7.8
& \\
Abell 1758a     & 13 32 45         & +50 32 31       & 0.280         & 11.7
& \\
Abell 1758b     & 13 32 29         & +50 24 42       & 0.280         & 7.3
& \\
Abell 2009      & 15 00 21         & +21 22 04       & 0.153         & 9.1
& \\
Abell 2111      & 15 39 40         & +34 26 00       & 0.229         & 10.9
& \\
Abell 2146      & 15 55 58         & +66 21 09       & 0.234         & 9.0
& \\
Abell 2218      & 16 35 45         & +66 13 07       & 0.171         & 9.3
& \\
Abell 2409      & 22 00 57         & +20 57 50       & 0.147         & 8.1
& \\
RXJ0142+2131    & 01 42 03         & +21 31 40       & 0.280         & 9.9
& \\
RXJ1720.1+2638  & 17 20 10         & +26 37 31       & 0.164         & 16.1
& \\
Zw0857.9+2107   & 09 00 39         & +20 55 17       & 0.235         & 10.8
& Z2089 \\
Zw1454.8+2233   & 14 57 15         & +22 20 34       & 0.258         & 13.2
& Z7160  \\ \hline
\end{tabular}
\end{table*}

\section{Instrument and observations}\label{sec:obs}

AMI consists of two aperture-synthesis interferometric arrays located
near Cambridge.
 The Small Array (SA) is optimized for SZ imaging while the Large Array (LA) is
used to observe radio sources
 that contaminate the SZ effect in the SA observations.
 AMI's $uv$-coverage is well-filled all the way down to $\approx 180\lambda$, corresponding to a maximum angular scale of $\approx 10\arcmin$.
 AMI is described in
detail in AMI Consortium: Zwart et al. (2008)\footnote{The observing frequency range given in AMI Consortium: Zwart et al. has been altered
slightly,
  as described in AMI Consortium: Franzen et al. 2010.}.

SA pointed observations of all the clusters were taken between 2007 and
2010 while LA raster observations, which were mostly 61+19\,pt hexagonal
rasters\footnote{A 61+19\,point raster observation consists of 61 pointings
with separations of 4$\arcmin$, of which the central 19 pointings have lower noise
levels, see e.g., \cite{hurley2011} for example LA maps.}
centred on the cluster X-ray position, were made between 2008 and 2010.
Typically,
each cluster was observed for 20-80 hours with the SA and for 10-25 hours with the LA.
The thermal noise levels for the SA ($\sigma_{\rm{SA}}$) and for the LA
($\sigma_{\rm{LA}}$) were obtained by
applying the \textsc{aips}\footnote{http://www.aips.nrao.edu} task {\sc{imean}}
on a
 section of the map far down the primary beam and free from any significant 
 contamination. In Tab. \ref{tab:mapnoise} we provide central thermal noise
estimates for the SA and LA observations; they reflect
the amounts of data remaining after flagging. A series of algorithms has been developed to 
remove (or `flag-out') bad data points arising from interference,
shadowing, hardware and other errors. This is a stringent process that typically results in $\approx$ 30-50$\%$ of
 data being discarded before the analysis.
 A primary-beam correction factor has been applied, as the thermal noise level is dependant on the distance 
 from the pointing centre.

The raw visibility files were put through our local data reduction
pipeline, {\textsc{reduce}}, described in detail in 
\cite{davies2010}, and exported in {\textsc{fits}} format.
Bi-daily observations of 3C286 and 3C84 were used for flux calibration while
interleaved calibrators
selected from the Jodrell Bank VLA Survey \citep{patnaik1992, browne1998,
wilkinson1998}
were observed every hour for phase calibration.

\section{Mapping and source detection and subtraction}\label{sec:maps}

Our LA map-making and source-finding procedures follow \cite{SHIMWELL}.  We
applied standard \textsc{aips} tasks to image the continuum and
individual-channel \textsc{uvfits} data output from {\textsc{reduce}}.

 At 16\,GHz, the dominant contaminants to the SZ decrements are radio sources.
In order to recover the SZ signal, the contribution of these radio sources to the data 
 need to be removed; this is done as follows.
\begin{itemize}
\item{ First, the {\textsc{clean}}ed LA continuum maps\footnote{The LA continuum maps were {\sc{clean}}ed down to $3\sigma_{\rm{LA}}$ with no boxes.} were put
 through the AMI-developed source-extraction software {\textsc{sourcefind}} \citep{franzen2010} to 
 identify and characterize radio sources on the LA maps above a certain signal-to-noise. {\textsc{sourcefind}} provides estimates for the
 right ascension $x_s$, declination $y_s$, flux density\footnote{We catalogue
the peak flux of the source,
 unless the source is extended, in which case we
 integrate the source surface brightness over its projected solid angle to give its integrated flux density (see e.g., \citealt{franzen2010}).}, 
$S_0$, and spectral index $\alpha$\footnote{We adopt the convention $S\propto \nu^{-\alpha}$.} at the
central frequency $\nu_0$ for identified radio sources. We impose a detection threshold such that 
we select only those radio sources with a flux density $\geq 4\sigma_{LA_p}$ on
the {\textsc{clean}}ed LA continuum maps, where $\sigma_{\rm{LA_p}}$ refers to
pixel values on the LA noise maps.
 The number of $\geq$4$\sigma_{LA_p}$ sources
 detected in our LA observations of each cluster is given in Tab.
\ref{tab:mapnoise}.}

\item{
Second, \emph{prior to any source subtraction}, we run our cluster-analysis software, which fits for the position, flux and spectral index of the 
{\textsc{sourcefind}}-detected radio sources using the source parameters obtained by {\textsc{sourcefind}} as priors. For some of the
 less contaminating radio sources, our cluster-analysis software uses delta-priors for the source parameters centred at the LA estimates (see Sec. \ref{analysis} for further details) .
 }

\item{
 Third, the source parameters given by the cluster analysis were used to perform source subtraction on the SA maps. This was done using in-house software,
{\textsc{muesli}}, which is an adaptation of the standard {\textsc{aips}} task
{\textsc{uvsub}} optimized for processing AMI data.
The flux-density contributions from detected radio sources were subtracted from each SA channel \textsc{uvfits} file using either
the mean values for their position, spectral index and flux density derived
from our Bayesian analysis, when these parameters are not given
delta-function priors, or, otherwise, using the LA estimates for these
source parameters. Details of the
priors assigned to each of the sources labelled on the SA maps can be found in
Sec. \ref{sourcepriors} and Tab. \ref{tab:sourcelabel}.}

\item{
Fourth, after source subtraction, the SA maps were {\textsc{clean}ed} with a
tight
box around the SZ signal. In contrast, the LA maps and SA maps before source subtraction
  were {\textsc{clean}ed} with a single box comprising the entire map. 
Both the SA and the LA maps were {\textsc{clean}ed} down to $3\sigma$.}

\end{itemize}

\begin{table*}
\caption{Observational details. SA and LA noise levels, $\sigma_{\rm{SA}}$ and $\sigma_{\rm{LA}} \geq 4\sigma$ and the number of sources detected above $4\sigma_{\rm{LA_p}}$ on the LA rasters for each cluster.
 Abell~1758b is not part of LoCuSS.}
 \label{tab:mapnoise}
\begin{tabular}{lccc}
\hline
Cluster   & $\sigma_{SA}$               & $\sigma_{LA}$   & Number of LA
$4\sigma_{LA}$ sources  \\
          & (mJy)                       & (mJy)           \\ \hline
Abell 586       & 0.17   & 0.09        & 23 \\
Abell 611       & 0.11   & 0.07        & 23 \\
Abell 621       & 0.11   & 0.09        & 13 \\
Abell 773       & 0.13   & 0.09        &  9 \\
Abell 781       & 0.12   & 0.07        & 24 \\
Abell 990       & 0.10   & 0.08        & 20 \\
Abell 1413      & 0.13   & 0.09        & 17 \\
Abell 1423      & 0.08   & 0.07        & 31 \\
Abell 1704      & 0.09   & 0.06        & 13 \\
Abell 1758a     & 0.12   & 0.08        & 14 \\
Abell 1758b     & 0.13   & 0.08        & 14 \\
Abell 2009      & 0.11   & 0.14        & 18 \\
Abell 2111      & 0.09   & 0.07        & 22 \\
Abell 2146      & 0.15   & 0.06        & 15 \\
Abell 2218      & 0.07   & 0.10        & 15 \\
Abell 2409      & 0.14   & 0.05        & 15 \\  
RXJ0142+2131    & 0.11   & 0.06        & 22 \\
RXJ1720.1+2638  & 0.08   & 0.10        & 17 \\
Zw0857.9+2107   & 0.13   & 0.12        & 13 \\
Zw1454.8+2233   & 0.10   & 0.10        & 16 \\ \hline
 \end{tabular}
\end{table*}

\section{ Analysis} \label{analysis}

We use our own Bayesian analysis package, {\sc{McAdam}}, to 
 estimate cluster parameters internal to $r_{500}$ and $r_{200}$ from AMI data in the presence
 of radio point sources, receiver noise and primordial CMB anisotropy.
The cluster and radio sources are parameterized in our analysis (see below) while the remaining 
 components are included in a generalized noise covariance matrix; we note that these are the 
 only significant noise contributions because large-scale emission from e.g., foreground galactic emission is resolved 
 out by our interferometric observations.

 {\sc{McAdam}} was originally
developed by \cite{marshall2003} and \cite{feroz2009} adapted it to work on AMI data. The latest
\textsc{McAdam} uses {\sc MultiNest} (\citealt{feroz2008}, \citealt{bridges2009}) as its inference engine to 
allow Bayesian evidence and posterior
distributions to be calculated efficiently, even for posterior distributions
with large (curved) degeneracies and/or mutiple peaks. This addition has been key to 
 our analysis since the posteriors of AMI data often have challenging dimensionalities, $>30$, primarily as a result
 of the presence of a large number of radio sources in the AMI observations.

\subsection{Model} \label{clusmod}

We have modelled the cluster density profile assuming
spherical symmetry using a $\beta$-model \citep{cavaliere1978}:
\begin{equation}
\rho_{\rm{g}}(r)=\frac{\rho_{\rm{g}}(0)}{\left[ 1 + \left( \frac{r}{r_{\rm{c}}}
\right)^2 \right]^{\frac{3\beta}{2}}},
\label{eq:den}
\end{equation}
where gas mass density $\rho_{\rm{g}}(r)=\mu n(r)$, $\mu=1.14m_{\rm{p}}$ is the gas mass per electron and $m_{\rm{p}}$ is
the proton mass.
The core radius $r_{\rm{c}}$ gives the density profile a flat top at low
$\frac{r}{r_{\rm{c}}}$ and $\rho_{\rm{g}}$ has a logarithmic slope of
$3\beta$ at large  $\frac{r}{r_{\rm{c}}}$. 

We choose to model the gas as isothermal, using the virial mass-temperature
relation and assuming that all kinetic energy is in gas internal energy:

\begin{align}
\rm{k}_{\rm{B}}T(r_{200}) &= \frac{\rm{G}\mu M_{\rm{T}}(r_{200})}{2r_{200}} \\
&=
\frac{\rm{G}\mu}{2\left(\frac{3}{4\pi\left(200\rho_{\rm{crit}}\right)}\right)^{1/3}}M_{\rm{T}}^{2/3}(r_{200})
\\
&= 8.2
\textrm{keV}\left(\frac{M_{\rm{T}}(r_{200})}{10^{15}h^{-1}\rm{M}_{\odot}}\right)^{2/3}\left(\frac{H(z)}{H_{0}}\right)^{2/3}.
\label{eq:virtemp}
\end{align}
$M_{\rm{T}}(r_{200})$ and $T(r_{200})$
refer to the total mass and gas temperature within $r_{200}$ (see e.g.,
\citealt{2005RvMP...77..207V}). This relation allows cluster parameters within
$r_{200}$ to be inferred without
 assuming hydrostatic equilibrium; note that, in our methodology, parameters
 describing the  cluster at smaller $r$ (e.g., $r_{500}$) do, however,
 assume hydrostatic equilibrium.
Further details of the cluster analysis can be found in \cite{carmen2010} and
\cite{olamaie2011}. The good agreement between mass estimates from weak-lensing and AMI data on 6 clusters in \cite{hurley2011}
 supports the use of this $M-T$ relation in our analysis.

\subsection{ Priors} \label{priors}

\subsubsection{ Cluster priors} \label{sec:cluspriors}

The cluster model parameters
$\vect{\Theta}_{\rm{c}}=(x_{\rm{c}}$, $y_{\rm{c}}$,
$M_{\rm{T}}(r_{200})$, $f_{\rm{g}}(r_{200})$, $\beta$, $r_{\rm{c}}$, $z$) have
priors that are assumed to be separable. $x_{\rm{c}}, y_{\rm{c}}$ are the cluster position (RA and Dec, respectively) and $f_{\rm{g}}$ is the gas fraction, which is defined as
\begin{equation}
f_{\rm{g}} =\frac{M_{\rm{g}}}{M_{\rm{T}}}.
\label{eq:fg}
\end{equation}
Further details on these priors are given in Tab. \ref{tab:Summary-of-cluster}.

This set of sampling parameters has proved sufficient for our cluster detection
algorithm (\citealt{SHIMWELL}) and to describe the physical cluster parameters.
We emphasize that this way of analysing the data is \textit{different} from the
way used traditionally, in which an X-ray spectroscopic temperature is used as
an input parameter. The difficulty with this use of an X-ray temperature is
that, in practice, the temperature measurement usually applies to gas relatively
close to the cluster centre (but any cooling flow is excised). By sampling from $M_{\rm{T}}$ and using the $M-T$ relation (Eq. \ref{eq:virtemp}), the 
\textit{temperature of each cluster is derived from SZ data only} and is
averaged over the angular scale of the SZ observation, which is typically larger
than the angular scale of the X-ray temperature measurement. This way, although our analysis
 does not yield $T(r)$, it gives and uses a temperature which is representative of the cluster volume
 we are investigating.

\begin{table*}
\caption{Summary of the priors for the sampling parameters in each
  model. The value for the redshift and position priors have not been
  included in this table since they are cluster specific. Instead, they are
  given in Tab. \ref{tab:clusdetails} for each cluster. }
\label{tab:Summary-of-cluster}
\begin{tabular}{cccc}
\hline
{ Parameter}  & { Prior Type} & { Values} & { Origin}\\

\hline
{{} $x_{\rm{c}},y_{\rm{c}} {''}$}  & { Gaussian at
  $\vect{x}_{\textrm{X-ray}},\sigma=60^{''}$} &
{cluster position}
& { \cite{ebeling2000}}\\
{ $\beta$}  &  {uniform } & { $0.3-2.5$} & {\cite{marshall2003}}\\
{ $M_{\rm{T}}(r_{200})/h^{-1}\rm{M}_\odot$} & { uniform in log} & {
$1\times10^{13.5}-5\times10^{15}$} & { physically reasonable, e.g., \citealt{vikhlinin2006}}\\
{ $r_{\rm{c}}/h^{-1}\textrm{kpc}$}  & { uniform} & { $10-1000$} & { physically
reasonable  e.g., \citealt{vikhlinin2006}}\\
{ $z$} & { delta} & {cluster redshift } & {\cite{ebeling2000}}\\
{ $f_{\rm{g}}(r_{200})/h^{-1}$} & { Gaussian, $\sigma= 0.0216$} & { $0.0864$} &
{ \cite{larson2010,zhang2010}}\\
\hline
\end{tabular}
\end{table*}

\subsubsection{ Source priors} \label{sourcepriors}

Radio sources detected on the LA maps using {\textsc{sourcefind}} are modelled
by four source parameters,
$\vect{\Theta}_{\rm{S}}$ = ($x_s$, $y_s$, $S_0$, $\alpha$). Priors on these
parameters are based on LA measurements,
discussed in Sec. \ref{sec:maps}.

Sources on the source-subtracted SA maps are labelled according to Tab.
\ref{tab:sourcelabel}. Delta-function priors on all the
source parameters tend to be given to those sources whose flux density
is $<4\sigma_{SA}$ and to those outside the $10\%$ radius of the SA power
primary
beam. The remaining sources are usually assigned a delta-function prior on
position and Gaussian priors on $\alpha$
and $S_0$. However, in a few cases we replace delta-function priors on the
 source parameters with Gaussian priors as this can increase the accuracy of
the source
 subtraction. These wider priors can be necessary to account for discrepancies
between the LA and SA measurements. Reasons for these differences include: a
poor fit of our
Gaussian model for the power primary beam far from the
pointing centre,  correlator artifacts, source variability and source
extension.

\begin{table}
\caption{Priors on position, spectral index and flux density given to detected
  sources. The symbols correspond to the labels in the SA
  source-subtracted maps. The Gaussian priors are centred
  on the LA measurements. $\sigma$ values for the Gaussian priors are assigned as follows:
 for the Gaussian prior on the flux-density of each radio source, $\sigma$ is set to 40\% of the source flux density;  
 for the spectral index $alpha$, $\sigma$ is set to the LA error on $\alpha$ and for 
the source position, $\sigma$ is set to $60\arcsec$.}
 \label{tab:sourcelabel}
\centering
\begin{tabular}{cccc}
\hline
{Symbol} & {$\Pi(S_0)$} & {$\Pi(\alpha)$} & {$\Pi(x_s, y_s)$}\\ \hline  
{+} & {delta}& {delta} & {delta}\\
{$\times$} & {Gaussian}& {Gaussian} & {delta}\\
{$\triangle$} & {Gaussian} & {Gaussian}& {Gaussian}\\
\hline
\end{tabular}
\end{table}

\section{Results and commentary} \label{results}


Out of the 20 clusters listed in Tab. \ref{tab:clusdetails}, we detect SZ
decrements towards 17. For these clusters we
 present  SA maps before and after source subtraction as well as 
marginalized
posterior distributions for some cluster and source parameters
 and mean values of 
selected cluster parameters (Tab. \ref{tab:clusparams}), with the exception of
 Abell~2409, which was found
to have a local environment which renders it unsuitable for robust
parameter estimation (see Sec. \ref{resultsA2409}). 
For the posterior distributions all ordinates and abscissae in these plots are linear,
 the $y$-axis for the 1-d marginals is the probability density and $h$ is short for $h_{100}$.
 It is important to note that, while the posterior
probability distributions for large-scale cluster parameters reflect the
uncertainty in the
{\textsc{McAdam}}-derived flux-density estimates, the radio source-subtracted maps
do not, as they simply use a single value (the mean) for each source parameter.
The effect of our priors on the results
has been thoroughly tested in a previous study by \cite{olamaie2011}, which
found that the priors used in this parameterization do not to lead
to any strong biases in the cluster parameter estimates. 

 The SA maps have labels indicating the position of detected radio
sources and their priors (Tab. \ref{tab:sourcelabel}); the square box in these plots
indicates the best-fit cluster position determined by
{\textsc{McAdam}}. No primary-beam
correction has been applied to the SA maps presented in this paper, unless stated otherwise. The contour levels on the SA maps, unless otherwise stated, start at $2\sigma_{SA}$ and 
increase linearly from 2 to $10\sigma_{SA}$. On radio-only images, positive contours are shown as solid lines and negative contours as dashed lines,
 but on radio+X-ray images, negative radio contours are shown as solid lines and X-ray shown as greyscale.  The bottom-left ellipses on the SA maps are the FWHMs of the synthesized beams.
  A $0.6$-k$\lambda$ taper
was applied to the SA source-subtracted maps to downweight
long-baseline visibilities with the purpose of increasing the
signal-to-noise of the large-scale structure; this typically leads to a
$\approx 20\%$ increase in the noise.
 The X-ray images are obtained from archive \emph{ROSAT} and {\sc{Chandra}} data.

We remind readers that when looking at a radio map -- necessarily with a particular $uv$-weighting -- a near-circular image does not mean that the SA failed to resolve the SZ signal. Investigating angular structure / size requires assessment in $uv$-space,
 which can be done with a selection of maps made over different $uv$ ranges but is optimally done here in $uv$-space with \textsc{McAdam}. In fact, all the SZ decrements in this paper \emph{are} resolved. 


\onecolumn
\begin{center}
\begin{longtable}{lllllllll}
\caption{{Source properties for detected sources within 5$\arcmin$ of the
SZ mean central position. The number next to each cluster name denotes the source number; this label is used in the plots showing the marginalized posterior distributions for the source fluxes. 
$S_0{\rm{McA}}$ is the {\sc{McAdam}}-derived best-fit source flux at 16\,GHz. $\alpha$ is the source spectral index estimated by {\sc{McAdam}} and centred at the {\sc{McAdam}}-derived mean frequency and
the last column contains the distance between the cluster SZ centroid (as determined by {\sc{McAdam}}) and the source.}}
\label{tab:source_info} \\
\hline Name     & Right ascension & Declination  &  $S_0{\rm{McA}}$
& $\alpha$ & Distance from SZ centroid \\          
 & (hh:mm:ss , J2000)      & ($^{\circ}$:$\arcmin$:$\arcsec$ , J2000)       &  mJy   &          & $\arcsec$ \\ \hline
\endhead
A586\_0         &   07:32:20.5 &  +31:38:02.8   &  0.26  &   1.20 &  28 \\
A586\_1         &   07:32:19.1 &  +31:40:25.6 &  0.86  &   0.28 &  171 \\
A586\_2         &   07:32:11.0 &  +31:39:47.6 &  0.86  &   1.20 &  178 \\
A586\_3         &   07:32:04.5 &  +31:39:09.6  &  0.41  &   1.53 &  222 \\
A586\_4         &   07:32:35.4 &  +31:35:35.5 &  1.03  &   -0.47 &  227 \\
A586\_5         &   07:32:21.2 &  +31:41:26.3 &  7.44  &   0.46 &  232 \\
A586\_6         &   07:32:42.7 &  +31:38:37.1 &  0.53  &   0.16 &  293 \\
\\
A611\_0         &   08:01:07.0 &  +36:02:18.9  &  0.32  &   -0.27 &  108 \\
A611\_1         &   08:00:52.6 &  +36:06:14.2  &  0.44  &   2.18 &  199 \\
A611\_2         &   08:01:17.0 &  +36:04:27.8  &  0.5   &   -0.71 &  229 \\
\\
A621\_3         &   08:11:12.8 &  +70:02:27.2  &  7.18  &   1.34 &  25 \\
A621\_4         &   08:11:19.3 &  +70:00:48.4  &  0.6   &   -1.31 &  127 \\
A621\_5         &   08:11:35.2 &  +70:04:25.6  &  0.16  &   0.13 &  166 \\
A621\_6         &   08:10:38.0 &  +70:04:09.3  &  0.09  &   0.01 &  181 \\
\\
A781\_0         &   09:20:24.7 &  +30:31:49.9 &  0.24  &   -0.04 &  4 \\
A781\_1         &   09:20:23.3 &  +30:29:49.3 &  8.97  &   0.97 &  126 \\
A781\_2         &   09:20:08.4 &  +30:32:15.8 &  1.49  &   -0.18 &  213 \\
A781\_3         &   09:20:14.0 &  +30:28:60.0 &  2.12  &   0.63 &  223 \\
\\
A990\_0         &   10:23:47.3 &  +49:11:25.5 &  0.44  &   -0.21 &  208 \\
A990\_1         &   10:24:02.1 &  +49:06:51.8  &  2.78  &   2.18 &  239 \\
\\
A1413\_0        &   11:55:15.4 &  +23:23:59.4 &  0.47  &   1.04 &  59 \\
A1413\_1        &   11:55:08.8 &  +23:26:16.6 &  3.1   &   0.98 &  222 \\
\\
A1423\_2        &   11:57:17.1 &  +33:36:30.6 &  0.54  &   -0.19 &  63 \\
A1423\_3        &   11:57:28.5 &  +33:35:31.0  &  0.26  &   2.16 &  134 \\
A1423\_4        &   11:57:19.7 &  +33:39:58.3 &  0.39  &   0.73 &  171 \\
A1423\_5        &   11:57:35.2 &  +33:37:21.8 &  0.19  &   0.58 &  176 \\
A1423\_6        &   11:57:40.5 &  +33:35:10.1 &  0.18  &   -0.40 &  270 \\
A1423\_7        &   11:57:20.5 &  +33:41:57.8  &  0.73  &   -0.22 &  290 \\
A1423\_8        &   11:57:39.0 &  +33:34:03.3   &  0.24  &   1.47 &  290 \\
\\
A1704\_0        &   13:14:02.3 &  +64:38:29.5 &  0.2   &   -0.03 &  273 \\
A1704\_1        &   13:14:52.6 &  +64:37:59.9 &  0.81  &   0.86 &  287 \\
\\
A1758a\_0 &         13:32:53.3 &  +50:31:40.6 &  7.08  &   0.5 &  54 \\
A1758a\_1 &         13:32:38.6 &  +50:33:37.7 &  0.77   &  0.36 &  150 \\
A1758a\_2 &         13:33:02.2 &  +50:29:26.4 &  1.43  &    0.28 & 190 \\
A1758a\_3 &         13:32:39.6 &  +50:34:31.1 &  0.3  &   1.45 &  192 \\
A1758a\_4 &         13:32:41.5 &  +50:26:47.7 &  0.46  &    -0.76 & 294 \\
A1758b\_0 &         13:32:33.1 &  +50:22:35.1 &  0.23  &    0.11 &  90 \\
A1758b\_1 &         13:32:41.5 &  +50:26:47.7 &  0.51  &    -0.68 & 196 \\
\\
A2009\_0        &   15:00:19.7 &  +21:22:12.6 &  1.85  &   3.14 &  58 \\
A2009\_1        &   15:00:28.6 &  +21:22:45.8 &  0.18  &   0.66 &  133 \\
A2009\_2        &  15:00:19.6 &   +21:22:11.3 &  1.97   &   2.64 &  77 \\
A2009\_3        &  15:00:28.6 &   +21:22:45.7 &  0.18   &   0.66 &  135 \\
\\
A2111\_0        &   15:39:30.1 &  +34:29:05.5   &  0.5   &   0.68 &  222 \\
A2111\_1        &   15:39:56.7 &  +34:29:31.8 &  0.81  &   -1.47 &  297 \\
\\
A2146\_0        &   15:56:04.2 &  +66:22:13.0 &  5.94  &   0.55 &  43 \\
A2146\_1        &   15:56:14.0 &  +66:20:53.5 &  1.82  &   1.03 &  59 \\
A2146\_2        &   15:56:15.4 &  +66:22:44.5 &  0.15  &   0.34 &  89 \\
A2146\_3        &   15:55:57.4 &  +66:20:03.1  &  1.67  &   -0.22 &  106 \\
A2146\_4        &   15:56:27.1 &  +66:19:43.8 &  0.1   &   0.64 &  164 \\
A2146\_5        &   15:55:25.7 &  +66:22:04.0  &  0.48  &   -0.22 &  249 \\
\\
A2218\_0        &   16:35:47.4 &  +66:14:46.1 &  2.86  &   0.07 &  100 \\
A2218\_1        &   16:35:21.8 &  +66:13:20.6 &  5.99  &   0.23 &  141 \\
A2218\_2        &   16:36:15.6 &  +66:14:24.0  &  1.77  &   0.72 &  200 \\
\\
A2409\_0        &   22:00:39.7 &  +20:58:55.0 &  0.75  &   1.9 &  241 \\
A2409\_1        &   22:01:11.2 &  +20:54:56.8 &  3.12  &   0.1 &  275 \\
\\
RXJ0142+2131\_0 &   01:42:09.2 &  +21:33:23.4 &  1.09  &   0.7 &  117 \\
RXJ0142+2131\_1 &   01:42:11.0 &  +21:29:45.3 &  1.16  &   1.52 &  156 \\
RXJ0142+2131\_2 &   01:42:23.3 &  +21:30:46.7 &  0.3   &   0.03 &  273 \\
\\
RXJ1720+2638\_0 &   17:20:10.0 &  +26:37:29.7 &  6.92  &   1.24 &  46 \\
RXJ1720+2638\_1 &   17:20:01.2 &  +26:36:32.3 &  2.05  &   0.57 &  105 \\
RXJ1720+2638\_2 &   17:19:58.4 &  +26:34:19.6 &  1.22  &   1.46 &  203 \\
RXJ1720+2638\_3 &   17:20:25.5 &  +26:37:57.2  &  0.88  &   0.89 &  234 \\
\\
Zw0857.9+2107\_0 &   09:00:36.9 &  +20:53:41.4  &  1.22  &   0.31 &  102 \\
Zw0857.9+2107\_1 &   09:00:55.5 &  +20:57:21.2 &  0.96  &   1.37 &  259 \\
Zw0857.9+2107\_2 &   09:00:52.8 &  +20:58:36.5 &  5.57  &   0.09 &  274 \\
\\
Zw1454.8+2233\_0 &   14:57:14.8 &  +22:20:34.2 &  1.64  &   0.28 &  14 \\
Zw1454.8+2233\_1 &   14:57:08.2 &  +22:20:08.6  &  1.55  &   1.89 &  108 \\
Zw1454.8+2233\_2 &   14:57:10.6 &  +22:18:45.6 &  1.49  &   0.94 &  137 \\
Zw1454.8+2233\_3 &   14:56:58.9 &  +22:18:49.6 &  8.36  &   0.17 &  258 \\
Zw1454.8+2233\_4 &   14:57:04.3 &  +22:24:11.9 &  0.83  &   -0.63 &  260 \\
Zw1454.8+2233\_5 &   14:57:24.8 &  +22:24:52.6 &  0.13  &   -0.25 &  281 \\
Zw1454.8+2233\_6 &   14:57:35.7 &  +22:19:46.8 &  1.04  &   1.67 &  285 \\ \hline
\end{longtable}
\end{center}
\twocolumn

\onecolumn
\begin{table}
\caption{Mean and 68$\%$-confidence uncertainties for some {\sc{McAdam}}-derived large-scale cluster parameters.}
\label{tab:clusparams}
\begin{tabular}{lcccccccccc}
\hline
Cluster name & $M_{T}(r_{200})$              & $M_{T}(r_{500})$               & $M_{g}(r_{200})$               & $M_{g}(r_{500})$              & $r_{200}$   & $r_{500}$                        & $T_{\rm{AMI}}$                            & $Y(r_{200})$                   & $Y(r_{500})$ \\
             &$\times10^{14}h_{100}^{-1}M_{\odot}$ & $\times10^{14}h_{100}^{-1}M_{\odot}$ & $\times10^{13}h_{100}^{-2}M_{\odot}$ & $\times10^{13}h_{100}^{-2}M_{\odot}$& $h_{100}^{-1}$\,Mpc & $\times10^{-1}h_{100}^{-1}$   Mpc      & keV                            & $\times 10^{-5}\rm{arcmin}^2$                    & $\times 10^{-5}\rm{arcmin}^2$\\ \hline
A586 & $5.1 \pm 2.1$ & $2.1 \pm 0.9$ & $4.3 \pm 1.7$ & $2.6 \pm 0.7$ & $1.2 \pm 0.2$ & $6.6 \pm 1.0$ & $5.2 \pm 1.4$ & $3.6_{-2.1}^{+2.0}$  & $2.7 \pm 1.4$  \\
A611 & $4.0_{-0.8}^{+0.7}$ & $2.0 \pm 0.5$ & $3.5 \pm 0.6$ & $2.8 \pm 0.3$ & $1.1 \pm 0.1$ & $6.3 \pm 0.5$ & $4.5 \pm 0.6$ & $2.2 \pm 0.5$  & $2.1 \pm 0.4$ \\
A621 & $4.8_{-1.8}^{+1.7}$ & $1.4 \pm 0.9 $ & $4.1 \pm 1.0$ & $1.5 \pm 0.8$ & $1.2_{-0.1}^{+0.2}$ & $5.3_{-0.1}^{+0.2}$  & $5.0 \pm 1.2$  & $3.1_{-1.6}^{+1.5}$ & $1.9 \pm 1.0$\\
A773 & $3.6 \pm 1.2$ & $1.7 \pm 0.6$ & $3.1_{-0.9}^{+1.0}$ & $2.1 \pm 0.4$ & $1.1 \pm 0.1$ & $6.0 \pm 0.7$ & $4.1_{-1.0}^{+0.9}$ & $1.9 \pm 0.9$  & $1.6_{-0.7}^{+0.6}$\\
A781 & $4.1 \pm 0.8$ & $2.0 \pm 0.5$ & $3.6 \pm 0.6$ & $2.9 \pm 0.4$ & $1.1 \pm 0.1$ & $6.3 \pm 0.5$ & $4.5 \pm 0.6$  & $2.3 \pm 0.6$ & $2.2 \pm +0.5$\\
A990 & $2.0_{-0.1}^{+0.4}$ & $1.1 \pm 0.2$ & $1.8 \pm 0.3$ & $1.6 \pm 0.2$ & $0.9 \pm 0.1$ & $5.5 \pm 0.3$  & $2.8 \pm 0.3$ & $0.7 \pm 0.2$ & $0.7 \pm 0.15$ \\
A1413 & $4.0 \pm 1.0$ & $1.9 \pm 0.6$ & $3.5 \pm 0.8$ & $2.7 \pm 0.4$ & $1.1 \pm 0.1$ & $6.6 \pm 0.6$ & $4.4 \pm 0.7$ & $2.2_{-0.8}^{+0.7}$ & $2.1 \pm 0.6$ \\
A1423 & $2.2 \pm 0.8$ & $1.1 \pm 0.4$ & $1.9 \pm 0.7$ & $1.5 \pm 0.4$ & $0.9 \pm 0.1$ & $5.3 \pm 0.6$  & $3.0_{-0.7}^{+0.8}$  & $0.9 \pm 0.5$ & $0.8 \pm 0.4$ \\
A1758a & $4.1_{-0.8}^{+0.7}$ & $2.5 \pm 0.4$ & $3.6 \pm 0.5$ & $3.4 \pm 0.4$ & $1.1 \pm 0.1$ & $6.8 \pm 0.4$  & $4.5 \pm 0.5$  & $2.3 \pm 0.5$ & $2.3 \pm 0.4$\\
A1758b & $4.4 \pm 1.9$ & $2.2 \pm 1.0$ & $3.7_{-1.5}^{+1.6}$ & $2.2 \pm 0.5$ & $1.1 \pm 0.2$ & $6.4_{-1.0}^{+1.1}$ & $4.6 \pm 1.4$ & $2.7 \pm 1.7$ & $1.6 \pm 0.6$ \\
A2009 & $4.6 \pm 1.5$ & $2.0_{-0.6}^{+0.2}$ & $3.9 \pm 0.7$ & $2.4_{-0.3}^{+0.2}$ & $1.2 \pm 0.1$ & $6.5_{-0.6}^{+0.4}$ & $4.8_{-0.6}^{+0.4}$  & $2.8_{-1.4}^{+1.3}$ & $2.2 \pm 0.9$\\
A2111 & $4.2 \pm 0.9$ & $1.8 \pm 0.5$ & $3.6 \pm 0.7$ & $2.5 \pm 0.3$ & $1.1 \pm 0.1$ & $6.2 \pm 0.6$ & $4.6 \pm 0.6$  & $2.4_{-0.7}^{+0.6}$ & $2.2 \pm 0.5$\\
A2146 & $5.0 \pm 0.7$ & $2.7 \pm 0.5$ & $4.4 \pm 0.5$ & $3.7 \pm 0.4$ & $1.2 \pm 0.1$ & $7.1 \pm 0.5$ & $5.2 \pm 0.5$  & $3.2 \pm 0.5$ & $3.1 \pm 0.4$\\
A2218 & $6.1 \pm 0.9$ & $2.7 \pm 0.6$ & $5.4_{0.7}^{+0.6}$ & $4.3 \pm 0.4$ & $1.3 \pm 0.1$ & $7.3 \pm 0.5$  & $5.9 \pm 0.6$ &  $4.5 \pm 0.7$ & $4.4 \pm 0.7$\\
RXJ0142+2131 & $3.7_{-1.2}^{+1.1}$ & $1.7 \pm 0.6$ & $3.1_{-1.0}^{+0.9}$ & $2.1 \pm 0.4$ & $1.0 \pm 0.1$ & $5.9_{0.7}^{+0.8}$  & $4.2 \pm 0.9$  & $2.0_{-0.9}^{+0.8}$ & $1.5_{-0.5}^{+0.6}$\\
RXJ1720+2638 & $2.0 \pm 0.4$ & $1.2 \pm 0.2$ & $1.7 \pm 0.3$ & $1.6 \pm 0.3$ & $0.9 \pm 0.1$ & $5.6 \pm 0.4$ & $2.8 \pm 0.4$ & $0.7 \pm 0.2$ & $0.7 \pm 0.2$ \\

\hline
\end{tabular}
\end{table}
\twocolumn

\subsection{Comparison with gNFW parameterizations}\label{sec:gnfw}

The adequacy of different profiles, such as the $\beta$, Navarro Frenk and White (NFW), generalized NFW (gNFW, \citealt{NFW}) and other
hybrid profiles (e.g., \citealt{mroc2011}, \citealt{allison2011}, \citealt{malakprof})
is still very much under debate.
We attempt to illustrate the impact that the choice of some of these profiles may
have on the parameter estimates by comparing the results obtained from five gNFW parameterizations and from our $\beta$ parameterization (see Sec.
\ref{analysis})
for two clusters: Abell~611 (see Sec. \ref{sec:A611}) and Abell~2111 (see Sec. \ref{resultsA2111}).

For this analysis we sample from the cluster position parameters ($x_{\rm{c}}, y_{\rm{c}}$), $\theta_S=r_S/D_{A}$,
and $Y_{\theta}=Y_{\rm{tot}}/D^2_A$. $r_s$ is the scale radius, $D_{A}$ the angular diameter distance and $Y_{\rm{tot}}=Y_{\rm{sph}}(5r_{500})$,
 where $Y_{\rm{sph}}$ is the integrated Compton $y$ parameter within $5r_{500}$ (\cite{arnaud2010} take $5r_{500}$ 
 as the radius where the pressure profile flattens). Assuming a spherical geometry, $Y_{\rm{sph}}$ is calculated by integrating the plasma pressure within a spherical volume of radius $r$:
 \begin{equation}
    Y_{\rm{sph}}(r) = \frac{\sigma_{\rm{T}}}{m_{\rm{e}}c^2} \int^{r}_0  P_{\rm{e}}(r')4\pi r'^2 \rm{d}r',
    \label{eq:ytotsph}
  \end{equation}
where $\sigma_{\rm{T}}$ is the Thomson scattering cross-section, $m_{\rm{e}}$ is the electron mass, $c$ is the speed of light and $P_{\rm{e}}(r)$ is
 the electron pressure at radius $r$.
The following priors were given for the sampling parameters: an exponential
prior between 1.3$\arcmin$ and 45$\arcmin$ for $\theta_S$ and a power law prior between
0.0005 and 0.2 arcmin$^2$ for $Y_{\rm{sph}}/D^2_A$, with a power law index of 1.6 . We note that for the purposes
of this exercise -- to show the
    different degeneracies for different, plausible sets of gNFW-profile
parameters -- such wide priors are acceptable; naturally, where appropriate, the
prior ranges can be refined.
We choose to use a pressure profile for this parameterization since
self-similarity
 has been shown to hold best for this quantity and pressure profiles
 have low cluster-to-cluster scatter (e.g., \citealt{nagai2007}). The gNFW
profile is given by
  \begin{equation}
    P_{\rm{e}}(r)= \frac{P_{\rm{e},i}}{(r/r_s)^{c}\left[ 1+ (r/r_s)^{a}
\right]^{\frac{b-c}{a}}}.
  \label{eq:pgNFW}
  \end{equation}
$P_{e,i}$, the overall normalisation coefficient of the pressure profile, 
 is calculated by computing Eq. (\ref{eq:ytotsph}) for $5r_{500}$; once $P_{e,i}$ has been found, $y$ can be obtained.
The shape of the gNFW profile is governed by $c$ in the inner cluster
regions ($r<<r_s$), by $a$ at intermediate radii ($r\approx r_s$), and
by $b$ on the cluster outskirts ($r>r_s$). These parameters, together
with the concentration parameter, $c_{500}$, are fixed in most analyses to some
best-fit values (e.g., \citealt{mroc2011}). With $c_{500}$, $r_s$ can be expressed in terms of $r_{500}$: $r_s =r_{500}/c_{500}$,
 which is a common reparameterization (see e.g., \citealt{arnaud2010} and \citealt{nagai2007}).

 We ran our analysis using a gNFW
profile with parameters defined by Nagai et al. as $\rm{gNFW}_N$, another
defined by Arnaud et al. as the `universal' profile $\rm{gNFW}_A$, and  three other
combinations for the slope parameters and
$c_{500}$ that were found to provide the best fit for some clusters in Arnaud
et al.; $\rm{gNFW}_1$, $\rm{gNFW}_2$ and $\rm{gNFW}_3$\footnote{The parameters for the three other gNFW profiles all lie within
3$\sigma$ of the average value of
each parameter obtained using the cluster sample in Arnaud et al. .}. The gNFW
parameters for our five choices are given in Tab. \ref{tab:valprof}.

\begin{table}
\begin{center}
\caption{Parameters for the gNFW pressure profile. The parameters for  gNFW$_A$
and gNFW$_N$ have been taken from \citealt{arnaud2010} and
  \citealt{mrocz2009}, respectively. The values in
\citealt{mrocz2009} are the corrected values for the results published by
\citealt{nagai2007}.}
\label{tab:valprof}
\begin{tabular}{lccccc}
\hline
Profile Label & $a$ & $b$ & $c$ & $c_{500}$ \\\hline
gNFW$_1$      & 1.37 & 5.49 & 0.035 & 2.16 \\
gNFW$_2$      & 0.33 & 5.49 & 0.065 & 0.17 \\
gNFW$_3$      & 2.01 & 5.49 & 0.860 & 1.37 \\
gNFW$_A$      & 1.0620 & 5.4807 & 0.3292 & 1.156 \\
gNFW$_N$      & 0.9 & 5.0 & 0.4 & 1.3 \\ \hline
\end{tabular}
\end{center}
\end{table}
The 2D-marginalized posterior distributions of $Y_{\rm{sph}}(r_{500})$
  against $r_{500}$ obtained for each of the five parameterizations, as well as
the $\beta$ parameterization from Sec. \ref{analysis},
  for Abell~611 and Abell~2111 are shown in
  Fig. \ref{fig:gNFWA2111} .
We test for possible biases in our results from the
choice of priors by running the analysis without data; the results indicate the constraints imposed by our priors.
We find no evidence for significant biases, as shown in Fig. \ref{fig:betanodata}.

When the shape parameters of the $\beta$ profile are fitted to the SZ data instead of being set to the X-ray value (typically derived 
for data sensitive to smaller scales than AMI data) we find the mean values for $Y_{\rm{sph}}(r_{500})$ and
$r_{500}$ derived from the $\beta$ analysis to be consistent (within 1-2$\sigma$) with those from $\rm{gNFW}_A$
 and $\rm{gNFW}_{N}$ -- the averaged gNFW profiles. For these two clusters we find all gNFW parameterizations
 yield lower values for $Y_{\rm{sph}}(r_{500})$ than for the $\beta$ analysis; this is not the case for $r_{500}$, for which
 no systematic difference is seen. The constraints on $Y_{\rm{sph}}(r_{500})$ are similar for most of the gNFW models (with 
 the exception of $\rm{gNFW}_2$) and the $\beta$ model, while those for $r_{500}$ appear to be tighter for the $\beta$ model. One striking 
 difference between the two types of parameterizations is the shape and orientation of the $Y_{\rm{sph}}-r_{500}$  degeneracy.

 The resolution and limited spatial dynamic range of 
the AMI data do not allow profile selection to be made robustly, as indicated by the small difference in evidence values between the different parameterizations (Tab. \ref{tab:evidences}). Hence,
 our $\beta$ parameterization provides a comparable fit to that of the commonly used, averaged gNFW profiles, $\rm{gNFW}_A$, and $\rm{gNFW}_N$.
 It is clear from Fig. \ref{fig:gNFWA2111} that the distribution for the
$Y_{\rm{sph}}(r_{500})-r_{500}$ degeneracies is very sensitive to the choice
for the slope parameters (and $c_{500}$ for gNFW). Cluster parameters for a cluster with a profile described by e.g., a $\rm{gNFW}_2$ recovered using a $\rm{gNFW}_A$ 
 paramaterization will be biased.

\begin{figure*}
\begin{center}
\centerline{\includegraphics[width=8.0cm,height=8.0cm,clip=,angle=0.]{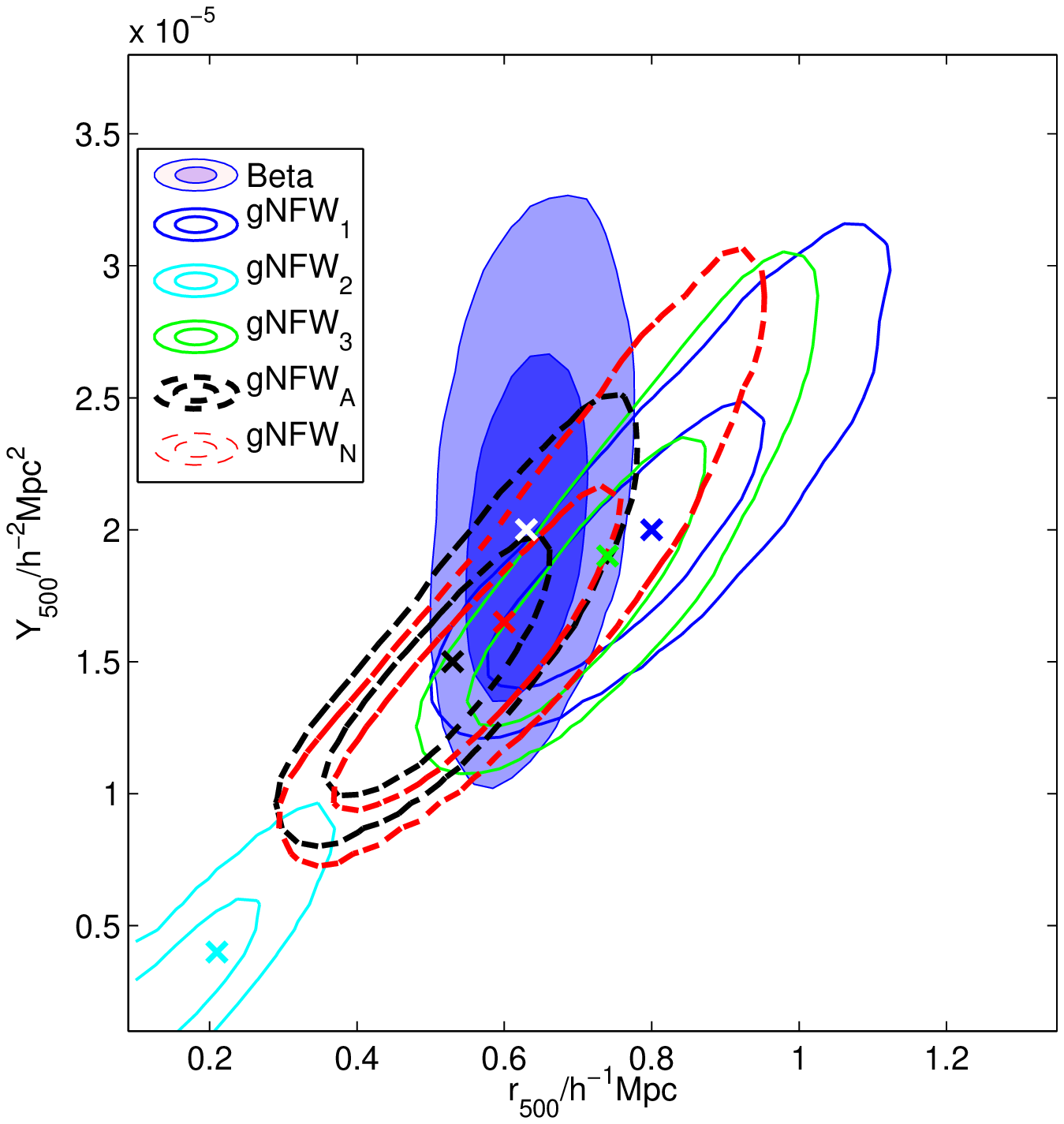}\qquad\includegraphics[width=8.0cm,height=8.0cm,clip=,angle=0.]{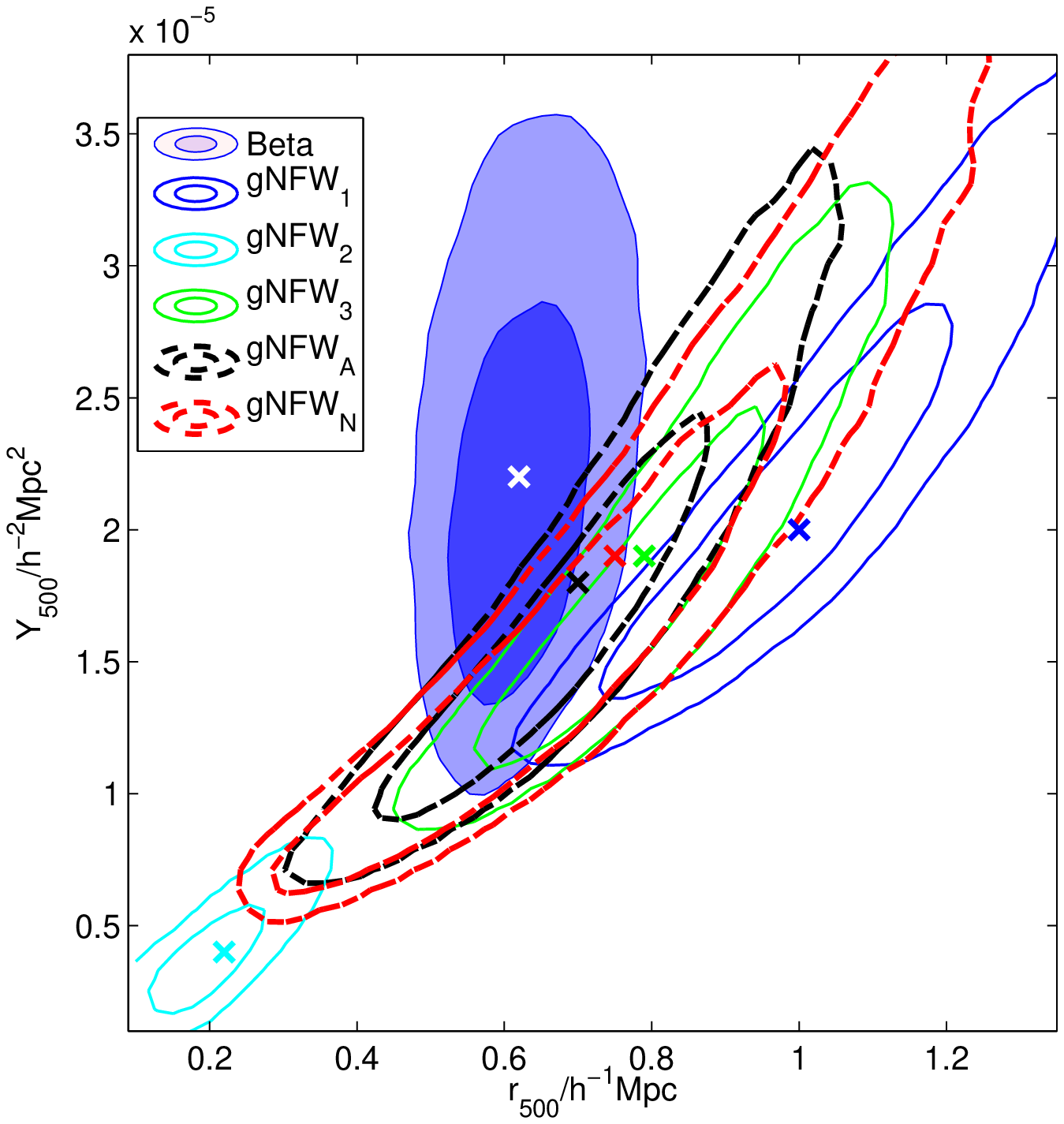}}
\caption{2-D marginalized distributions for $Y_{\rm{sph}}(r_{500})$ against
$r_{500}$ obtained using the $\beta$-based cluster parameterization
  and five gNFW-based cluster
  parameterizations with slope parameters and $c_{500}$ given in
  Tab. \ref{tab:valprof}.
 The crosses denote the {\sc{McAdam}}-derived mean values.
 The results are for Abell~611
  (left) and Abell~2111 (right). The blue filled
ellipses
 show the results of the $\beta$ parameterization.}
\label{fig:gNFWA2111}
\end{center}
\end{figure*}

\begin{figure}
\begin{center}
\centerline{\includegraphics[height=6.0cm,clip=,angle=0.]{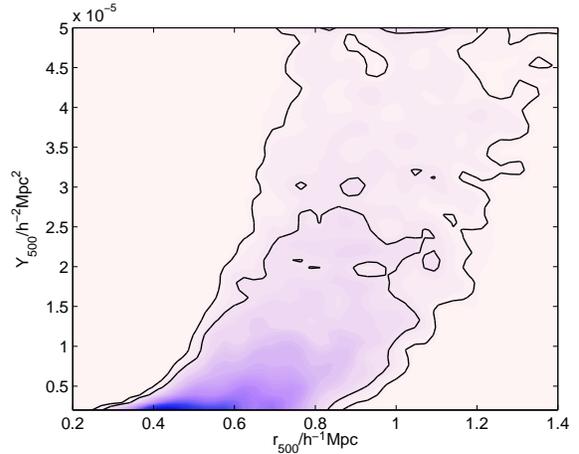}}
\caption{2-D marginalized distribution for $Y_{\rm{sph}}(r_{500})$ against
$r_{500}$ obtained using the $\beta$-based cluster parameterization
  without any data.}
\label{fig:betanodata}
\end{center}
\end{figure}

\begin{table}
\begin{center}
\caption{Log$_e$ evidences for five cluster parameterizations applied to Abell~2111 and Abell~611}
\label{tab:evidences}
\begin{tabular}{lccc}
\hline
                & Abell~2111 & Abell~611 \\ \hline
Profile Label &              &  \\
gNFW$_1$      & 23198.88 & 21114.08 \\
gNFW$_2$      & 23199.51 & 21114.40 \\
gNFW$_3$      & 23198.64 & 21114.74 \\
gNFW$_A$      & 23198.94 & 21114.50 \\
gNFW$_N$      & 23198.76 & 21114.65 \\
$\beta$       & 23194.92 & 21112.05 \\ \hline
\end{tabular}
\end{center}
\end{table}

\subsection{Abell~586}

Results for Abell~586 are given in Figs. \ref{fig:A586} and \ref{fig:A586_2}.
 This cluster has a complex source environment, with 7 sources within $5\arcmin$ from the cluster SZ centroid, which include two radio sources
of $\approx 260$ and $744$\,$\mu$Jy at $0.5\arcmin$ and
$4\arcmin$ from the pointing centre.
 After source-subtraction there are only $\approx 1\sigma$
residuals left on the map. Uncertainties in the source fluxes are carried
through into the posterior distributions for the cluster parameters. From
Fig. \ref{fig:A586_2},
it can be seen that there is no strong degeneracy between the source flux
densities and the cluster mass. 

Abell~586 has been studied extensively in the X-ray band (e.g., \citealt{allen2000}
and \citealt{white2000}). A recent
analysis of the temperature profile \citep{cypriano2004} shows how the
temperature falls from $\approx9$\,keV at the cluster centre to
$\approx 5.5$\,keV at a radius $\approx 280\arcsec$. Cypriano et al. have used
the Gemini Multi-Object Spectrograph together with X-ray data taken from the
{\sc{Chandra}} archive to measure the properties of Abell~586.
 They compare mass estimates derived from the velocity distribution and from
the X-ray temperature profile and find that both give very similar results,
$M_{\rm{g}}\approx 0.48\times10^{14}M_{\odot}$ (for $h_{70}=1$) within 1.3$h_{70}^{-1}$\,Mpc.
 They suggest that the cluster is spherical and relaxed with no recent mergers.
It is less clear whether this cluster has a cool core or not, with \cite{allen1998}
reporting its existence and \cite{marrone2011} saying otherwise.
The peak X-ray and SZ emissions are consistent with each other and the AMI
SZ decrement shows some signs of being extended towards the SW (Fig. \ref{fig:A586} B and C); there are no contaminating sources in the vicinity of this SZ-`tail'.

The SZ effect from Abell~586 has previously been observed with OVRO/BIMA by
\cite{laroque2006} and \cite{bona_chandra}. LaRoque et al. apply an isothermal
$\beta$-model to SZ and {\sc{Chandra}} X-ray observations
 and find $M_{\rm{g}}(r_{2500})=2.49\pm{0.32}\times10^{13}M_{\odot}$
 and
$M_{\rm{g}}(r_{2500})=2.26^{+0.13}_{-0.11}\times10^{13}M_{\odot}$
, respectively (using
$h_{70}=1$ and excising the inner 100\,kpc from the X-ray data).
 In addition, they determine an X-ray
 spectroscopic temperature of the cluster gas of $\approx 6.35$\,keV between a
radius of 100\,kpc and $r_{2500}$. In comparison, \cite{okabe2010} use
\emph{Subaru} to calculate the cluster mass from weak lensing by applying
a Navarro, Frenk \& White (NFW; \citealt{NFW}) profile. They find
$M_{\rm{T}}(r_{2500})=2.41^{+0.45}_{-0.41} \times10^{14}M_{\odot}$
 and 
 $M_{\rm{T}}(r_{500})=4.74^{+1.40}_{-1.14} \times10^{14}M_{\odot}$ (using
$h_{72}=1$). 
In this work, we find  $M_{\rm{T}}(r_{500})=3.0\pm 1.3
\times10^{14}M_{\odot}$, where $r_{500}=0.94\pm 0.14$\,Mpc and $h_{70}=1$. Note
that the fluxes of the radio sources $S_1$ and $S_5$ are degenerate in our
analysis of Abell~586 (see Fig. \ref{fig:A586_2}); this is because these sources
are separated by only 66$\arcsec$ and their individual fluxes cannot be
disentangled in the analysis of the AMI SA data.

  \begin{figure*}
\centerline{\huge{Abell~586}}
\centerline{{A}\includegraphics[width=7.5cm,height=7.5cm,clip=,angle=0.]{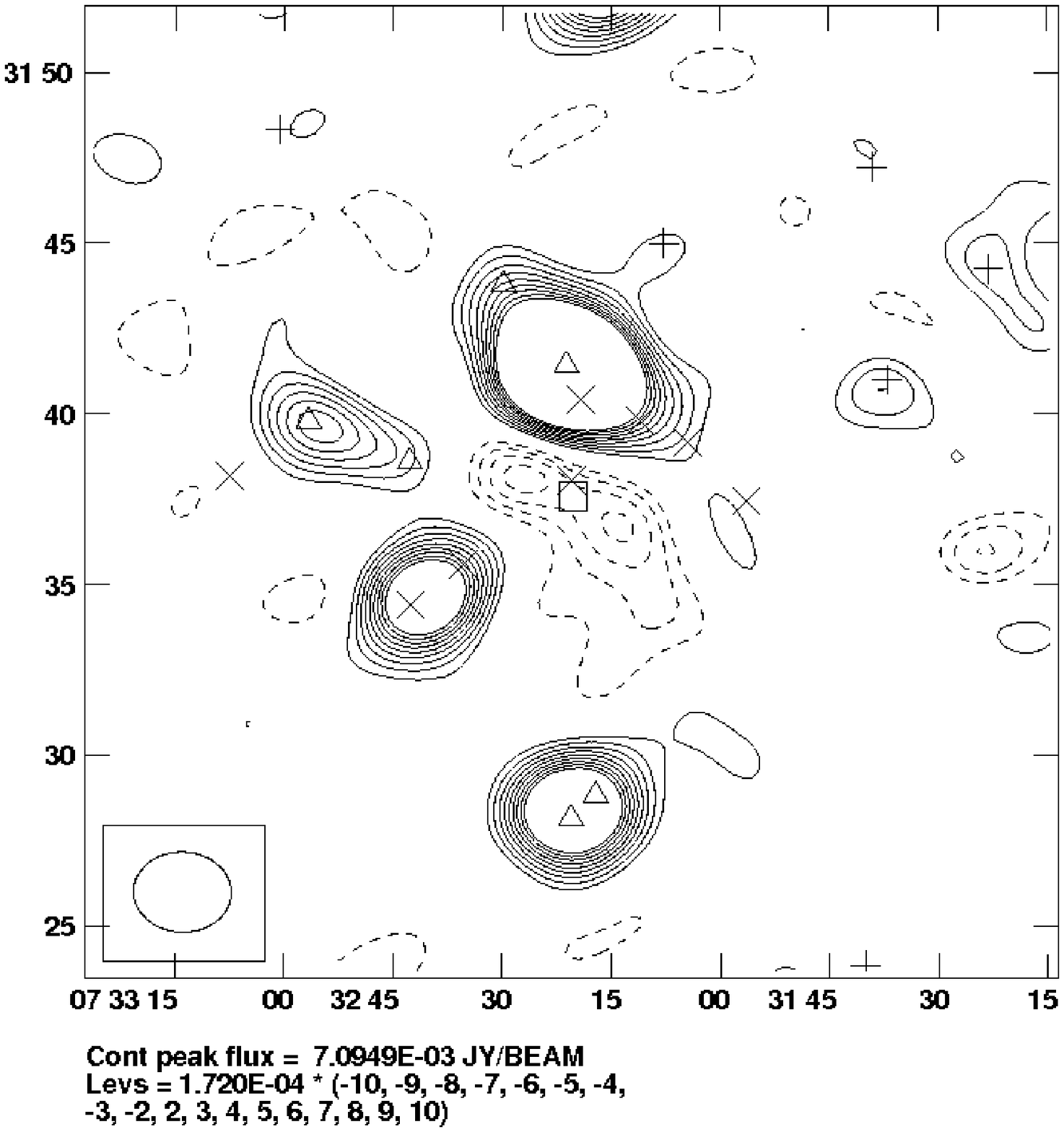}\qquad{D}\includegraphics[width=7.5cm,height=7.5cm,clip=,angle=0.]{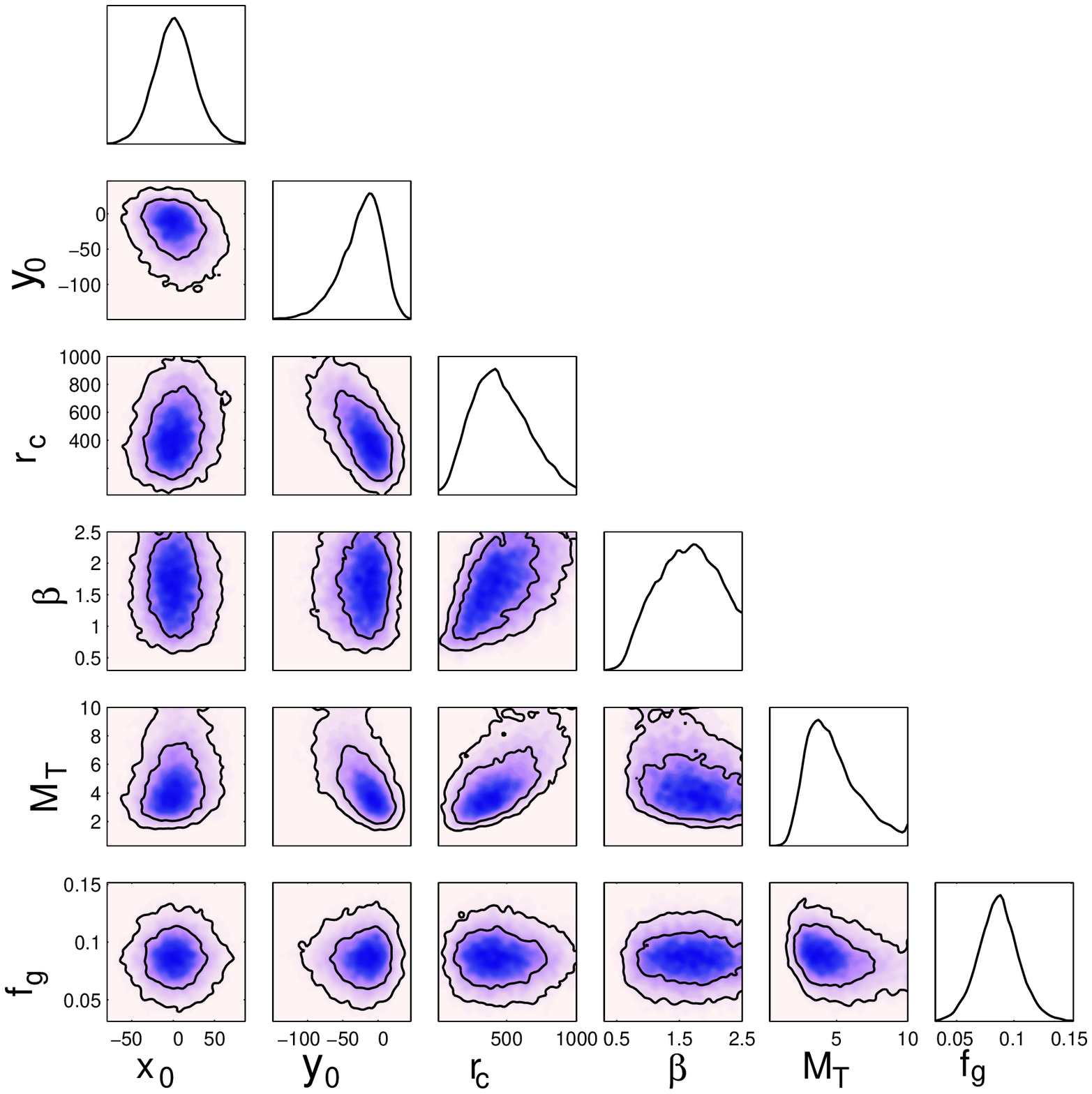}}
\centerline{{B}\includegraphics[width=7.5cm,height=7.5cm,clip=,angle=0.]{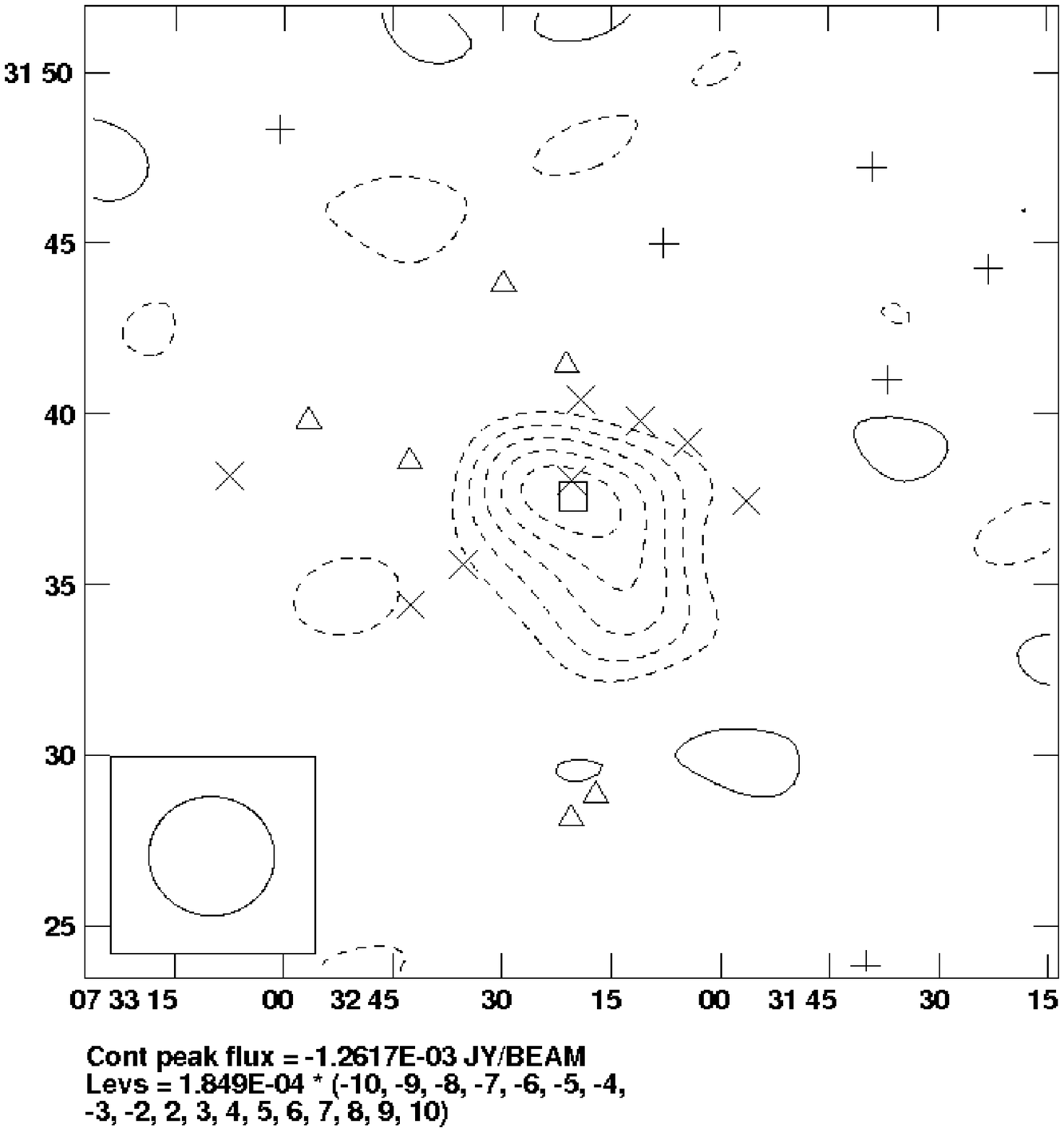}\qquad{E}\includegraphics[width=7.5cm,height=7.5cm,clip=,angle=0.]{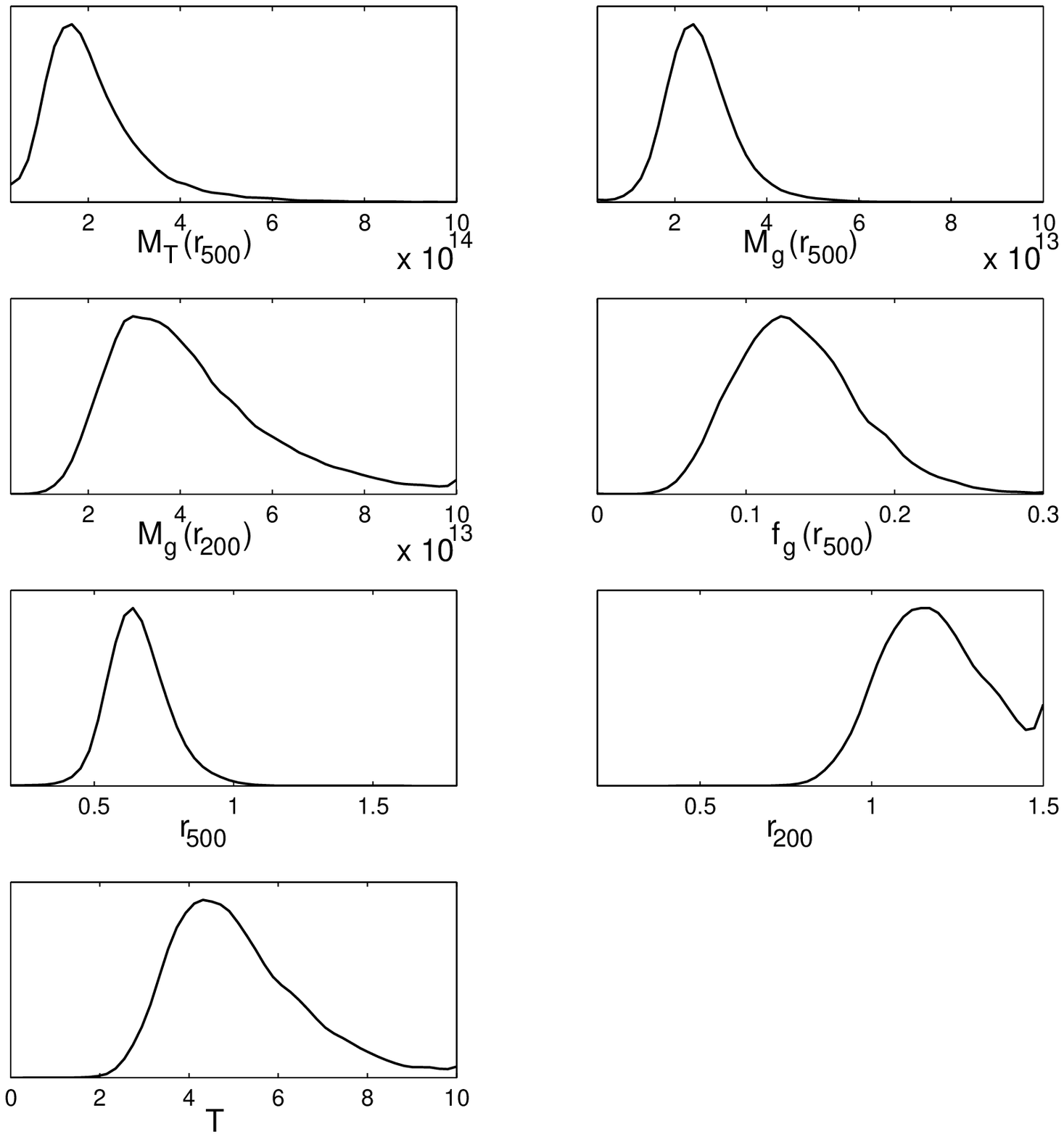}}
\centerline{{C}\includegraphics[width=7.5cm,height=6.5cm,clip=,angle=0.]{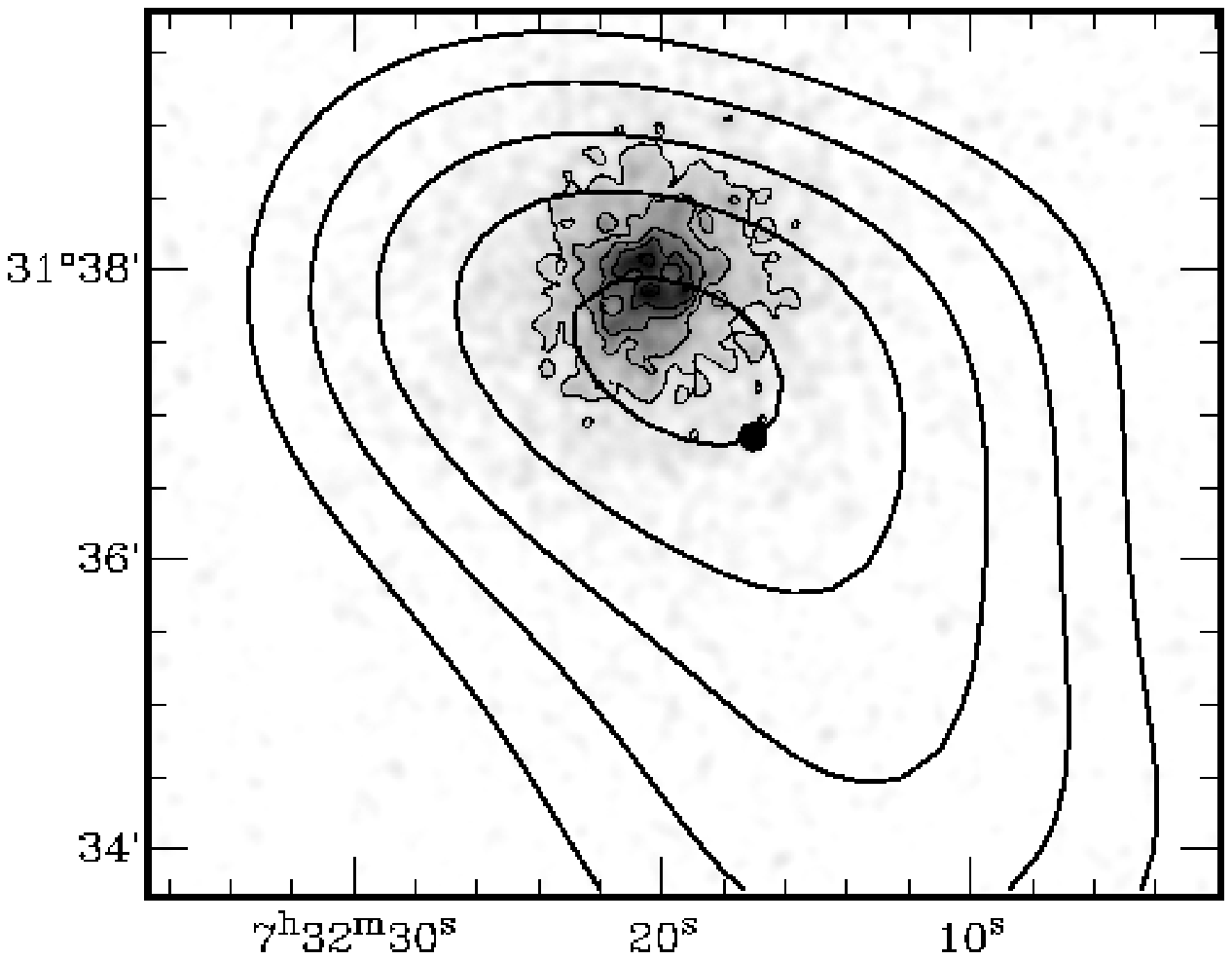}}
 
\caption{Results for Abell~586. Panels A and B show the SA map before and after source subtraction, respectively; a 0.6\,k$\lambda$ taper has been applied in B. The box in panels A and B indicates the cluster SZ centroid, for the other symbols see Tab. \ref{tab:sourcelabel}. Panel C shows the smoothed {\sc{Chandra}} X-ray map overlaid with contours from B. D and E show the marginalized posterior distributions for the cluster sampling and derived parameters, respectively. The $y$-axis for the 1-d marginals is the probability density and for all the posterior distributions plots in this paper $h$ refers to $h_{100}$. In D $M_{\rm{T}}$ is given in units of $h_{100}^{-1}\times 10^{14}M_{\rm{\odot}}$ and $f_{\rm{g}}$ in $h_{100}^{-1}$; both parameters are estimated within $r_{200}$. In E $M_{\rm{g}}$ is in units of $h_{100}^{-2}M_{\odot}$, $r$ in $h_{100}^{-1}$Mpc and $T$ in KeV. The slight rise in the distribution for $r_{200}$ at large $r$ is a result of a binning artifact and, in fact,
this distribution does tail off smoothly, as expected.}
\label{fig:A586}
\end{figure*}

\begin{figure}
\centerline{\includegraphics[width=8.4cm,clip=,angle=0.]{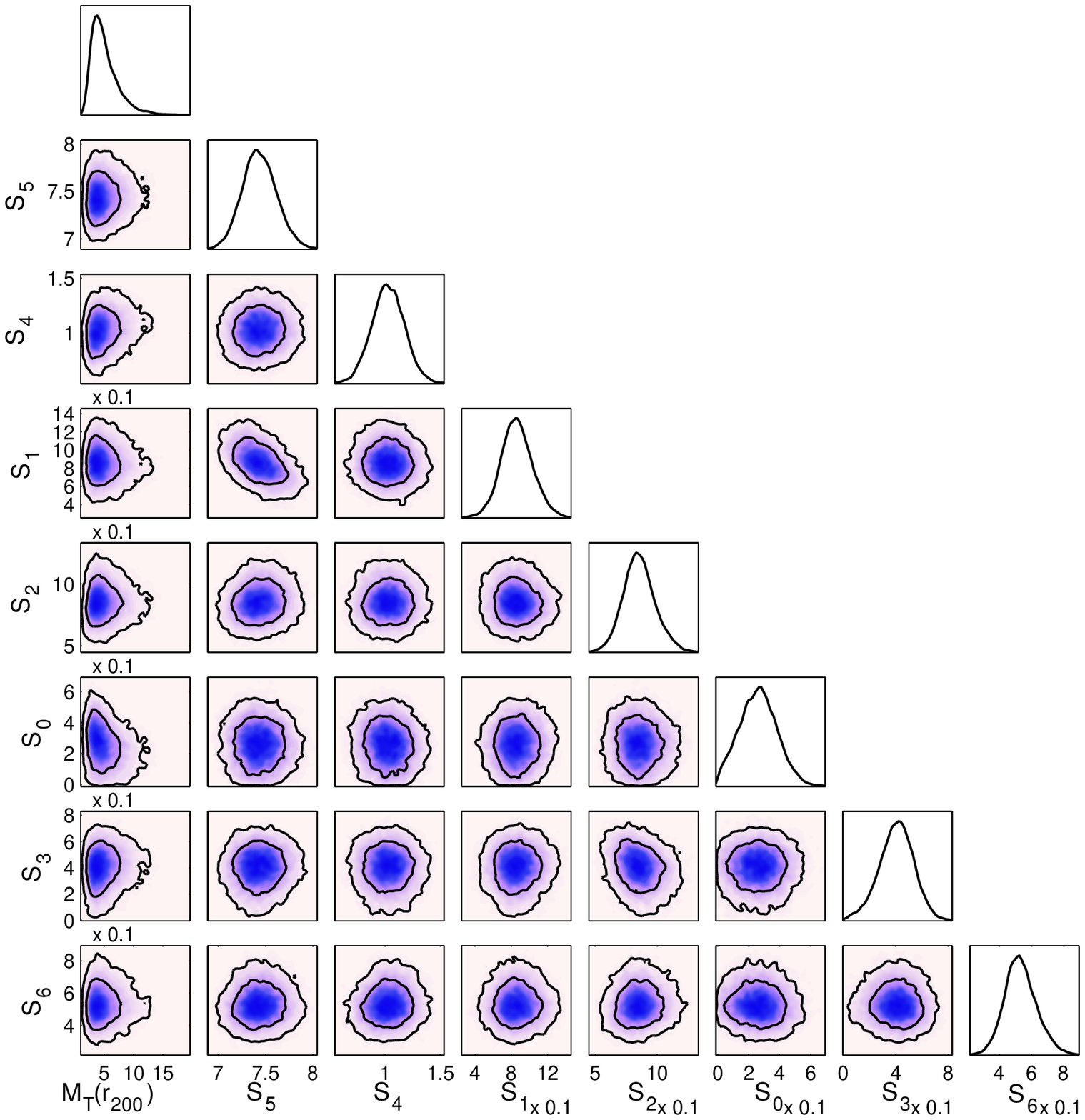}}
\caption{1 and 2-D marginalized posterior distributions for the flux densities, in Jys, of sources detected within $5\arcmin$ of the SZ centroid of Abell~586 (see Tab. \ref{tab:source_info}) and $M_{\rm{T}}(r_{200})$, in units of $h_{100}^{-1}M_{\odot}\times 10^{14}$.}
\label{fig:A586_2}
\end{figure}

\subsection{Abell~611}\label{sec:A611}

 Results for Abell~621 are presented in Fig. \ref{fig:A611}. Our methodology is able to model the radio sources + cluster enviroment well,
as demonstrated by the good constraints on the mass and other parameters and
the lack of degeneracies between the sources closest to the cluster and the
cluster mass (Fig. \ref{fig:A611} D, E and F). We do not expect any significant contamination from
radio sources nor from extended emission since GMRT observations by
\cite{GMRT_HALO} found no evidence for a radio halo associated with Abell~611
at 610\,MHz.
The decrement on the source-subtracted maps appears to be circular, in
agreement with the X-ray surface brightness from the {\sc{Chandra}} achive data
shown in Fig. \ref{fig:A611} C, which also appears to be
smooth and whose peak is close to the position of the brightest cluster
galaxy and the SZ peak. These facts might be taken to imply the cluster is relaxed but, 
 it does not seem to have a cool core \citep{marrone2011}. Abell~611 has also previously been observed in the SZ at 15\,GHz by
\cite{RT_Abell 611}, AMI Consortium: Zwart et al. (2010) and \cite{hurley2011}, and at 30\,GHz by \cite{OCRO_Abell 611}, \cite{bona_chandra} and \cite{laroque2006}.

 From the analysis in \cite{Xray_lense_Abell 611_2} 
 the cluster mass was estimated to be 9.32--11.11$\times 10^{14} \rm{M_{\odot}}$ (within a radius of 1.8$\pm$0.5\,Mpc) by fitting different cluster models
 to X-ray data and between 4.01--6.32$\times 10^{14} \rm{M_{\odot}}$ (within a radius of 1.5$\pm$0.2\,Mpc) when fitting different models to the
lensing data; all estimates use $h_{70}=1$.
 Several other analyses of {\sc{Chandra}}
data produce comparable mass estimates (e.g., \citealt{schmidt_A611},
\citealt{MORANDI_1_Abell 611}, \citealt{MORANDI_2_Abell 611} and
\citealt{sanderson2009}).
\cite{romano_a611} perform a weak-lensing analysis of Abell~611 using data from
the Large Binocular Telescope; with an NFW profile they estimate
$M_{\rm{T}}(r_{200})$ = 4--7$\times 10^{14} \rm{M_{\odot}}$ for $h_{70}=1$.

These are in agreement with the values
obtained from \emph{Subaru} weak lensing observations by Okabe et al.. AMI Consortium: Hurley-Walker et al. estimate the total mass for this system within $r_{200}$.
 Using lensing data they find it is $4.7\pm 1.2 \times 10^{14} h_{70}^{-1}M_{\odot}$ and using AMI SZ data they find it is $6.0\pm 1.9 \times 10^{14} h_{70}^{-1}M_{\odot}$.
 We find $M_{\rm{T}}(r_{200})$ = $5.7\pm 1.1\times 10^{14}\rm{M_{\odot}}$, where
$r_{200}=1.6\pm 0.1$ and $h_{70}=1$; this value is significantly smaller than the result given
in AMI Consortium: Zwart et al. (2010); this is due to their mass measurements being biased
high, as they said, and is further discussed in \cite{olamaie2011}.

\begin{figure*}
\centerline{\huge{A611}}
\centerline{{A}\includegraphics[width=7.5cm,height=7.5cm,clip=,angle=0.]{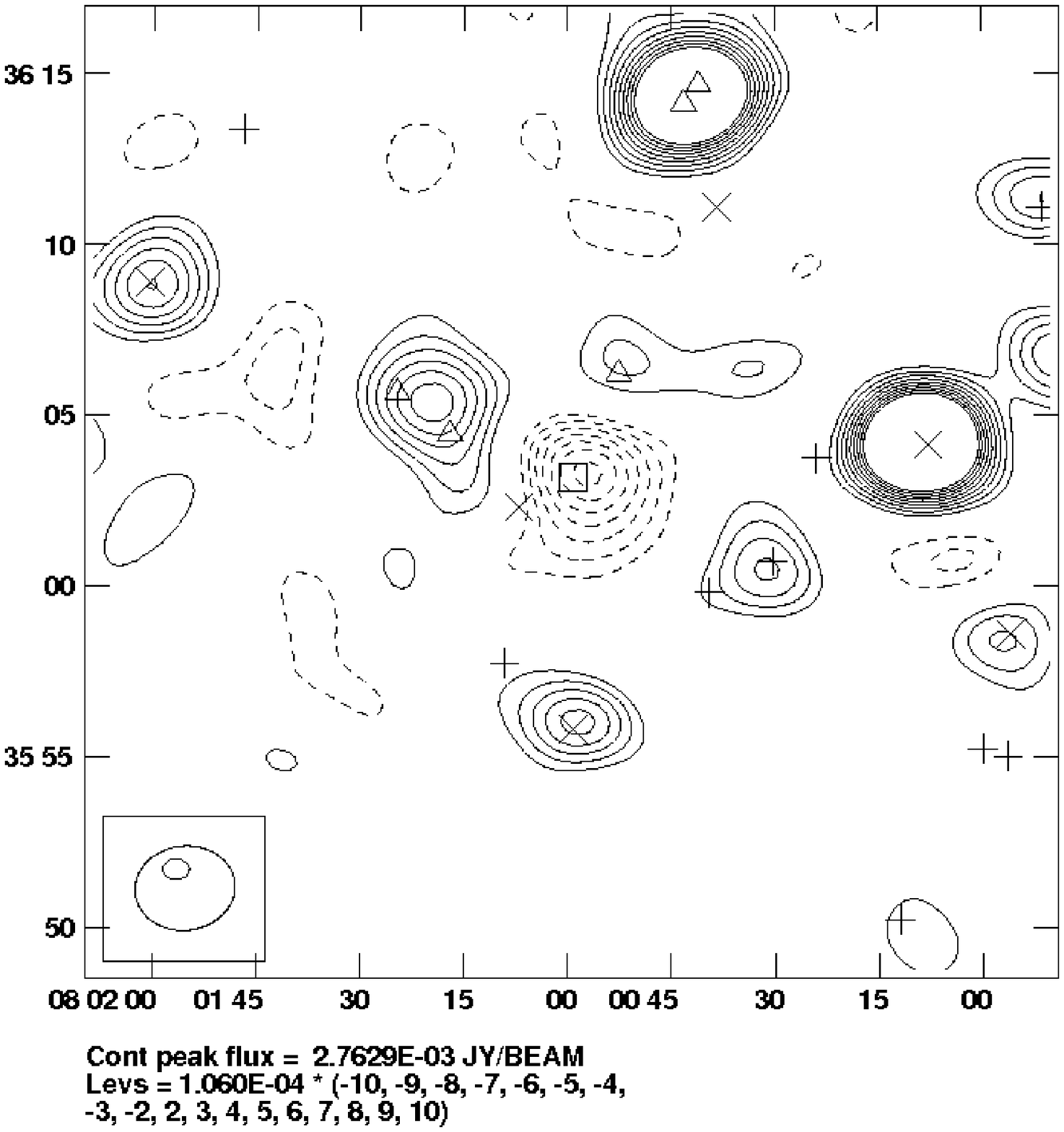}\qquad{D}\includegraphics[width=7.5cm,height=7.5cm,clip=,angle=0.]{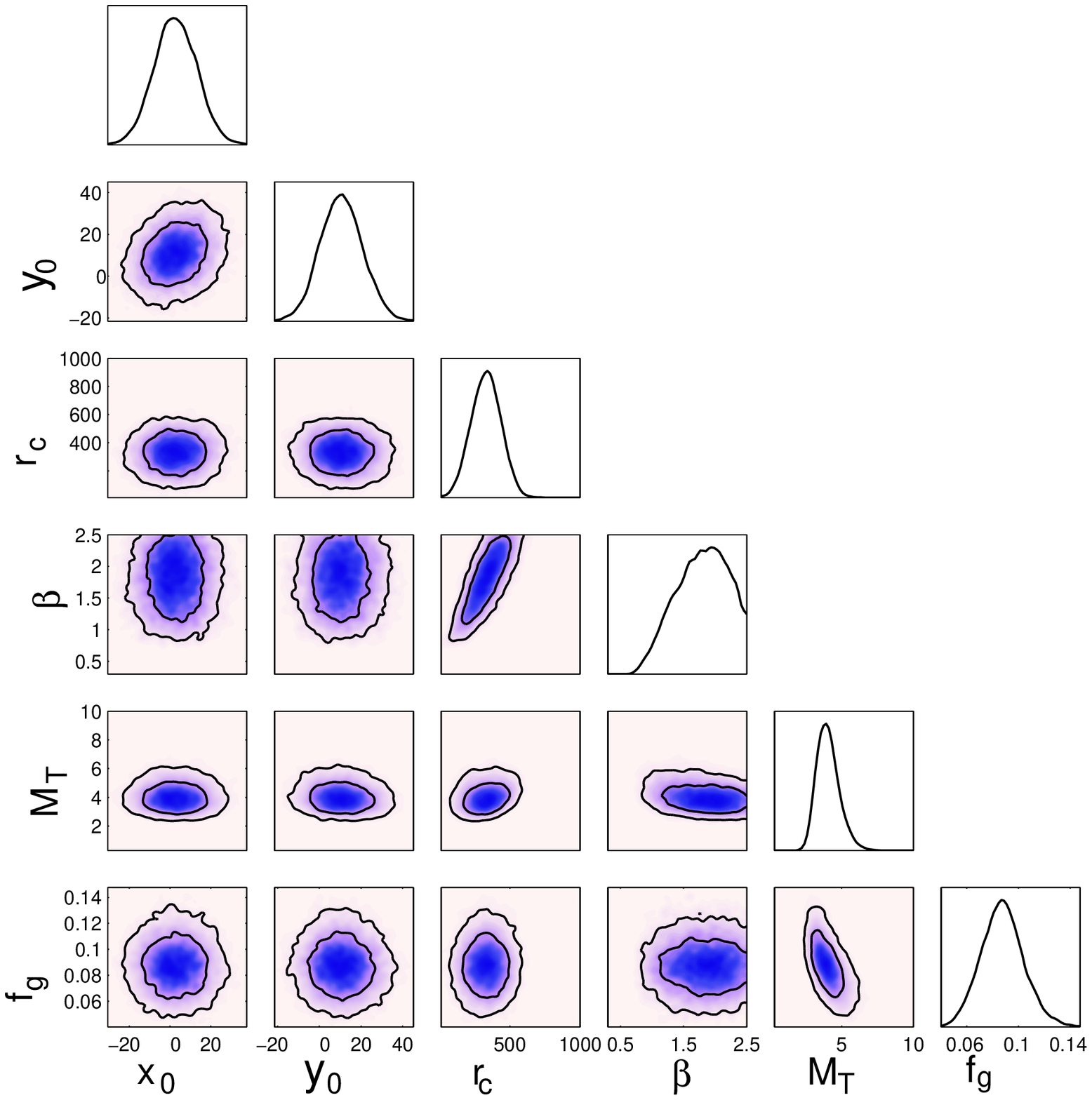}}
 \centerline{{B}\includegraphics[width=7.5cm,height=7.5cm,clip=,angle=0.]{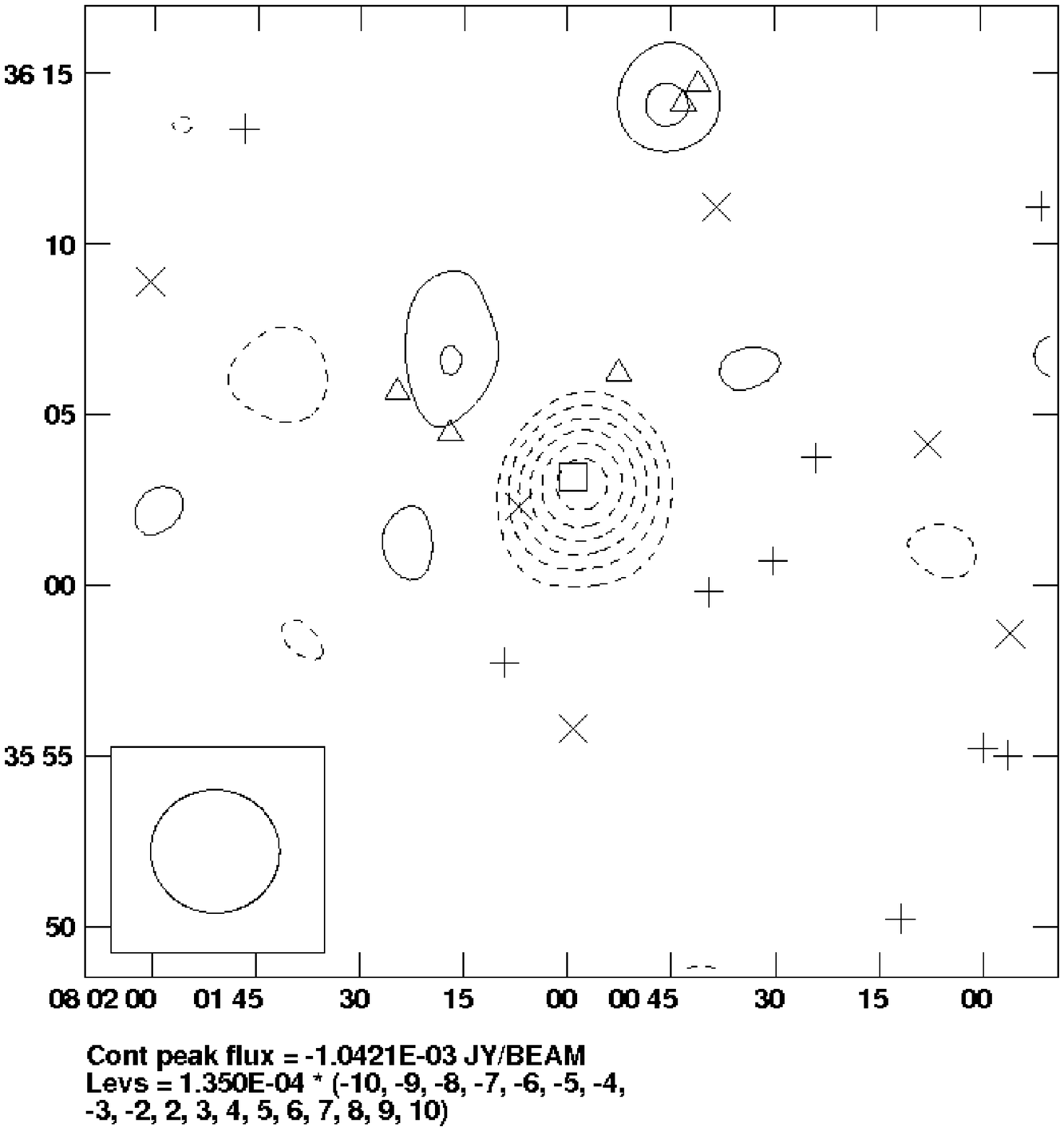}\qquad{E}\includegraphics[width=7.5cm,height=7.5cm,clip=,angle=0.]{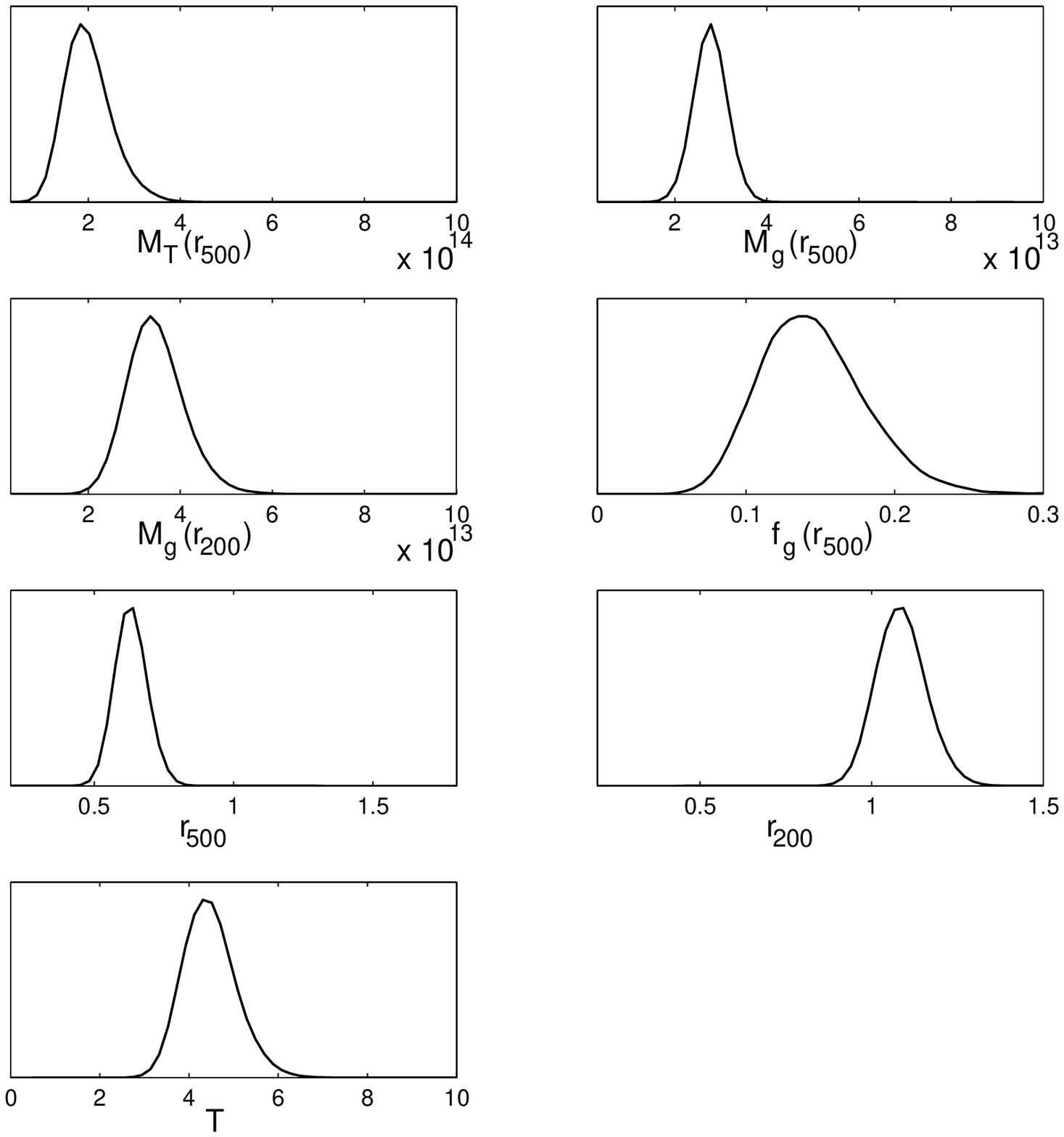}}
 \centerline{{C}\includegraphics[width=7.5cm,height=6.5cm,clip=,angle=0.]{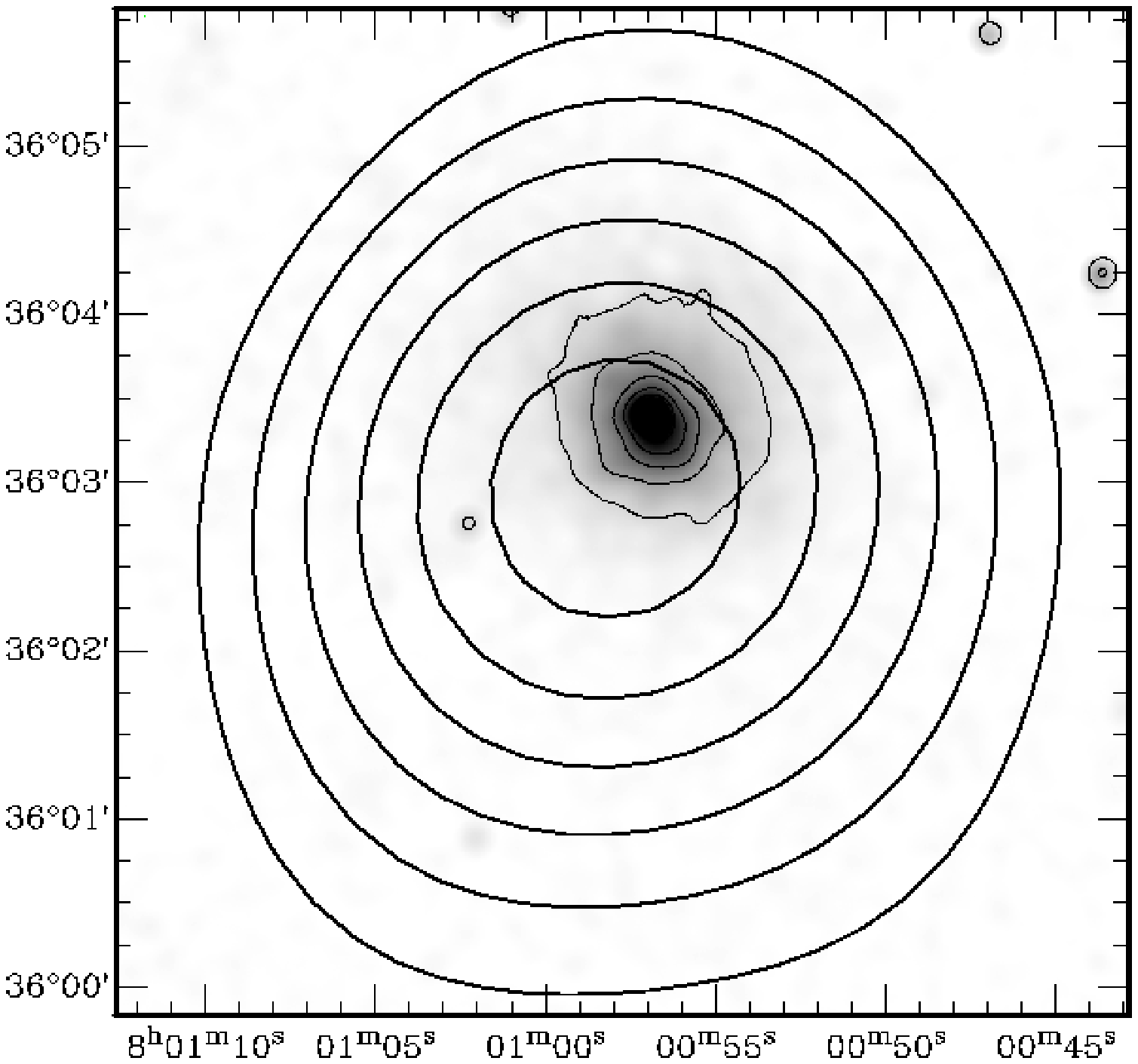}\qquad{F}\includegraphics[width=7.5cm,height=6.5cm,clip=,angle=0.]{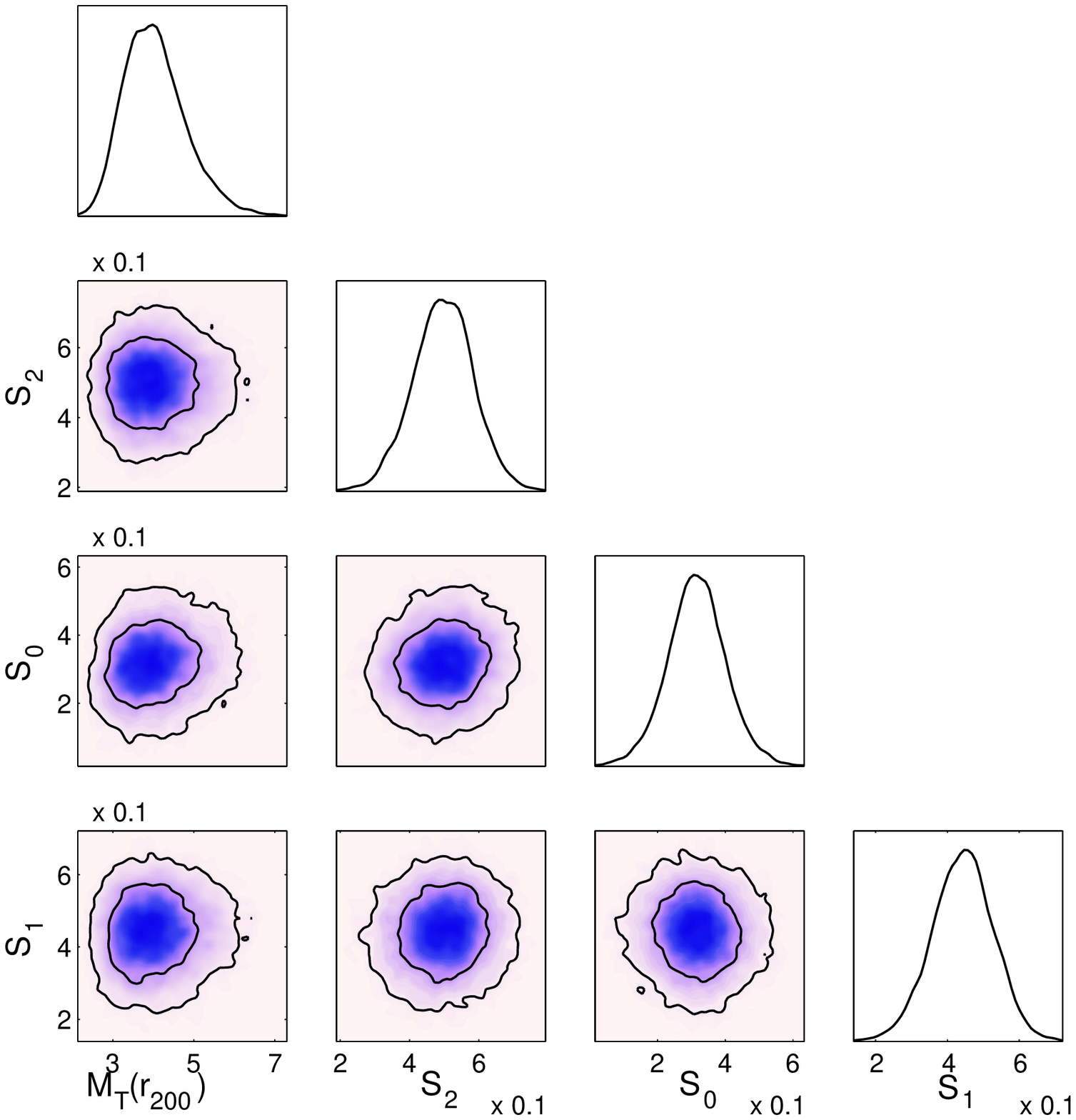}}

\caption{Results for Abell~611. Panels A and B show the SA map before and after source subtraction, respectively; a $0.6$\,k$\lambda$ taper has been applied to B. The box in panels A and B indicates the cluster SZ centroid, for the other symbols see Tab. \ref{tab:sourcelabel}. C shows the smoothed {\sc{Chandra}} X-ray map overlaid with contours from B. Panels D and E show the marginalized posterior distributions for the cluster sampling and derived parameters, respectively. F shows the 1 and 2-D marginalized posterior distributions for source flux densities, in Jys, (within $5\arcmin$ of the cluster SZ centroid, see Tab. \ref{tab:source_info}) and $M_{\rm{T}}(r_{200})$, in $h_{100}^{-1}\times 10^{14}M_{\odot}$. In panel D $M_{\rm{T}}$ is given in units of $h_{100}^{-1}\times10^{14}M_{\odot}$ and $f_{\rm{g}}$ in $h_{100}^{-1}$; both parameters are estimated within $r_{200}$. In E $M_{\rm{g}}$ is in units of $h_{100}^{-2}M_{\odot}$, $r$ in $h_{100}^{-1}$Mpc and $T$ in KeV.}

\label{fig:A611}
\end{figure*}

\subsection{ Abell~621} \label{resultsA621}

Fig. \ref{fig:A621} contains our results for Abell~621. 
Out of the 13 radio sources detected on the LA raster for Abell~621, three lie
near the edge of the cluster decrement in the source-subtracted map and one, which has a flux
density $\approx 7$\,$\rm{mJy}$, is coincident with the best-fit cluster
position, as indicated
by the box in Fig. \ref{fig:A621} A. 
However, whatever reasonable source subtraction we try makes almost no difference to the inferred cluster mass.
 The \emph{ROSAT HRI} X-ray image presented in Fig. \ref{fig:A621} C appears to
be
quite uniform and circular and the offset between the X-ray and SZ cluster
centroids is small. We find the cluster mass to be $M_{\rm{T}}(r_{200})$ =
4.8$^{+1.7}_{-1.8}$$\times 10^{14}h_{100}^{-1}\rm{M_{\odot}}$ from our analysis;
 at $6\sigma$, this is one of our less significant
detections.

The data for the probability distributions
in Fig.  \ref{fig:A621} E have been binned relatively finely to avoid misleading
features, in particular towards the lower limits of our plots. As a result,
the noise in these bins is higher, which makes the distributions appear
less smooth. For some combinations of cluster parameters, there is nowhere in
the
cluster density estimation that the density of the gas
reaches $500\rho_{\rm{crit}}$. In these cases, where there is no physical
solution
for $r_{500}$, we set $r_{500}=0$. This leads to
sharp, meaningless peaks at small radius in the distributions for some cluster parameters at $r_{500}$ (Fig.
\ref{fig:A621} E). These features have also been discussed in \cite{zwart2010}.

\begin{figure*}
\begin{center}
\begin{tabular}{m{8cm}cm{8cm}}
\multicolumn{3}{c}{\huge{Abell~621}}\\
{A}\includegraphics[width=7.5cm,height=7.5cm,clip=,angle=0.]{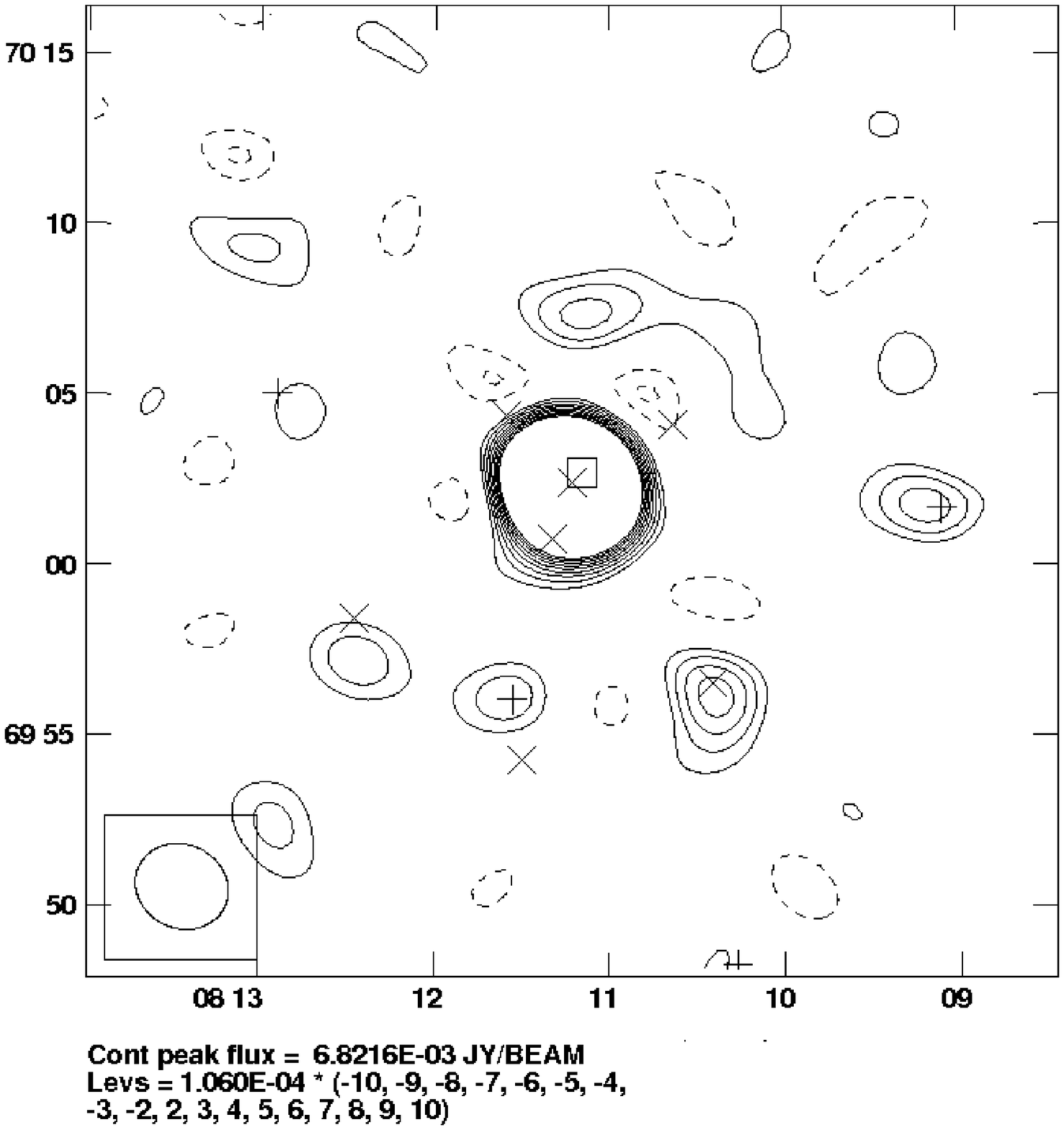}
& \quad &
{D}\includegraphics[width=7.5cm,height=7.5cm,clip=,angle=0.]{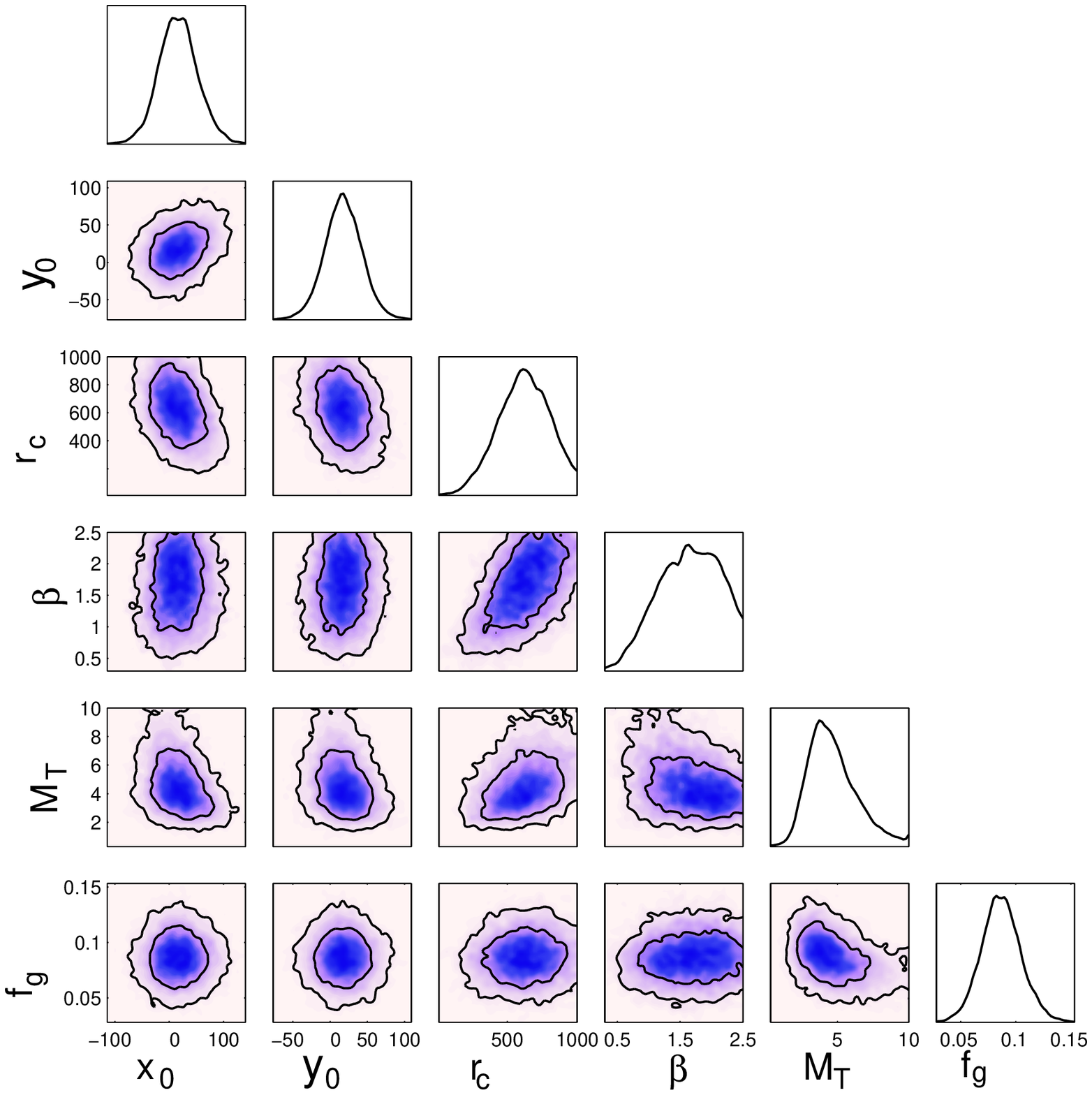}
\\
{B}\includegraphics[width=7.5cm,height=7.5cm,clip=,angle=0.]{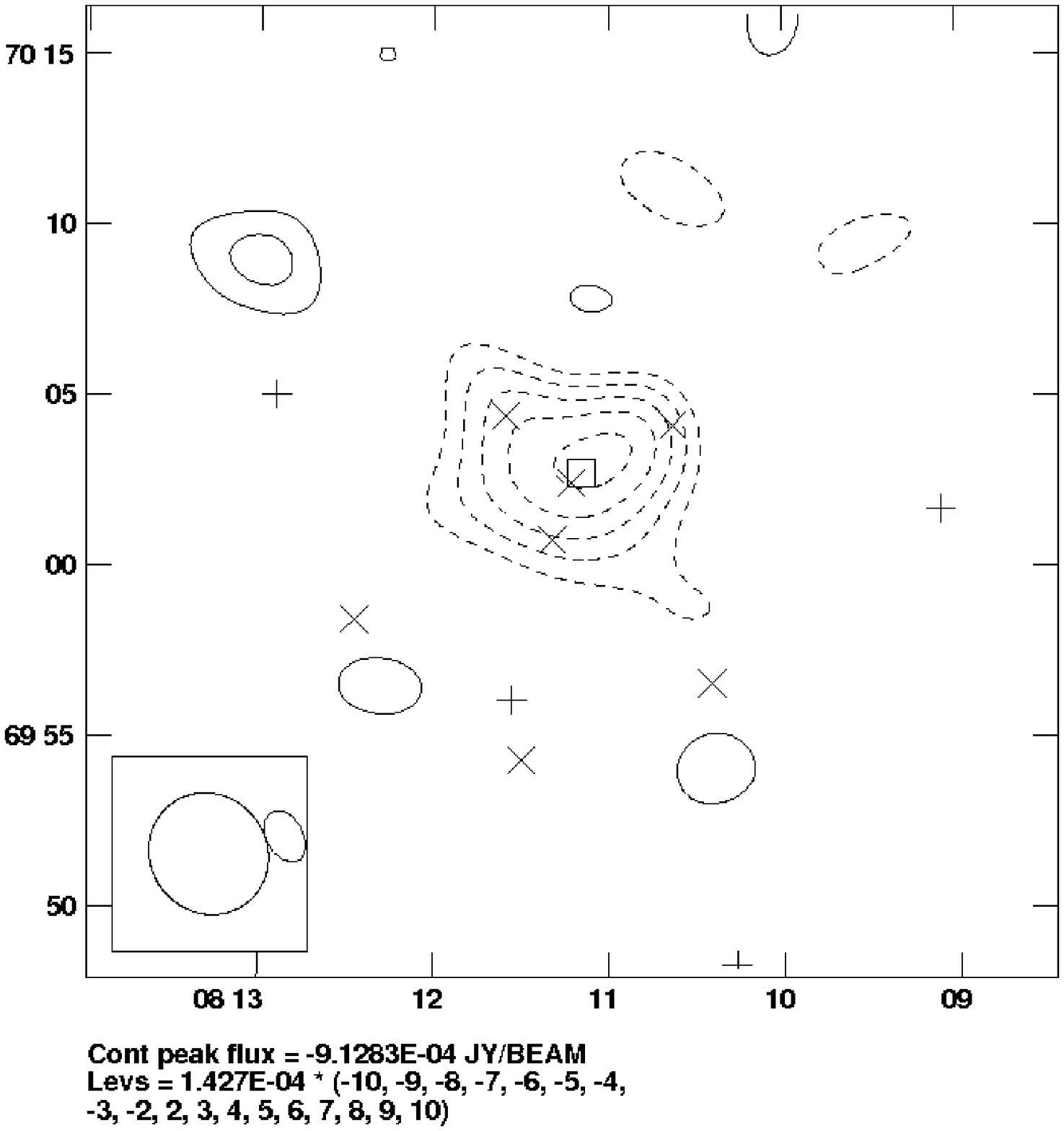}
& \quad &
{E}\includegraphics[width=7.5cm,height=7.5cm,clip=,angle=0.]{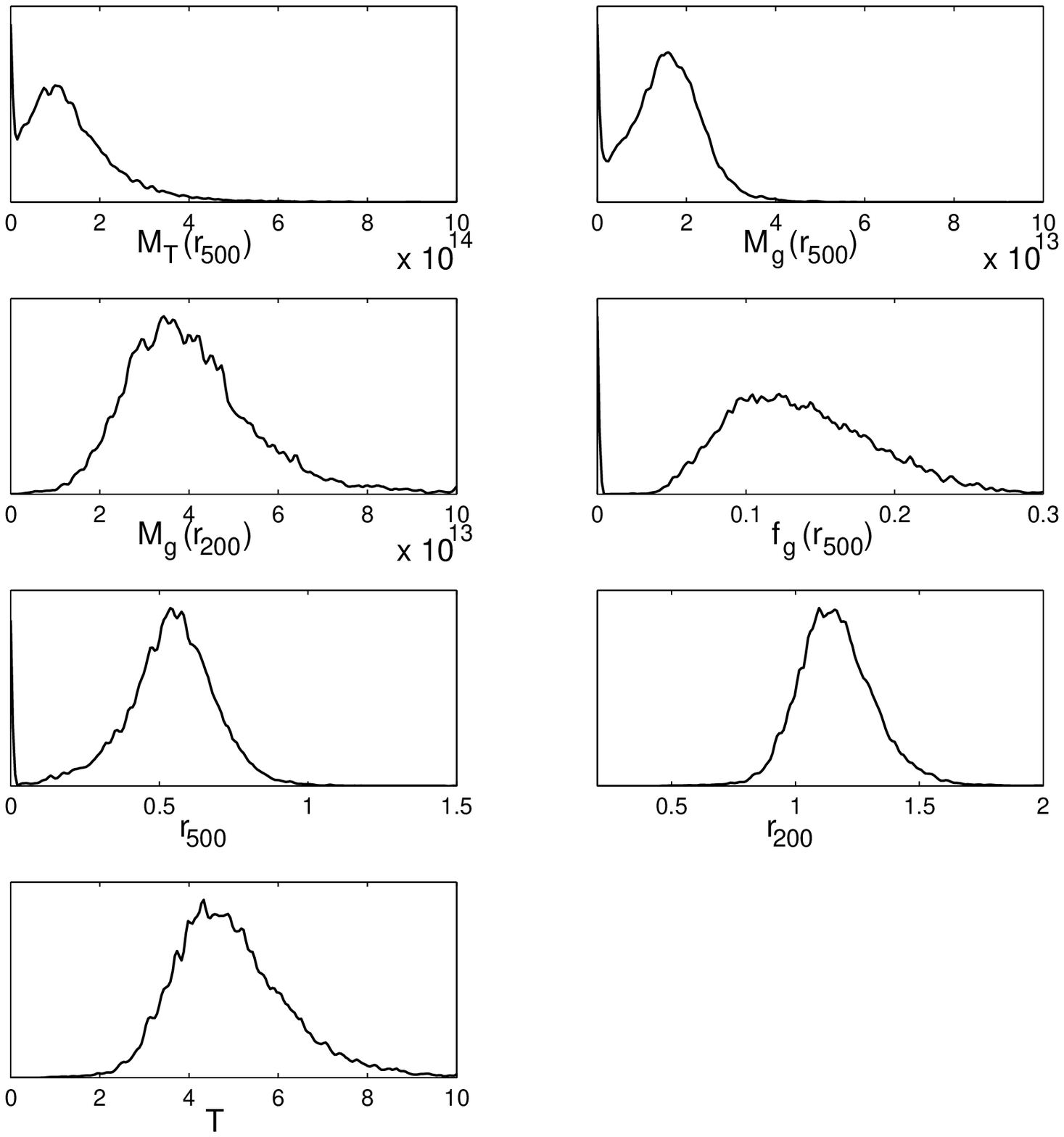}
\\
{C}\includegraphics[width=6.5cm,height=6.5cm,clip=,angle=0.]{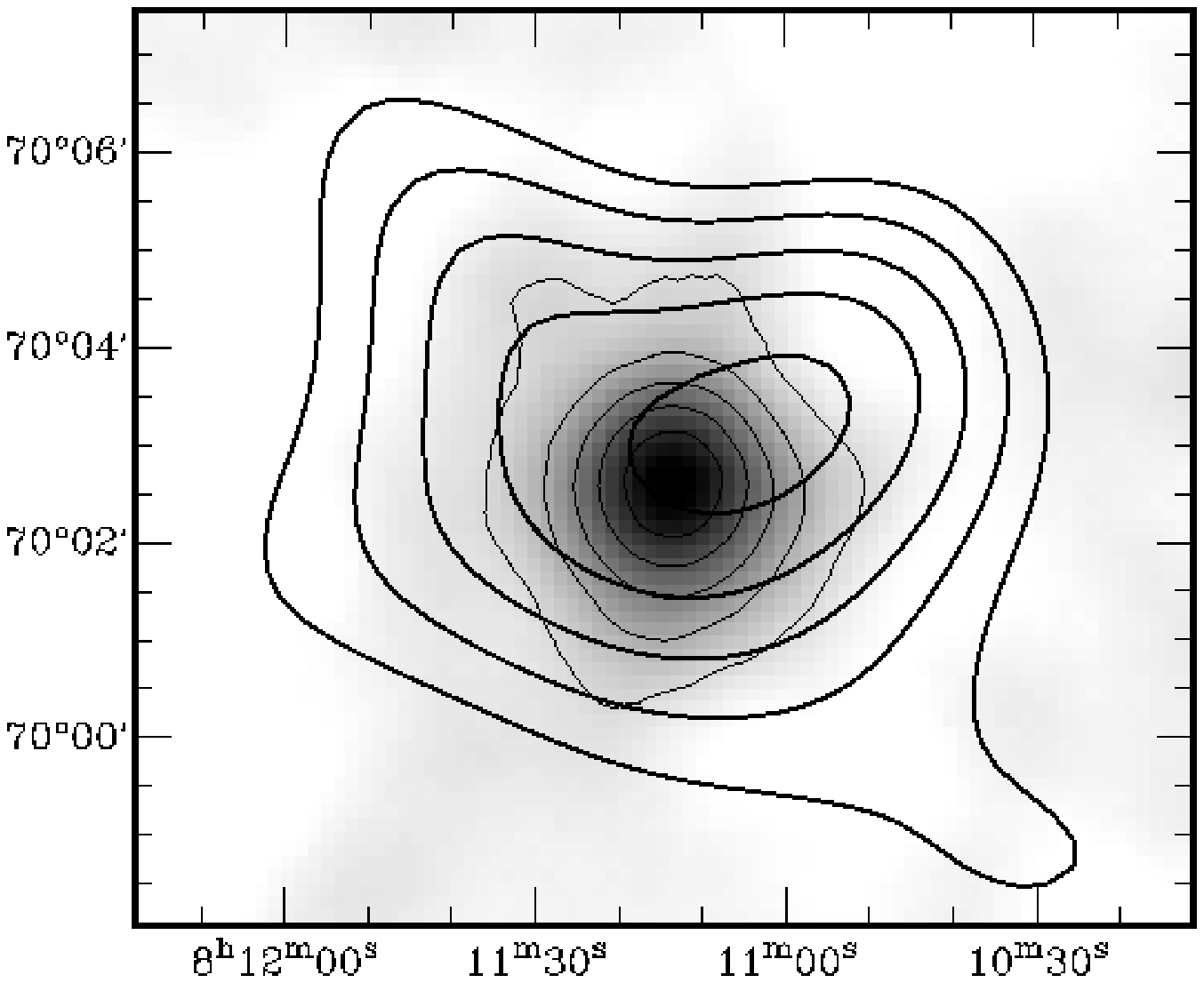}
& \quad &
{F}\includegraphics[width=7.0cm,height=6.5cm,clip=,angle=0]{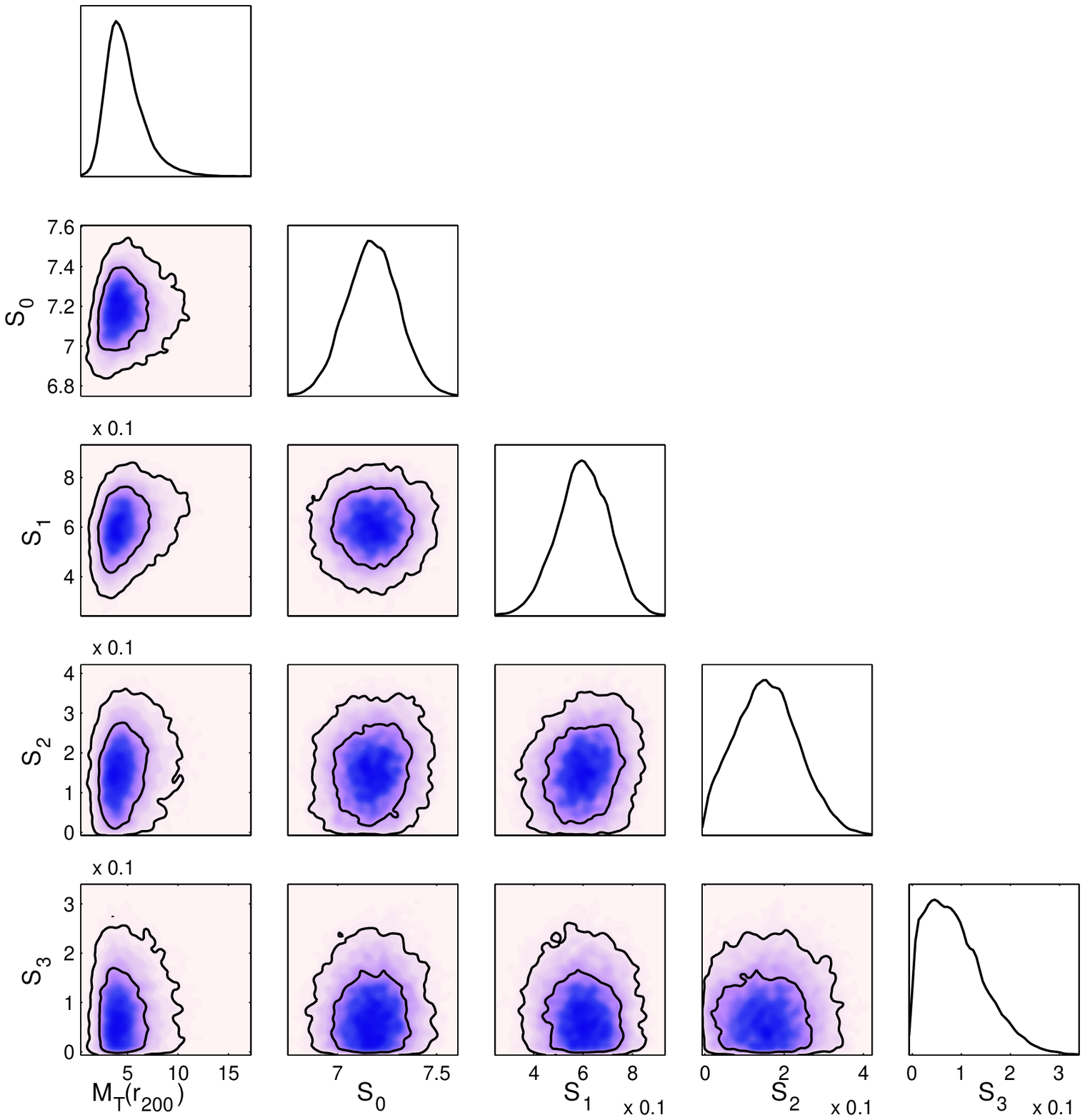}\\

\end{tabular}
\caption{Results for Abell~621. Panels A and B show the SA map before and after source-subtraction, respectively; a $0.6$\,k$\lambda$ taper has been applied to B. The box in panels A and B indicates the cluster SZ centroid, the other symbols are in Tab. \ref{tab:sourcelabel}. The smoothed {\emph{ROSAT HRI}} X-ray map overlaid with contours from B is given in panel C. Panels D and E show the marginalized posterior distributions for the cluster sampling and derived parameters, respectively. F shows the 1 and 2-D marginalized posterior distributions for source flux densities (in Jys) within $5\arcmin$ of the cluster SZ centroid (see Tab. \ref{tab:source_info}) and $M_{\rm{T}}(r_{200})$ in $h_{100}^{-1}\times 10^{14}M_{\odot}$. In D $M_{\rm{T}}$ is given in units of $h_{100}^{-1}\times10^{14}M_{\odot}$ and $f_{\rm{g}}$ in $h_{100}^{-1}$; both parameters are estimated within $r_{200}$. In E $M_{\rm{g}}$ is in units of $h_{100}^{-2}M_{\odot}$, $r$ in $h_{100}^{-1}$Mpc and $T$ in KeV.}
\label{fig:A621}
\end{center}
\end{figure*}


\subsection{Abell~773}\label{sec:A773}

Results for Abell~773 are shown in Fig. \ref{fig:A773}. 
Abell~773 has few associated radio sources, all of which are $\gtrsim
10\arcmin$ away from the pointing centre, weak ($\lesssim
3$\,mJy), and are subtracted well from our data (Fig.\ref{fig:A773} B).
 We do not find any evidence for extended positive emission in our maps. Observations by
\cite{Gio_A773_halo} revealed
the presence of a radio halo with a luminosity of $2.8\times
10^{24}$\,WHz$^{-1}$ at 1.4\,GHz; this result has been confirmed with
the VLA by \cite{govoni2001}. Given the typical steep spectral index of radio
halos, we do not expect our SZ signal to be affected at 16\,GHz.

Our observations clearly show the SZ image is extended along the NW-SE direction, contrary to the
X-ray image from {\sc{Chandra}} observations, which appears to be elongated in an approximately perpendicular
 direction. As might be expected from a
 disturbed system, Abell~773 appears to not have a cool core \citep{allen1998}.

\cite{2007A&A...467...37B} present a comprehensive study of Abell~773 from
the Telescopio Nazionale Galileo (TNG) telescope
 and X-ray data from the {\sc{Chandra}} data archive. They
find two peaks in the velocity distribution of the cluster members which are
separated
 by 2$\arcmin$ along the E-W direction. Two peaks can also be seen in the
X-ray, although these are along the NE-SW direction. Barrena et al. estimate
 the  virial mass of the main cluster to be
$M_{\rm{T}}(r_{\rm{vir}})=1.0-2.5\times 10^{15}$\,$h_{70}^{-1}\rm{M_{\odot}}$
and
 $M_{\rm{T}}$ = 1.2-2.7$\times 10^{15}h_{70}^{-1}\rm{M_{\odot}}$ for the entire
system, using the virial theorem, dispersion velocity measurements and a galaxy King-like 
distribution. Assuming an NFW profile they estimate the mass for the system to be
$M_{\rm{T}}(< r=1 h_{70}^{-1}\rm{Mpc})=5.9-11.1\times 10^{14} h_{70}^{-1}M_{\odot}$.
A further analysis of {\sc{Chandra}} data by \cite{govoni2004} yielded a mean
temperature of 7.5$\pm$0.8\,keV within a radius of 800\,kpc ($h_{70}=1$).
Another X-ray study of this cluster by \cite{zhang2008} using {\sc{XMM-Newton}}
found $M_{\rm{T}}(r_{500})$ = 8.3 $\pm$2.5 $\times 10^{14}\rm{M_{\odot}}$
 assuming isothermality, spherical symmetry and
$h_{70}=1$.

 The SZ effect associated with Abell~773 has been observed several times
(\citealt{1993MNRAS.265L..57G},
 \citealt{carlstrom1996}, \citealt{2003MNRAS.341..937S},  \citealt{bona_chandra}, LaRoque et al. 2006). Most recently,
 AMI Consortium: Zwart et al. (2010) observed the cluster and found a cluster mass of
$M_{\rm{T}}(r_{200})$ = 1.9$^{+0.3}_{-0.4}$$\times 10^{15}\rm{M_{\odot}}$
 using $h_{100}=1$; however, their  $M_{\rm{T}}$ estimates are biased high, as they say, 
 and we find $M_{\rm{T}}(r_{200}) = 3.6 \pm 1.2 \times 10^{14} h_{100} \rm{M_{\odot}}$.

\begin{figure*}
\centerline{\huge{Abell 773}}
\centerline{{A}\includegraphics[width=7.5cm,height=
7.5cm,clip=,angle=0.]{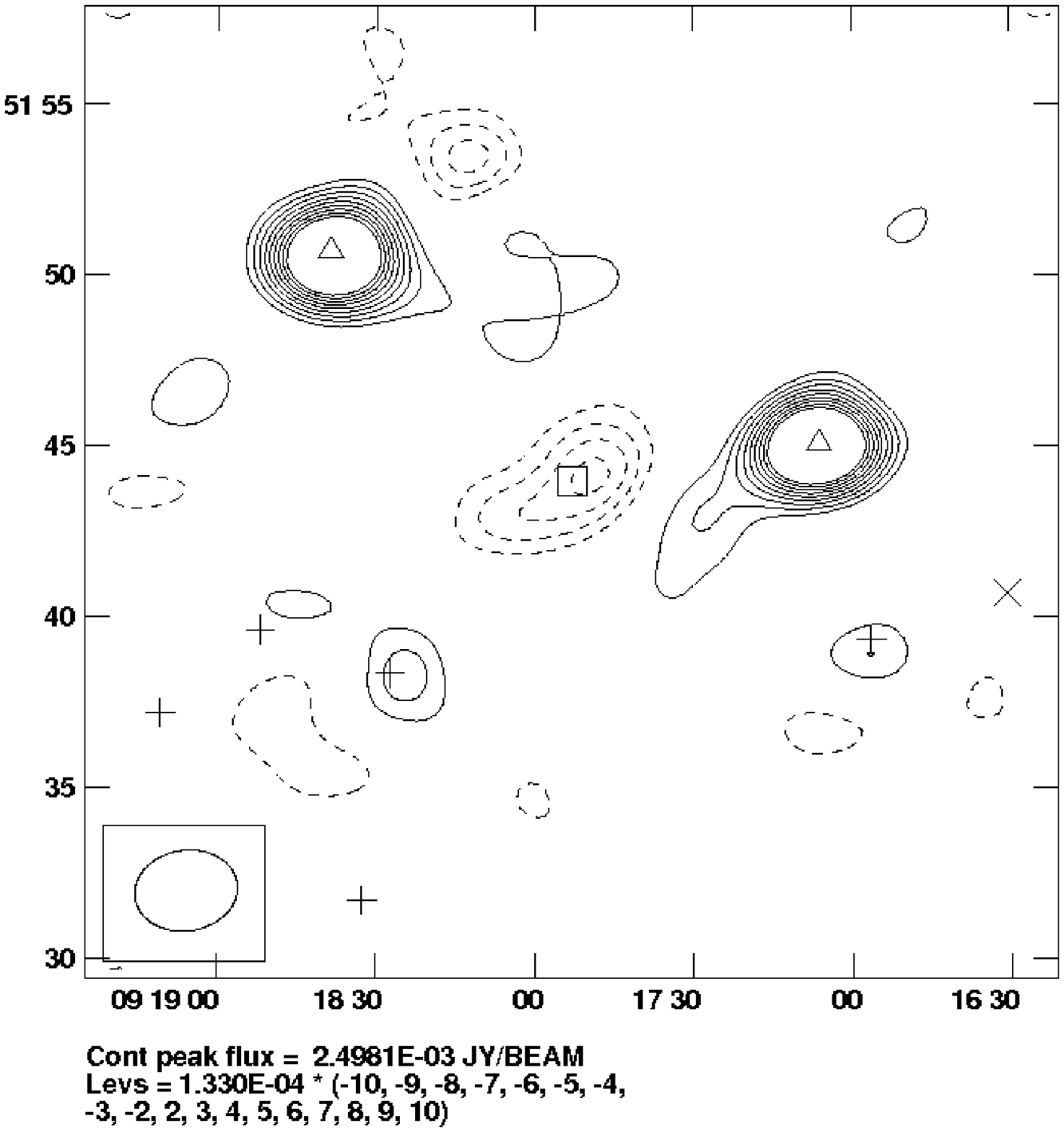}\qquad{D}\includegraphics[width=7.5cm,height=
7.5cm,clip=,angle=0.]{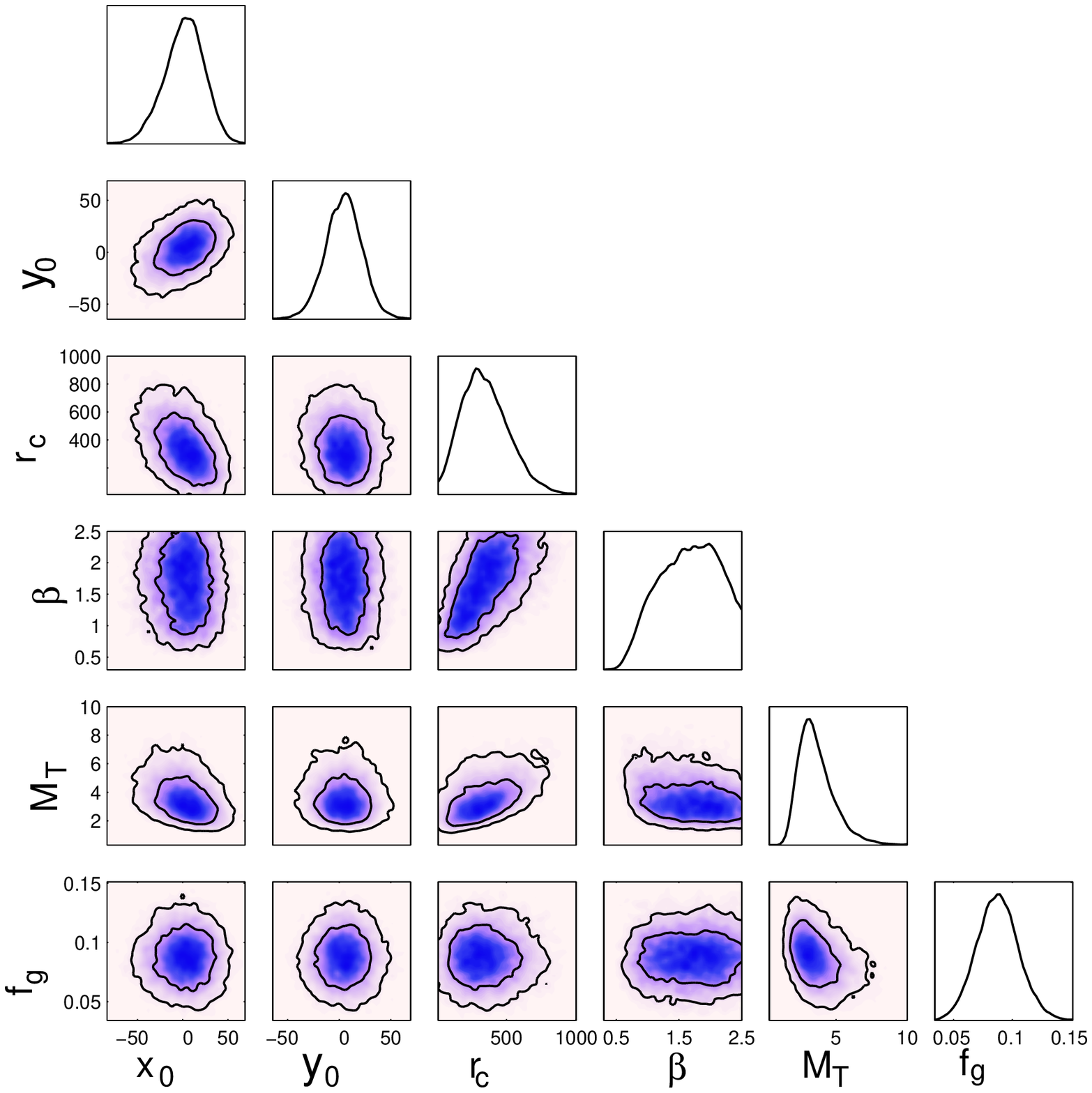}}
 \centerline{{B}\includegraphics[width=7.5cm,height=
7.5cm,clip=,angle=0.]{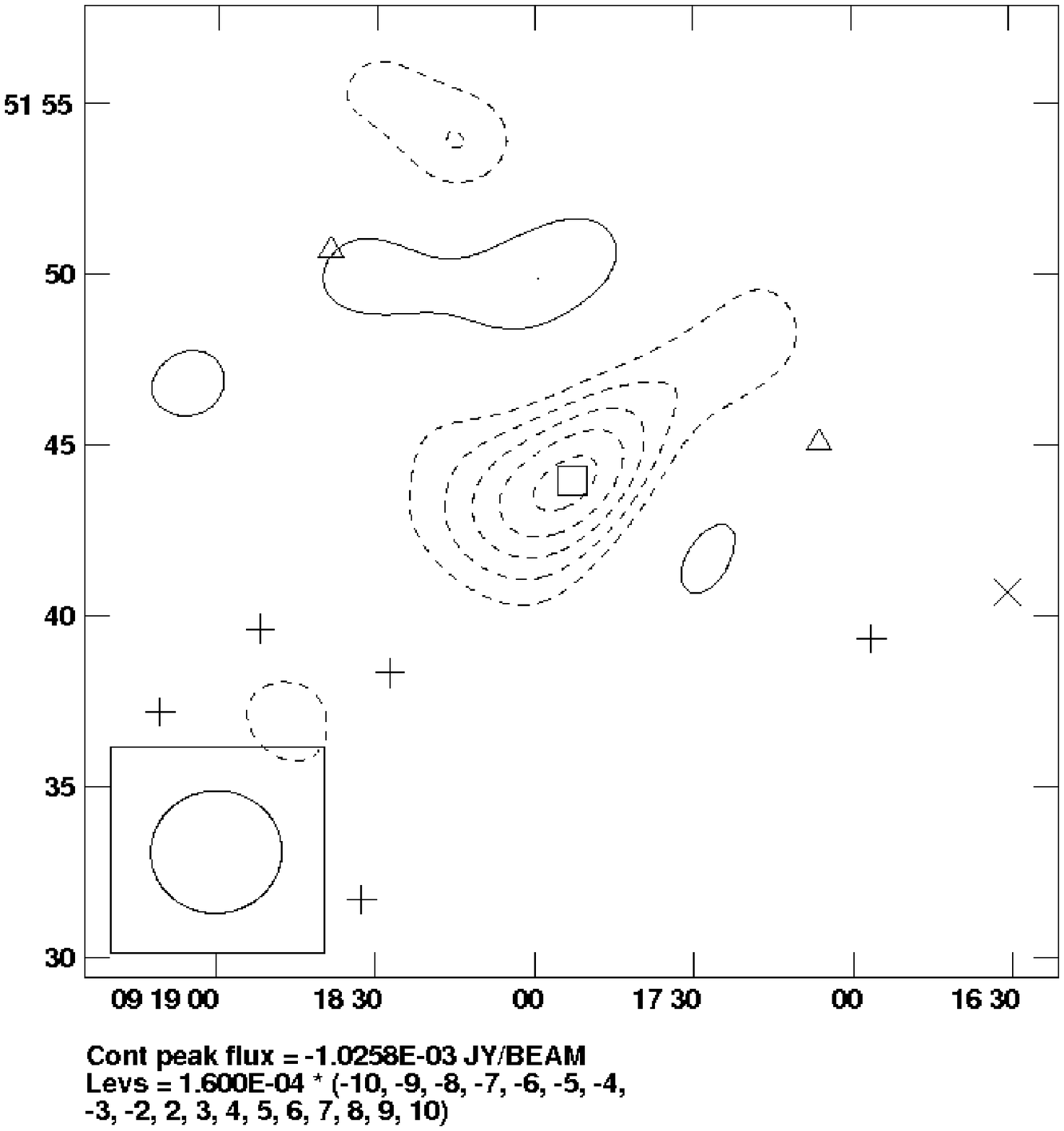}\qquad{E}\includegraphics[width=7.5cm,height=
7.5cm,clip=,angle=0.]{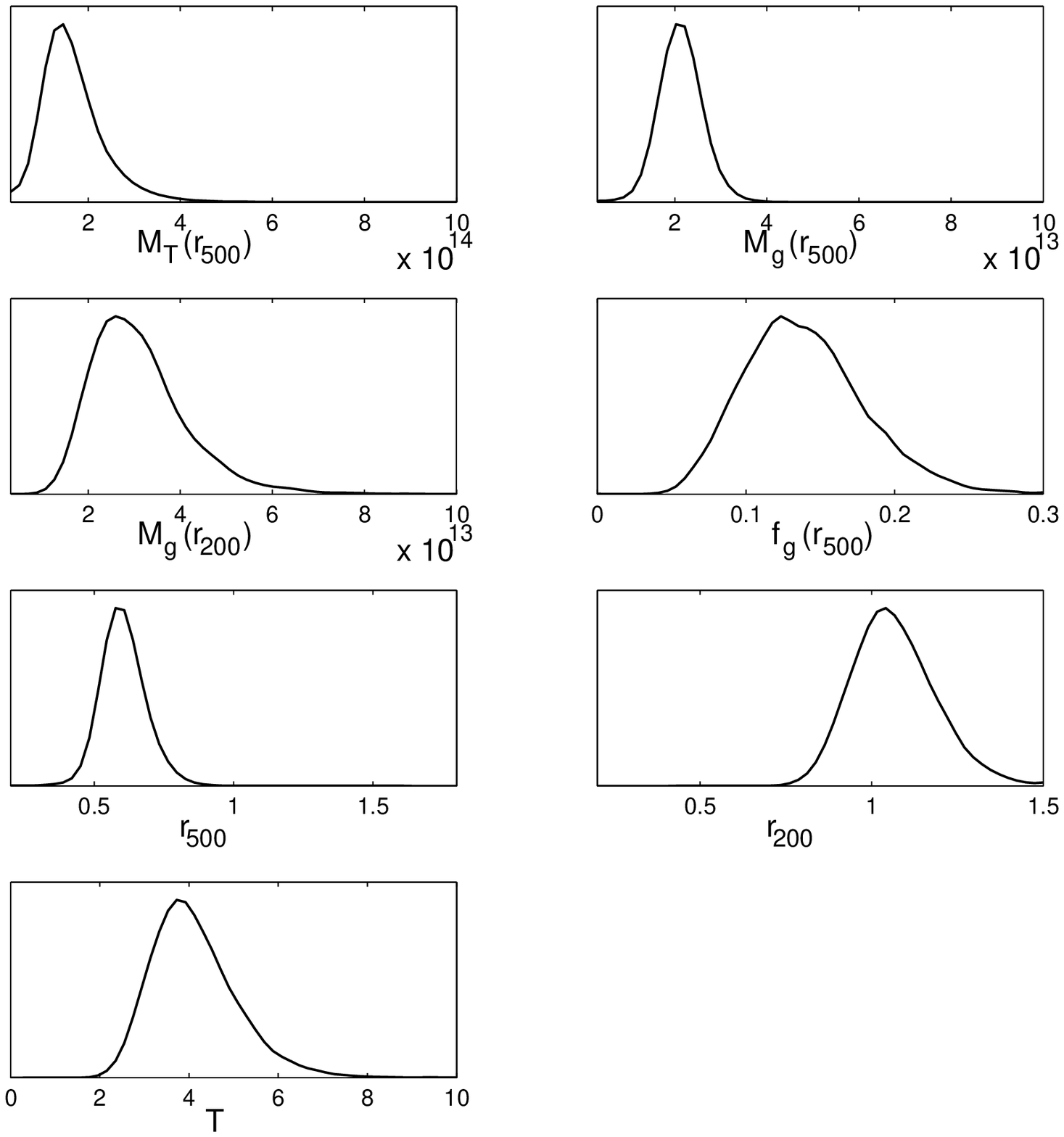}}
\centerline{{C}\includegraphics[width=7.5cm,height=
6.5cm,clip=,angle=0.]{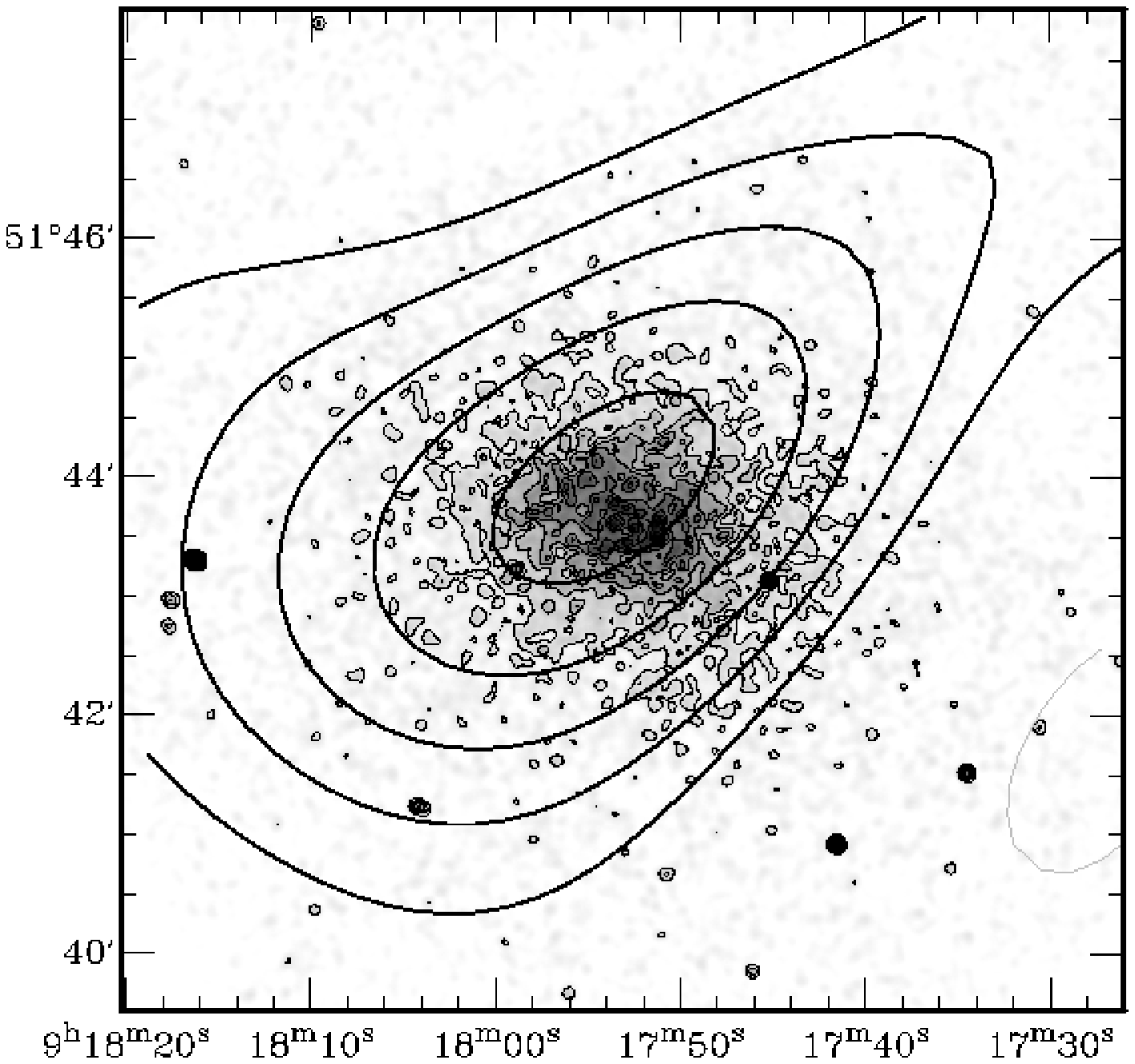}}

\caption{Results for Abell~773. Panels A and B show the SA map before and after source subtraction, respectively; a $0.6$\,k$\lambda$ taper has been applied to B. The box in panels A and B indicates the cluster SZ centroid, for the other symbols see Tab. \ref{tab:sourcelabel}. The smoothed {\sc{Chandra}} X-ray map overlaid with contours from B is given in image C. Panels D and E show the marginalized posterior distributions for the cluster sampling and derived parameters, respectively. In panel D $M_{\rm{T}}$ is given in units of $h_{100}^{-1}\times10^{14}M_{\odot}$ and $f_{\rm{g}}$ in $h_{100}^{-1}$; both parameters are estimated within $r_{200}$. In E $M_{\rm{g}}$ is in units of $h_{100}^{-2}M_{\odot}$, $r$ in $h_{100}^{-1}$Mpc and $T$ in KeV.
No plots of the degeneracy between cluster mass and source flux densities are shown 
 since all detected sources are $>5'$ from the cluster SZ centroid and thus should not have a strong impact on the marginalized distribution for the cluster mass.}
\label{fig:A773}
\end{figure*}

\subsection{Abell~781}

Fig. \ref{fig:A781} contains our results for Abell~781.
It is evident from inspection of Figs. \ref{fig:A781} A and F and Tab. \ref{tab:source_info} that 
there is strong emission from radio sources lying on the decrement. One of the
sources with a flux density of $9$\,mJy lies on top of the {\sc{McAdam}} best-fit cluster position. The 
difficulty of accurately disentangling the signal contributions from this source and the cluster is translated into a
degeneracy between the source's flux density and the cluster mass:  Fig. \ref{fig:A781} F.
 No extended emission was detected on the LA
maps and, after source subtraction, the residuals on the maps are $\lesssim
2\sigma$ (Fig. \ref{fig:A781} B). \cite{rudnick2009}
 have found evidence in WENSS data at 327\,MHz of diffuse emission from a radio galaxy and some other
unknown source
 with a flux  within a radius of 500\,kpc of 40\,mJy, while
\cite{GMRT_HALO} estimate diffuse emission at the centre to be $\approx
15-20$\,mJy
 using 325-MHz GMRT data. Assuming a typical steep spectral index
 for radio halos, in the range of $1.2-1.4$ (e.g., \citealt{hanisch1980}), even as far as 16\,GHz, we
would expect to find an $\approx 170$\,$\mu$Jy signal around
the cluster and $\lesssim 85$\,$\mu$Jy at the centre. The GMRT contour map in
Venturi et al. identifies the relic at a similar location to that of some
unsubtracted positive
emission in our maps at $\approx$ RA 09:30:00, Dec 30:28:00.

 X-ray observations with {\sc{Chandra}} and {\sc{XMM-Newton}}
(\citealt{Abell 781_XMM}) imply that Abell~781 is a
 complex cluster merger: the main cluster is surrounded by three smaller
clusters, two to the East of the main cluster and one to
the West. Sehgal et al. estimate the  mass of Abell~781 within $r_{500}$
assuming a NFW matter density profile to be
$5.2^{+0.3}_{-0.7}$$\times$$10^{14}\rm{M_{\odot}}$ from  X-ray data
and $2.7^{+1.0}_{-0.9}$$\times$$10^{14}\rm{M_{\odot}}$ from the Kitt Peak
Mayall 4-m telescope lensing observations, using $h_{71}=1$.
 Further
results from {\sc{XMM-Newton}} by Zhang et al. yield
 $M_{\rm{T}}(r_{500})$ = 4.5$\pm$1.3 $\times 10^{14}\rm{M_{\odot}}$
 assuming isothermality, spherical symmetry and $h_{70}=1$. We obtain
 $M_{\rm{T}}(r_{500})$ =
2.9$\pm$0.6$\times 10^{14}\rm{M_{\odot}}$ and 
$M_{\rm{T}}(r_{200})$ = 5.9$\pm$1.1$\times 10^{14}\rm{M_{\odot}}$ for $h_{70}=1$.

\begin{figure*}
\centerline{\huge{A781}}
\centerline{{A}\includegraphics[width=7.5cm,height=7.5cm,clip=,angle=0.]{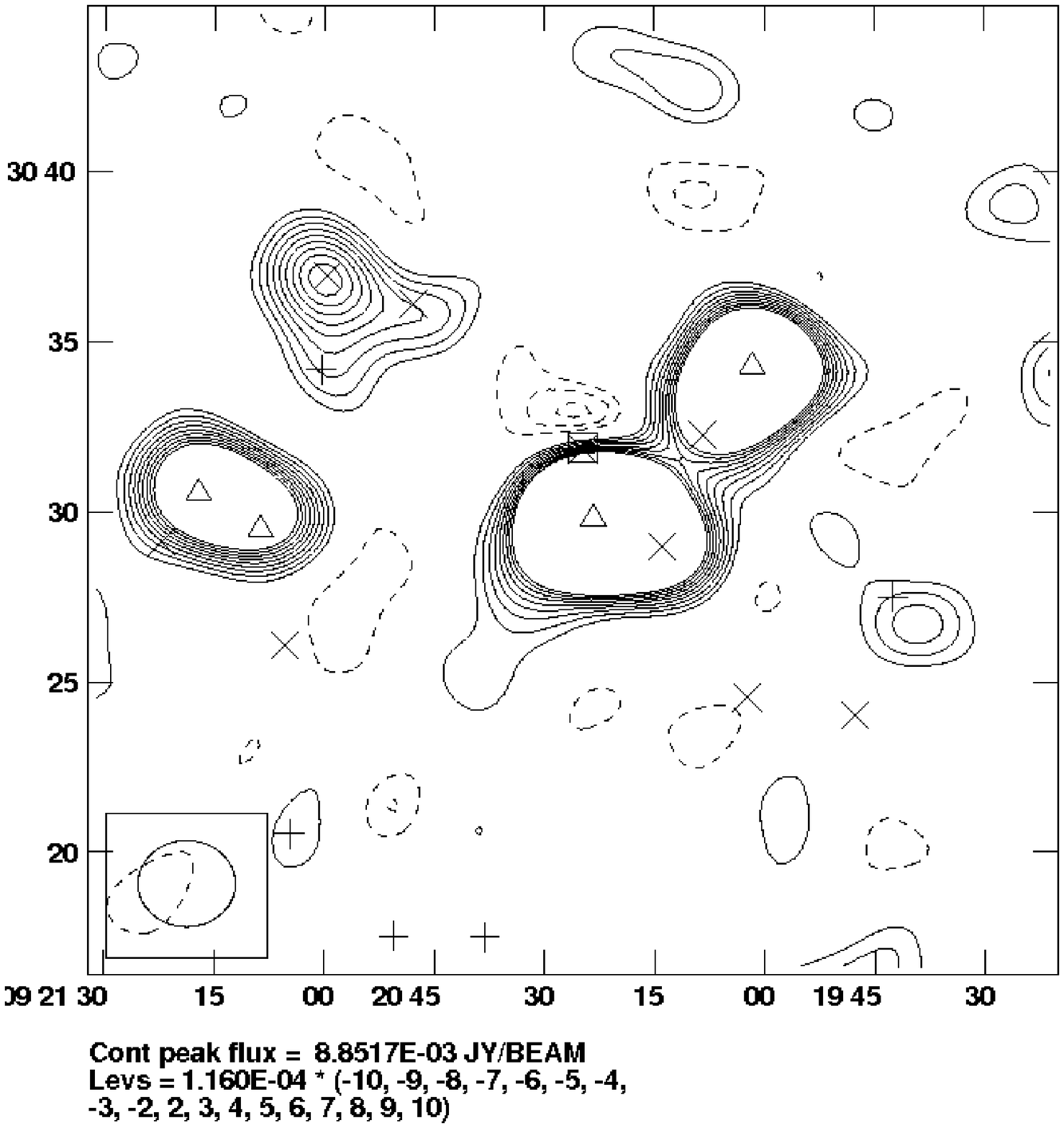}\qquad{D}\includegraphics[width=7.5cm,height=7.5cm,clip=,angle=0.]{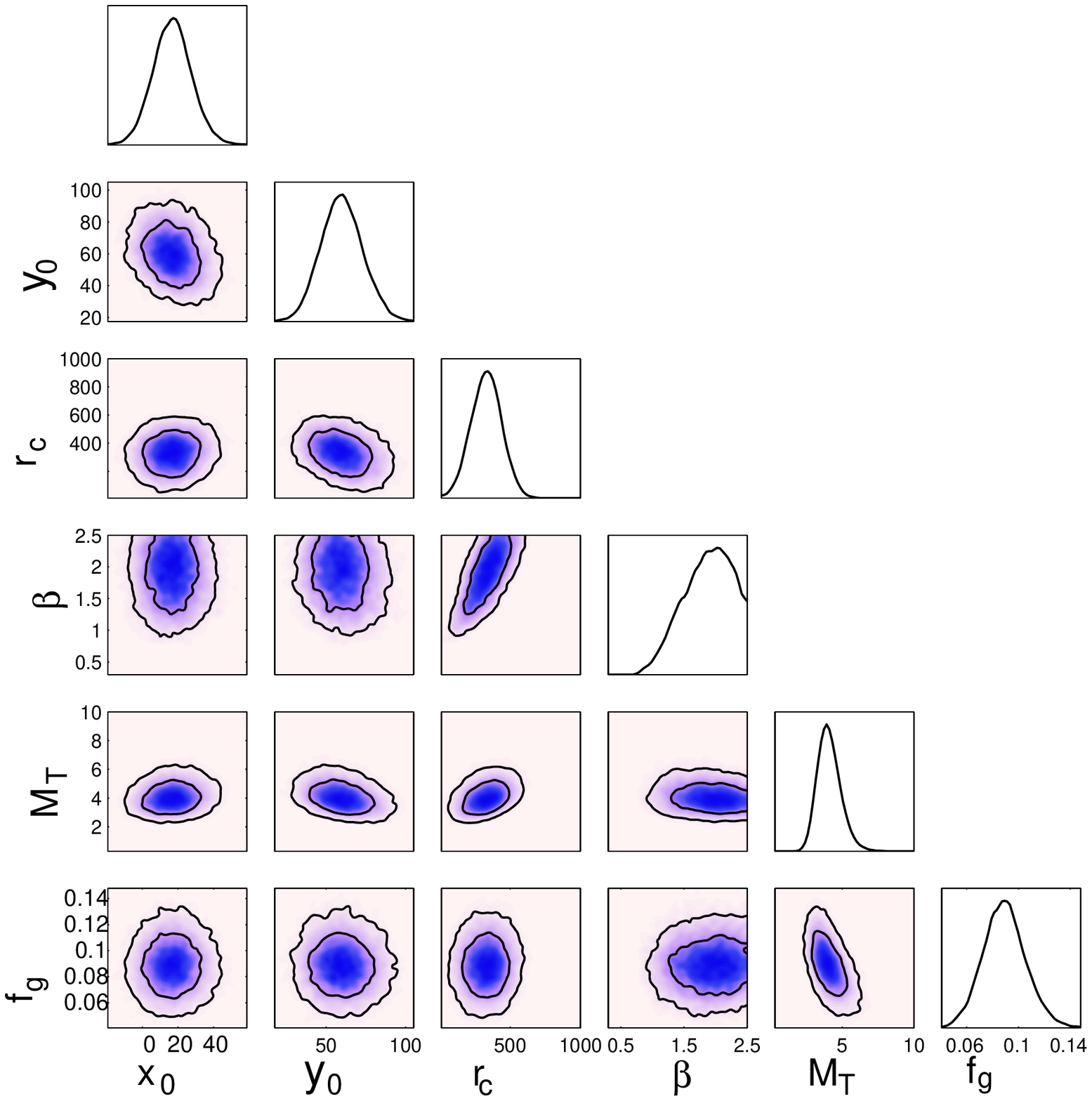}}
 \centerline{{B}\includegraphics[width=7.5cm,height=7.5cm,clip=,angle=0.]{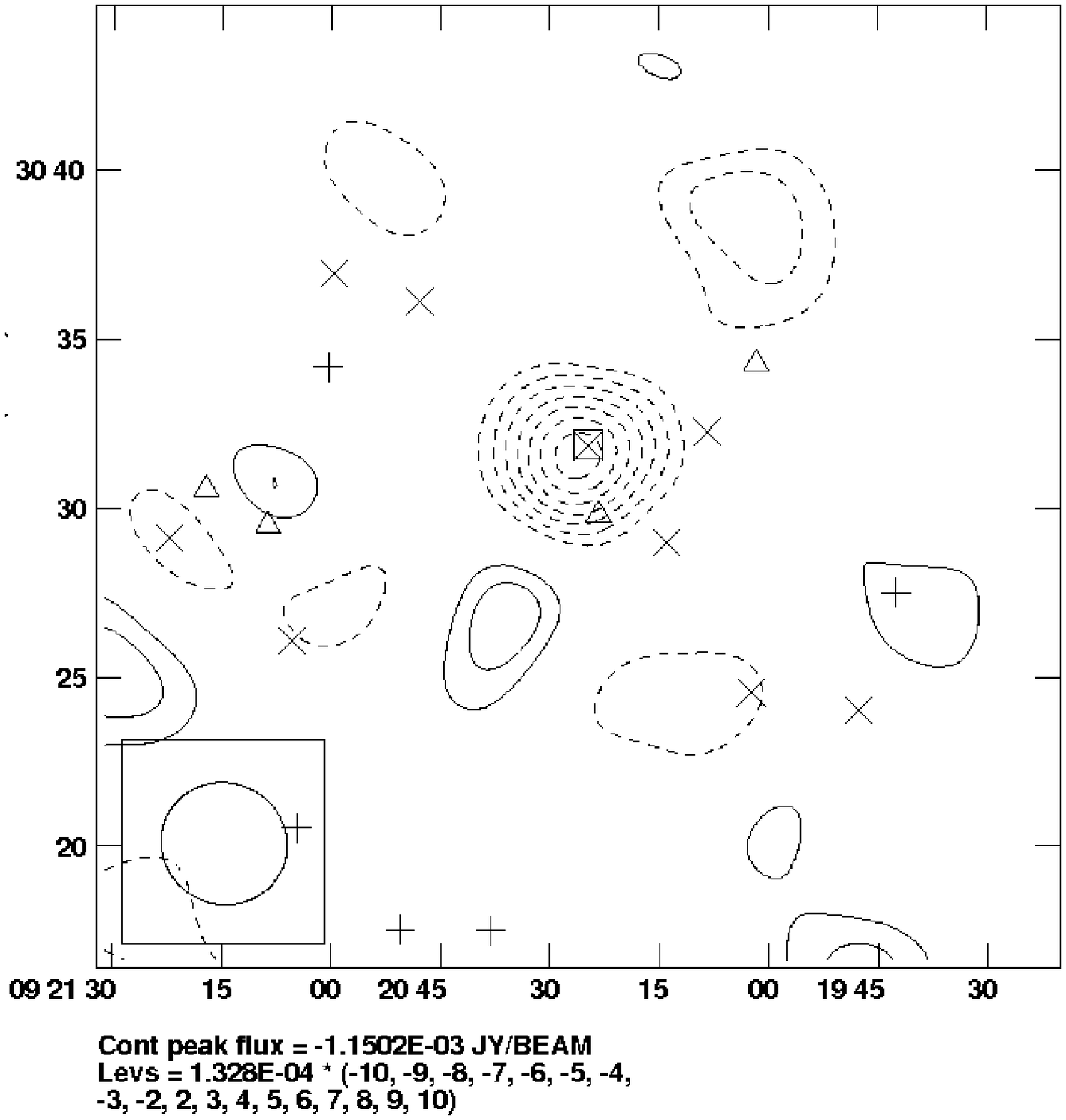}\qquad{E}\includegraphics[width=7.5cm,height=7.5cm,clip=,angle=0.]{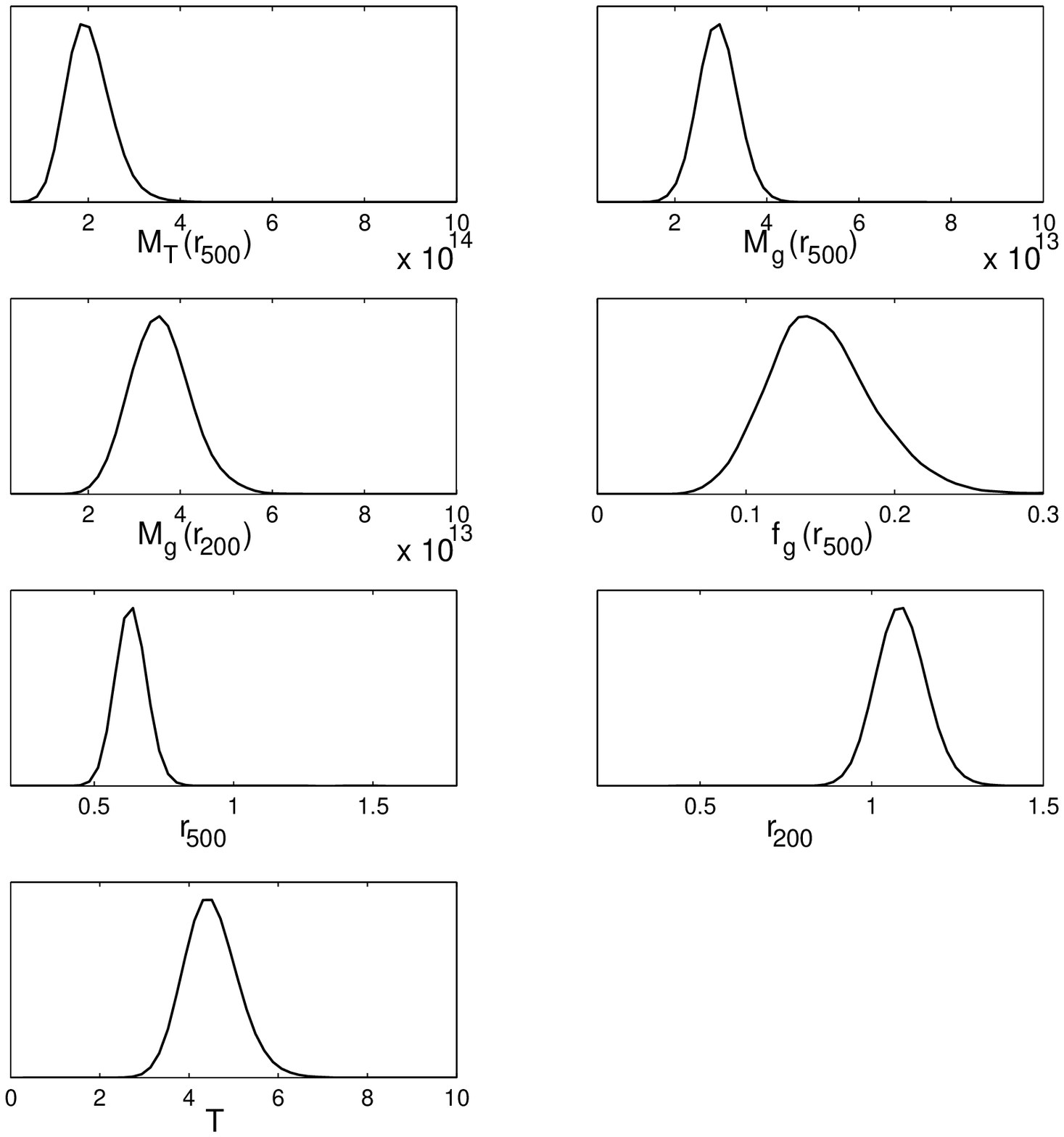}}
 \centerline{
{C}\includegraphics[width=7.5cm,height=6.5cm,clip=,angle=0.]{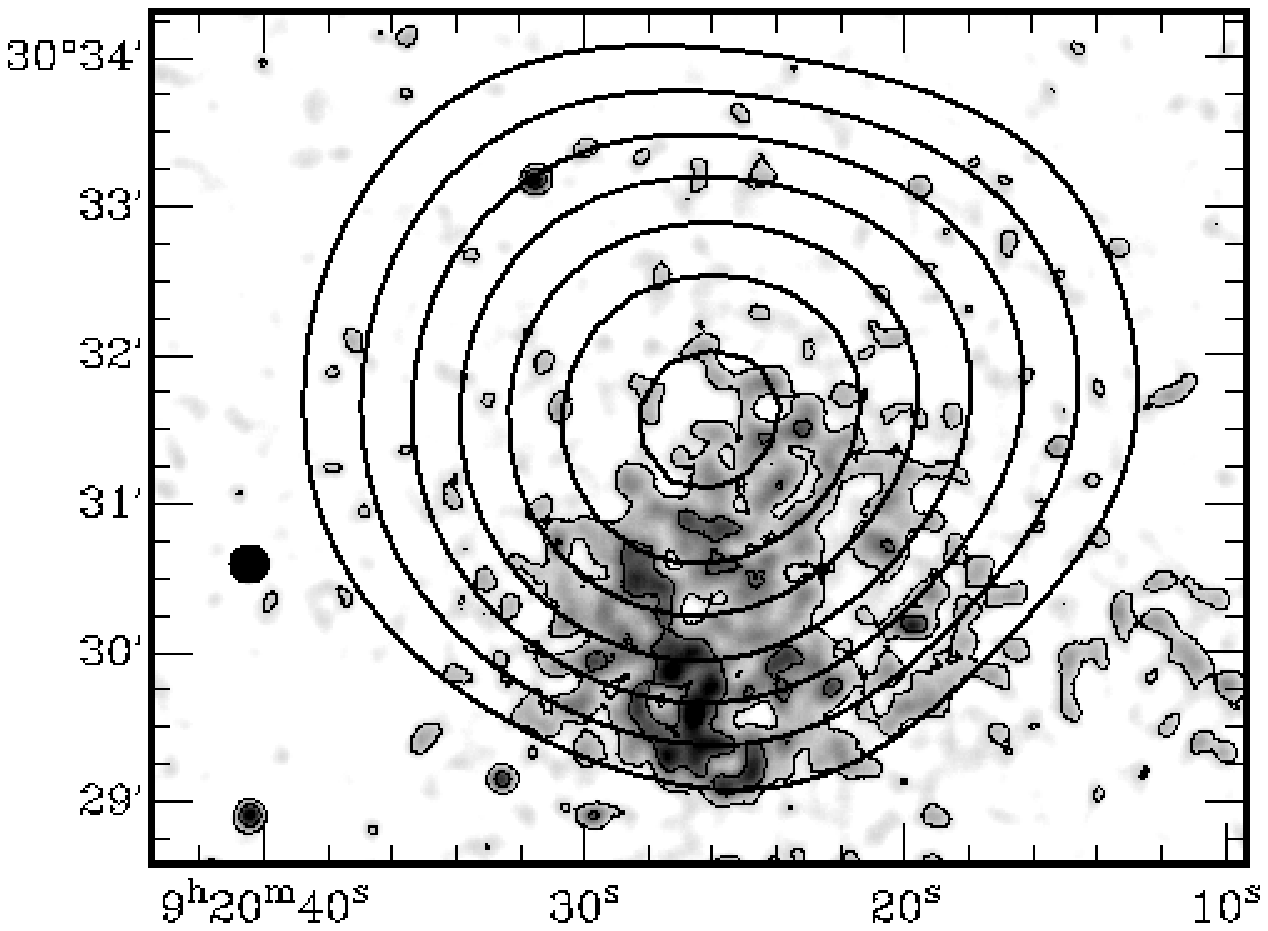}\qquad{F}\includegraphics[width=7.0cm,height=6.5cm,clip=,angle=0.]{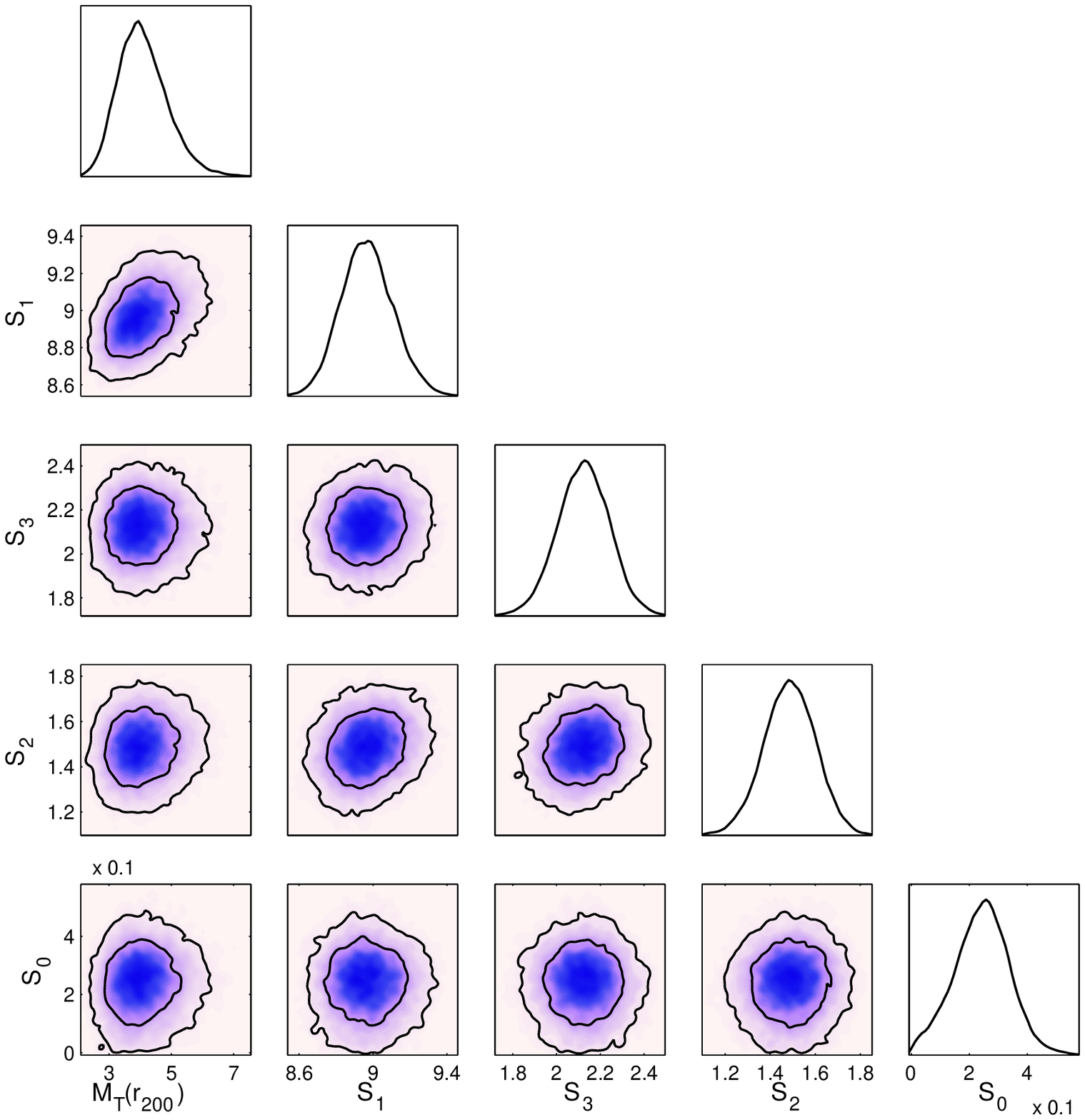}}

\caption{Results for Abell~781. Panels A and B show the SA map before and after source-subtraction, respectively; a $0.6$\,k$\lambda$ taper has been applied to B. The box in panels A and B indicates the cluster SZ centroid, other symbols are in Tab. \ref{tab:sourcelabel}. The smoothed {\sc{Chandra}} X-ray map overlaid with contours from B is given in image C. Panels D and E show the marginalized posterior distributions for the cluster sampling and derived parameters, respectively. F shows the 1 and 2-D marginalized posterior distributions for source flux densities (in Jys) within $5\arcmin$ of the cluster SZ centroid (see Tab. \ref{tab:source_info}) and $M_{\rm{T}}(r_{200})$ (in $h_{100}^{-1}\times 10^{14}M_{\odot}$). In D $M_{\rm{T}}$ is given in units of $h_{100}^{-1}\times10^{14}M_{\odot}$ and $f_{\rm{g}}$ in $h_{100}^{-1}$; both parameters are estimated within $r_{200}$. In E $M_{\rm{g}}$ is in units of $h_{100}^{-2}M_{\odot}$, $r$ in $h_{100}^{-1}$Mpc and $T$ in KeV.}
\label{fig:A781}
\end{figure*}

\subsection{ Abell~990} \label{resultsA990}

Results for Abell~990 are given in Fig. \ref{fig:A990}. We detected 20 sources towards Abell~990. Those
 detected above $4\sigma_{\rm{LA}}$ within $10\arcmin$ from the pointing
centre
were found to have flux densities $<2.8$\,mJy, not to be extended
with respect to the LA synthesized beam (Tab. \ref{tab:source_info}),
 and none to lie on the SZ decrement, as seen in the source-subtracted map (Fig.
\ref{fig:A990} B).
 The subtraction
has worked well and there are only low-level ($\approx 1-2\sigma$) residuals. \cite{rudnick2009} do not detect any
significant amount of diffuse emission within a radius of 500\,kpc in 327\,MHz WENSS data; given the
steep falling spectrum associated with this emission, we do not expect it to
contaminate our SZ signal.
 The imaged decrement is fairly circular but extended along the NE-SW direction 
coincident with the distribution of the X-ray signal. Our spherical cluster
model provides a good fit and the parameter distributions are tightly
constrained.
The low resolution X-ray map shown in Fig. \ref{fig:A990} C
provides tentative evidence that the X-ray emitting cluster gas
has a clumpy distribution.


\begin{figure*}
\begin{center}
\begin{tabular}{m{8cm}cm{8cm}}
\multicolumn{3}{c}{\huge{Abell 990}}\\
{A}\includegraphics[width=7.5cm,height=7.5cm,clip=,angle=0.]{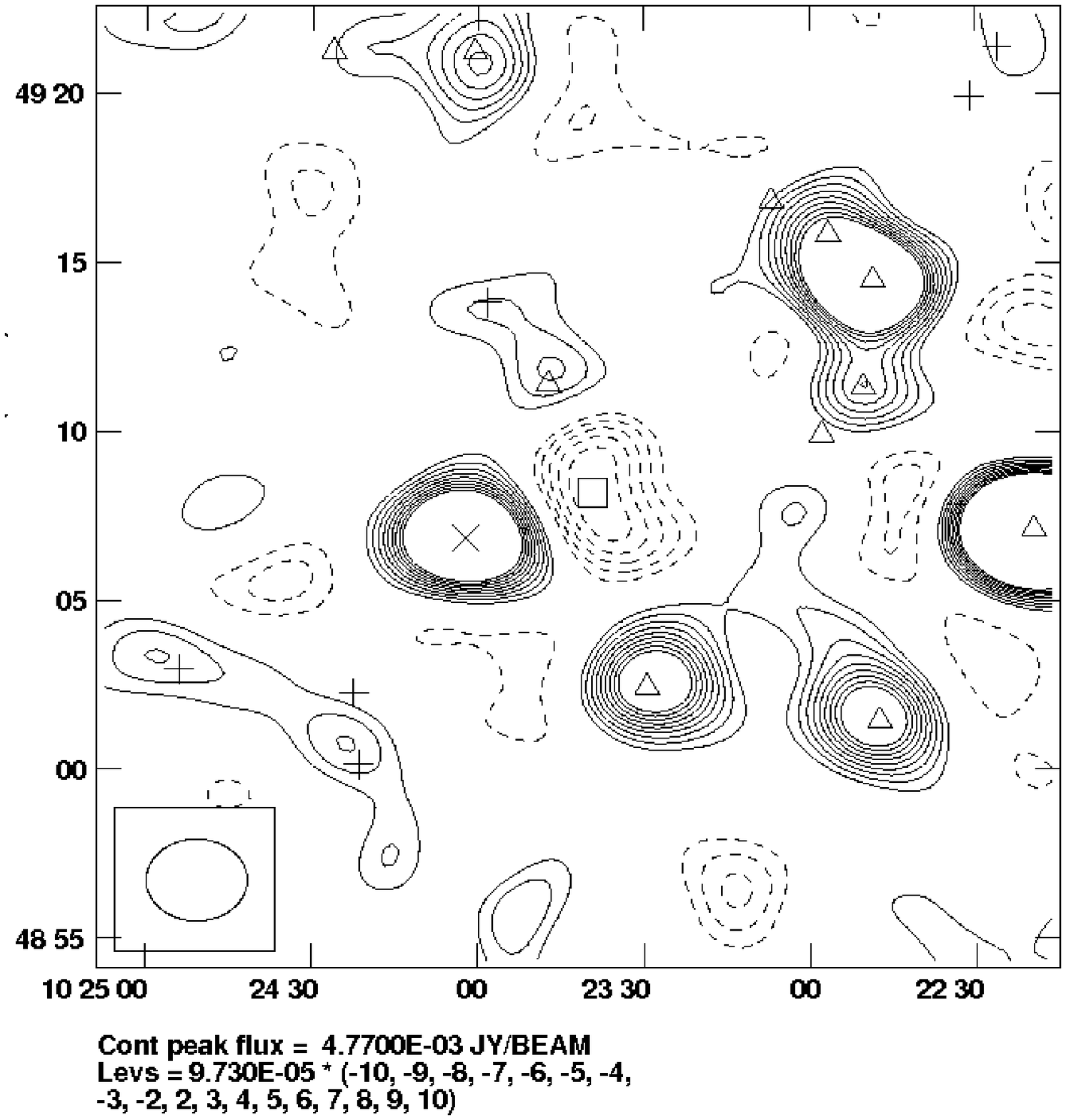}
& \quad &
{D}\includegraphics[width=7.5cm,height=7.5cm,clip=,angle=0.]{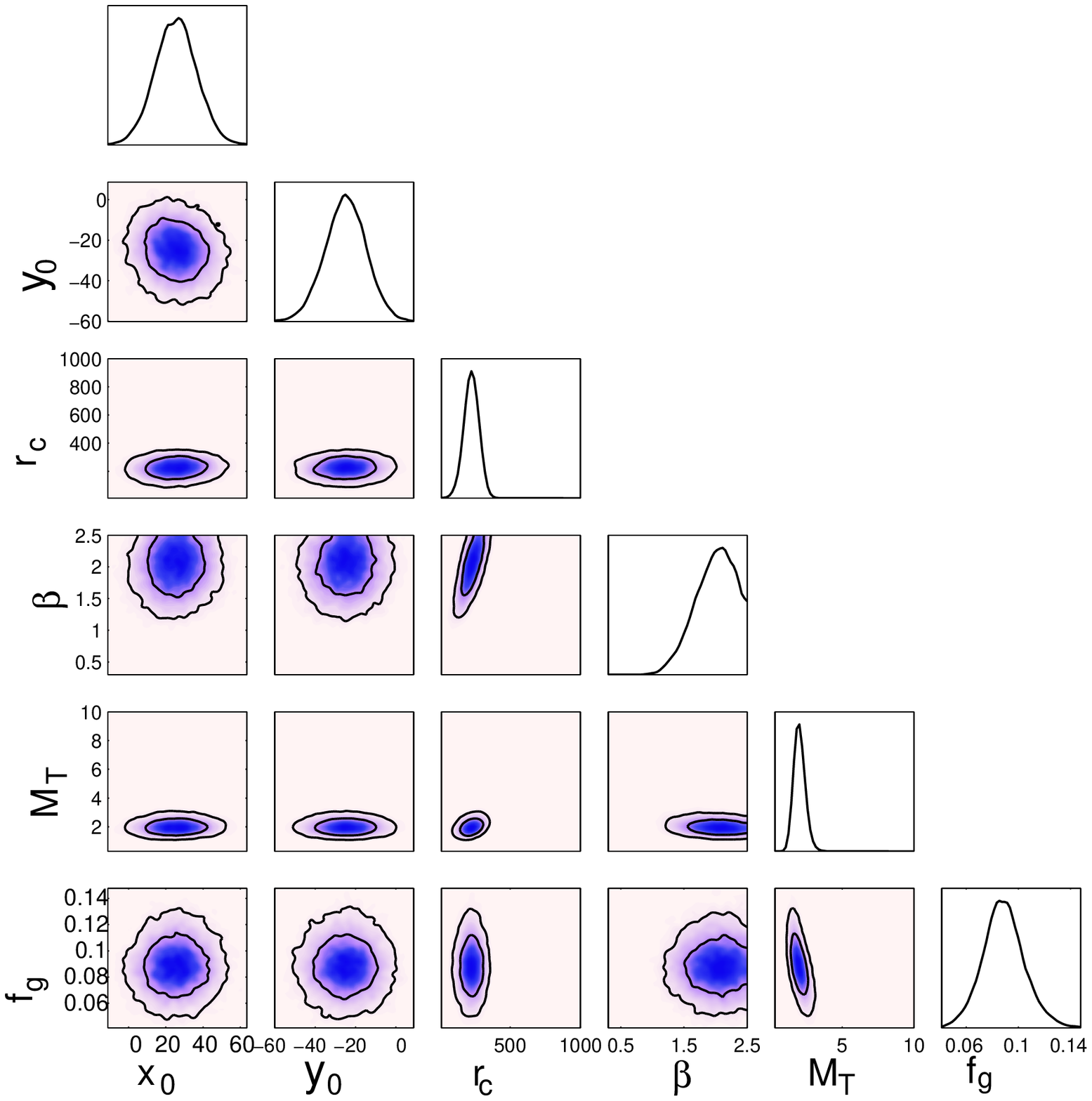}
\\
{B}\includegraphics[width=7.5cm,height=7.5cm,clip=,angle=0.]{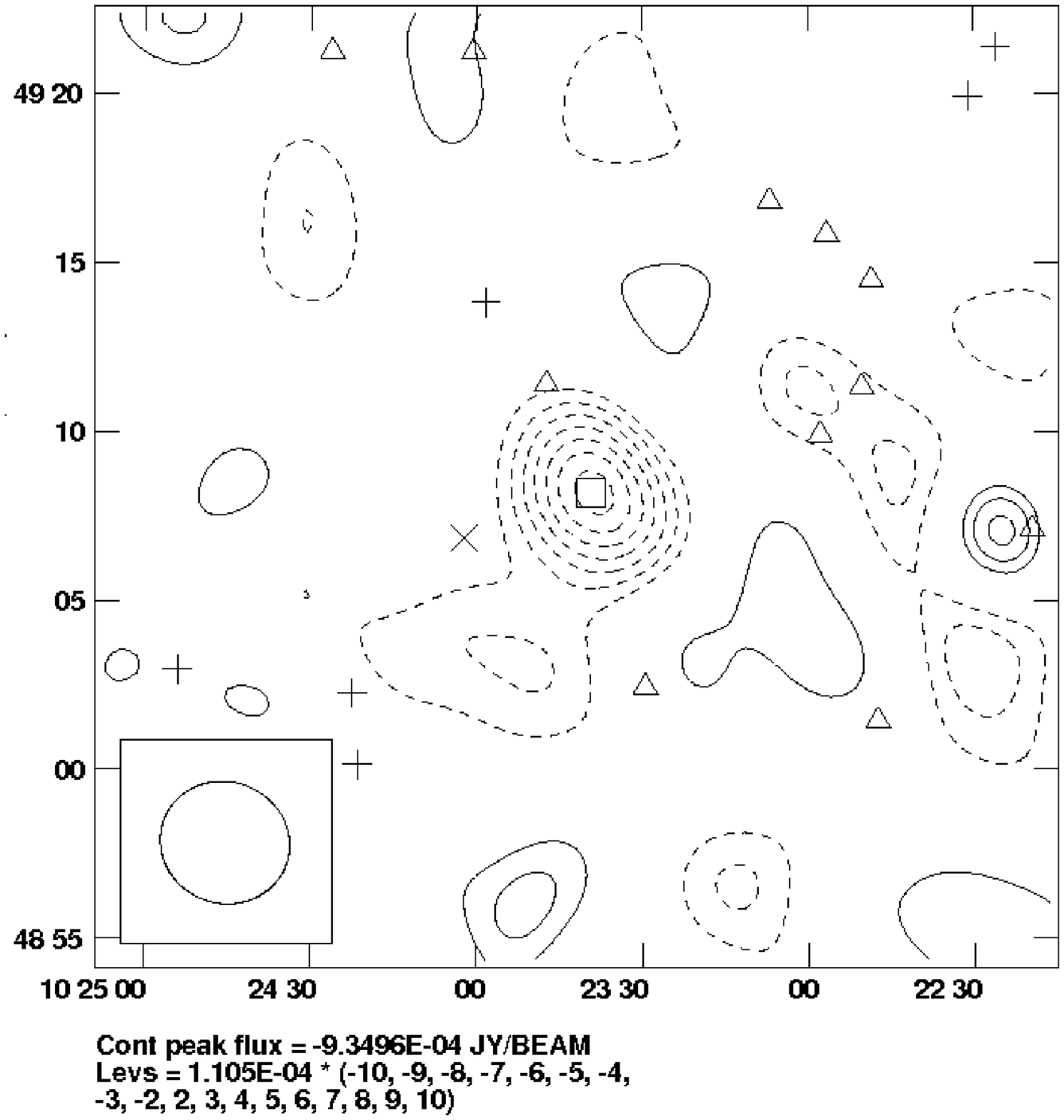}

& \qquad &
{E}\includegraphics[width=7.5cm,height=7.5cm,clip=,angle=0.]{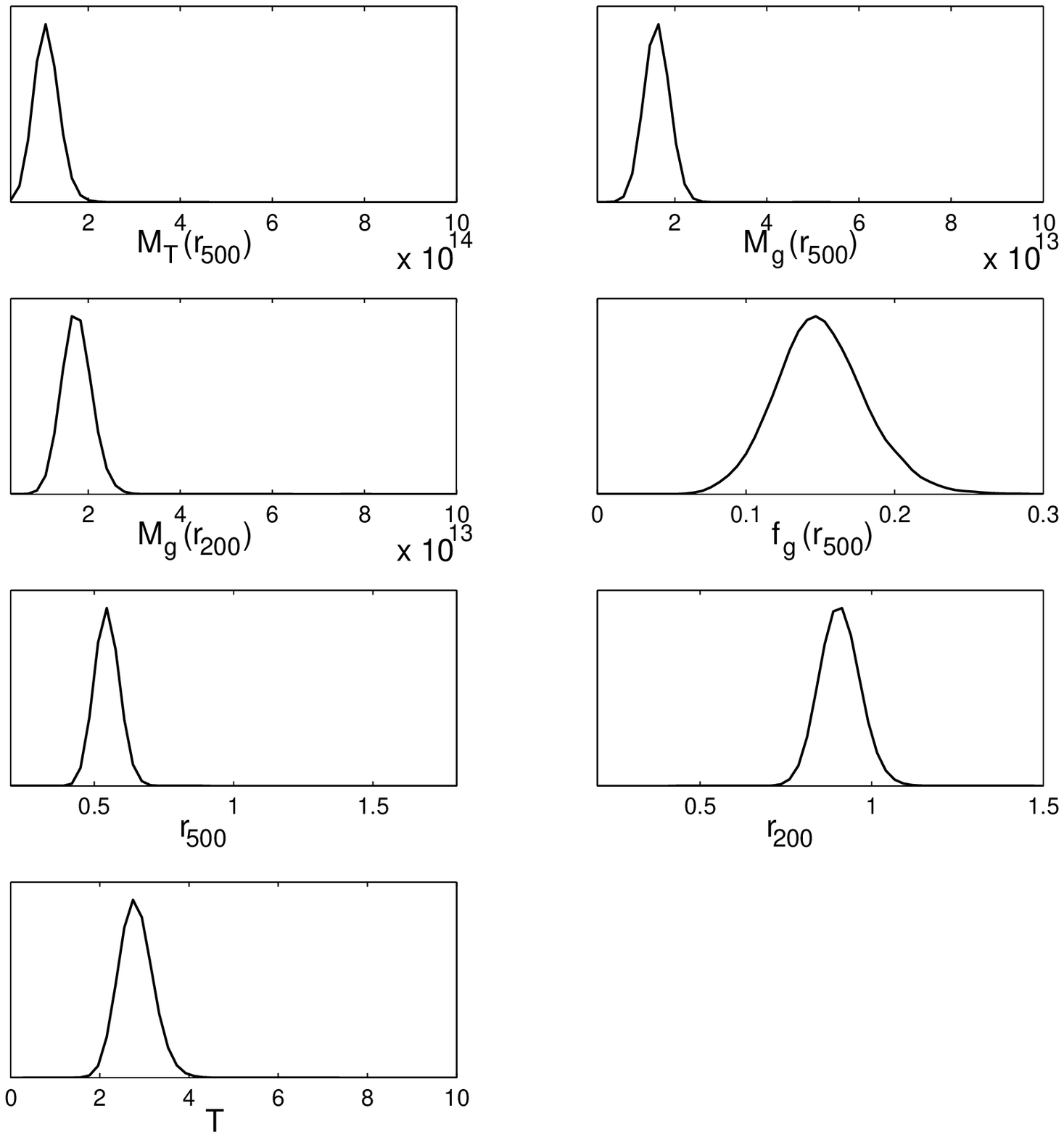}
\\
{C}\includegraphics[height=6.0cm,clip=,angle=0.]{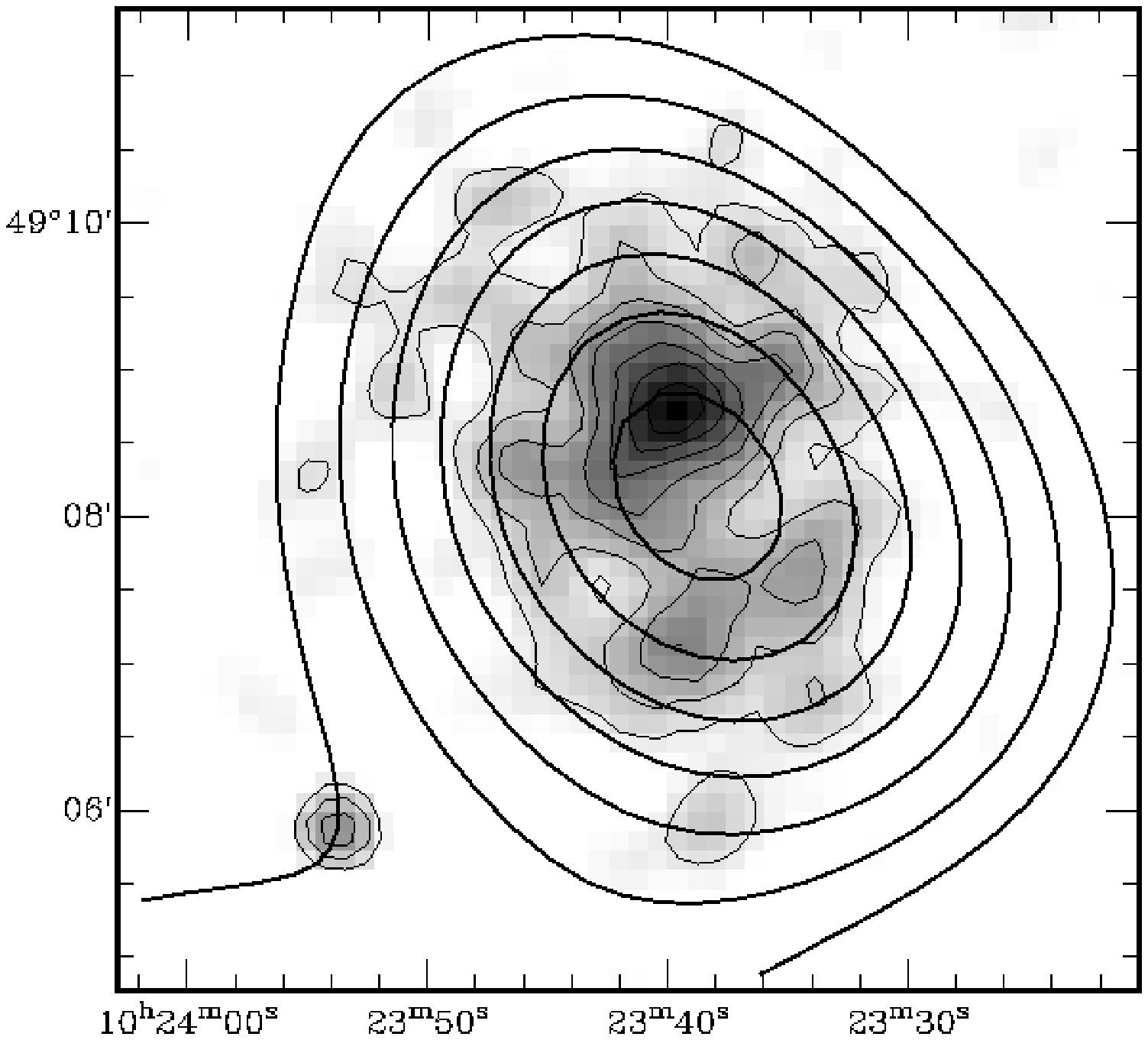}

& \qquad &
{F}\includegraphics[height=6.0cm,clip=,angle=0.]{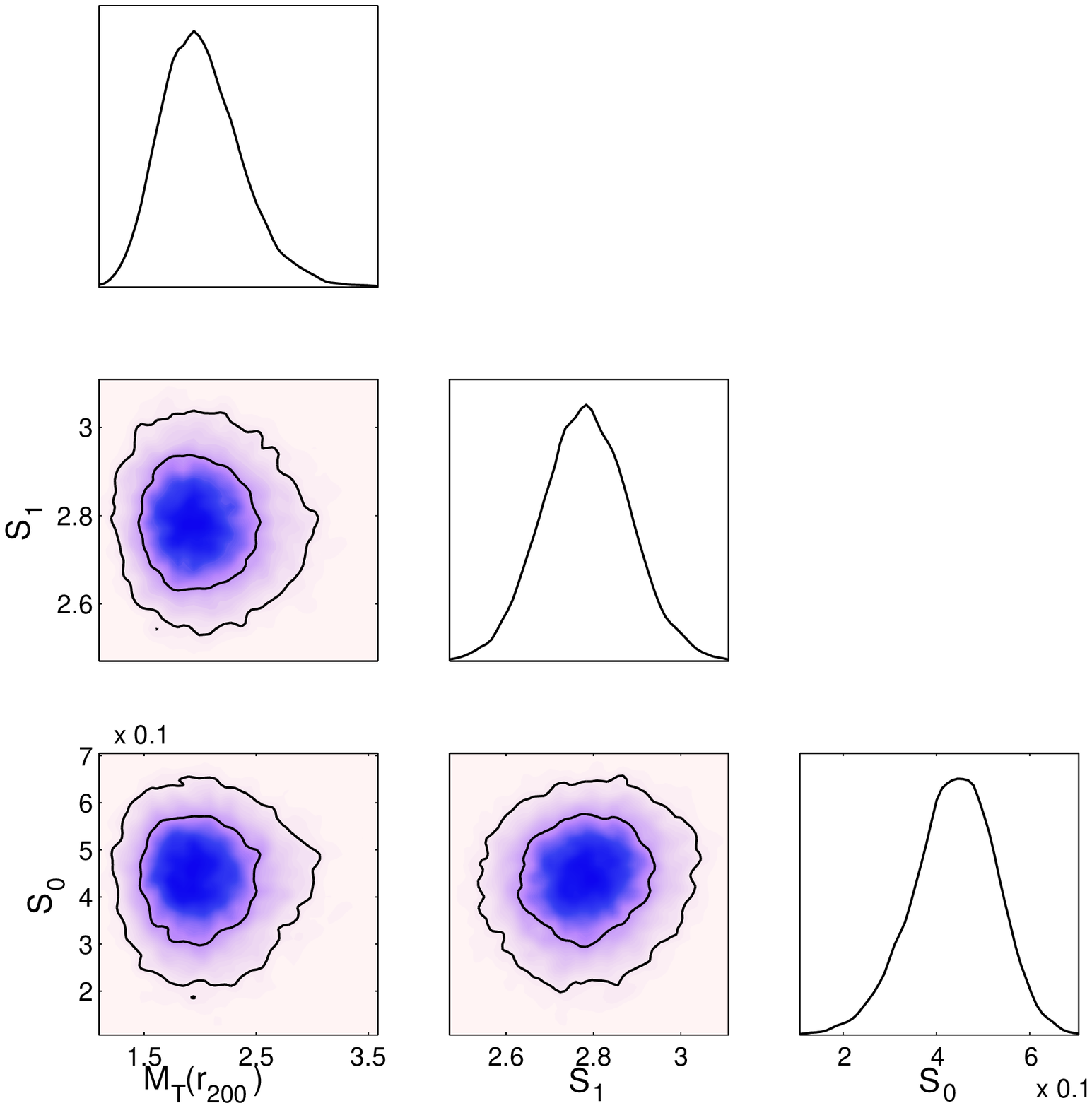}
\end{tabular}
\caption{Results for Abell~990. Panels A and B show the SA map before and after source-subtraction, respectively; a $0.6$\,k$\lambda$ taper has been applied to B. The box in panels A and B indicates the cluster SZ centroid, for the other symbols see Tab. \ref{tab:sourcelabel}. The smoothed {\emph{ROSAT HRI}} X-ray map overlaid with contours from B is shown in image C. Panels D and E show the marginalized posterior distributions for the cluster sampling and derived parameters, respectively. F shows the 1 and 2-D marginalized posterior distributions for source flux densities in Jy (see Tab. \ref{tab:source_info}) and $M_{\rm{T}}(r_{200})$ (in $\times 10^{14}M_{\odot}$). In panel D $M_{\rm{T}}$ is given in units of $h_{100}^{-1}\times10^{14}M_{\odot}$ and $f_{\rm{g}}$ in $h_{100}^{-1}$; both parameters are estimated within $r_{200}$. In E $M_{\rm{g}}$ is in units of $h_{100}^{-2}M_{\odot}$, $r$ in $h_{100}^{-1}$Mpc and $T$ in KeV. }

\label{fig:A990}
\end{center}

\end{figure*}

\subsection{Abell~1413}
\label{sec:A1413}

In Fig. \ref{fig:A1413} we present results for Abell~1413.
It can be seen from Figs. \ref{fig:A1413} A and B 
 that there are two of sources on the decrement
with flux densities of $0.47$ and $3.1$\,mJy (in Tab. \ref{tab:source_info}). The brightest source in
our LA maps has a flux density of 14\,mJy
but, since it is 700$''$ from the cluster X-ray centre, it does not contaminate
our SZ signal.
Some residual flux is seen on the source-subtracted SA maps; the strongest
residuals
 are not associated with sources in the LA data,
suggesting they could be extended emission resolved out from the LA maps.
\cite{govoni2009} find tentative ($\approx 3\sigma$)
evidence in FIRST data at 1.4\,GHz for a weak mini halo  with a luminosity of
$1.0\times 10^{23}$\,W Hz$^{-1}$.
 The peak signal from this mini halo is offset to the East with respect to the
central cD
galaxy, similarly to our SZ peak, which is slightly offset to the SE of the
X-ray centroid. Abell~1413 does seem to be a relaxed cluster; this is
supported by the
smooth X-ray distribution, the good agreement between the X-ray and SZ
centroids, the circular appearance of the projected SZ signal and the presence
 of a cool core \citep{allen1998}.
 We therefore expect our model to provide a good fit to the AMI data towards this cluster.

 Abell~1413 has been observed in the X-ray by {\sc{XMM-Newton}} (e.g.,
\citealt{XMM-A1413}), {\sc{Chandra}}
(e.g., \citealt{Chandra-A1413} and \citealt{bona_chandra}) and most recently by the {\sc{Suzaku}}
 satellite \citep{hoshino2010}; SZ images have been made with
the Ryle Telescope at 15\,GHz
(\citealt{RT_A1413}) and with OVRO/BIMA at 30\,GHz (LaRoque et al. and
\citealt{bona_chandra}). These analyses
 indicate that Abell~1413 seems indeed to be a relaxed cluster with no evidence of recent
merging. Different temperature and
density profiles obtained from X-ray data are in good agreement out
to half the virial radius. Hoshino et al. measure the variation of
temperature with radius, finding a
 temperature of 7.5\,keV near the centre and of 3.5\,keV at $r_{200}$; they
assume spherical symmetry, an NFW density profile
 and hydrostatic equilibrium to calculate $M_{\rm{T}}(r_{200})$ =
6.6$\pm$2.3$\times 10^{14}$\,$h_{70}^{-1}\rm{M_{\odot}}$.
 Zhang et al. use
{\sc{XMM-Newton}} and find
$M_{\rm{T}}(r_{500})$ = 5.4 $\pm$1.6 $\times 10^{14}\rm{M_{\odot}}$
 ; they assume isothermality,
 spherical symmetry and $h_{70}=1$.
We determine $M_{\rm{T}}(r_{200})$ to be $5.7\pm 1.4\times 10^{14}$\,$\rm{M_{\odot}}$ for $h_{70}=1$.


\begin{figure*}
\centerline{\huge{Abell~1413}}
\centerline{{A}\includegraphics[width=7.5cm,height=7.5cm,clip=,angle=0.]{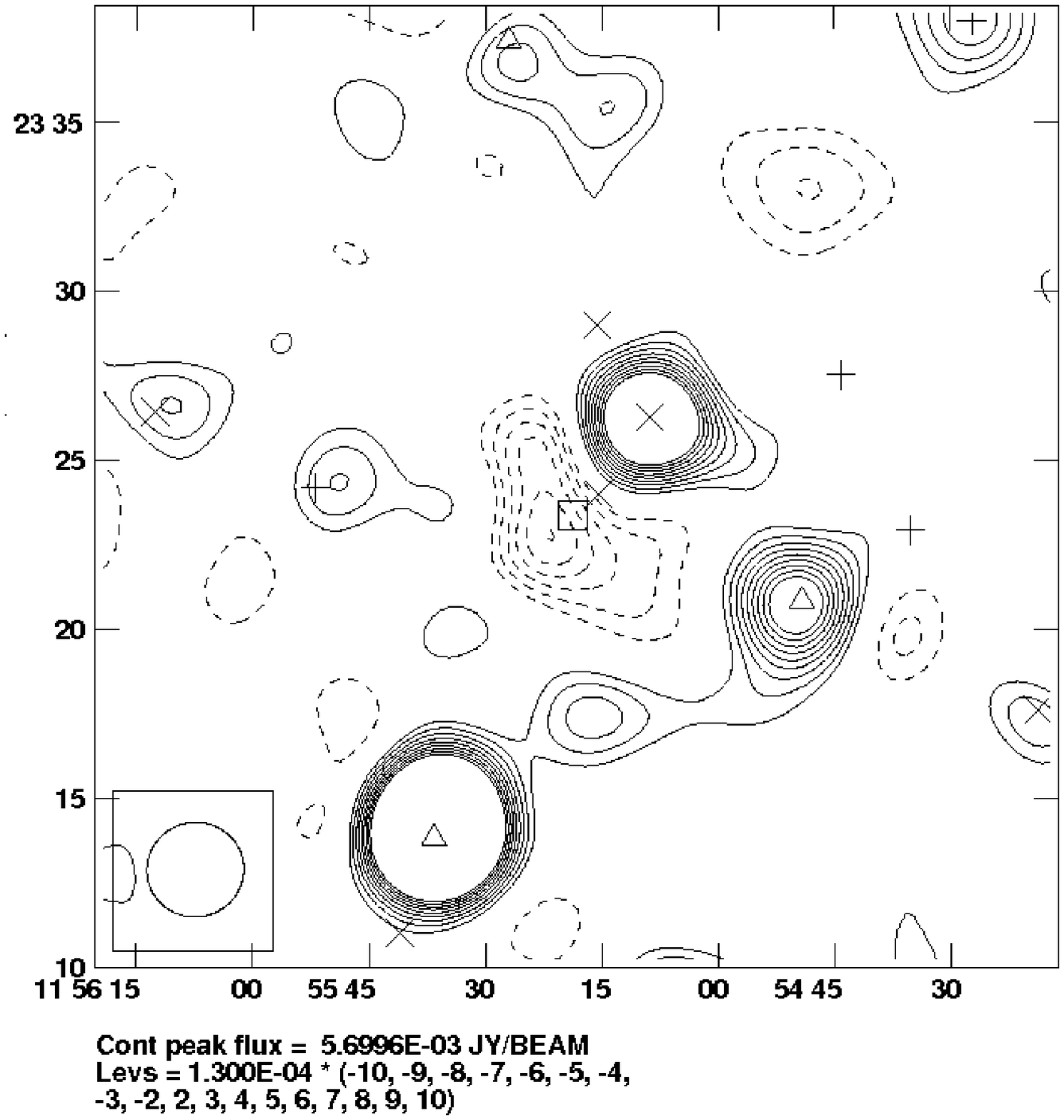}\qquad{D}\includegraphics[width=7.5cm,height=7.5cm,clip=,angle=0.]{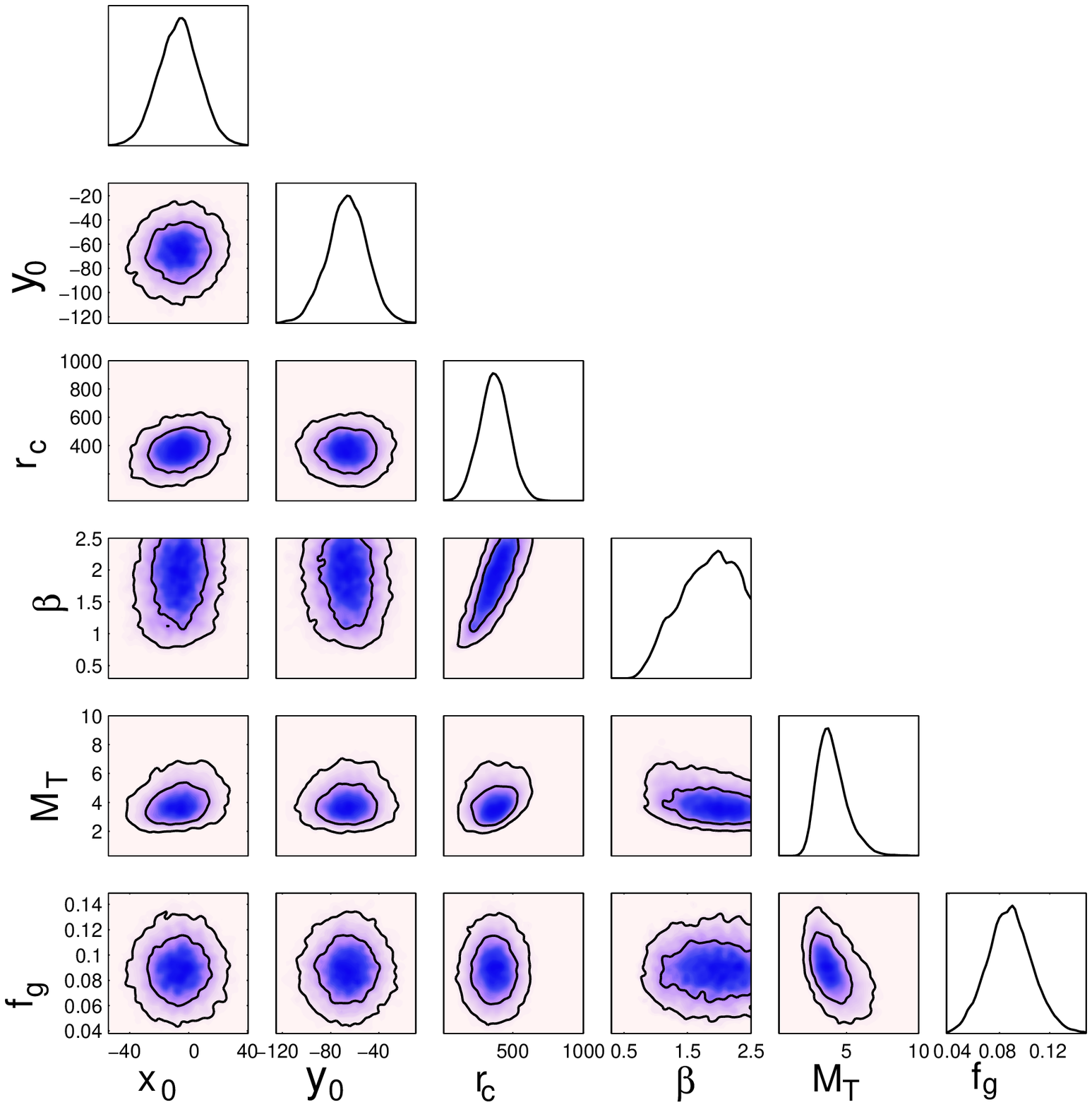}} 
\centerline{{B}\includegraphics[width=7.5cm,height=7.5cm,clip=,angle=0.]{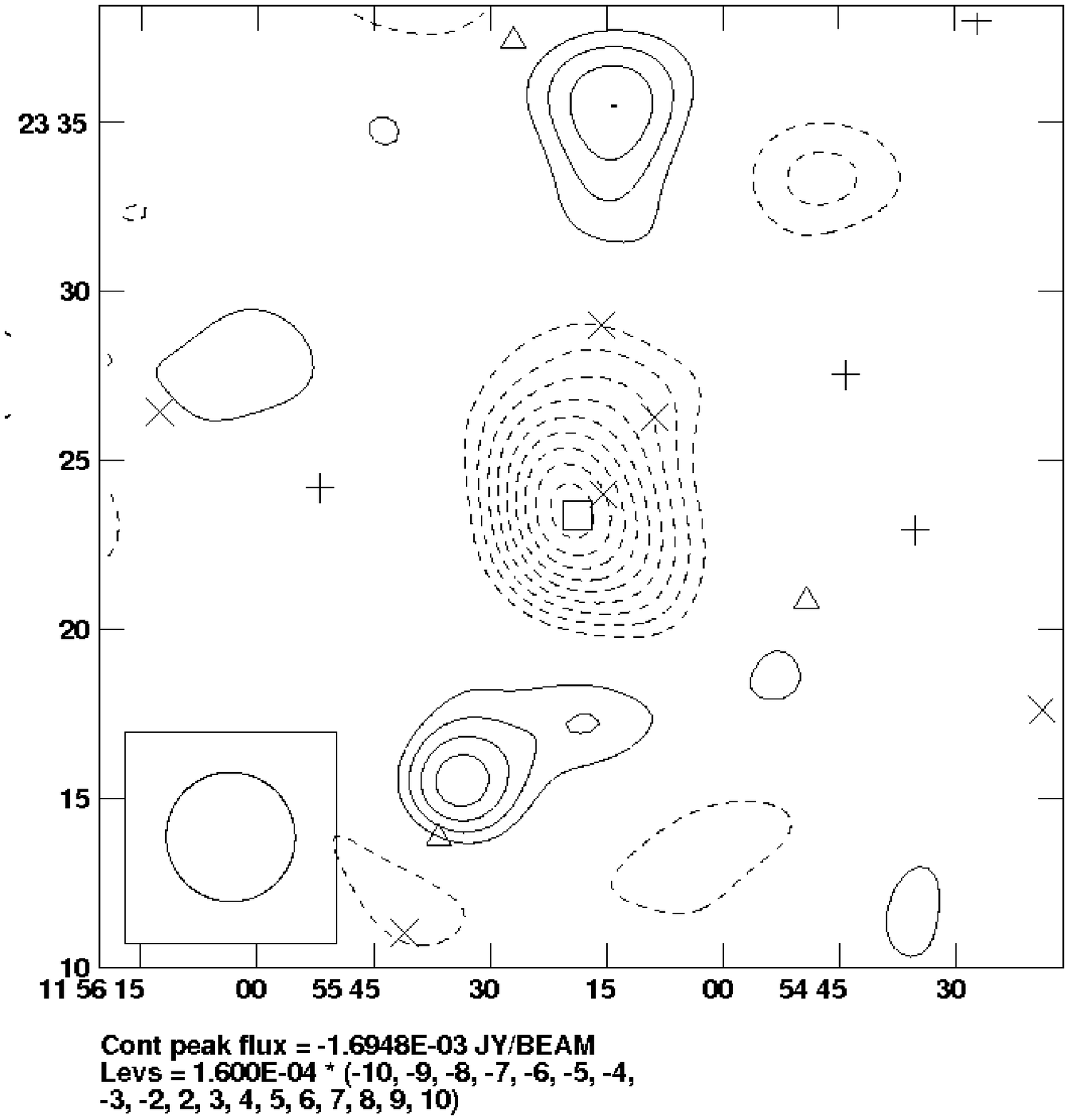}\qquad{E}\includegraphics[width=7.5cm,height=7.5cm,clip=,angle=0.]{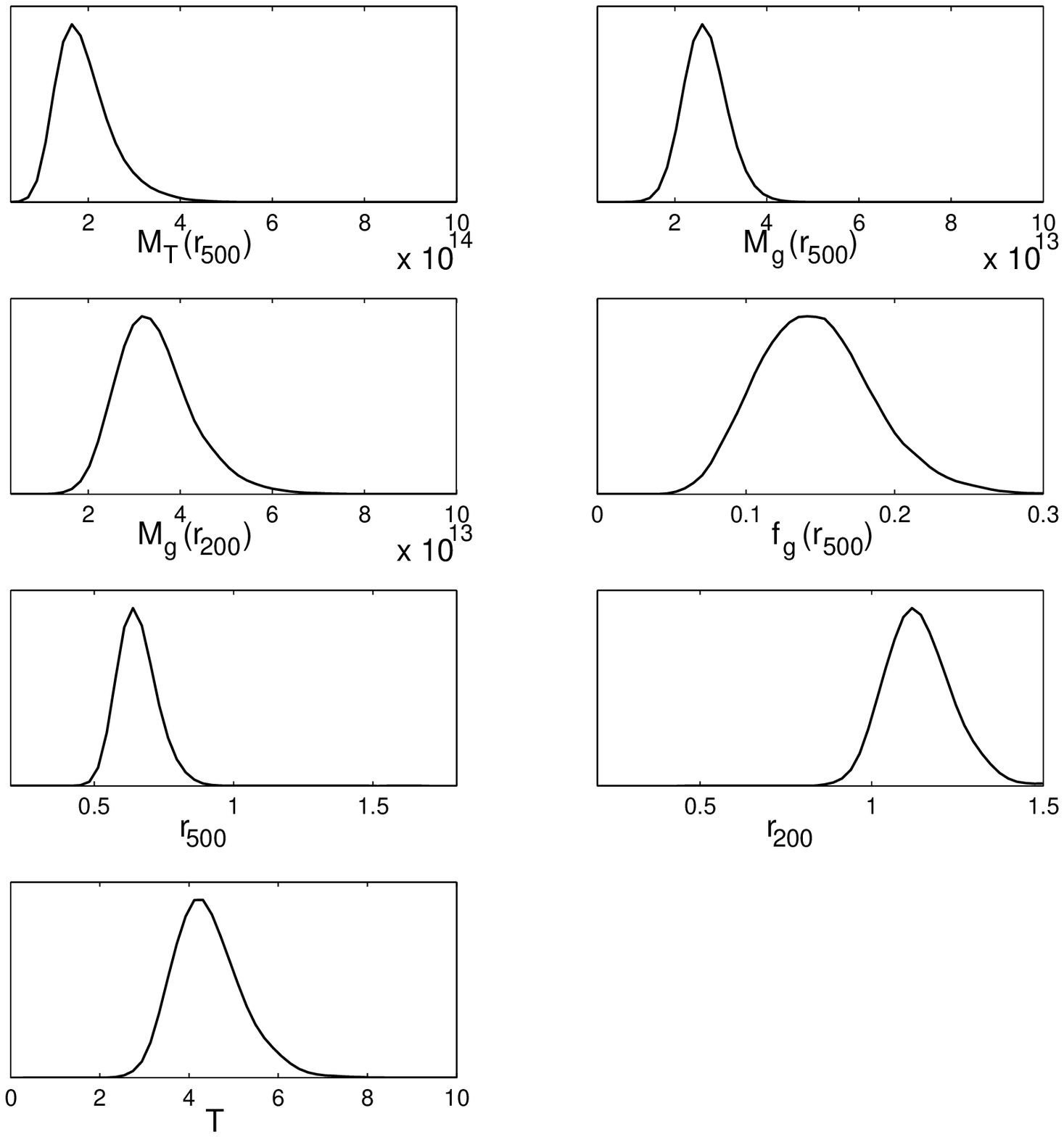}}
 \centerline{{C}\includegraphics[width=7.5cm,height=6.5cm,clip=,angle=0.]{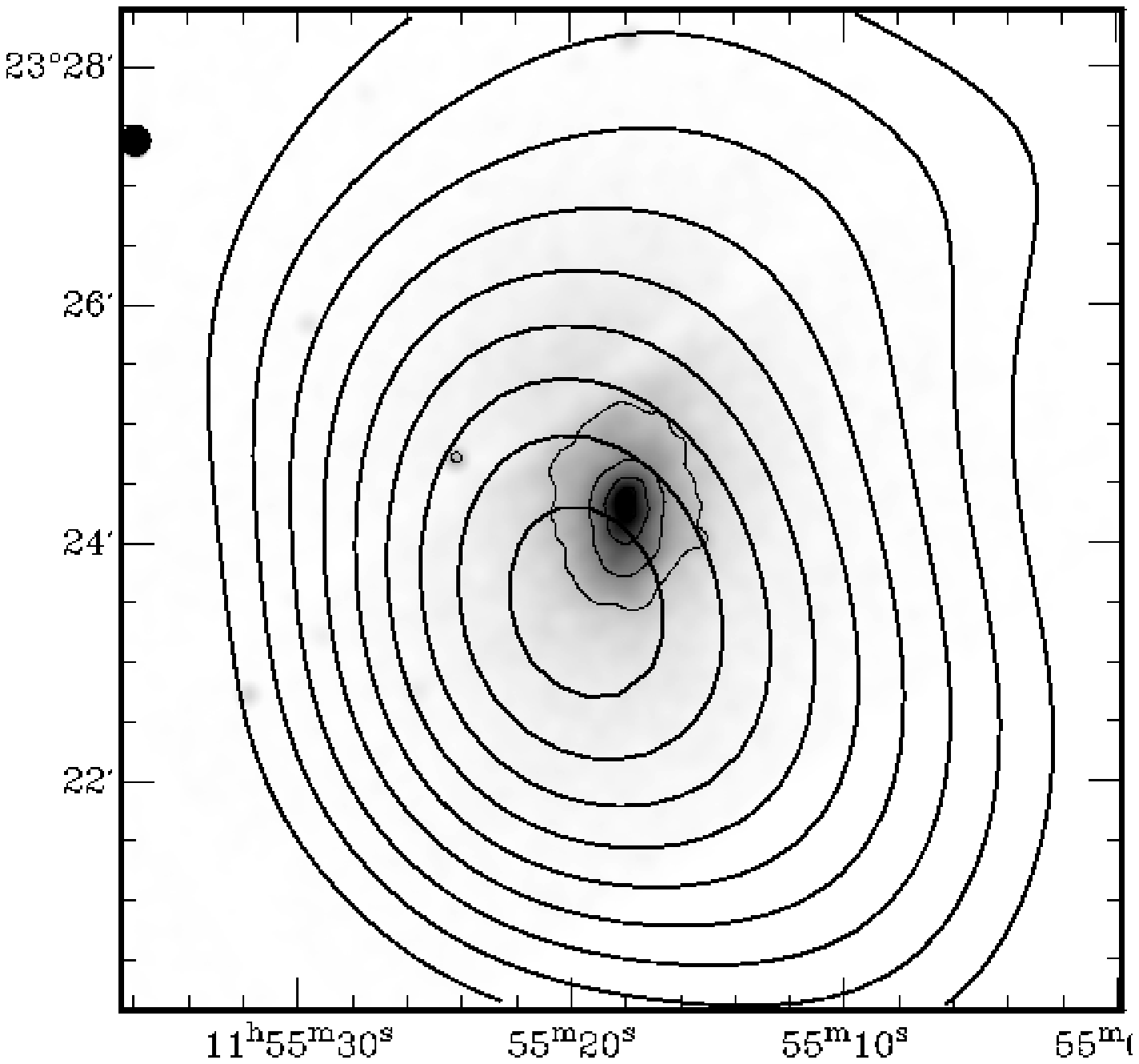}\qquad{F}\includegraphics[width=6.5cm,height=6.5cm,clip=,angle=0.]{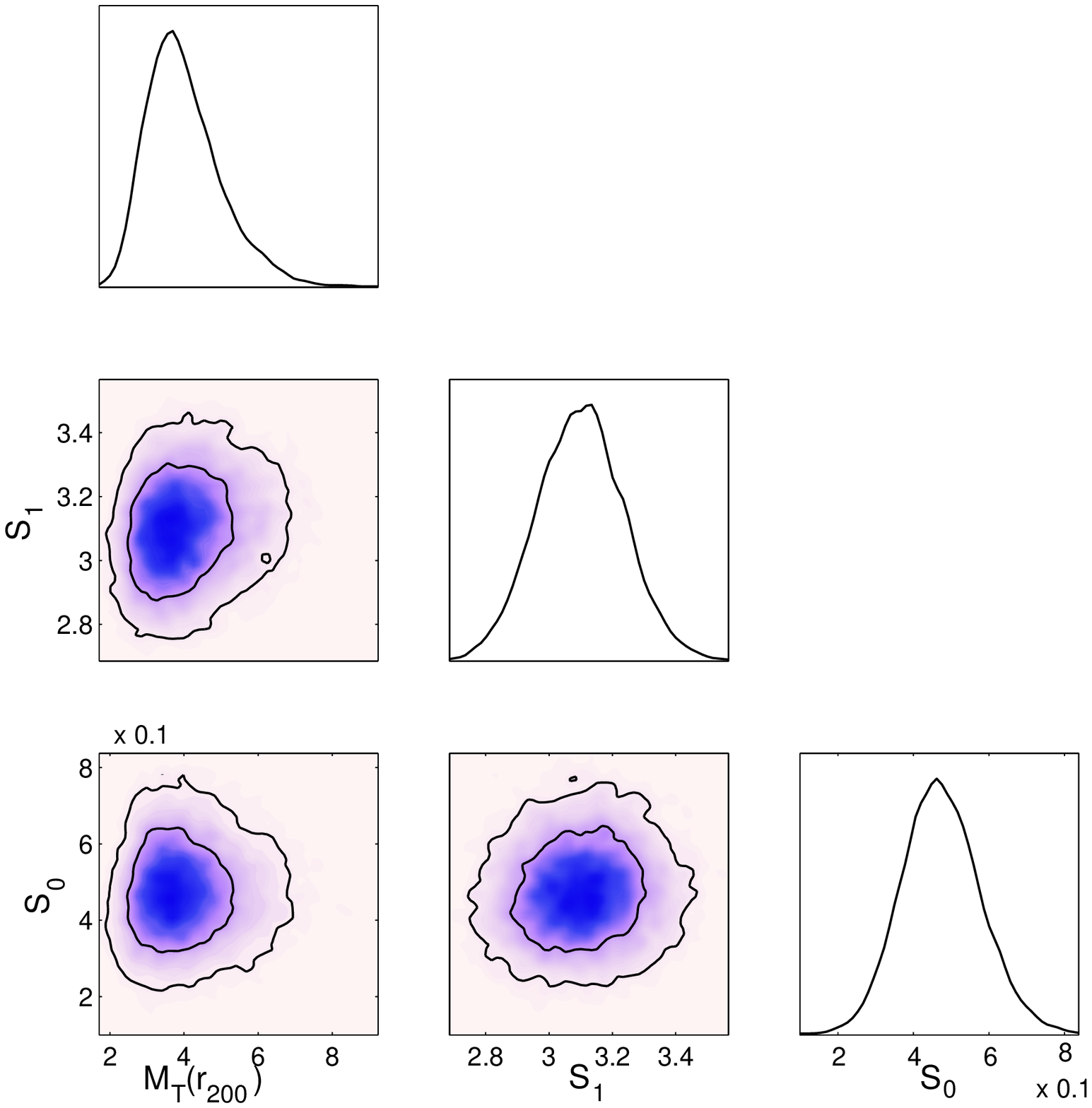}}

\caption{Results for Abell~1413. Panels A and B show the SA map before and after source-subtraction, respectively; a $0.6$\,k$\lambda$ taper has been applied to B. The box in panels A and B indicates the cluster SZ centroid, for the other symbols see Tab. \ref{tab:sourcelabel}. The smoothed {\sc{Chandra}} X-ray map overlaid with contours from B in presented in image C. Panels D and E show the marginalized posterior distributions for the cluster sampling and derived parameters, respectively. F shows the 1 and 2-D marginalized posterior distributions for source flux densities (in Jys) within $5\arcmin$ of the cluster SZ centroid (see Tab. \ref{tab:source_info}) and $M_{\rm{T}}(r_{200})$ (in $h_{100}^{-1}\times 10^{14}M_{\odot}$). In panel D $M_{\rm{T}}$ is given in units of $h_{100}^{-1}\times10^{14}M_{\odot}$ and $f_{\rm{g}}$ in $h_{100}^{-1}$; both parameters are estimated within $r_{200}$. In E $M_{\rm{g}}$ is in units of $h_{100}^{-2}M_{\odot}$, $r$ in $h_{100}^{-1}$Mpc and $T$ in KeV. }
\label{fig:A1413}
\end{figure*}

\subsection{ Abell~1423} \label{resultsA1423}

Results for Abell~1423 are shown in Fig. \ref{fig:A1423}.
The source environment for Abell~1423 is challenging (see Fig. \ref{fig:A1423} A)
-- 23 sources have been detected
within 10$\arcmin$ of the X-ray cluster centroid, of which 4 lie on the
decrement, as seen from the source-subtracted map. We find
no evidence for extended emission, in agreement with the lack of
diffuse emission towards this cluster at 327\,MHz reported by
\cite{rudnick2009} and
the results in \cite{rossetti2011}.
The sources closest to the cluster all have flux densities $<
1.3$\,mJy (Tab. \ref{tab:source_info}) and only small positive residuals remain after
source subtraction. As shown in Fig. \ref{fig:A1423} F, the flux densities for
some of the sources close to the cluster centroid
manifest degeneracies with the cluster mass. 

The details on the dynamics of Abell~1423 are largely unknown. The lack of
strong radio halo emission is indicative of a system without very significant
 dynamical activity \citep{buote2001}, as is the
good agreement between the X-ray and SZ emission peak positions.
 On the other hand, the X-ray data in Fig.  \ref{fig:A1423} C shows signs of
substructure and our SZ image is be elongated along the SE--NW direction. \cite{sanderson2009} find that the
logarithmic gradient for the gas density profile of Abell~1423
 at $0.04r_{500}$ is  $\alpha \approx -0.98$ --
 a key signature of cooling core clusters (\cite{vikhlinin2006} suggest
$\alpha<-0.7$ for strong cooling flows). In their study clusters with small
 offsets at $r_{500}$ between the X-ray and the Brightest Cluster Galaxy (BCG)
are tightly correlated with large, negative spectral indices, an indication
that the strength of cooling cores tends to drop in more disturbed systems, but
Abell~1423 is an unsual outlier in this trend with a small offset
and a steep $\alpha$.

\subsection{ Abell~1704 } \label{resultsA1704}

Abell~1704 has been observed with \emph{ROSAT HRI} and
\emph{PSPC} \citep{rizza1998}. These observations show a shift in
position
between the peak emission and the cluster centroid and distinct signs of
elongations in the gas distribution. Further analysis of X-ray observations
suggest the presence of a cooling flow \citep{allen1998}.
\cite{carlstrom1996} attempted to
detect an SZ effect using the OVRO array at 30\,GHz towards this cluster but found no 
 convincing SZ signal.

The NVSS map at $1.4$\,GHz shows complex, extended emission (Fig.
\ref{fig:A1704}). These
features are detected on our SA maps but a significant portion of the
emission is resolved out on our LA maps (see Tab. \ref{tab:source_info} for
more details on these sources).
 Our model is not sophisticated enough to deal
properly with extended structure and significant residual emission can
be seen in the source-subtracted SA map, Fig. \ref{fig:A1704}. Consequently, we
are not able
to convincingly detect an SZ effect towards Abell~1704.

\begin{figure*}
\begin{center}
\begin{tabular}{m{8cm}cm{8cm}}
\multicolumn{3}{c}{\huge{Abell 1423}}\\
{A}\includegraphics[width=7.5cm,height=7.5cm,clip=,angle=0.]{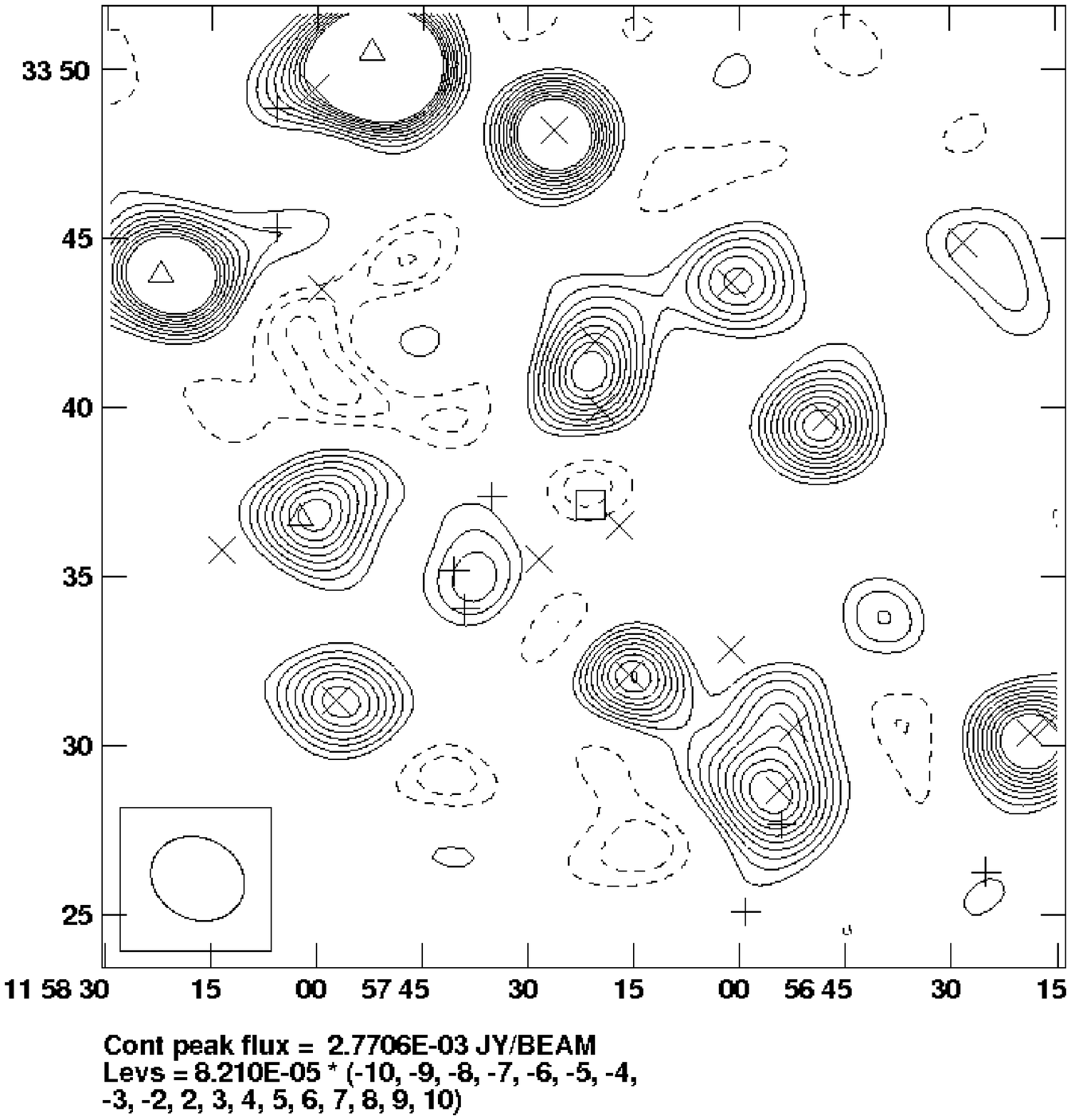}
& \quad &
{D}\includegraphics[width=7.5cm,height=7.5cm,clip=,angle=0.]{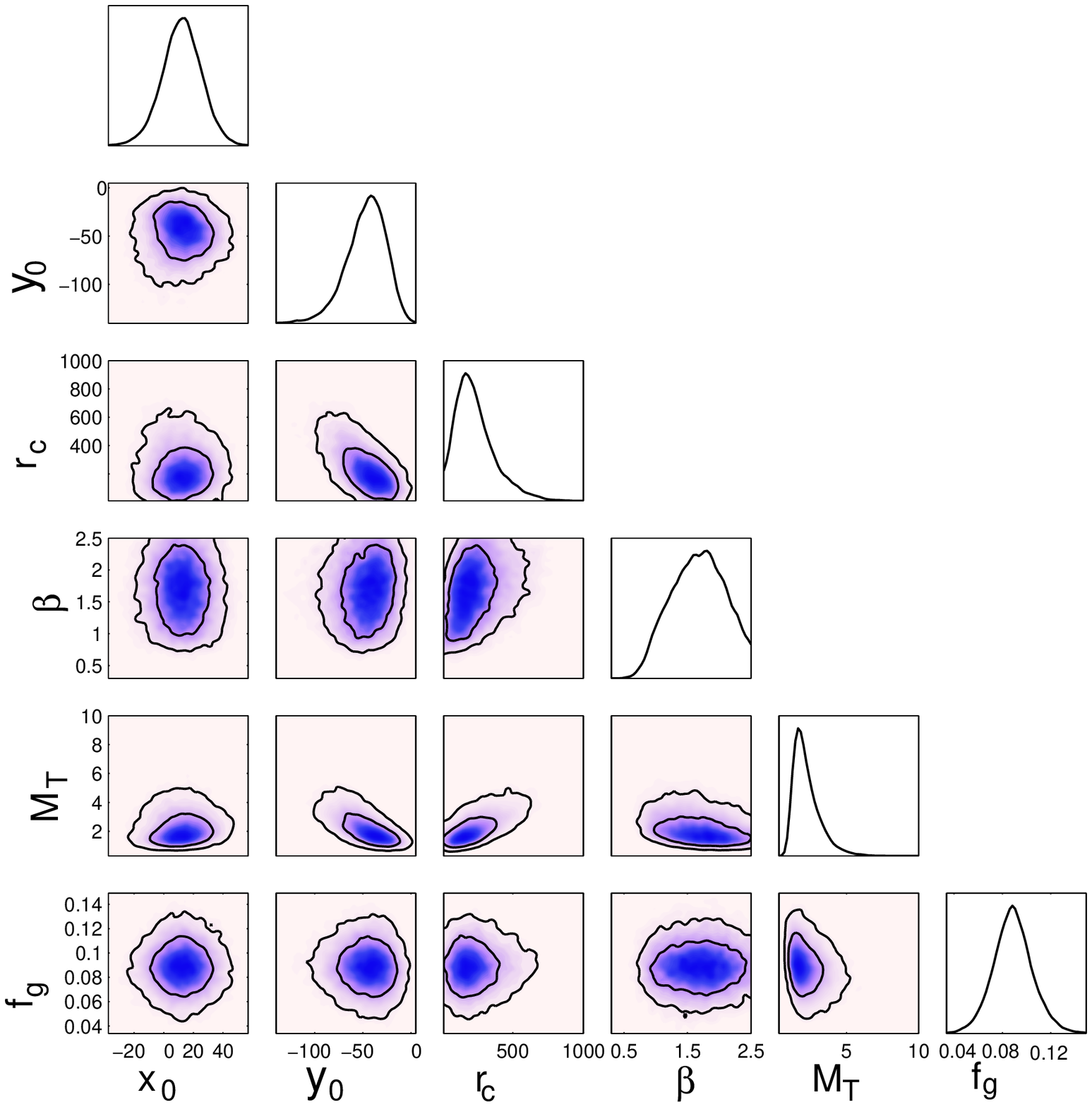}\\
{B}\includegraphics[width=7.5cm,height=7.5cm,clip=,angle=0.]{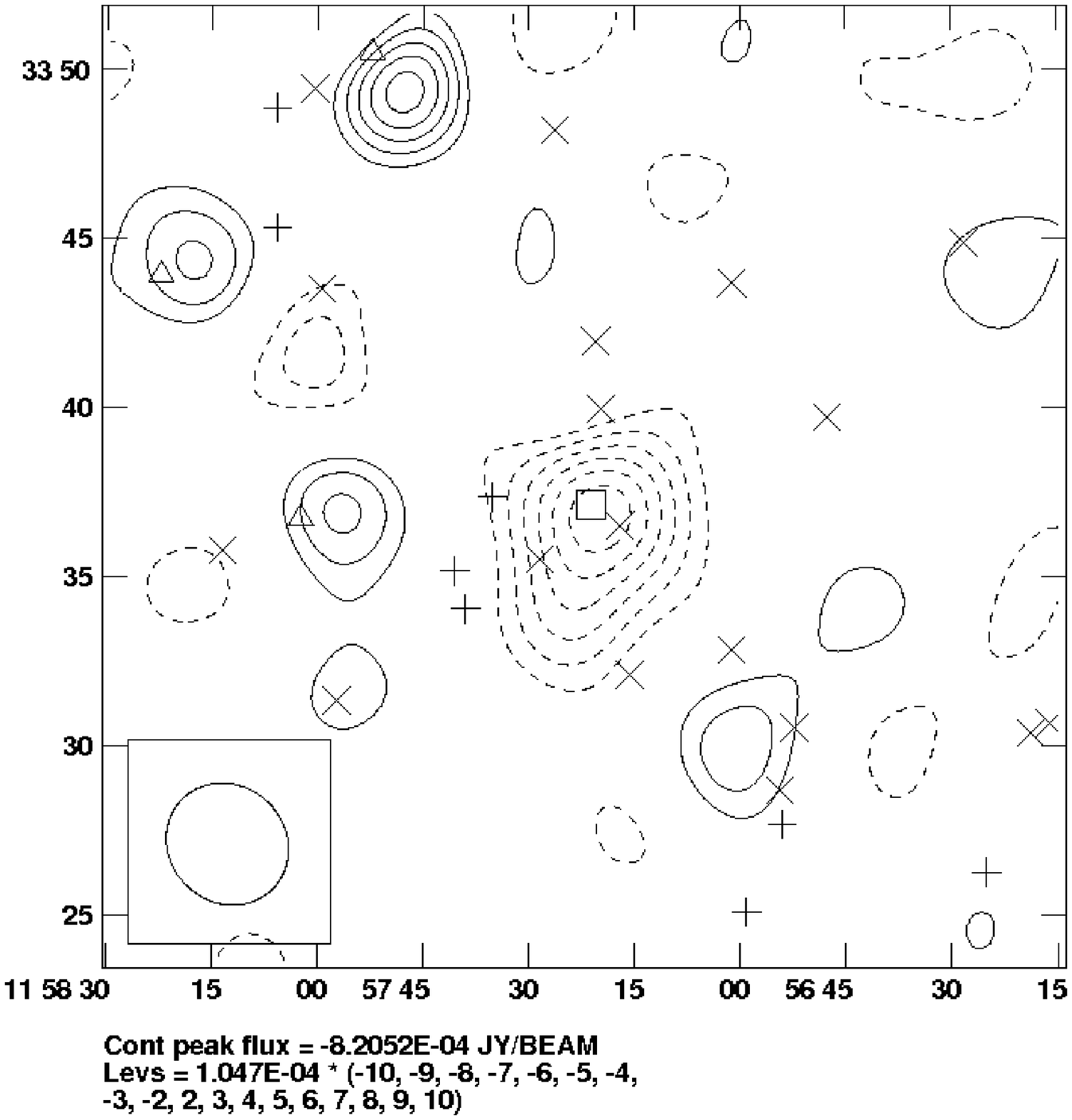}
& \quad &
{E}\includegraphics[width=7.5cm,height=7.5cm,clip=,angle=0.]{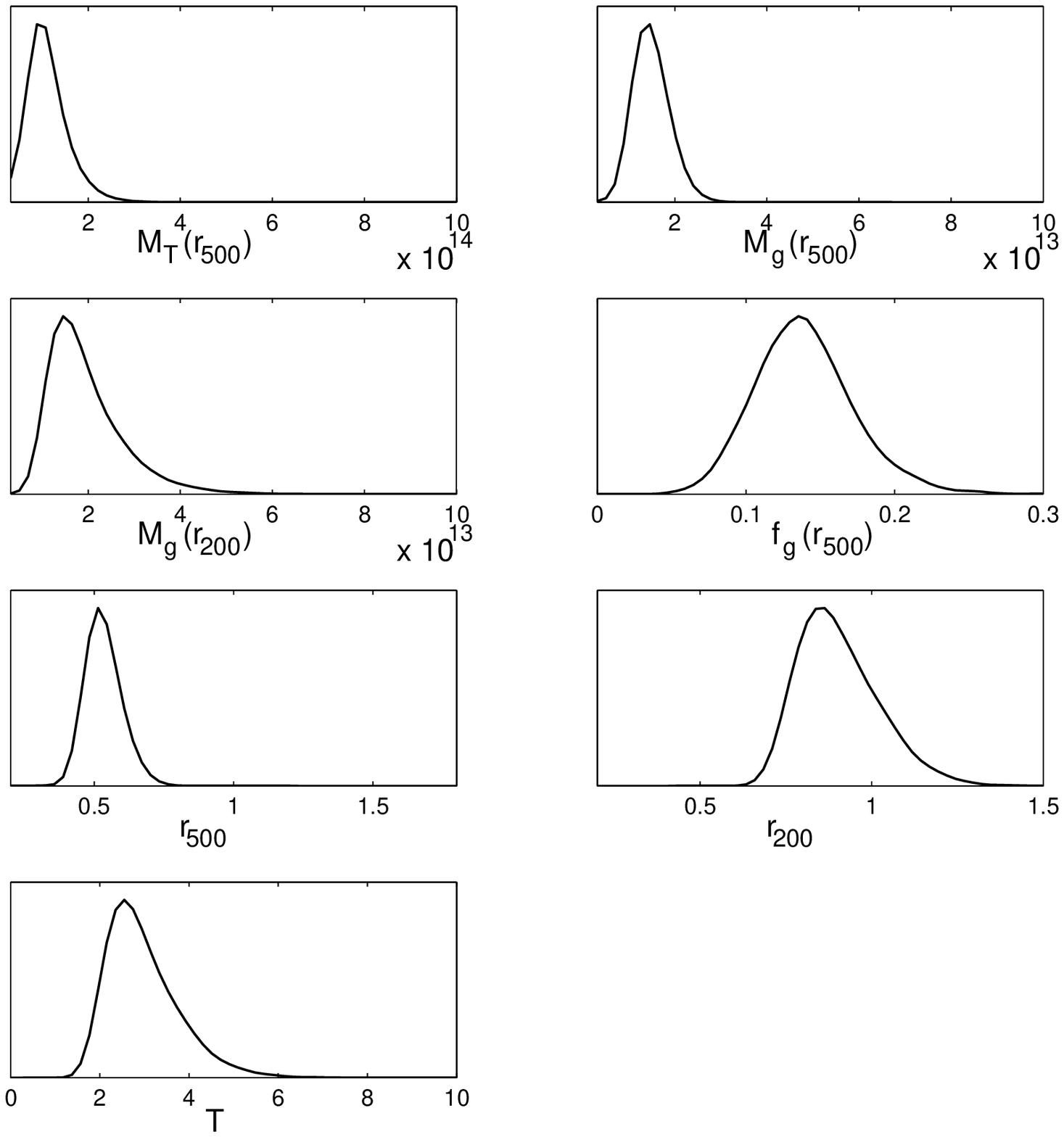}
\\
{C}\includegraphics[width=7.0cm,height=6.5cm,clip=,angle=0.]{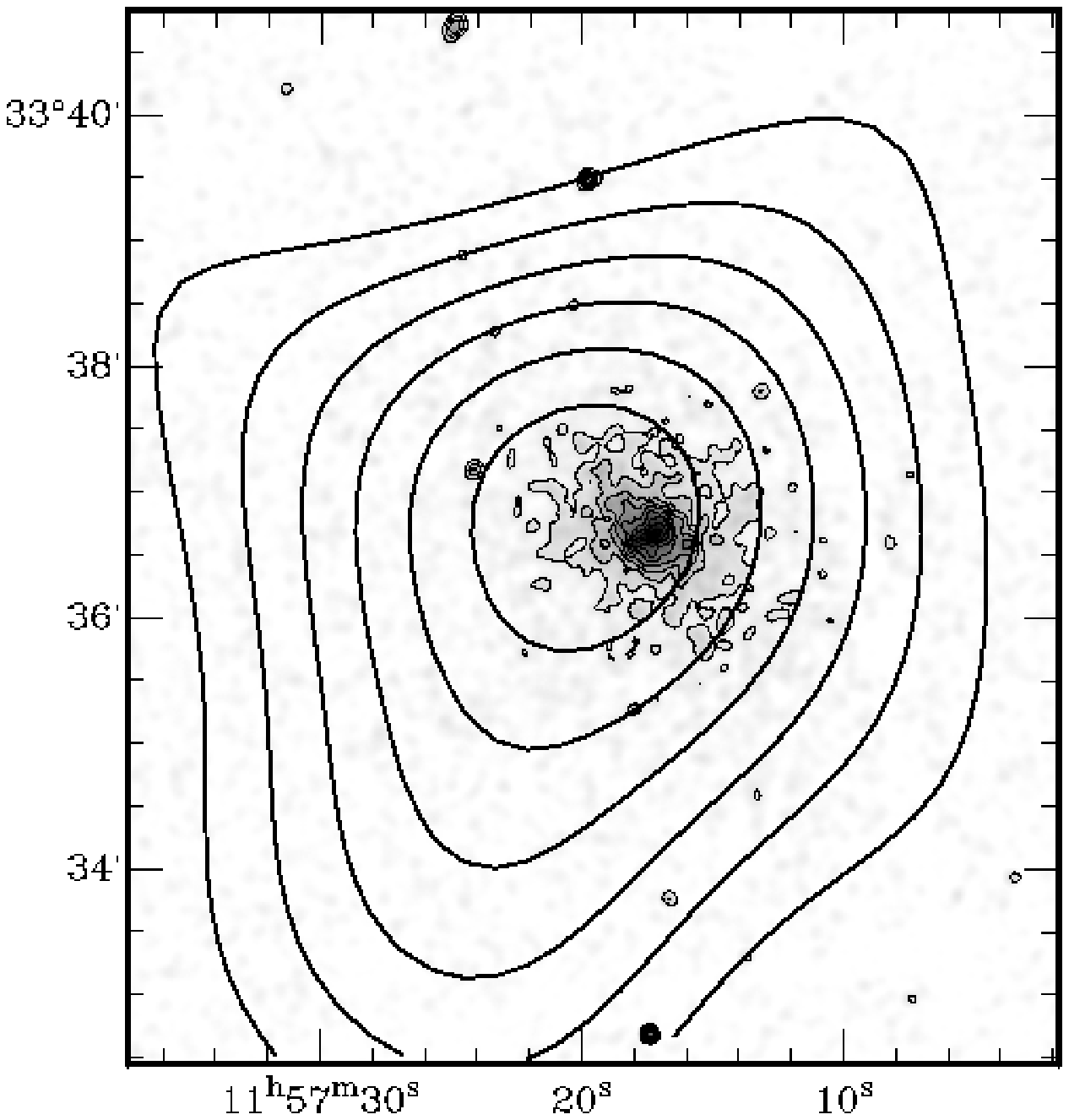}
& \quad &
{F}\includegraphics[width=7.0cm,height=6.5cm,clip=,angle=0.]{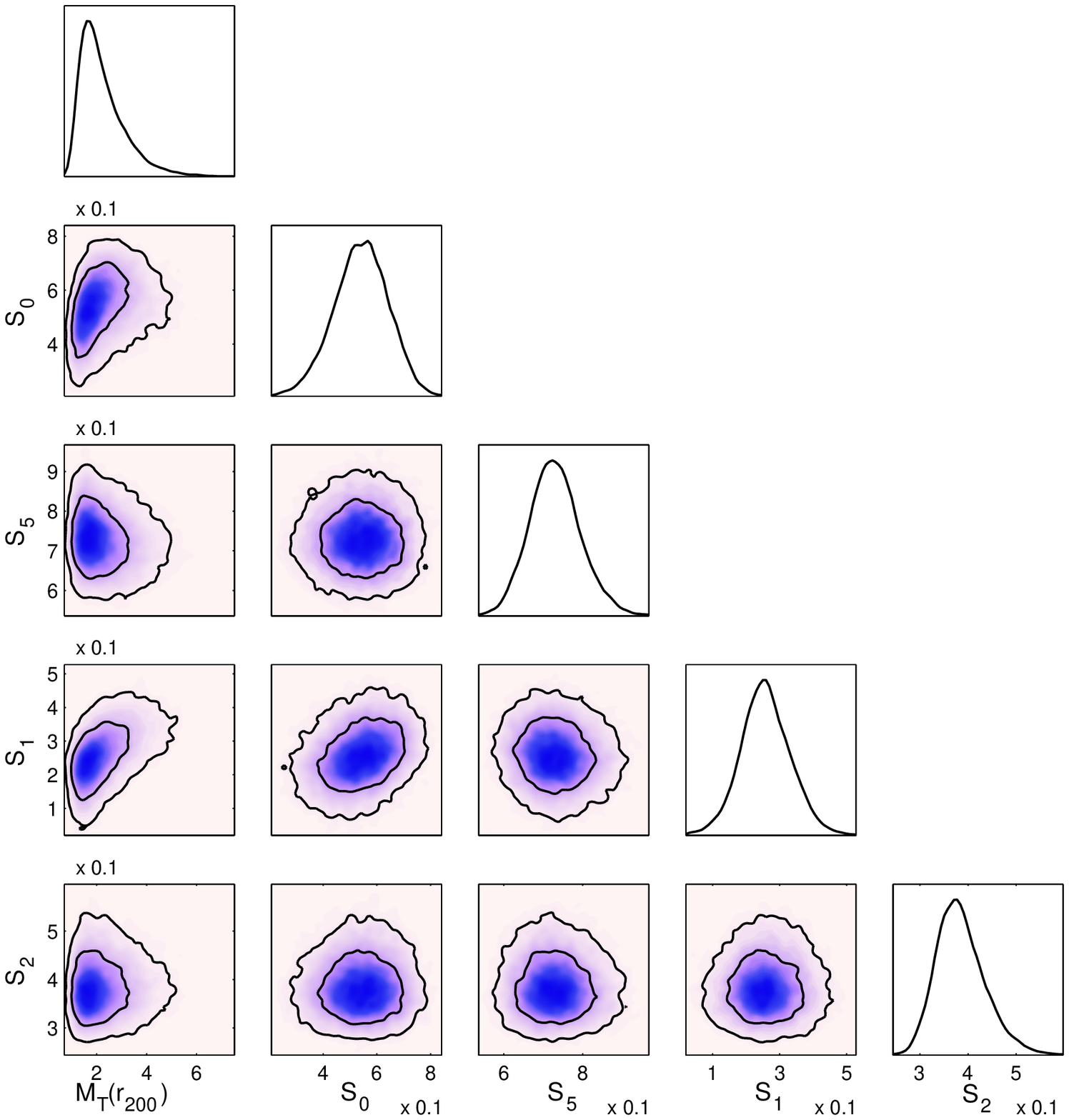}
\end{tabular}
\caption{Results for Abell~1423. Panels A and B show the SA map before and after surce-subtraction, respectively; a $0.6$\,k$\lambda$ taper has been applied to B. The box in panels A and B indicates the cluster SZ centroid, for the other symbols see Tab. \ref{tab:sourcelabel}. The smoothed {\sc{Chandra}} X-ray map overlaid with contours from B is presented in image C. Panels D and E show the marginalized posterior distributions for the cluster sampling and derived parameters, respectively. F shows the 1 and 2-D marginalized posterior distributions for source flux densities (in Jys) within $5\arcmin$ of the cluster SZ centroid (see Tab. \ref{tab:source_info}) and $M_{\rm{T}}(r_{200})$ (in $h_{100}^{-1}\times 10^{14}M_{\odot}$). In panel D $M_{\rm{T}}$ is given in units of $h_{100}^{-1}\times10^{14}M_{\odot}$ and $f_{\rm{g}}$ in $h_{100}^{-1}$; both parameters are estimated within $r_{200}$. In E $M_{\rm{g}}$ is in units of $h_{100}^{-2}M_{\odot}$, $r$ in $h_{100}^{-1}$Mpc and $T$ in KeV.}
\label{fig:A1423}
\end{center}
\end{figure*}

\begin{figure*}
\begin{center}
\begin{tabular}{m{8cm}cm{8cm}}
\multicolumn{1}{c}{\huge{Abell 1704}}\\
{A}\includegraphics[height=7.0cm,clip=,angle=0.]{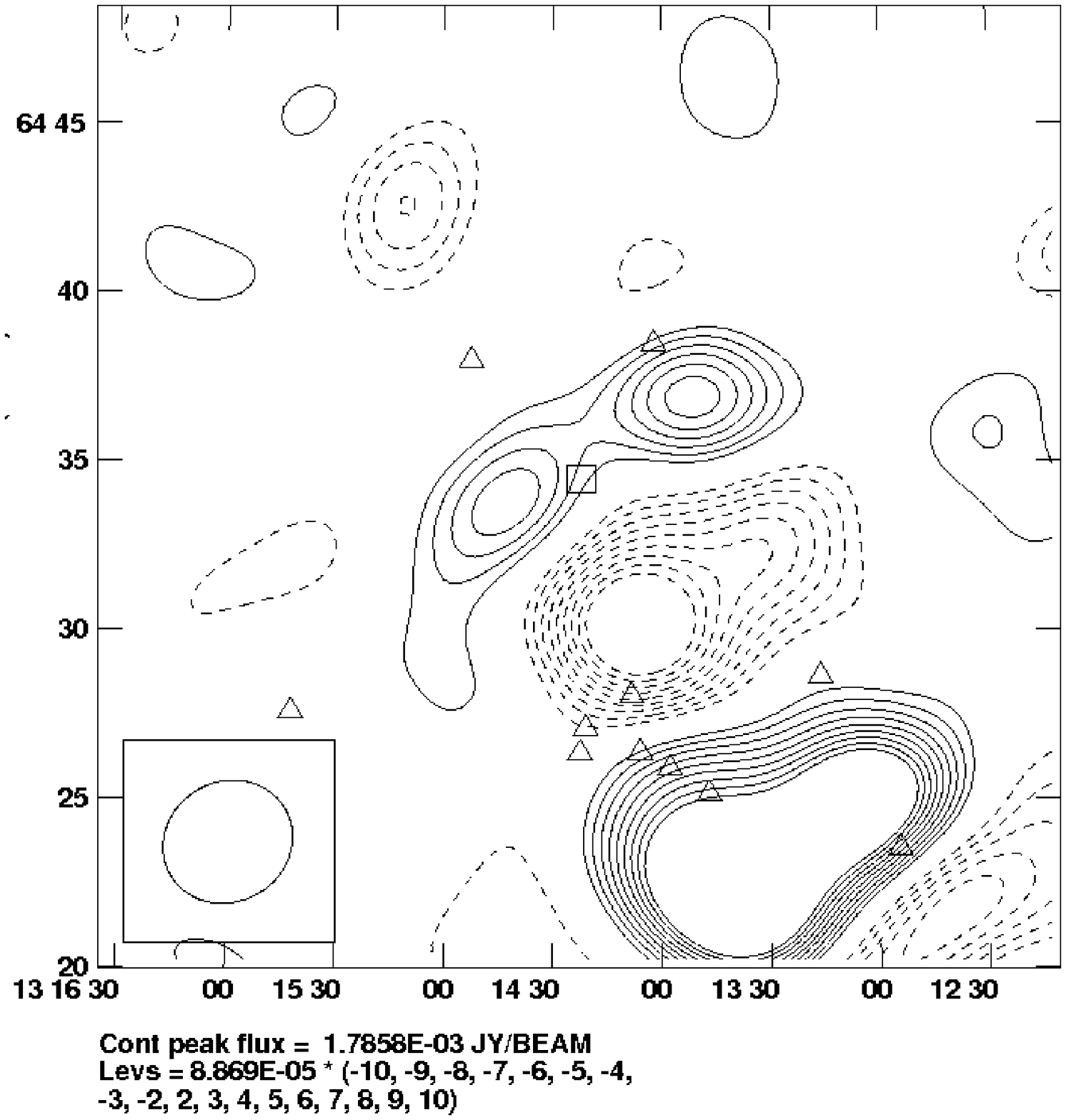}\\
{B}\includegraphics[height=7.0cm,clip=,angle=270.]{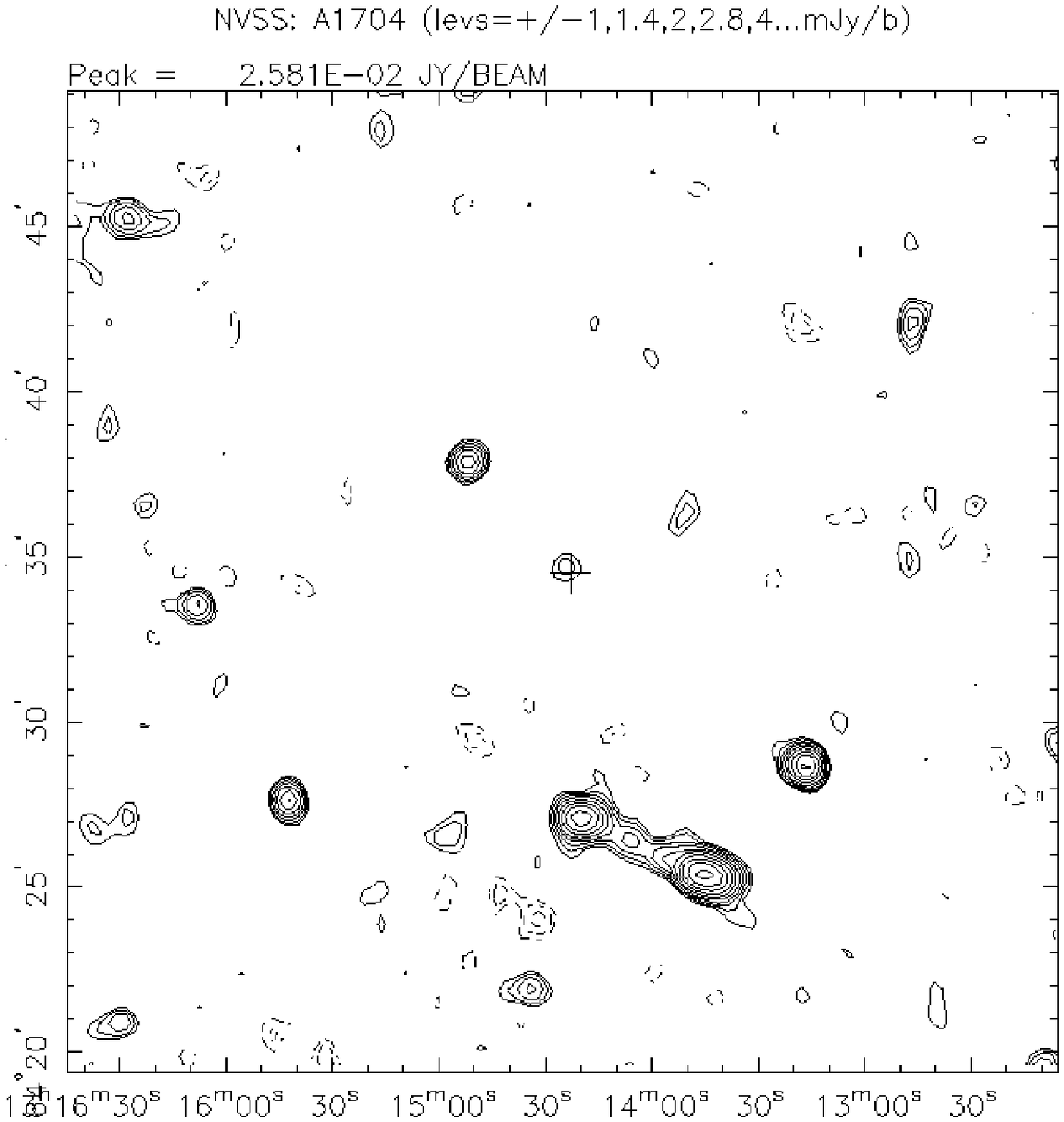}
\\
\end{tabular}
\caption{A: source-subtracted SA map produced using a
  0.6-k$\lambda$ taper. The contours increase linearly in units of
  $\sigma_{SA}$. B: 1.4-GHz NVSS map towards Abell 1704.}
\label{fig:A1704}
\end{center}
\end{figure*}

\subsection{Abell~1758}

Results for Abell1758a and b are given in Fig. \ref{fig:A1758A}-Fig \ref{fig:A1758A_2}.
It is clear from Fig. \ref{fig:A1758A} A, B and C that Abell~1758 is a complex system
comprising two gravitationally-bound main clusters,
 Abell~1758a and Abell~1758b, separated by $8\arcmin$ (\citealt{rizza1998} and
\citealt{2004ApJ...613..831D}).
 David \& Kempner find no conclusive evidence for interaction between
these two main clusters, yet each of them is undergoing major mergers --
Abell~1758a between two 7-keV clusters and Abell~1758b between two 5-keV
clusters; since
both sets of mergers are between clusters of approximately equal mass, provided
each of the primary clusters was virialized pre-merger,
we might expect the average temperature to be higher by some $25\%$ when all the
gas mass of the subcluster has merged with that of the primary cluster.

To map the full extent of this system we took raster observations with the SA, which
 are presented in Fig. \ref{fig:A1758A} B and C.
From \ref{fig:A1758A} Cii it can be seen that the SZ signal follows the X-ray
emission but there seems to be a hint of an SZ signal connecting these two clusters; 
note that the clusters have identical redshifts. No connecting X-ray signal would be
 expected and indeed none is seen.
A recent analysis of Spitzer/MIPS 24$\mu$m data by \cite{LoCuSS_A1758}
classifies Abell~1758 as the most
active system they have observed at that wavelength. They also identify
numerous smaller mass peaks and filamentary structures, which
are likely to indicate the presence of infalling galaxy groups, in support of
the David \& Kempner observations.

 For Abell~1758a we obtain
 $M_{\rm{T}}(r_{500})$ = 2.5 $\pm 0.4\times 10^{14}h_{100}\rm{M_{\odot}}$
and $M_{\rm{T}}(r_{200})$ = $4.1^{0.7}_{0.8}\times 10^{14}h_{100}\rm{M_{\odot}}$.
Zhang et al. studied Abell~1758a using {\sc{XMM-Newton}} and found
$M_{\rm{T}}(r_{500}) = 1.1\pm 0.3\times 10^{15}\rm{M_{\odot}}$
; they assumed isothermality, spherical symmetry and $h_{70}=1$.

\begin{figure*}
\centerline{\huge{Abell 1758a}}
\centerline{{A}\includegraphics[width=7.5cm,height=7.5cm,clip=,angle=0.]{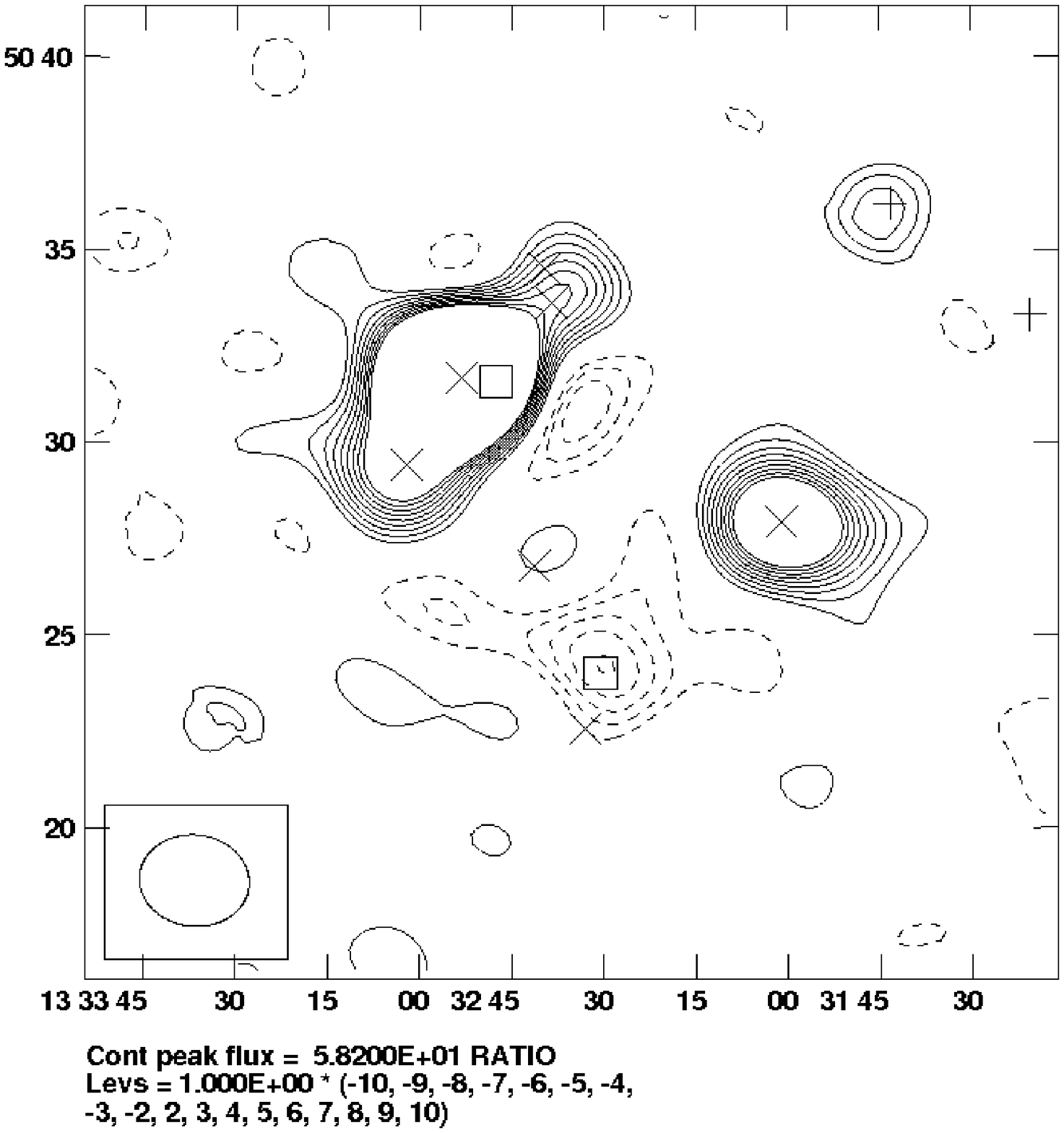}\qquad{D}\includegraphics[width=7.5cm,height=7.5cm,clip=,angle=0.]{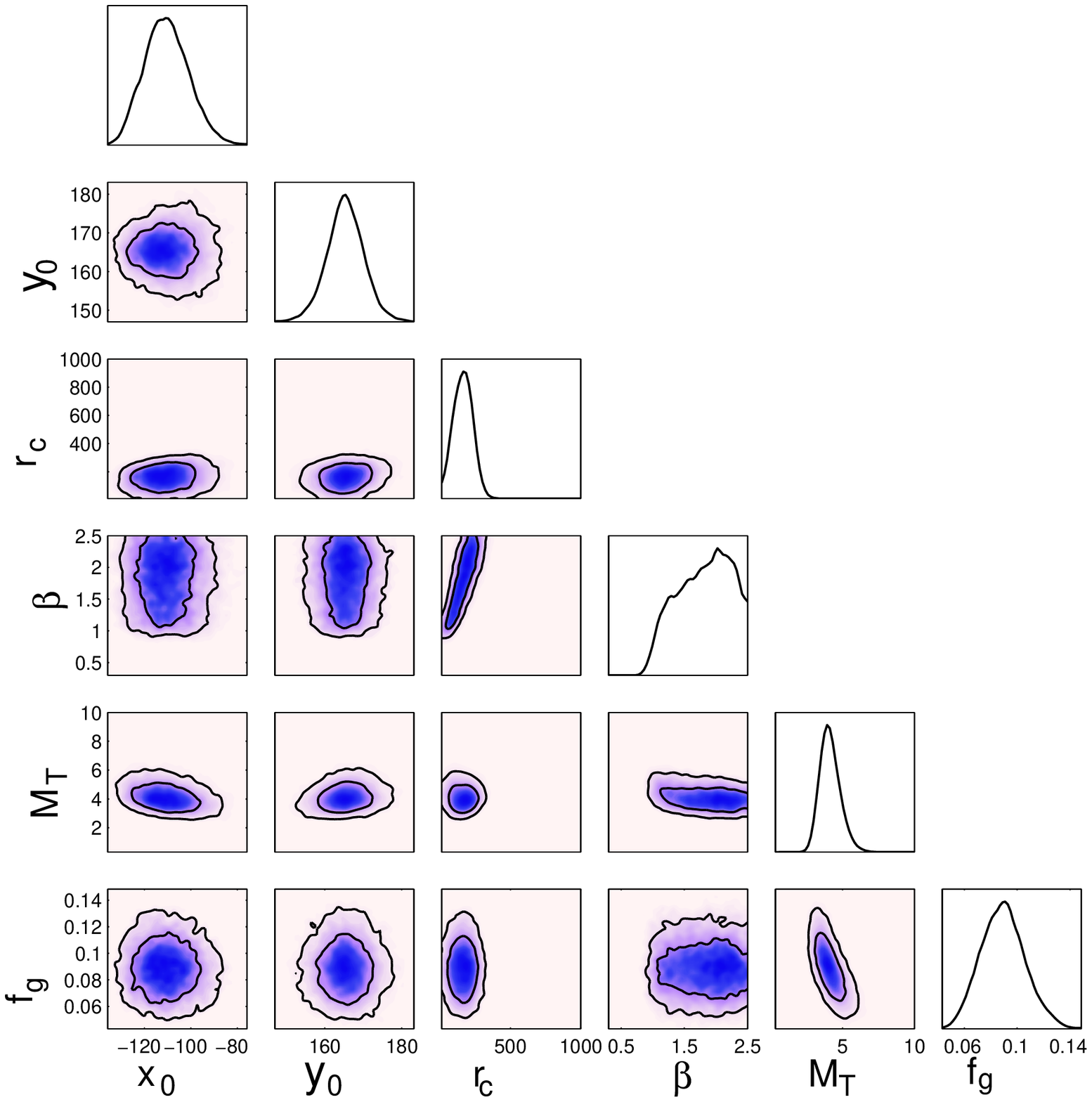}}
 \centerline{{B}\includegraphics[width=7.5cm,height=7.5cm,clip=,angle=0.]{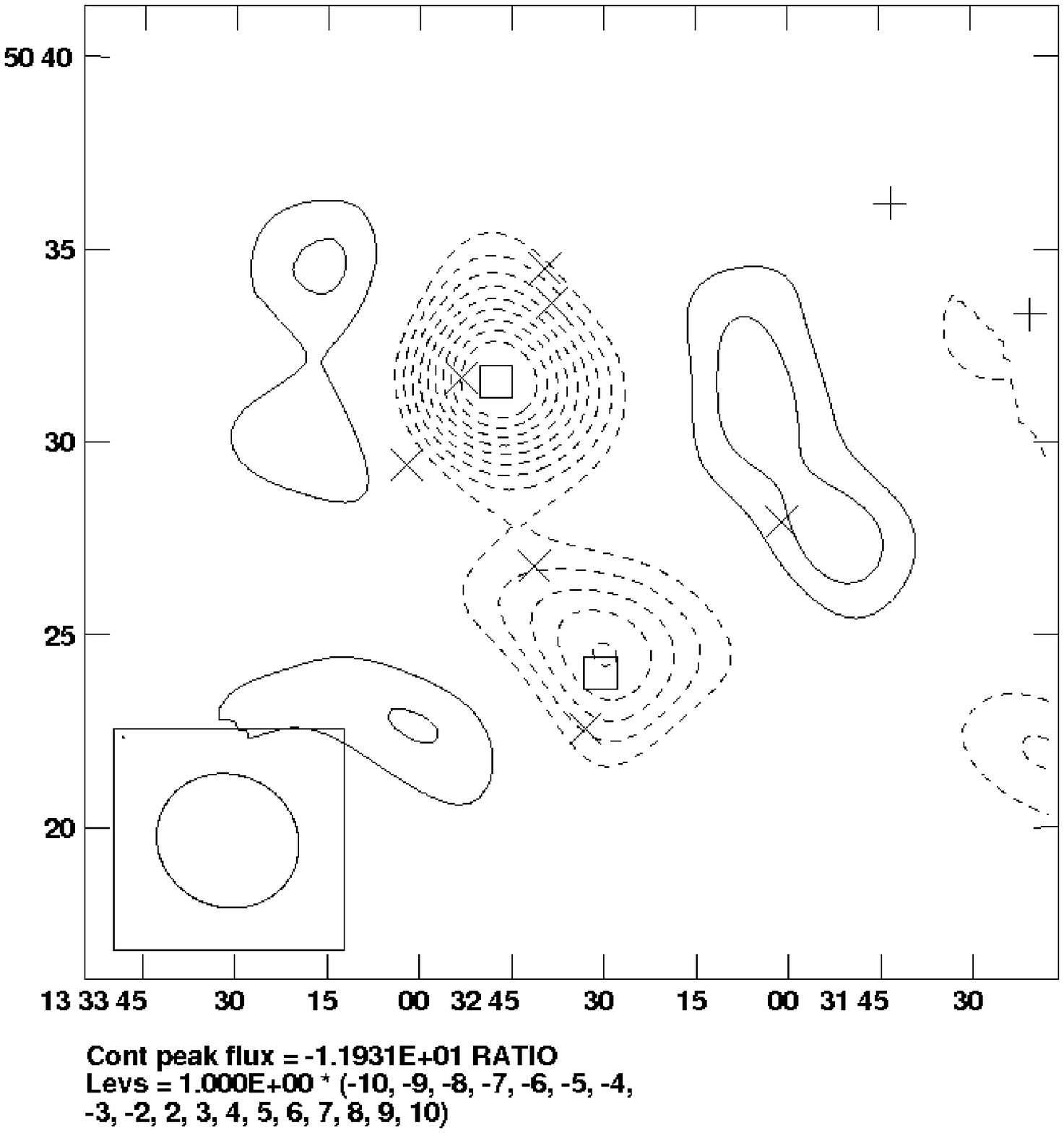}\qquad{E}\includegraphics[width=7.5cm,height=7.5cm,clip=,angle=0.]{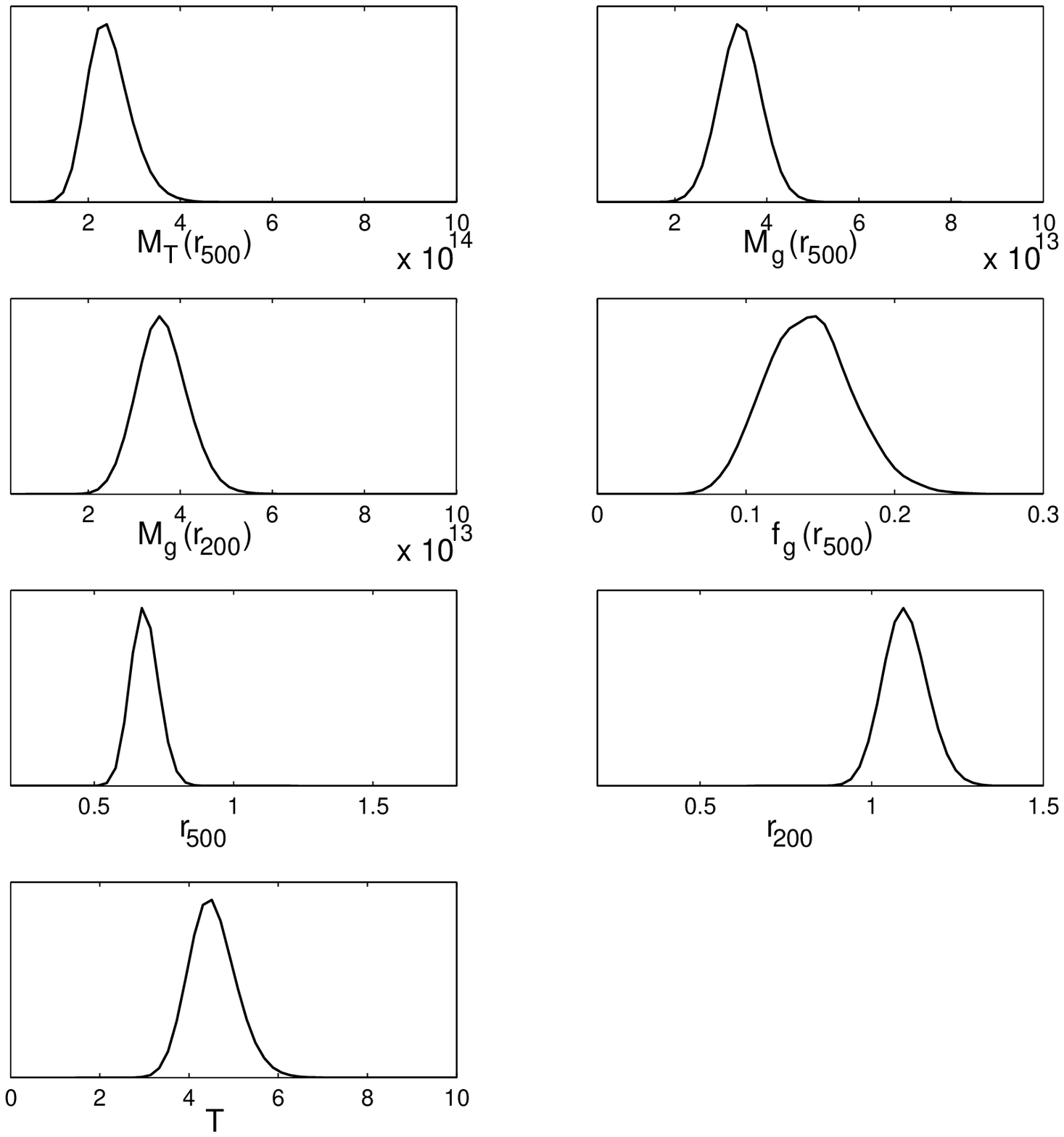}}
 \centerline{{C.i}\includegraphics[width=7.5cm,height=6.5cm,clip=,angle=0.]{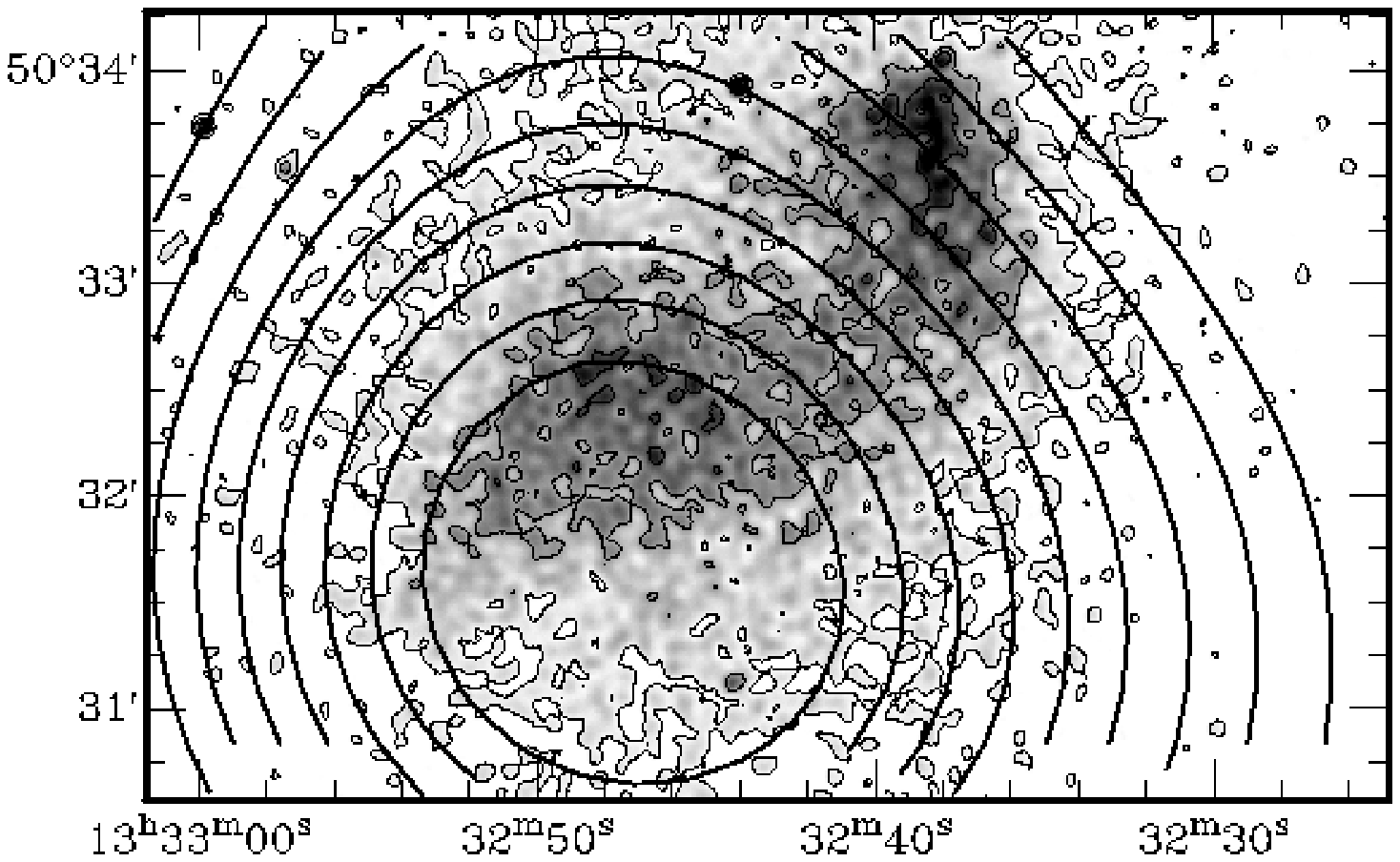}\qquad{C.ii}\includegraphics[width=7.5cm,height=6.5cm,clip=,angle=0.]{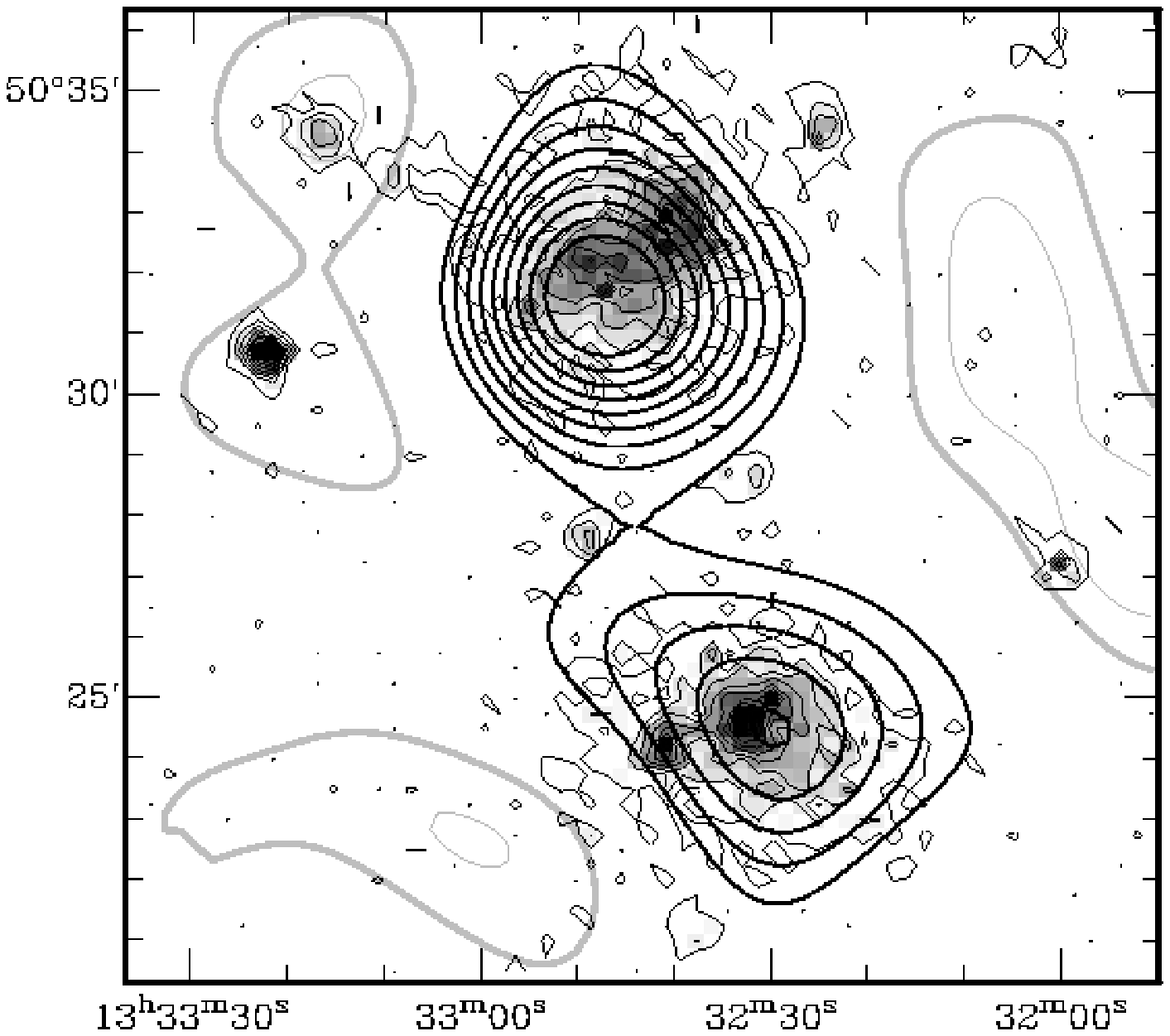}}
\caption{Panels A. and B. show the SA map before and after source subtraction
  (the latter map has had $0.6$-k$\lambda$ taper applied to it). The boxes in panels A and B indicates SZ centroid for each cluster, for the other symbols see Tab. \ref{tab:sourcelabel}.
 The maps shown here are primary beam corrected signal-to-noise maps cut off at 0.3 of the primary beam. The noise
level is $\approx$ 115$\mu$Jy towards the upper cluster (Abell 1758a) and
$\approx$ 130$\mu$Jy towards the lower cluster (Abell 1758b).
 The source-subtracted SA maps from B are overlaid with the the {\sc{Chandra}} map in Ci and with \emph{ROSAT} PSPC X-ray map in Cii.
 D and E show the marginalized posterior distributions for sampling and derived parameters, respectively. In panel D, $M_{\rm{T}}$ is given in units of $h_{100}^{-1}\times10^{14}M_{\odot}$ and $f_{\rm{g}}$ in $h_{100}^{-1}M_{\odot}$; both parameters are estimated within $r_{200}$. In E $M_{\rm{g}}$ is in units of $h_{100}^{-2}M_{\odot}$, $r$ in $h_{100}^{-1}$Mpc and $T$ in KeV.}
\label{fig:A1758A}
\end{figure*}

\begin{figure*}
\centerline{\huge{Abell 1758b}}
 \centerline{\includegraphics[width=7.5cm,height=7.5cm,clip=,angle=0.]{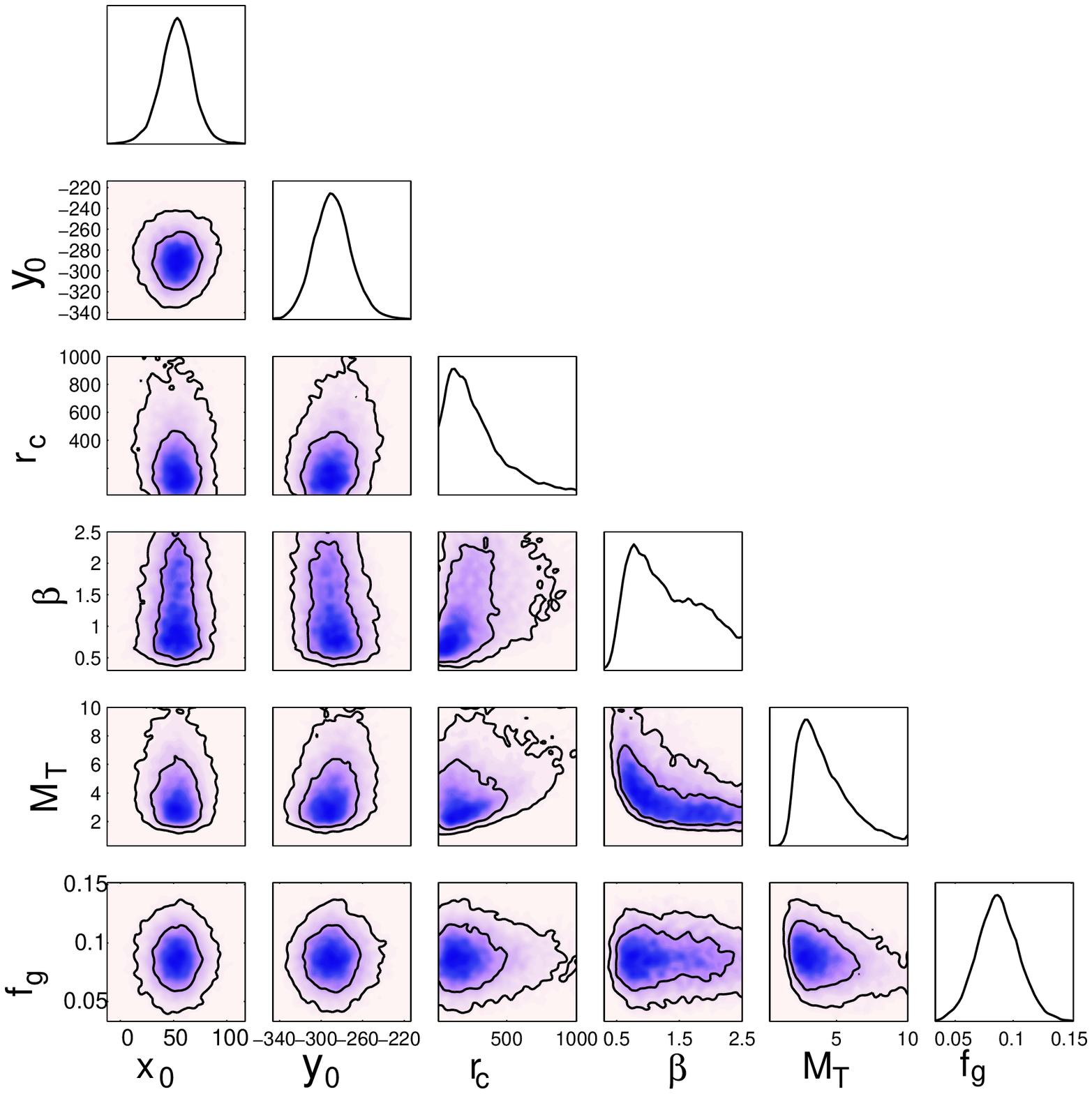}\qquad\includegraphics[width=7.5cm,height=7.5cm,clip=,angle=0.]{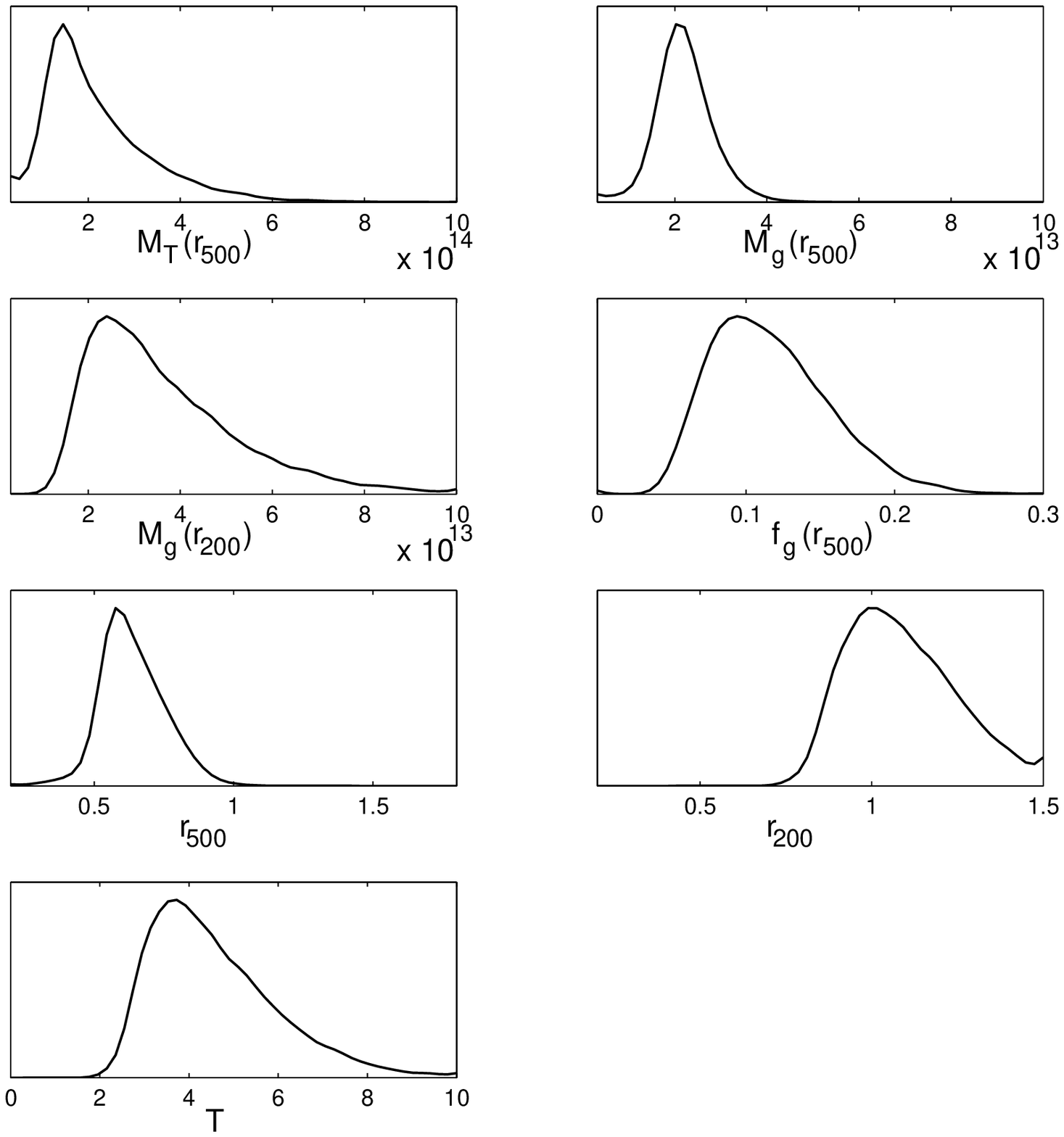}}
\caption{Abell~1758b. Left panel: 1 and 2-D marginalized posterior distributions for the cluster sampling parameters. $M_{\rm{T}}$ is given in units of $h_{100}^{-1}\times10^{14}M_{\odot}$ and $f_{\rm{g}}$ in $h_{100}^{-1}$; both parameters are estimated within $r_{200}$. Right panel: 1-D marginalized
posterior distributions for the cluster derived parameters. $M_{\rm{g}}$ is in units of $h_{100}^{-2}M_{\odot}$, $r$ in $h_{100}^{-1}$Mpc and $T$ in KeV.}
\label{fig:A1758B}
\end{figure*}

\begin{figure}
\includegraphics[width=8.0cm,height=8.0cm,clip=,angle=0.]{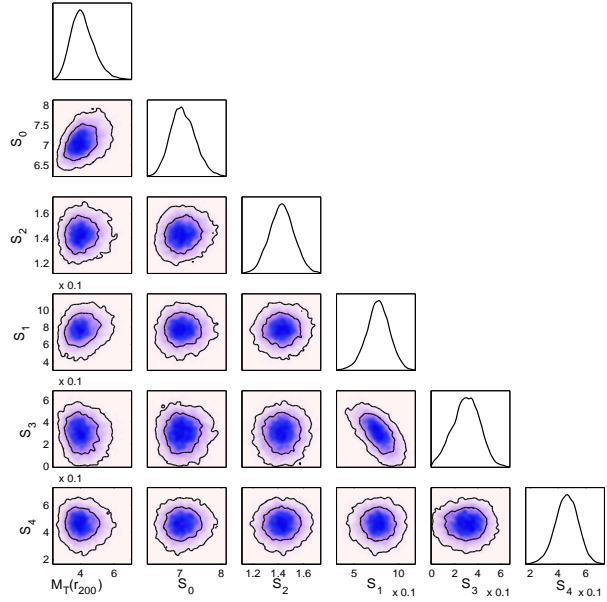}
\caption{1 and 2-D marginalized posterior distributions for
$M_{\rm{T}}(r_{200})$ and sources within $5\arcmin$ from the cluster X-ray
centroid for Abell~1758a. Source flux densities are given in units of Jys and $M_{\rm{T}}(r_{200})$ in units of $h_{100}^{-1}M_{\odot}\times10^{14}$.}
\label{fig:A1758A_2}
\end{figure}

\subsection{ Abell~2009} \label{resultsA2009}

Results for Abell~2009 are given in Fig. \ref{fig:A2009}.
Eighteen sources were detected above $4\sigma_{\rm{LA}}$ in our LA maps. Given
that all of the sources, except one, are further away than one arcminute from
the pointing centre and have
flux densities $<2$\,mJy, the source environment should not
 significantly contaminate the SZ signal on the SA maps. The source-subtraction has worked well and there are only 2$\sigma$ residuals
 (Fig. \ref{fig:A2009}, B); the most
prominent residual is likely to be associated with some extended emission seen in
the SA map before source subtraction.

We find the SZ image is extended in an approximately NS direction.
 \cite{okabe2010} fit an NFW profile to weak lensing data from the
\emph{Subaru/Suprime-Cam} and find
 $M_{\rm{T}}(r_{110})=3.86^{+1.20}_{-0.93}\times10^{14}h_{72}^{-1}\rm{M}_{\odot}$
(with $h_{72}=1.0$). 
 We find $M_{\rm{T}}(r_{200})=4.6 \pm 1.5\times10^{14}h_{100}^{-1}\rm{M}_{\odot}$.

The misleading sharp peaks at small radius in the distributions for  cluster parameters at
$r_{500}$ (Fig. \ref{fig:A2009} F)
are discussed in Sec. \ref{resultsA621}.

\begin{figure*}
\begin{center}
\begin{tabular}{m{8cm}cm{8cm}}
\multicolumn{3}{c}{\huge{Abell 2009}} \\
{A}\includegraphics[width=7.5cm,height=7.5cm,clip=,angle=0.]{./A2009.ps}
& \quad &
{D}\includegraphics[width=7.5cm,height=7.5cm,clip=,angle=0.]{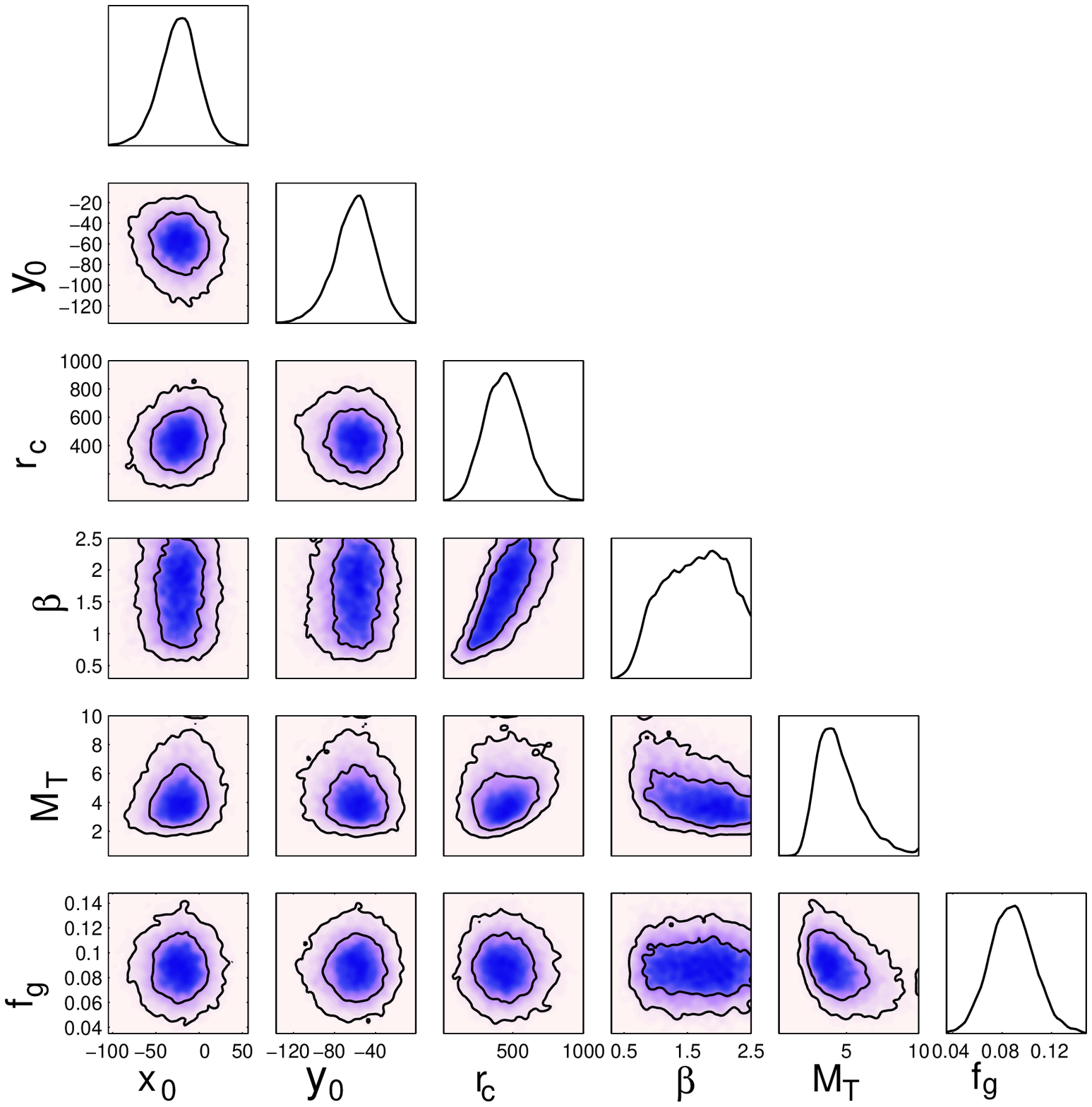}
\\
{B}\includegraphics[width=7.5cm,height=7.5cm,clip=,angle=0.]{./A2009_box.ps}
& \quad &
{E}\includegraphics[width=7.5cm,height=7.5cm,clip=,angle=0.]{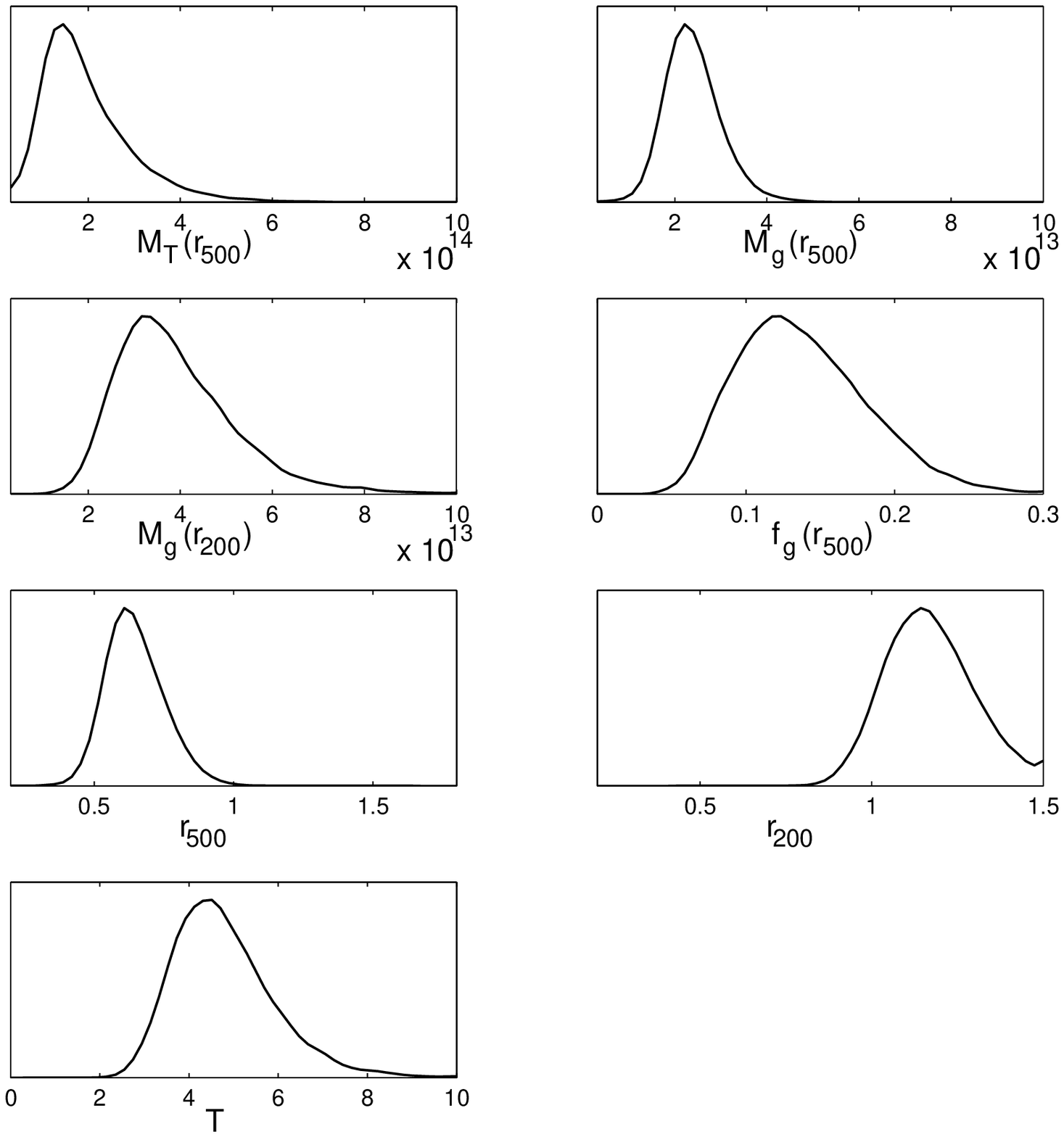}
\\
{C}\includegraphics[width=7.0cm,height=6.5cm,clip=,angle=0.]{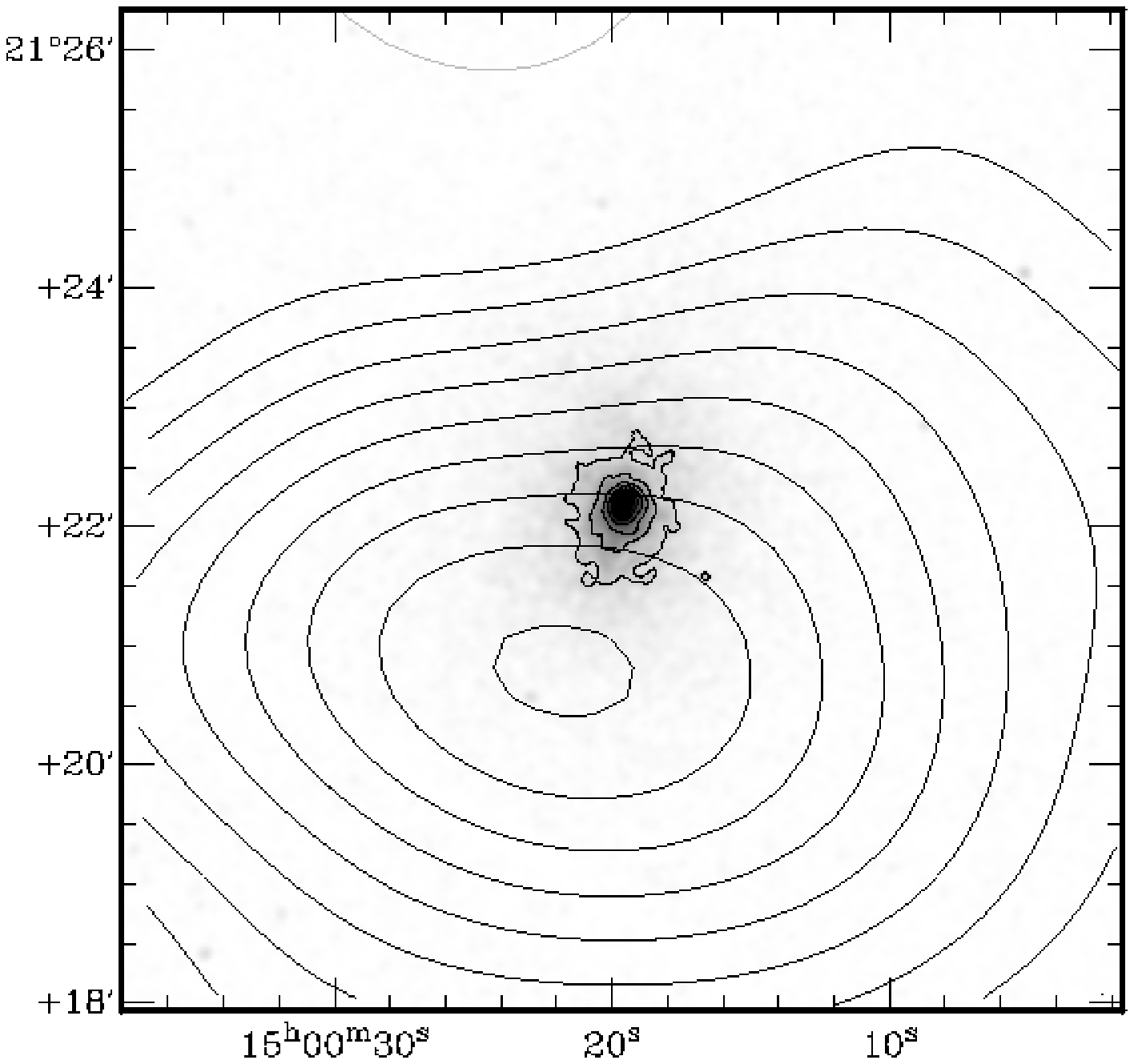}
& \quad &
{F}\includegraphics[width=7.0cm,height=6.5cm,clip=,angle=0.]{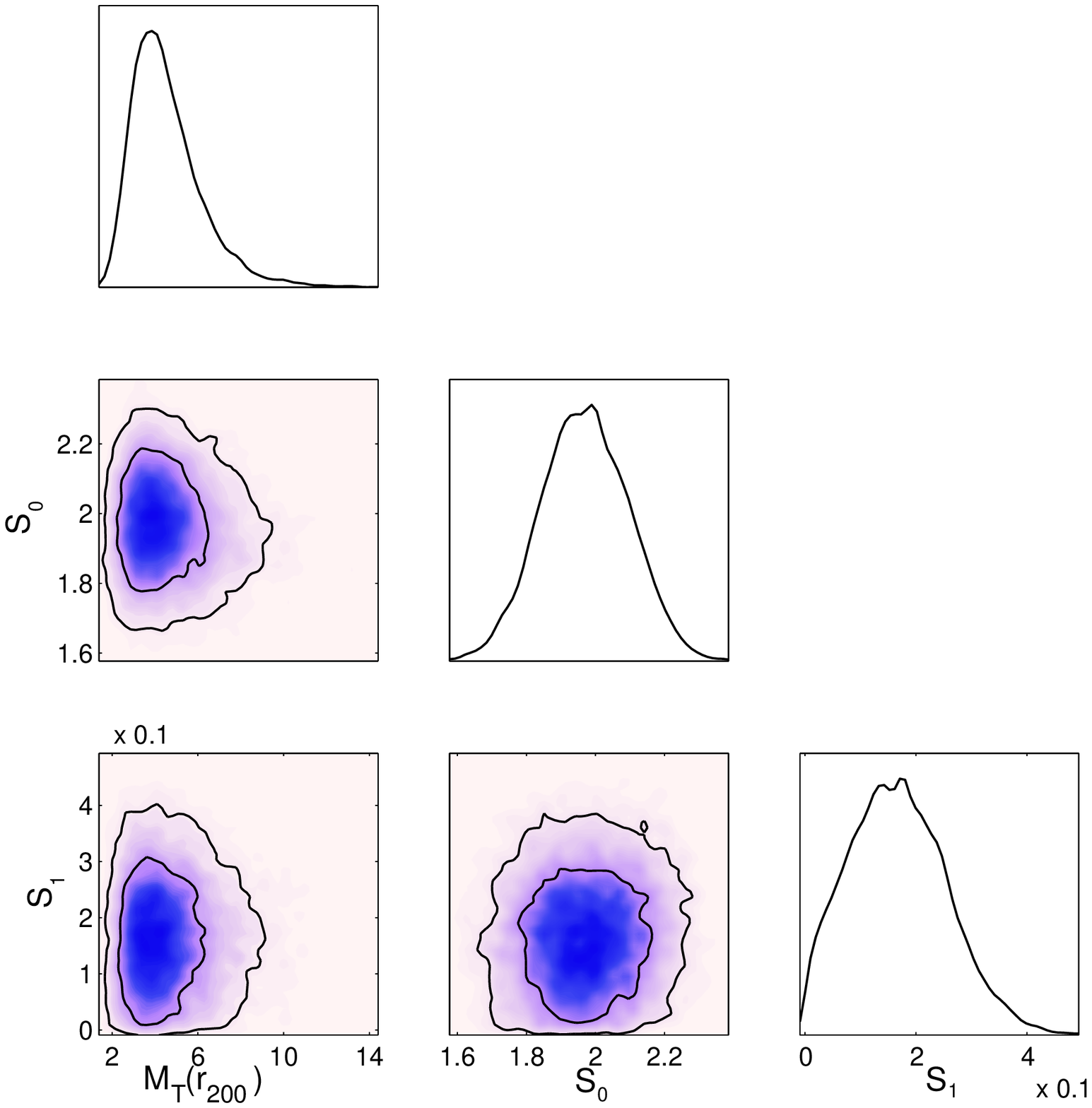}
%
\end{tabular}
\caption{Results for Abell~2009. Panels A and B show the SA map before and after source-subtraction, respectively; a $0.6$\,k$\lambda$ taper has been applied to B. The box in panels A and B indicates the cluster SZ centroid, for the other symbols see Tab. \ref{tab:sourcelabel}. The smoothed {\sc{Chandra}} X-ray map overlaid with contours from B is presented in image C. Panels D and E show the marginalized posterior distributions for the cluster sampling and derived parameters, respectively. F shows the 1 and 2-D marginalized posterior distributions for source flux densities (in Jys) given in Tab. \ref{tab:source_info} and $M_{\rm{T}}(r_{200})$ (in $h_{100}^{-1}\times 10^{14}M_{\odot}$). In panel D $M_{\rm{T}}$ is given in units of $h_{100}^{-1}\times10^{14}M_{\odot}$ and $f_{\rm{g}}$ in $h_{100}^{-1}$; both parameters are estimated within $r_{200}$.  In E $M_{\rm{g}}$ is in units of $h_{100}^{-2}M_{\odot}$, $r$ in $h_{100}^{-1}$Mpc and $T$ in KeV}
\label{fig:A2009}
\end{center}
\end{figure*}

\subsection{ Abell~2111} \label{resultsA2111}

Results for Abell~2111 are presented in Fig. \ref{fig:A2111}.
The source environment in the vicinity of Abell~2111 does not present a problem
in our analysis:
all the sources are located on the edge of the decrement or beyond and have
flux densities $\lesssim 3$\,mJy (Fig. \ref{fig:A2111} A and Tab, \ref{tab:source_info}).
Some residual flux with a peak surface brightness $\approx 700$\,$\mu$Jy\,beam$^{-1}$
remains in our source-subtracted map but is sufficiently far ($\approx
45\arcsec$) that it has a negligible effect on our SZ detection (Fig. \ref{fig:A2111} B).

X-ray studies of \emph{ROSAT PSPC} and \emph{HRI} data by \cite{wang1997}
reveal Abell~2111 has substructure on small scales but
 appears to be reasonably relaxed on larger scales away from the core.
Wang et al. identify a main X-ray emitting
component and a hotter subcomponent and conclude that Abell~2111 is most likely to be a head-on
merger between two subclusters; this is supported by \cite{henriksen1999}
using ASCA data. A disturbed nature of Abell~2111
 might also be indicated by the apparently clumpy X-ray emission and X-ray-SZ offset seen in Fig. \ref{fig:A2111} C.

Recent investigations by \cite{rines2010} find the virial mass for Abell~2111
to be
$M_{\rm{T}}(r_{100})=4.01\pm0.41\times10^{14}M_{\odot}$ using $h_{70}=1.0$ from
an
average of 90 member redshifts within $r_{100}$. \cite{maughan2008}
fit a modified version of the standard 1D isothermal $\beta$-model to {\sc{Chandra}} data with $h_{70}=1.0$ to compute
$M_{\rm{g}}(r_{500})=7.44^{+0.10}_{-0.05}\times10^{13}M_{\odot}$.
 We
obtain a value of $M_{g}(r_{500})= 2.5 \pm
0.3\times10^{13}h_{70}^{-2}M_{\odot}$.
Previously, \cite{laroque2006} fitted an isothermal $\beta$-model to {\sc{Chandra}} data (excising the
$r<100$\,kpc
from the core) and OVRO/BIMA data and found a gas mass
$M_{\rm{g}}(r_{2500})=2.15\pm{0.42}\times10^{13}M_{\odot}$ (for
$h_{70}=1.0$); they also found an X-ray
spectroscopic temperature of $\approx8.2$\,keV. On larger scales, at
$r_{200}$, we obtain a lower temperature, $4.6\pm 0.6$\,keV, which suggests
the average cluster temperature falls with
radius. Moreover, Henriksen et al. report a radially decreasing
temperature structure for Abell~2111 and parameterize it by a polytropic
index $\gamma\approx1.45$. On larger scales \cite{hurley2011} estimate $M_{\rm{T}}(r_{200})=6.9\pm 1.1\times10^{14}h_{70}^{-1}M_{\odot}$
from lensing data and $M_{\rm{T}}(r_{200})=6.3 \pm 2.1\times 10^{14}h_{70}^{-1}M_{\odot}$ from AMI SZ data; they also find that a circular geometry
is a slightly better fit to the data than an elliptical geometry. Our results,
$M_{\rm{T}}(r_{200})=4.2\pm 0.9\times10^{14}h_{100}^{-1}M_{\odot}$ are in very good agreement.

 \begin{figure*}
\begin{center}
\begin{tabular}{m{8cm}cm{8cm}}
\multicolumn{3}{c}{\huge{Abell 2111}} \\
{A}\includegraphics[width=7.5cm,height=7.5cm,clip=,angle=0.]{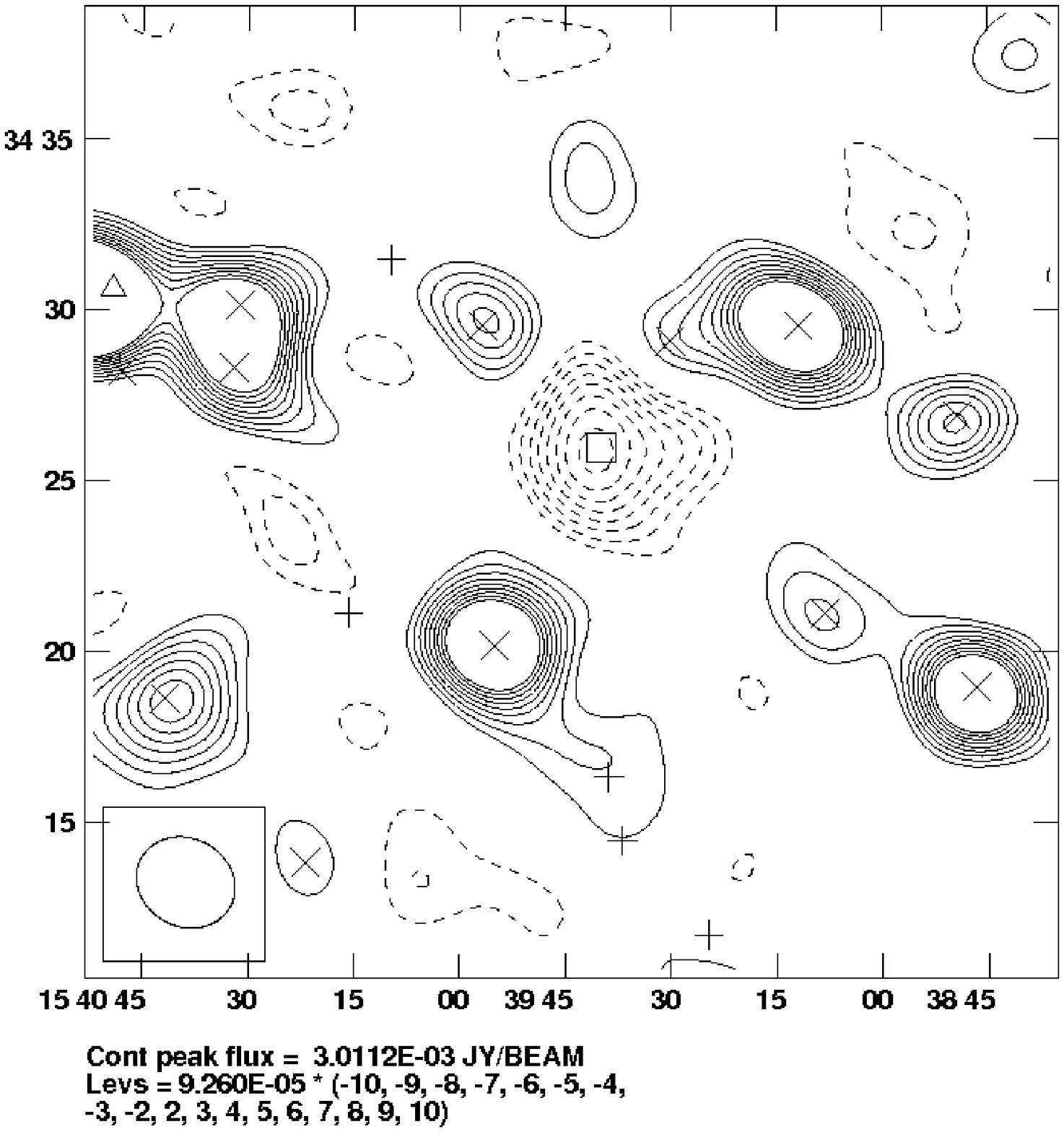}
& \quad &
{D}\includegraphics[width=7.5cm,height=7.5cm,clip=,angle=0.]{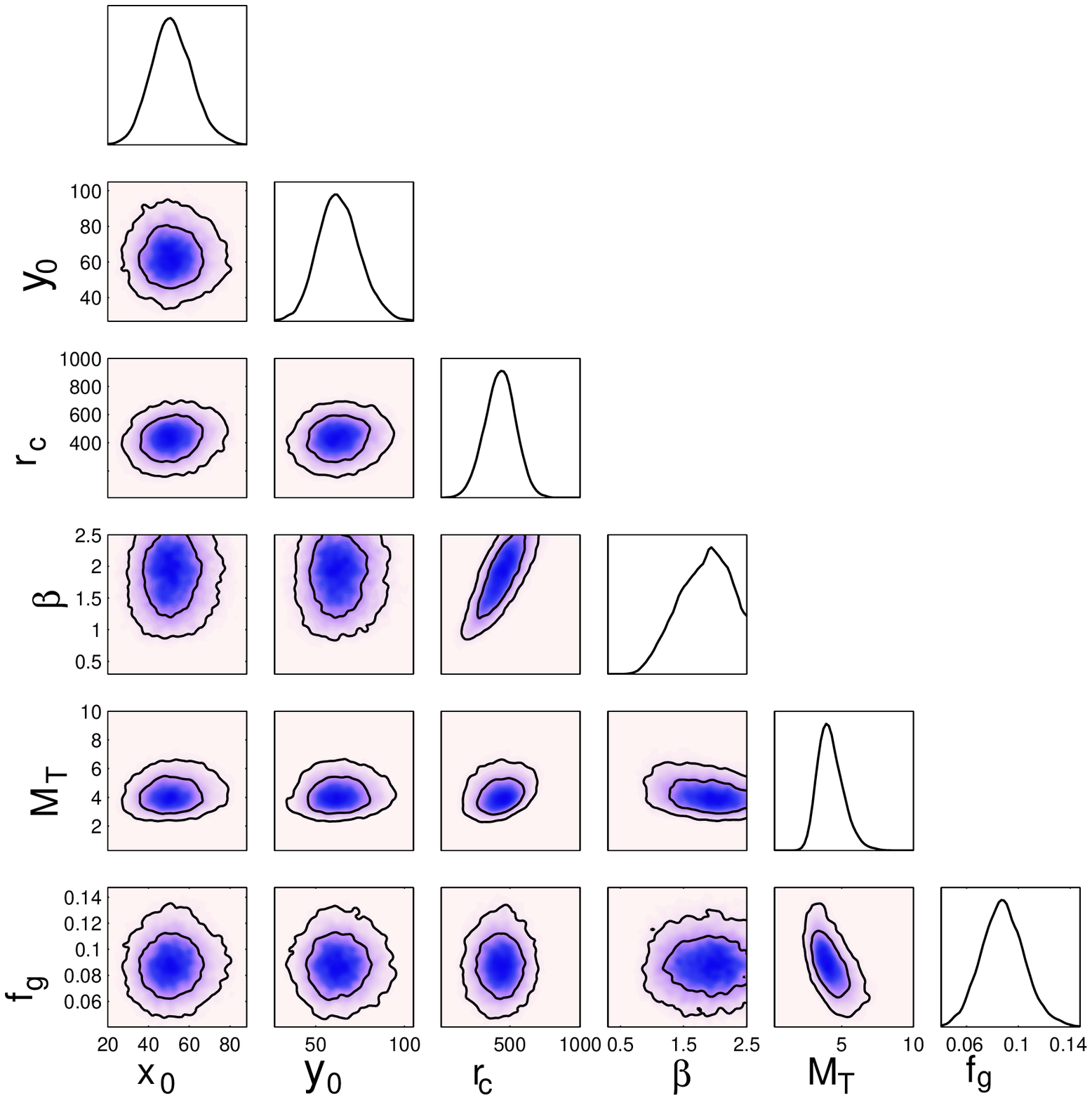}\\
{B}\includegraphics[width=7.5cm,height=7.5cm,clip=,angle=0.]{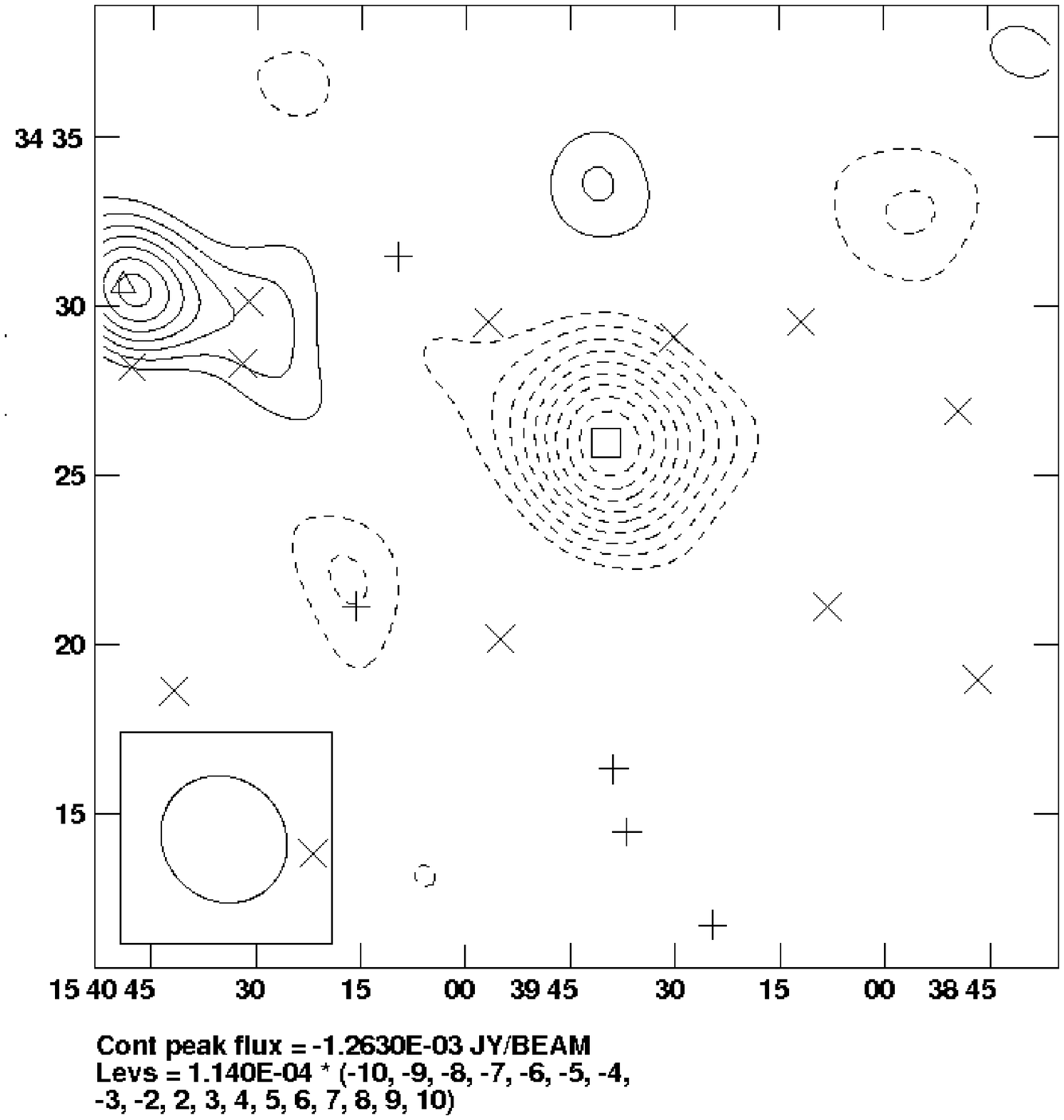}
& \quad &
{E}\includegraphics[width=7.5cm,height=7.5cm,clip=,angle=0.]{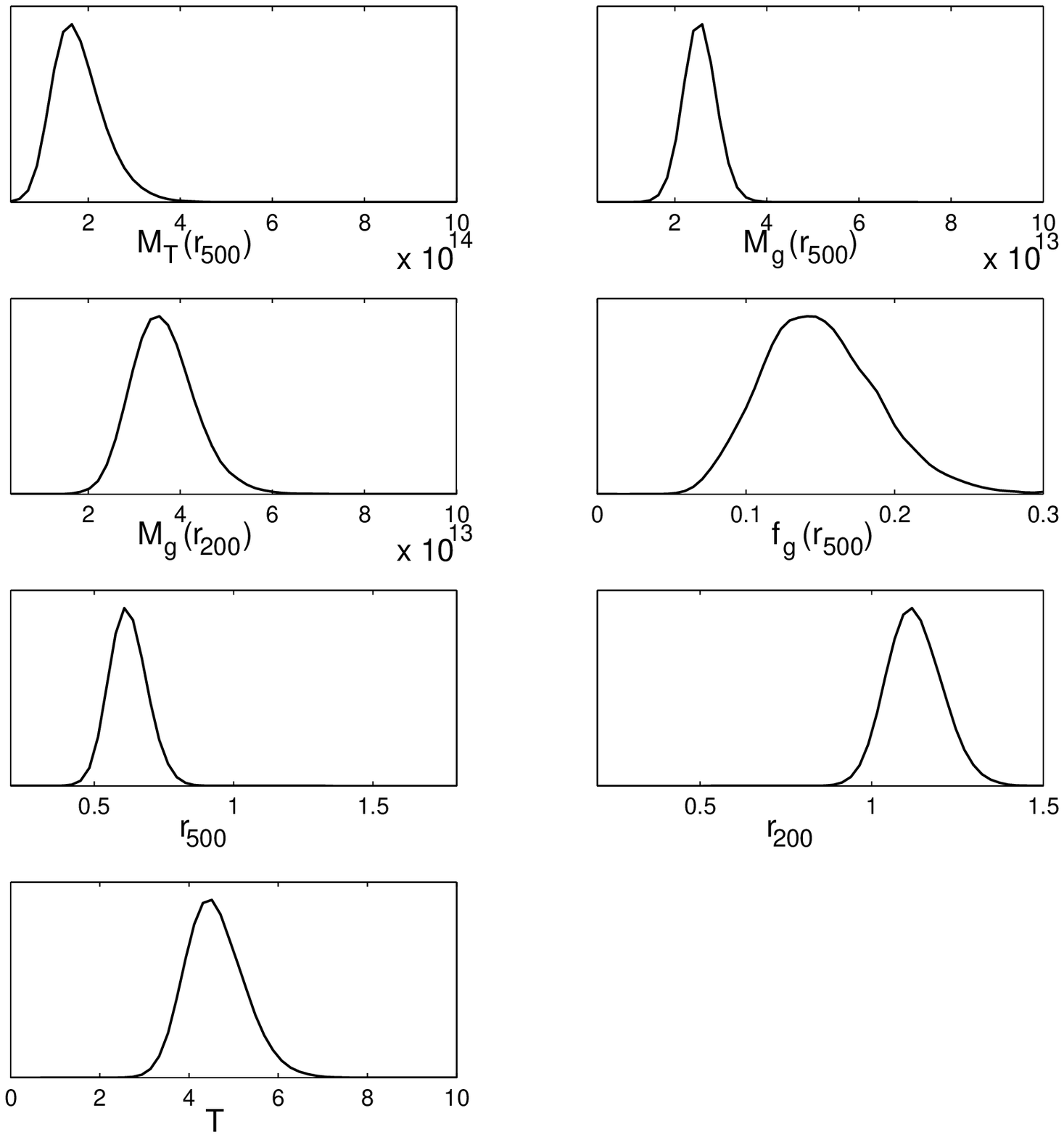}\\

{C}\includegraphics[width=6.5cm,height=6.0cm,clip=,angle=0.]{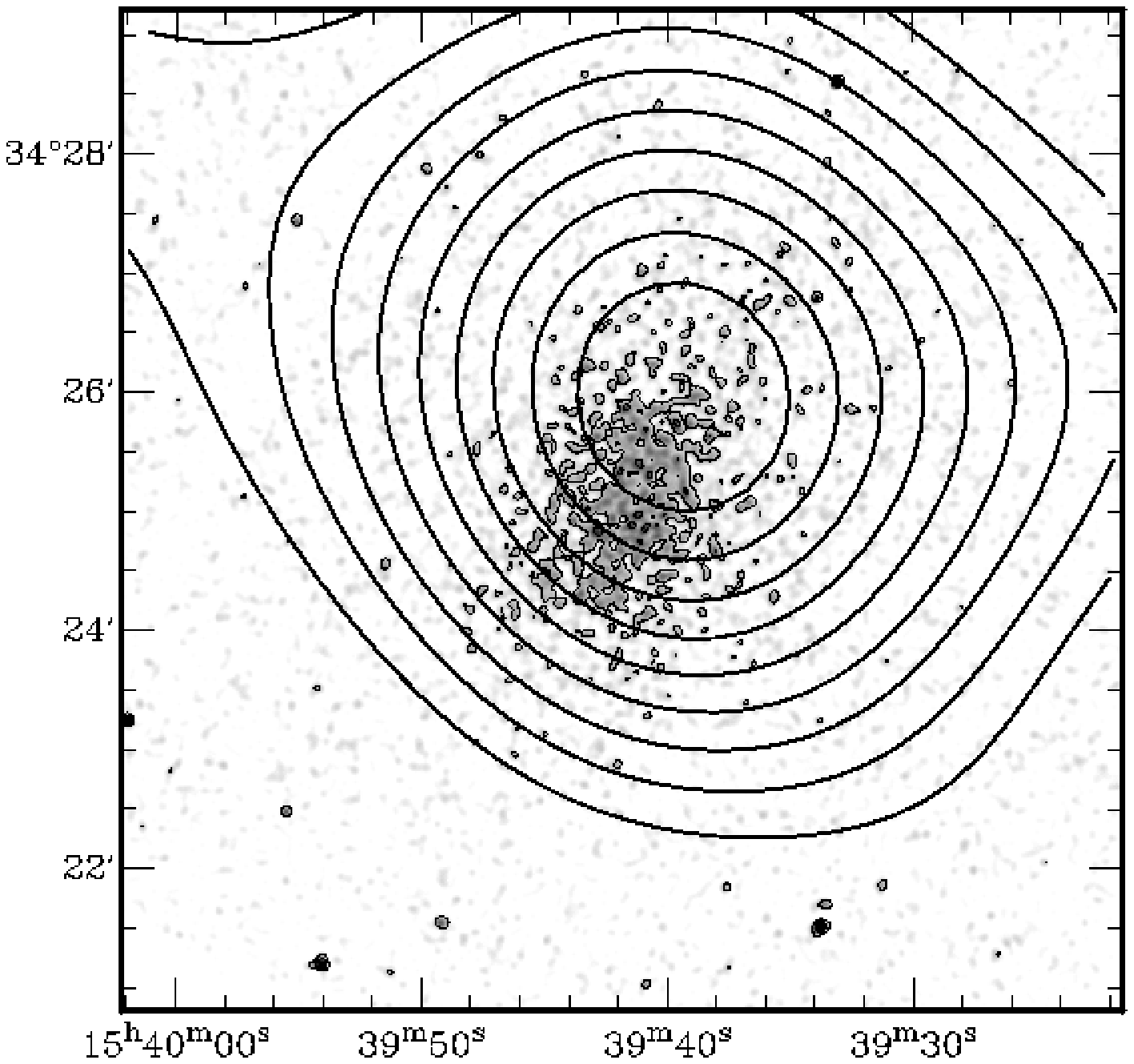}
& \quad &
{F}\includegraphics[width=6.5cm,height=6.0cm,clip=,angle=0.]{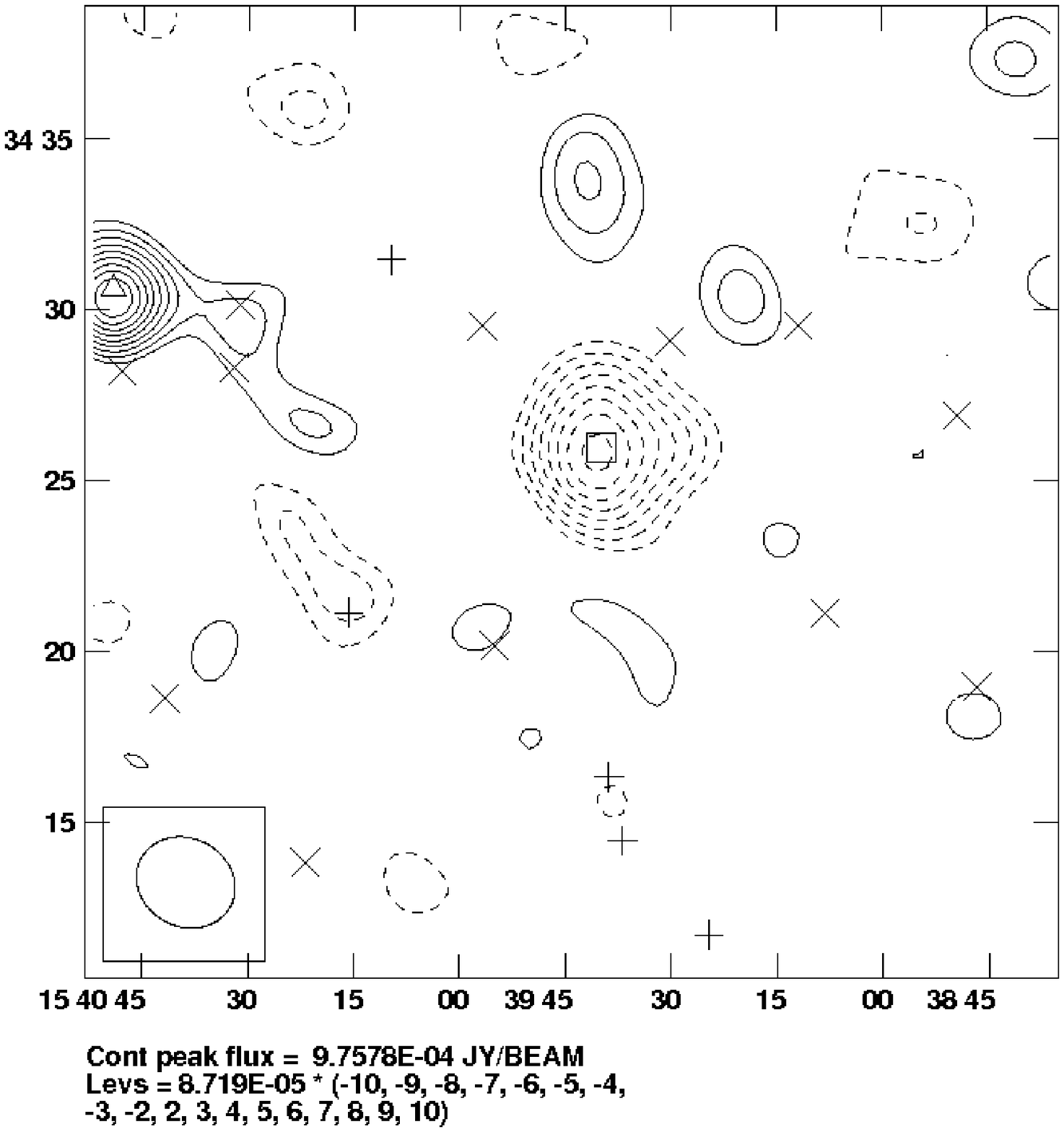}

\end{tabular}
\caption{Results for Abell~2111. Panels A and B show the SA map before and after source-subtraction, respectively; a $0.6$\,k$\lambda$ taper has been applied to B. The box in panels A and B indicates the cluster SZ centroid, for the other symbols see Tab. \ref{tab:sourcelabel}. The smoothed {\sc{Chandra}} X-ray map overlaid with contours from B is presented in image C. Panels D and E show the marginalized posterior distributions for the cluster sampling and derived parameters, respectively. F shows the higher-resolution source-subtracted map (no taper). In panel D $M_{\rm{T}}$ is given in units of $h_{100}^{-1}\times10^{14}$ and $f_{\rm{g}}$ in $h_{100}^{-1}$; both parameters are estimated within $r_{200}$. In E $M_{\rm{g}}$ is in units of $h_{100}^{-2}M_{\odot}$, $r$ in $h_{100}^{-1}$Mpc and $T$ in KeV}
\label{fig:A2111}
\end{center}
\end{figure*}

\begin{figure}
\includegraphics[width=8.0cm,height=8.0cm,clip=,angle=0.]{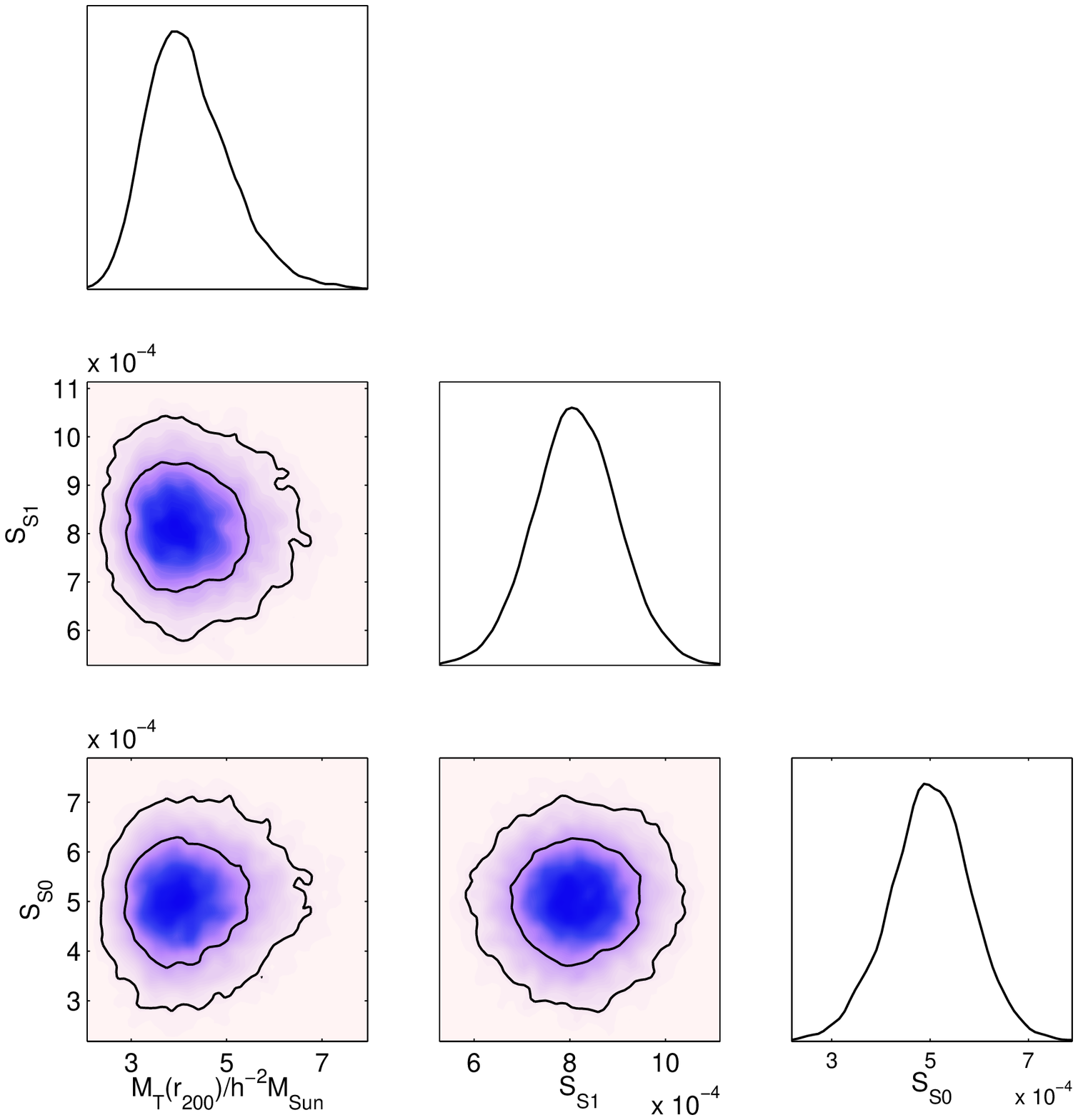}
\caption{2-D and 1-D marginalized posterior distributions for $M_{\rm{T}}(r_{200})$ (in  $h_{100}^{-1}M_{\odot}\times 10^{14}$) and source flux densities (in Jys) within $5\arcmin$ from the cluster X-ray centroid for Abell 2111.}
\label{fig:A2111_2}
\end{figure}

\subsection{ Abell~2146} \label{resultsA2146}

We have re-analysed the AMI data used in \cite{carmen2010}
with the cluster parameterization described in Sec. \ref{clusmod},
which is slightly different to theirs; our results are presented in Fig.
\ref{fig:A2146}.
 They obtain $M_{\rm{g}}(r_{200})= 4.9 \pm
0.5 \times 10^{13} h_{100}^{-2}\rm{M}_\odot$
and $T=4.5 \pm 0.5$\,keV while our results give
$M_{\rm{g}}(r_{200})= 4.4 \pm 0.6
  \times 10^{13} h_{100}^{-2}\rm{M}_\odot$
and $T=5.2 \pm 0.5$\,keV. Given the similarities between the
two analyses and the fact the same data were used for both, we would indeed expect this good
agreement
between these sets of results. We have further investigated the effect of
sources in this cluster
and have found a slight degeneracy between the cluster mass and the flux
density of the source lying closest to the cluster centre -- see Fig.
\ref{fig:A2146_2} F -- 
which had not been seen for the brighter, $\approx 6$\,mJy\,beam$^{-1}$
 source lying a few arcseconds away from the cluster centroid.

 {\sc{Chandra}} data analysed by \cite{russel09} have revealed that Abell~2146
is undergoing a rare merger event
 similar to that of ``Bullet-cluster'' \citep{markevitch2002}, with  two
shock fronts with Mach numbers $M\approx2$, and strong non-uniformities in the temperature
profile. Note the different -- essentially 90$^{\circ}$ -- orientaions  between the X-ray and the SZ
 extensions. To understand this  we have to consider collision geometry, mass ratio and, especially, time
 of snapshot since the merger start -- see Sec. \ref{discussion}.

 \begin{figure*}
\begin{center}
\begin{tabular}{m{8cm}cm{8cm}}
\multicolumn{3}{c}{\huge{Abell~2146}}\\

{A}\includegraphics[width=7.5cm,height=7.5cm,clip=,angle=0.]{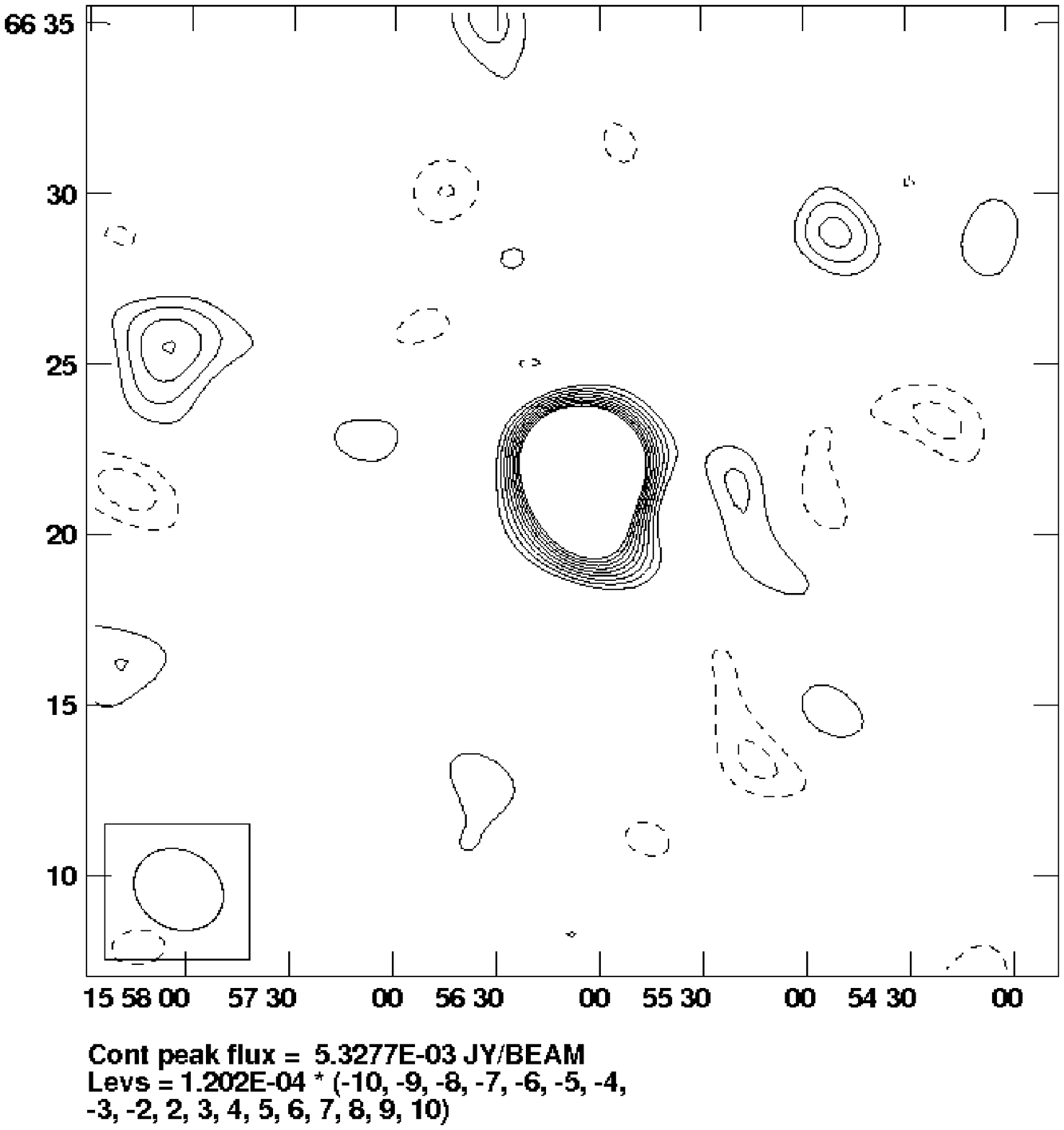}
& \quad &
{D}\includegraphics[width=7.5cm,height=7.5cm,clip=,angle=0.]{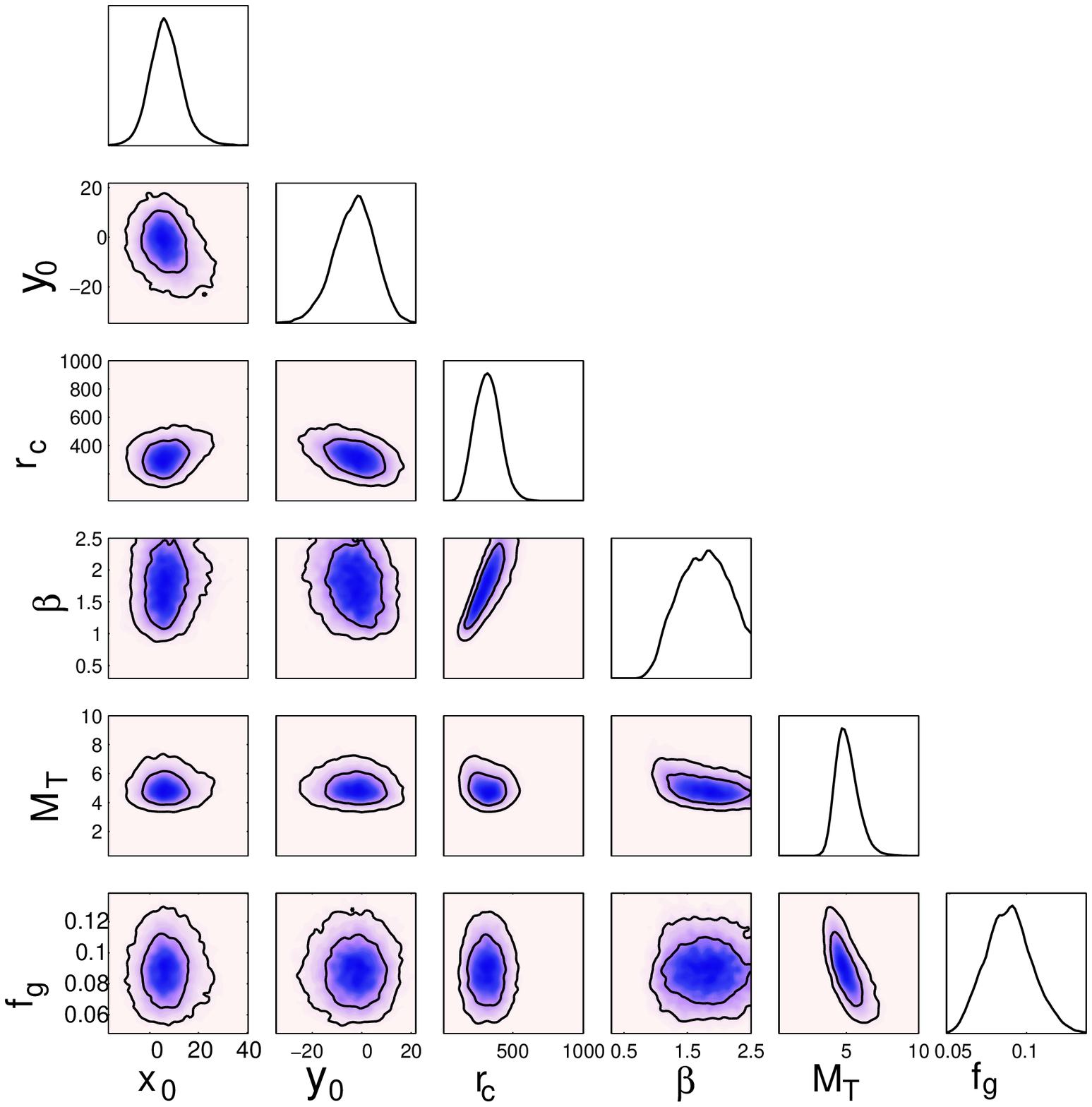}
\\
{B}\includegraphics[width=7.5cm,height=7.5cm,clip=,angle=0.]{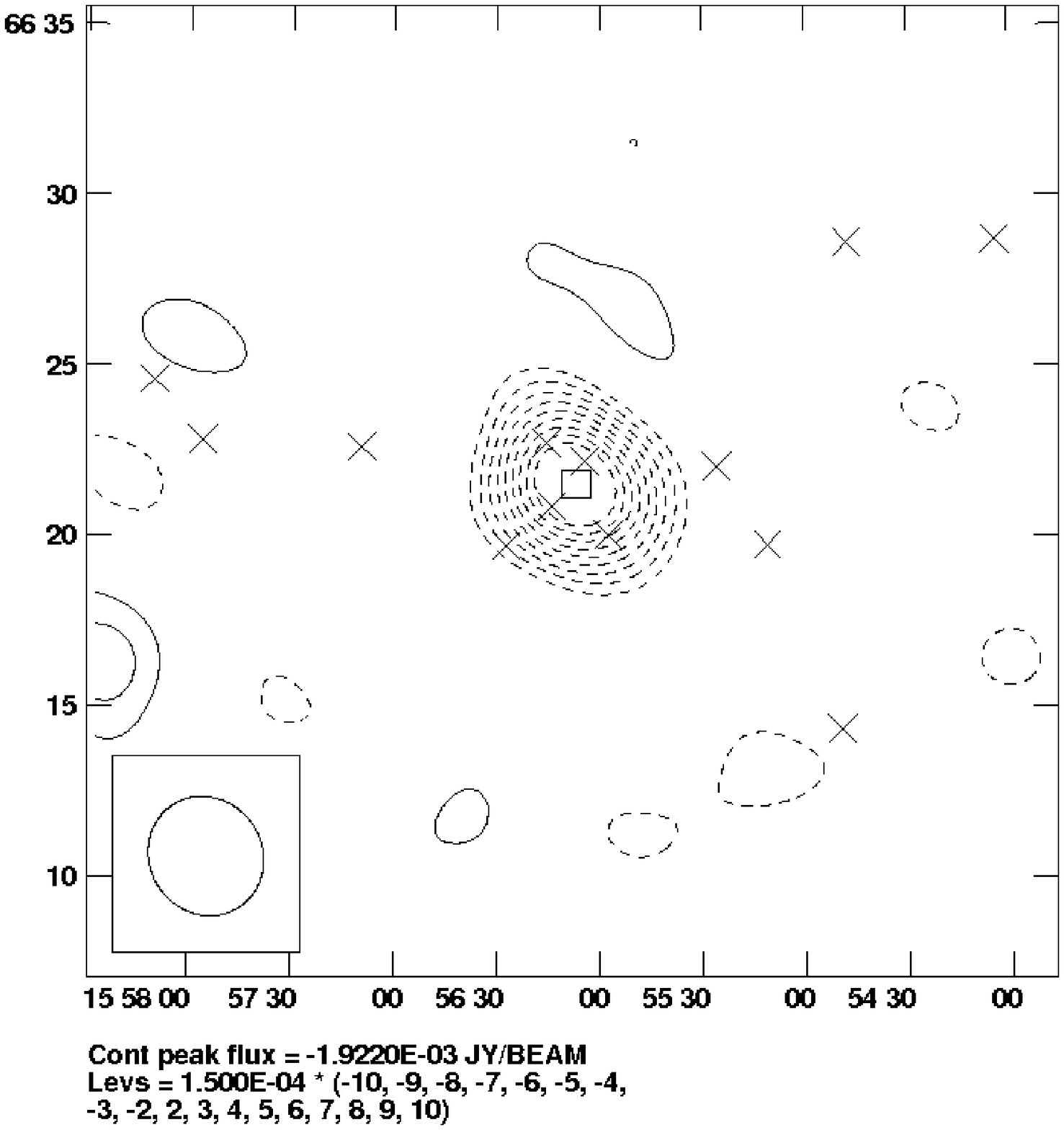}
& \quad &
{E}\includegraphics[width=7.5cm,height=7.5cm,clip=,angle=0.]{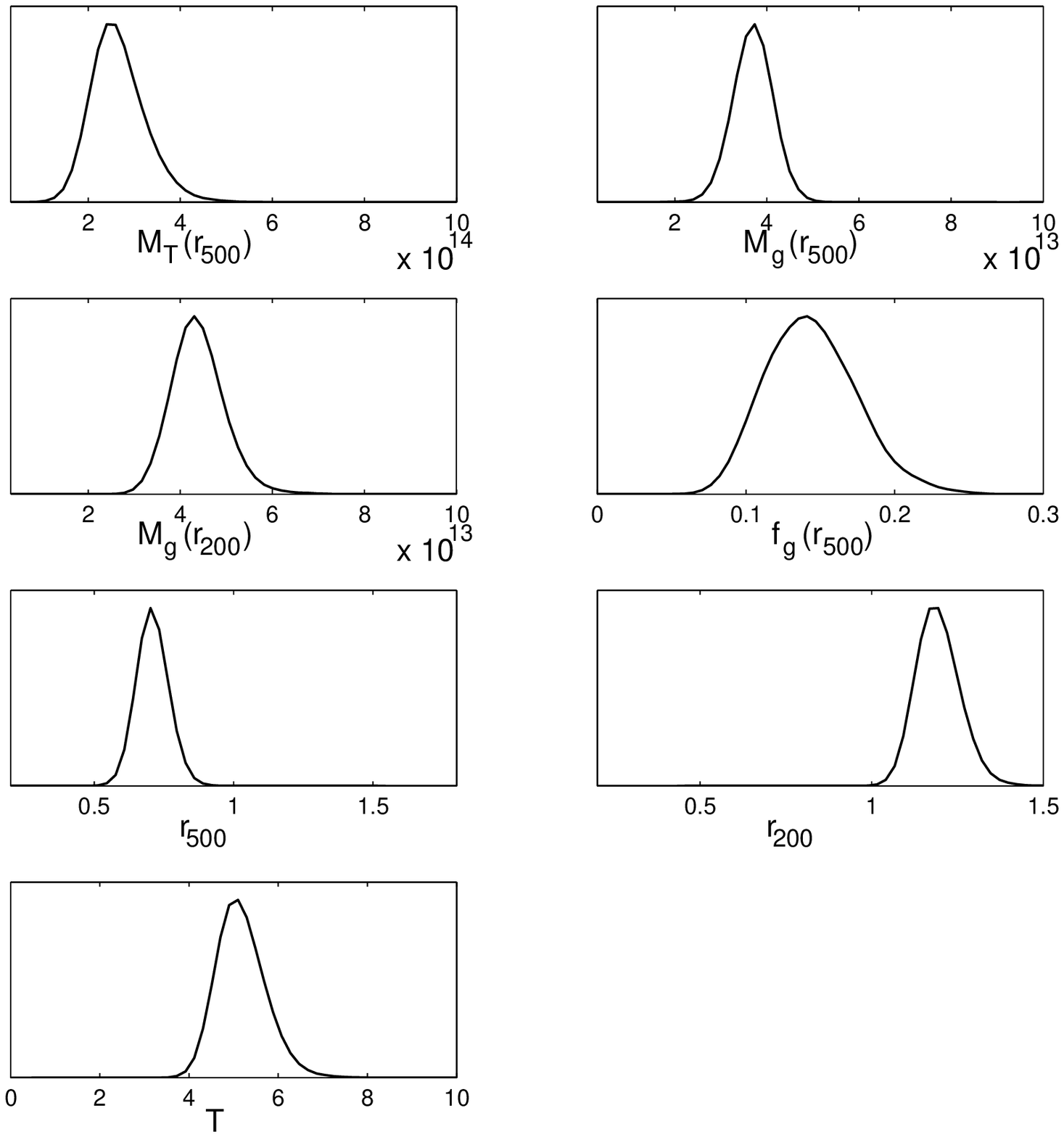}
\\
\centerline{{C}\includegraphics[width=7.5cm,height=7.5cm,clip=,angle=0.]{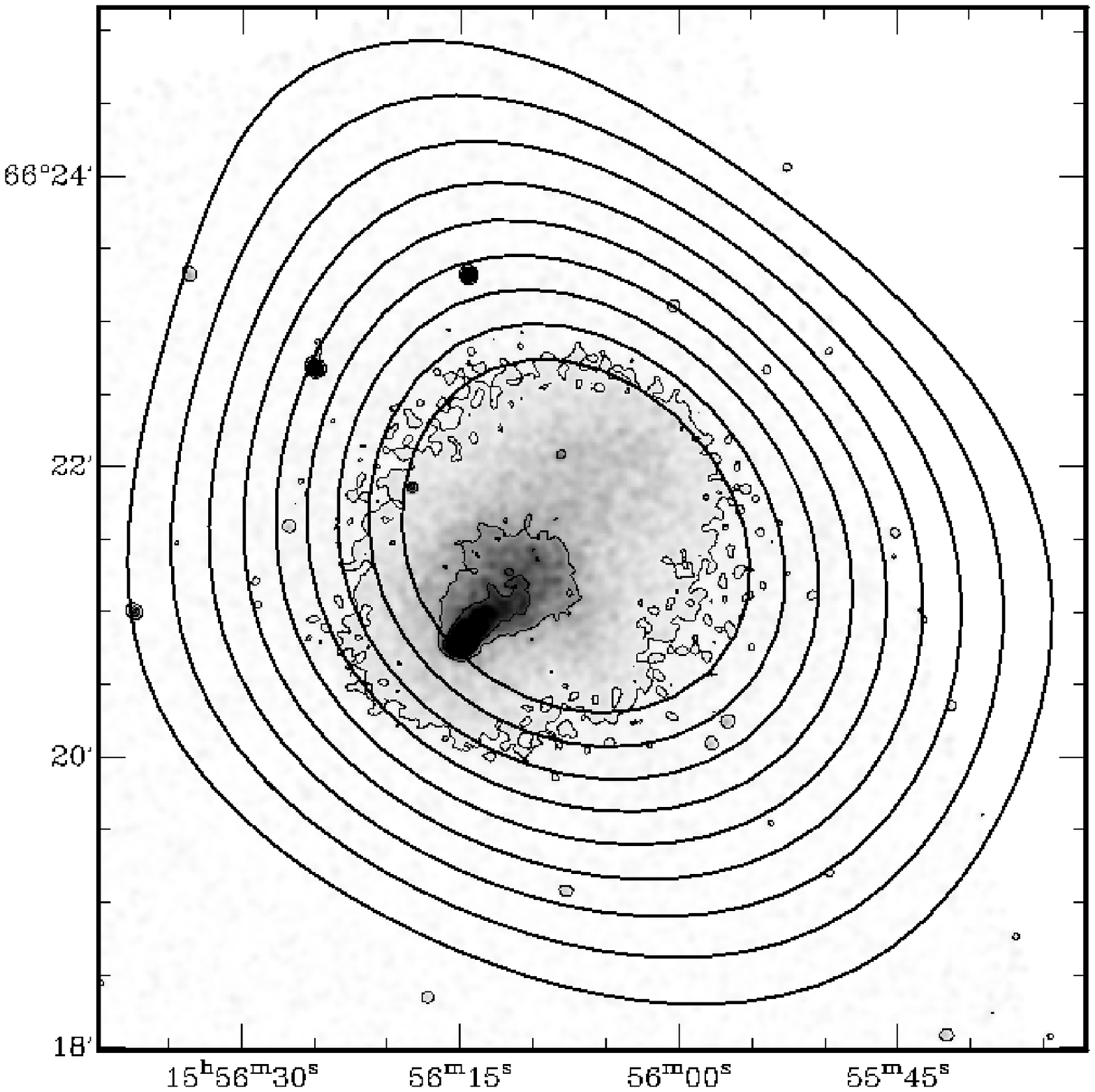}}
& \quad &
\caption{Results for Abell~2146. Panels A and B show the SA map before and after source subtraction, respectively; a $0.6$\,k$\lambda$ taper has been applied to B. The box in panels A and B indicates the cluster SZ centroid, for the other symbols see Tab. \ref{tab:sourcelabel}. The smoothed {\sc{Chandra}} X-ray map overlaid with contours from B is presented in image C. Panels D and E show the marginalized posterior distributions for the cluster sampling and derived parameters, respectively. In panel D $M_{\rm{T}}$ is given in units of $h_{100}^{-1}\times10^{14}M_{\odot}$ and $f_{\rm{g}}$ in $h_{100}^{-1}$; both parameters are estimated within $r_{200}$. In E $M_{\rm{g}}$ is in units of $h_{100}^{-2}M_{\odot}$, $r$ in $h_{100}^{-1}$Mpc and $T$ in KeV.}
\end{tabular}
%
\label{fig:A2146}
\end{center}
\end{figure*}

\begin{figure}
\includegraphics[width=8.0cm,height=8.0cm,clip=,angle=0.]{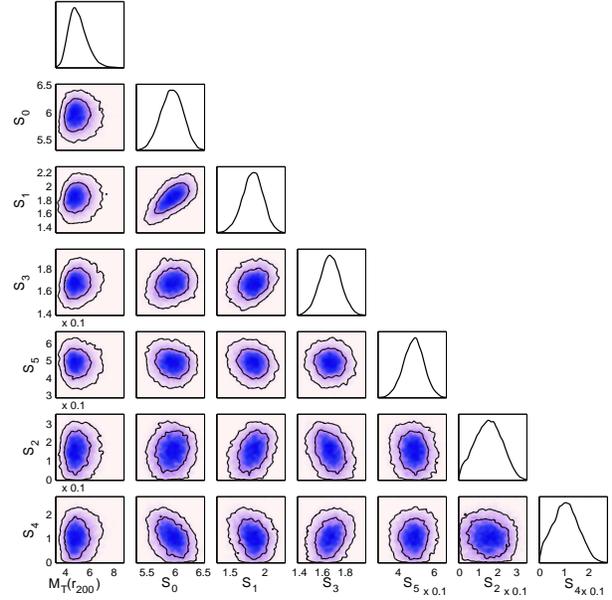}
\caption{2-D and 1-D marginalized posterior distributions for $M_{\rm{T}}(r_{200})$ (in  $h_{100}^{-1}\times 10^{14}M_{\odot}$) and source flux densities (in Jys; see Tab. \ref{tab:source_info}) within $5\arcmin$ from the cluster X-ray centroid for Abell~2146.}
\label{fig:A2146_2}
\end{figure}

\subsection{ Abell~2218} \label{A2218}

Results for Abell~2218 are shown in Fig. \ref{fig:A2218}.
There is substantial radio emission towards Abell~2218, most of which is
subtracted from our maps to leave a $470$\,$\mu$Jy\,beam$^{-1}$ positive feature to
the West of the decrement, which could be extended emission. Rudnick et al. detect diffuse emission from a radio halo with a flux of $0.05$\,Jy within
a 500\,kpc radius at 327\,MHz, from which one might expect a $\leq
200$\,$\mu$Jy signal at 16\,GHz (for a typical halo spectral index, see e.g.,
\cite{hanisch1980}.

Several observations in the X-ray (e.g., \citealt{markevitch1997,
govoni2004, machacek2002}), optical (e.g., \citealt{girardi2001}), SZ (e.g.,
\citealt{jones2005}) and
lensing (e.g., \citealt{squires1996} and \citealt{smith2004}) have suggested that Abell~2218 is
a complex,
disturbed system. High-resolution \emph{ROSAT} \citep{markevitch1997} and
{\sc{Chandra}} \citep{govoni2004, machacek2002} data
show signs of substructure, particularly on
small scales.  Moreover, lensing studies by \cite{squires1996} and \cite{smith2004} have
revealed a bi-modal
mass distribution and associated
elongated structures in the mass distribution. Abell~2218 also shows signs of
strong temperature
variations (\citealt{govoni2004} and \citealt{pratt2004}). All of these results
are indicative that the cluster is not relaxed.

SZ observations towards Abell~2218 have been made with the Ryle Telescope
\citep{jones2005} at 15\,GHz, at 36\,GHz using the Nobeyama Telescope
\citep{tsuboi1998}
and with OCRA-p at 30\,GHz \citep{lancaster2008}. Earlier SZ
observations towards this cluster include \cite{birkinshaw1981},
\cite{birkinshaw1984}, \cite{partridge1987}, \cite{klein1991}, \cite{jones1993} and \cite{birkinshaw1994}.

Pratt et al. find from XMM-Newton data
that $T(r)$ falls from 8\,keV near the centre to 6.6\,keV at
700\,kpc. \cite{zhang2008} calculate a cluster mass estimate from the
XMM-Newton data; using
$h_{70}=1.0$, they obtain
$M_{\rm{T}}(r_{500})=4.2\pm1.3\times10^{14}M_{\odot}$
and $f_{\rm{g}}(r_{500})=0.15\pm0.09$. We find $M_{\rm{T}}(r_{500})=2.7 \pm
0.6\times10^{14}h_{100}^{-1}M_{\odot}$.

The {\sc{Chandra}} X-ray image shown in Fig. \ref{fig:A2218} C appears to be
extended
along the N--S direction on arcminute scales and along the
$\approx$SE--NW direction on scales
$\approx2\arcmin$. On the other hand, the distribution of the X-ray signal on
larger
scales, $\approx3\arcmin$, tends to be more circular.
On the untapered, source-subtracted SA map, Fig. \ref{fig:A2218} F, the SZ
signal towards Abell~2218 is clearly extended.
There is  no significant degeneracy between the cluster mass and the source flux densities, as seen from Fig. \ref{fig:A2218_2}.

\begin{figure*}
\begin{center}
\begin{tabular}{m{8cm}cm{8cm}}
\multicolumn{3}{c}{\huge{Abell 2218}} \\

{A}\includegraphics[width=7.5cm,height=7.4cm,clip=,angle=0.]{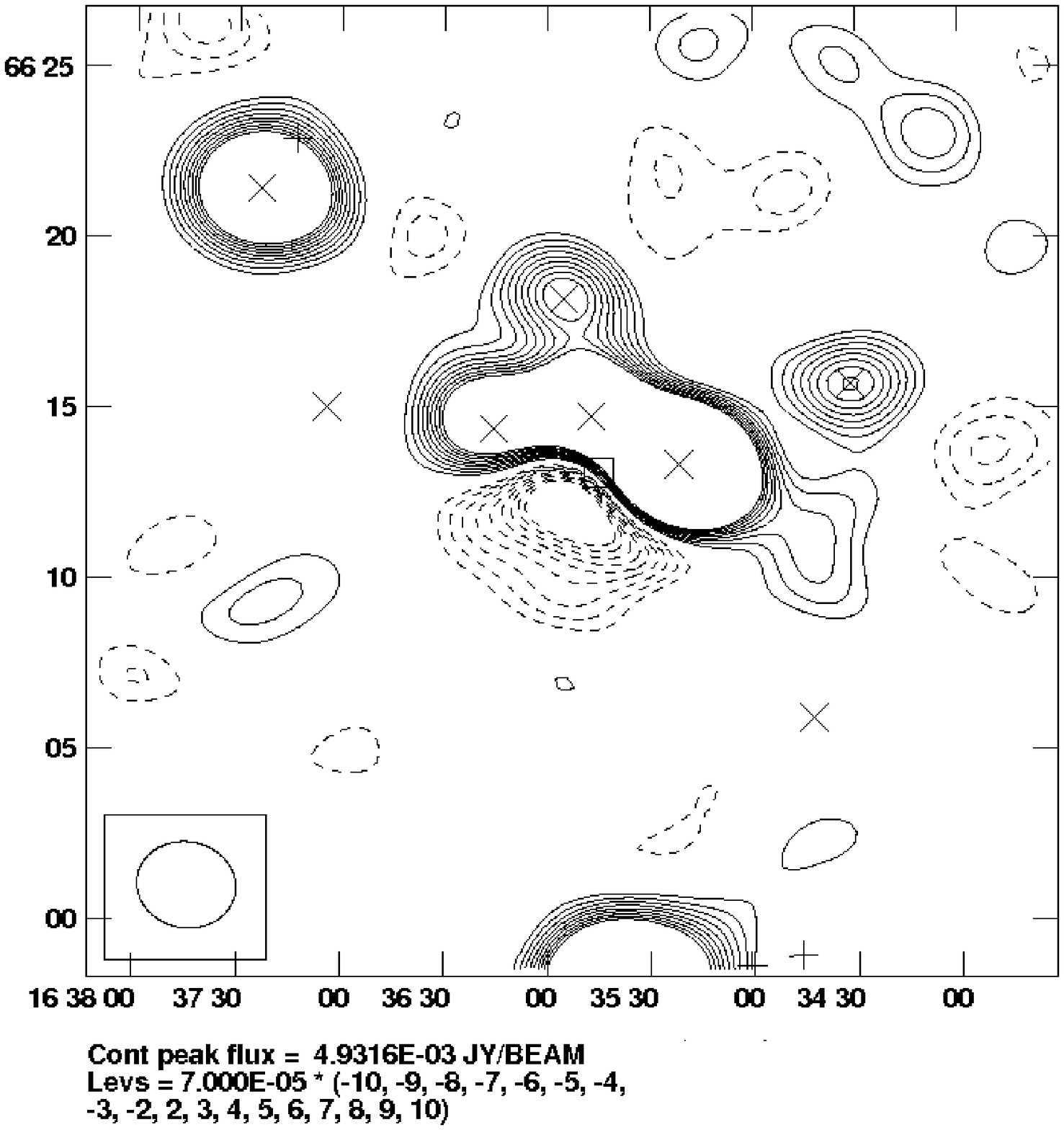}
& \quad &
{D}\includegraphics[width=7.5cm,height=7.4cm,clip=,angle=0.]{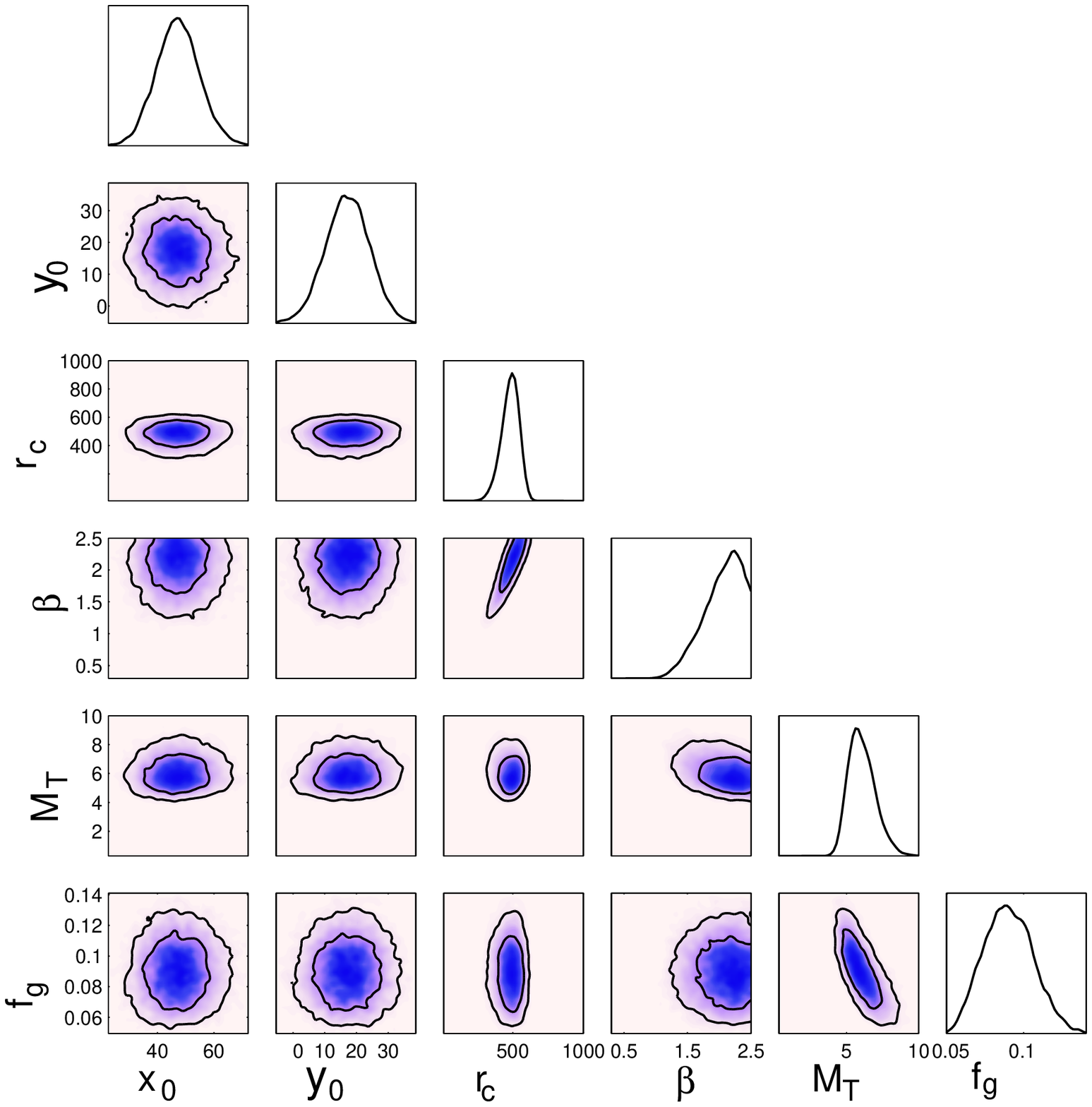}
\\
{B}\includegraphics[width=7.5cm,height=7.4cm,clip=,angle=0.]{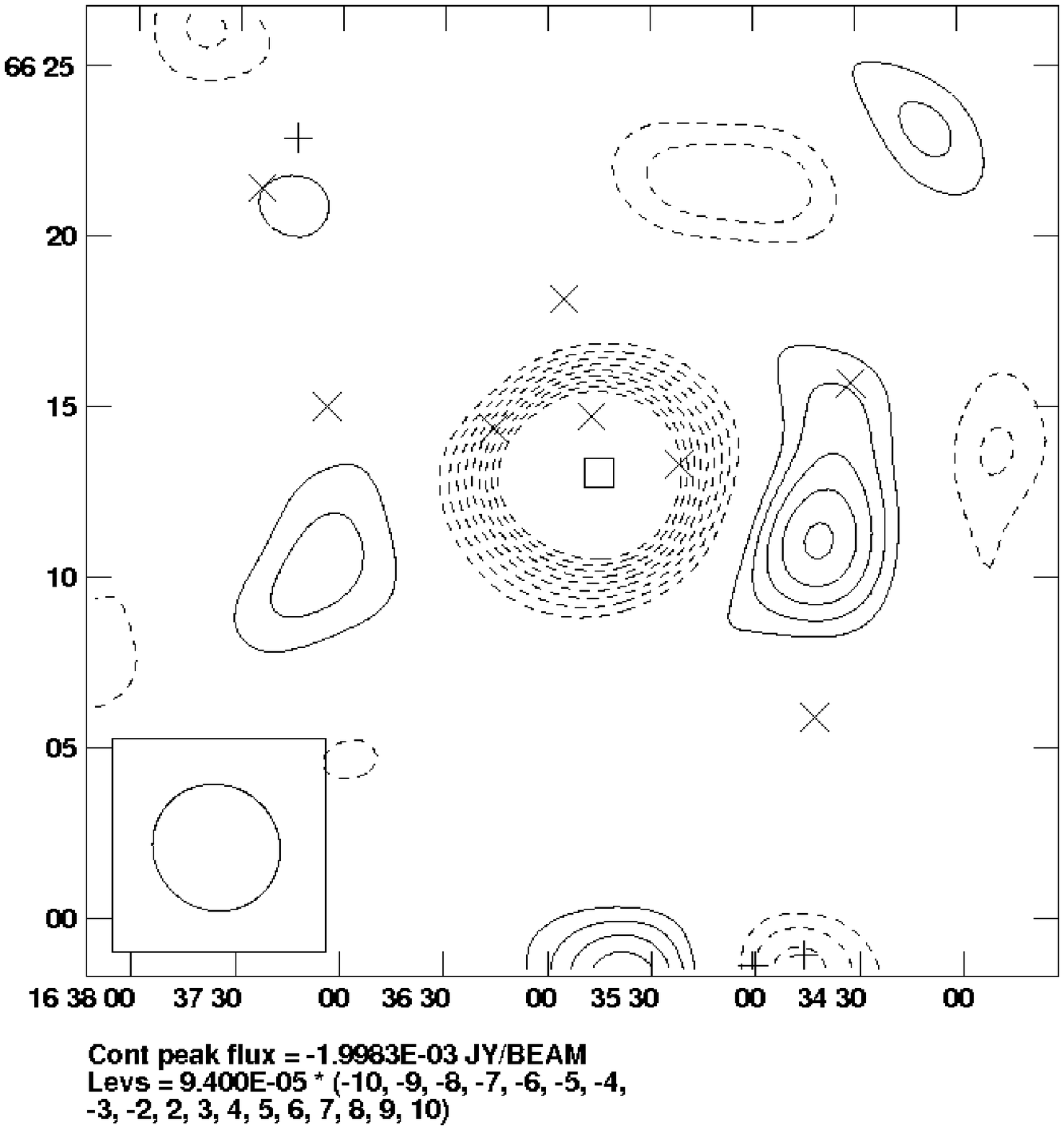}
& \quad &
{C}\includegraphics[width=7.5cm,height=7.4cm,clip=,angle=0.]{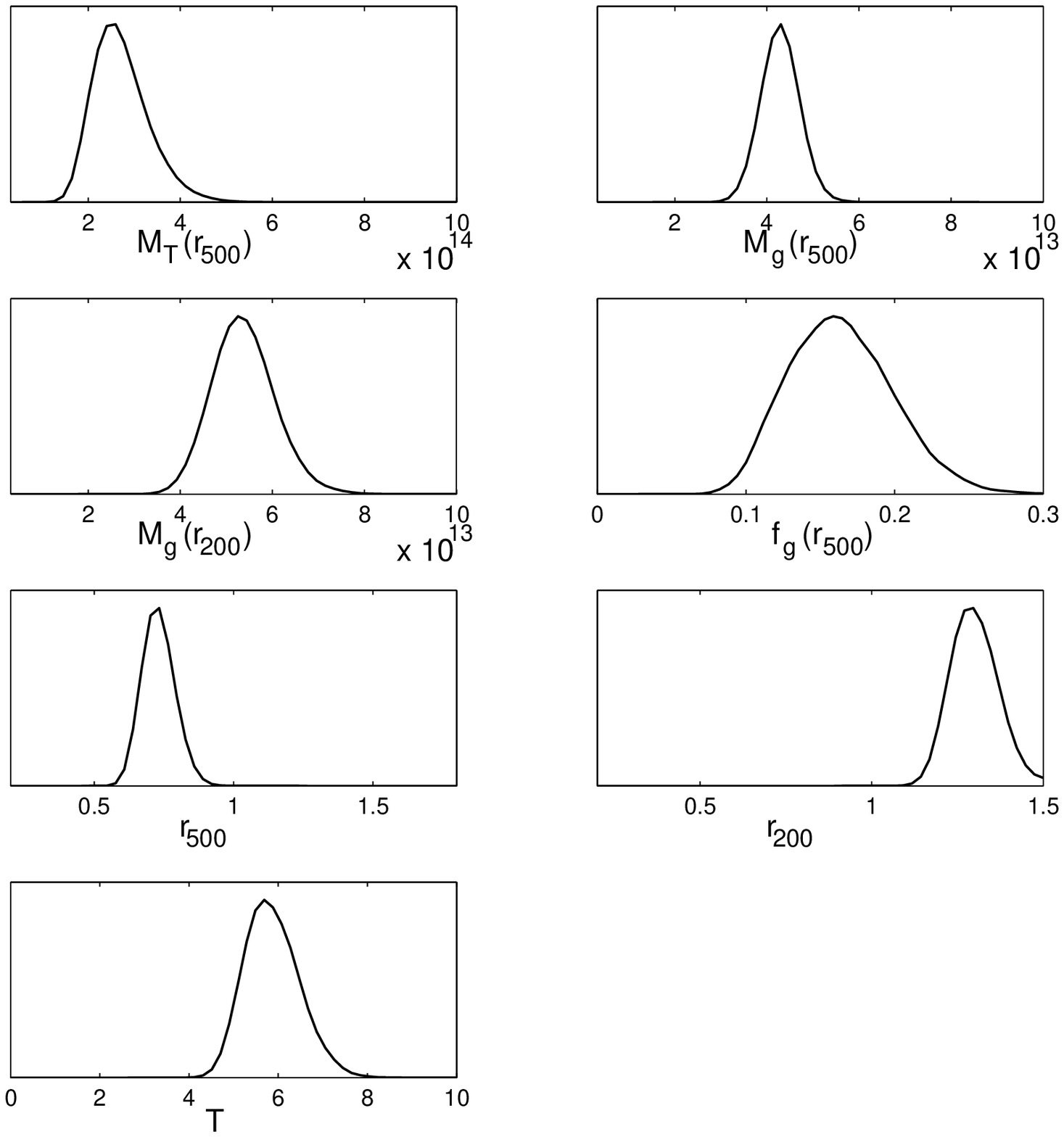}
\\
{E}\includegraphics[width=6.5cm,height=6.5cm,clip=,angle=0.]{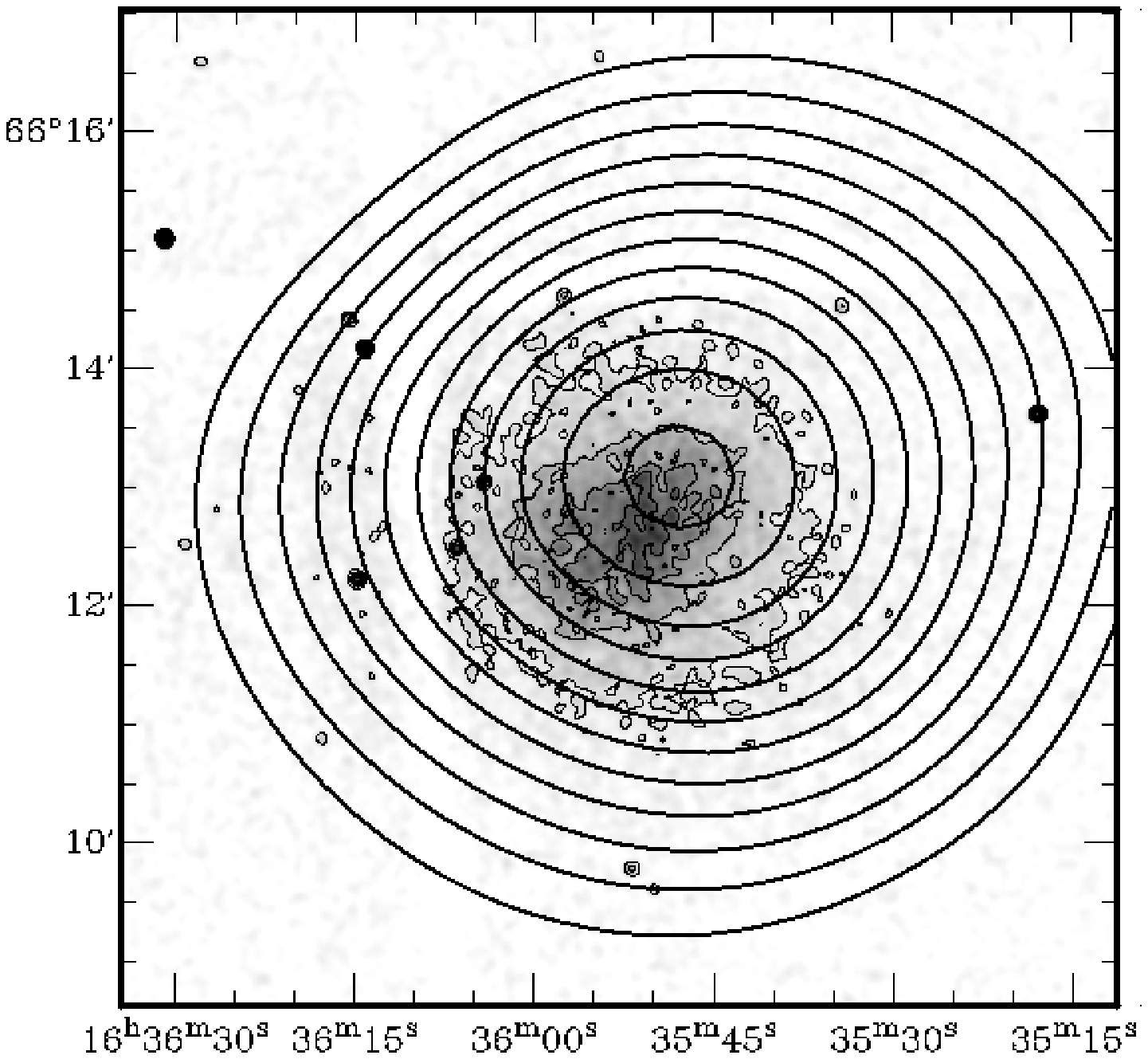}
& \quad &
{F}\includegraphics[width=6.5cm, height=6.5cm,clip=,angle=0.]{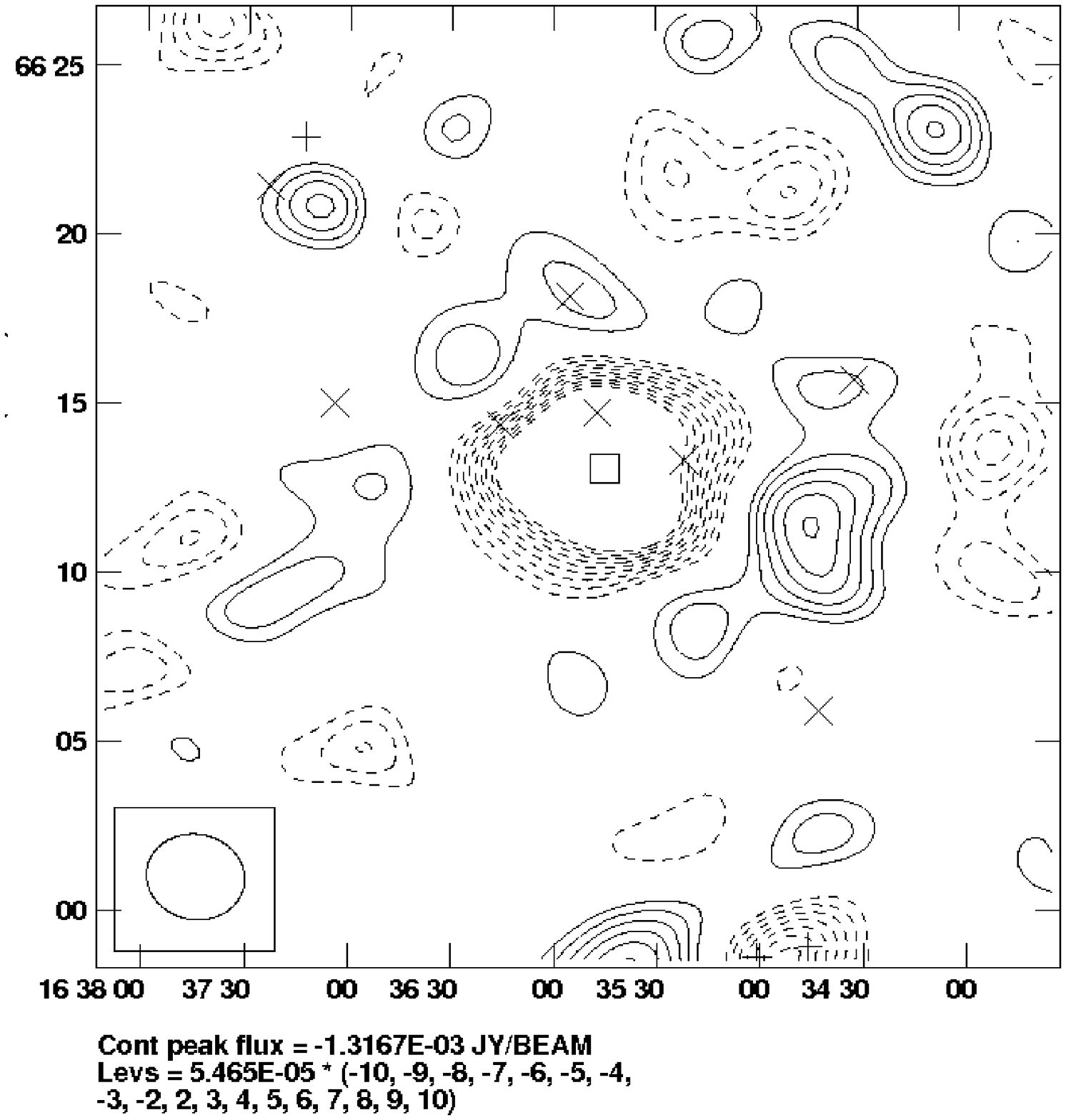}


%
\end{tabular}
\caption{Results for Abell~2218. Panels A and B show the SA map before and after source subtraction, respectively; a $0.6$\,k$\lambda$ taper has been applied to B. The box in panels A and B indicates the cluster SZ centroid, for the other symbols see Tab. \ref{tab:sourcelabel}. The smoothed {\sc{Chandra}} X-ray map overlaid with contours from B is presented in image C.  Panels D and E show the marginalized posterior distributions for the cluster sampling and derived parameters, respectively.  In panel D $M_{\rm{T}}$ is given in units of $h_{100}^{-1}\times10^{14}M_{\odot}$ and $f_{\rm{g}}$ in $h_{100}^{-1}$; both parameters are estimated within $r_{200}$.  In E $M_{\rm{g}}$ is in units of $h_{100}^{-2}M_{\odot}$, $r$ in $h_{100}^{-1}$Mpc and $T$ in KeV. F shows the higher-resolution source-subtracted map (no taper). Contours of the map in panel C are not the same as in panel B.; they range from -1.388 to
-0.188 mJy$\,$beam$^{-1}$ in steps of $+$0.15 mJy$\,$beam$^{-1}$. }
\label{fig:A2218}
\end{center}
\end{figure*}

\begin{figure}
\includegraphics[width=8.0cm,height=8.0cm,clip=,angle=0.]{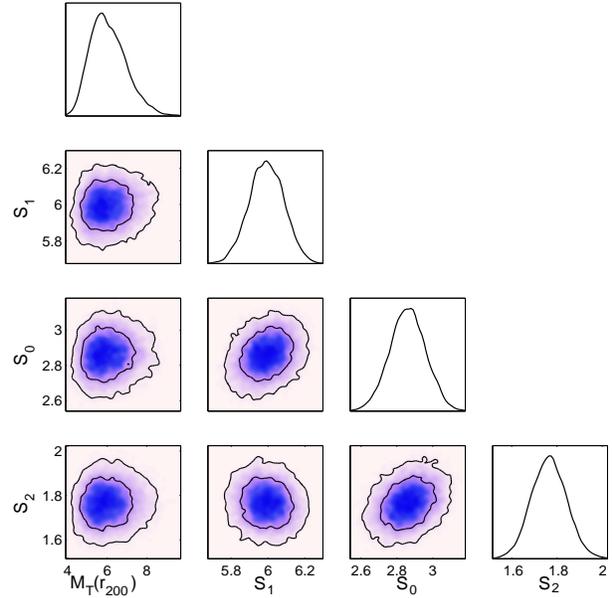}
\caption{2-D and 1-D marginalized posterior distributions for
$M_{\rm{T}}(r_{200})$ (in $h_{100}^{-1}\times 10^{14}M_{\odot}$) and source flux densities (in Jys) within $5\arcmin$ from the cluster X-ray
centroid for Abell 2218 (Tab. \ref{tab:source_info}).}
\label{fig:A2218_2}
\end{figure}

\subsection{ Abell~2409} \label{resultsA2409}

We detect a $12\sigma_{SA}$ SZ effect towards Abell 2409 in the
tapered, source-subtracted SA maps,
Fig. \ref{fig:A2409} B. Despite the high SNR we
are not able to obtain sensible parameter estimates for this
cluster. As shown in Fig. \ref{fig:A2409} A, the effect of some
 emission close to the
pointing centre is to give the decrement a shape that cannot be
well approximated by a spherical $\beta$-profile with free shape parameters. The
parameter estimates from {\textsc{McAdam}} are thus not reliable and we
present only the AMI SA map. Fixing the shape of the profile can improve the
fit to this cluster. Cluster parameters
for Abell~2409 from AMI data have been obtained using a
 gNFW parameterization -- see \cite{planckami}.

 The nature of the residual emission around the cluster is
uncertain. Pointed LA observations towards the location of these
sources of positive flux were made in an attempt to detect possible
sources lying just below our detection threshold. Despite the noise at these
locations on the LA map
reaching $\approx 50\mu$Jy beam$^{-1}$, no additional sources were detected; it
seems likely that
this is (at this resolution) extended emission with relatively low surface brightness. However, no
evidence for extended
emission was found in either the NVSS 1.4\,GHz or in the VLSS 74\,MHz maps.

  \begin{figure*}
\centerline{\huge{Abell 2409}}
\centerline{{A}\includegraphics[width=8.0cm,height=8.0cm,clip=,angle=0.]{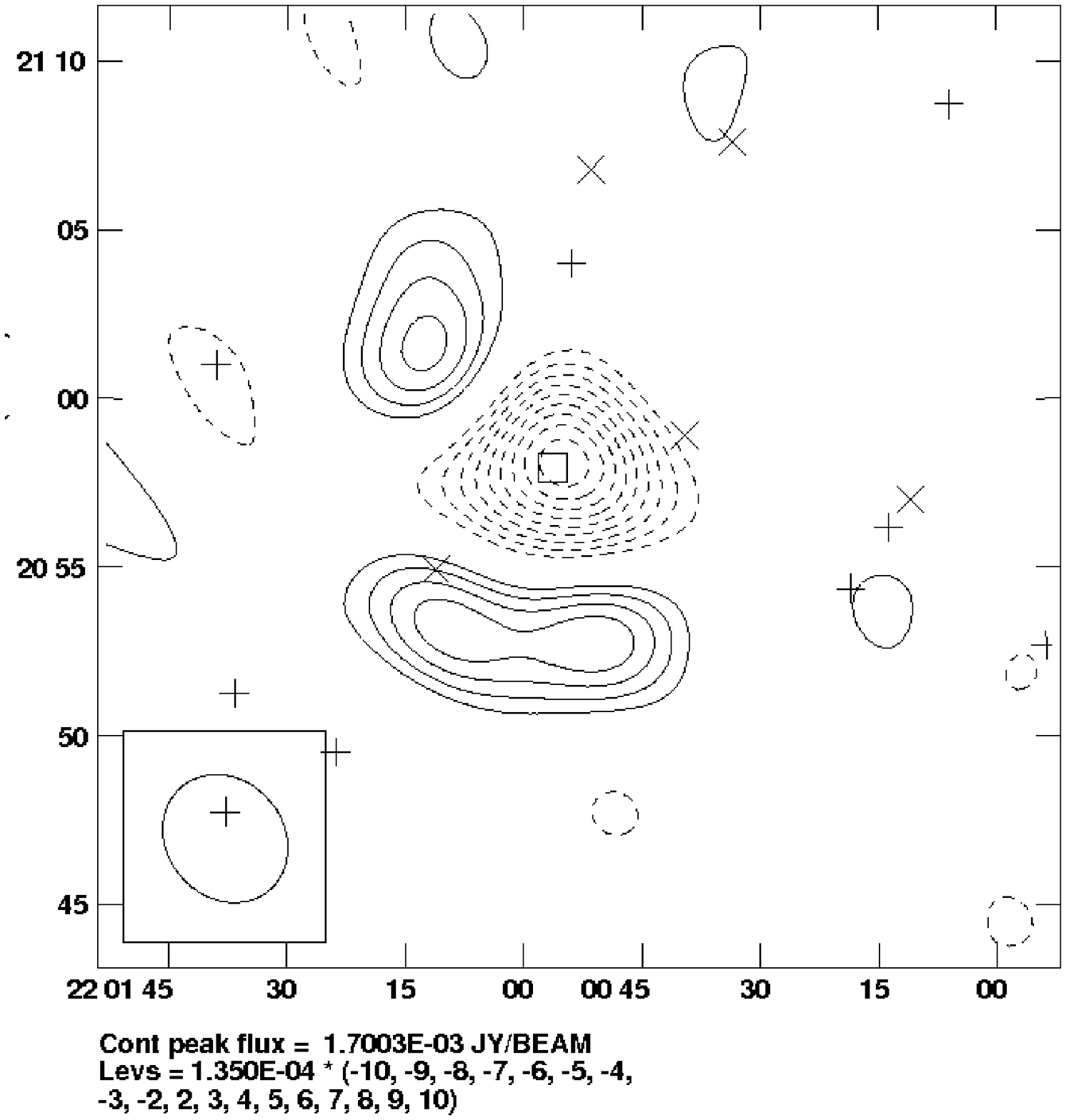}
\qquad{B}\includegraphics[width=8.0cm,height=8.0cm,clip=,angle=0.]{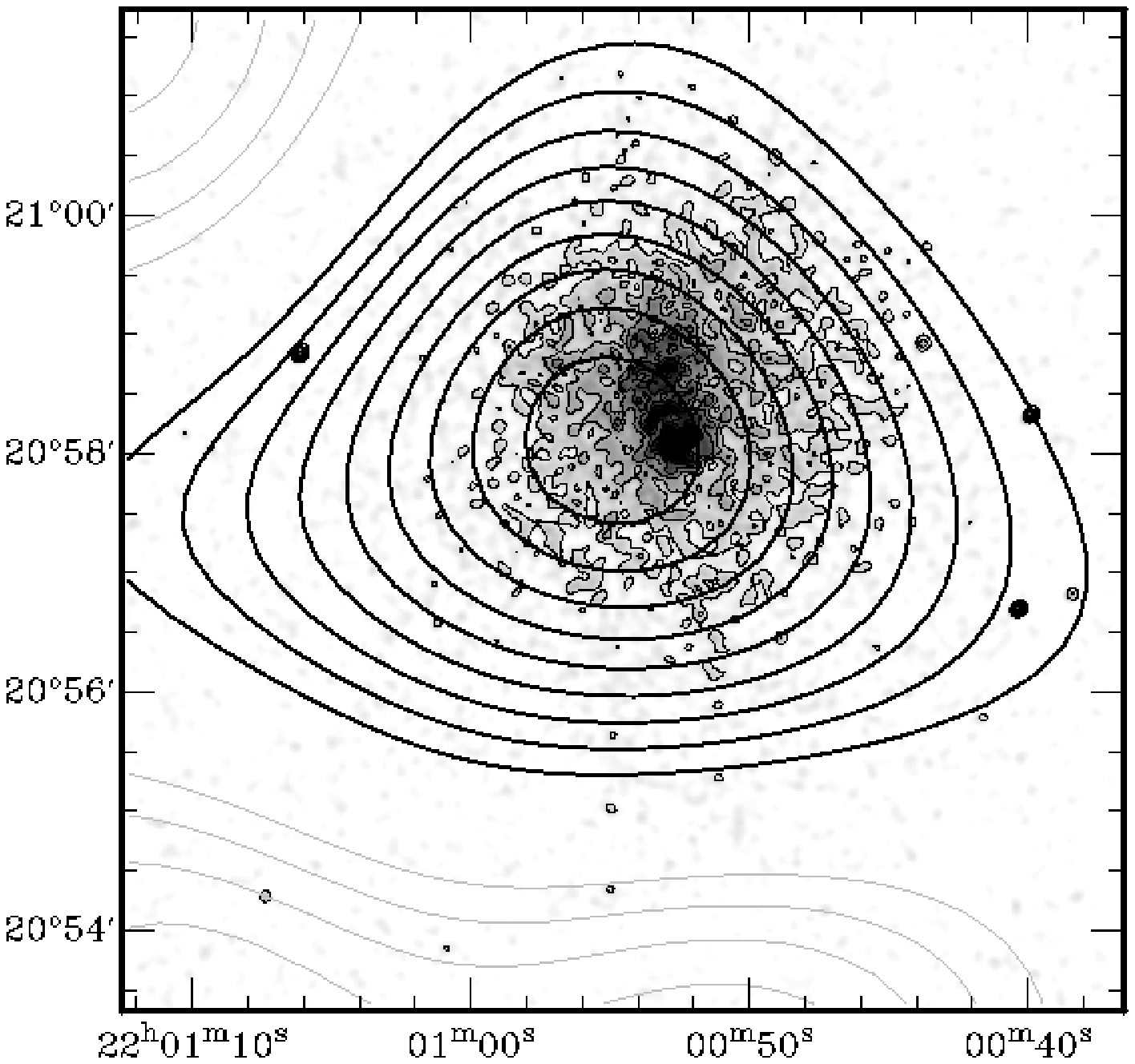}}

\caption{Results for Abell~2409. A: Source-subtracted SA map produced using a 0.6-k$\lambda$
  taper. B: SA map from A. overlaid onto {\sc{Chandra}} X-ray image. Contours increase linearly in units of $\sigma_{SA}$. }
\label{fig:A2409}
\end{figure*}

\subsection{ RXJ0142+2131} \label{resultsRXJ0142+2131}
 The maps and parameters for  RXJ0142+2131 are presented in Fig.\,\ref{fig:xrayszrxj0}. The source
environment
is not expected to contaminate our SZ detection, with the brightest source
having a flux density
of $\approx 2$\,mJy and lying several arcminutes away from the
cluster centroid; residual
emission after source subtraction is seen on the SA maps at the $1\sigma$
 level.
 The composite image of the SZ and X-ray data
  reveals good agreement between the emission peaks of these two datasets.

 A photometric and spectroscopic study of RXJ0142+2131 by
 \cite{baars2005} finds the velocity dispersion of this cluster
 ($\sigma_x=1278\pm134$\,km\,s$^{-1}$) to be surprisingly large, given its
 X-ray luminosity (Tab. \ref{tab:clusdetails}). This study indicates that galaxies
in this cluster have older
 luminosity-weighted mean ages than expected, which could be explained
 by a short increase in the star formation rate, possibly due to a
 cluster-cluster merger. Moreover, RXJ0142+2131 shows signs of not
 being fully virialized since the brightest cluster galaxy was
 found to be displaced by 1000\,km\,s$^{-1}$ from the systemic
 cluster redshift.
 \cite{okabe2010} fitted an NFW profile for the mass density to
\emph{Subaru/Suprime-Cam} data and assumed a spherical geometry for the cluster
to derive
 $M_{\rm{T}}(r_{500})=2.85^{+0.60}_{-0.53}\times10^{14}h_{72}^{-1}\rm{M}_{\odot}$
 and
 $M_{\rm{T}}(r_{200})=3.86^{+0.98}_{-0.82}\times10^{14}h_{72}^{-1}\rm{M}_{\odot}$ (using $h_{72}=1.0$).
 From our analysis, we find $M_{\rm{T}}(r_{500})=1.7 \pm
0.6\times10^{14}h_{100}^{-1}\rm{M}_{\odot}$ and $M_{\rm{T}}(r_{200})=3.7^{+1.1}_{-1.2}\times10^{14}h_{100}^{-1}\rm{M}_{\odot}$.


\begin{figure*}
\begin{center}
\begin{tabular}{m{8cm}cm{8cm}}
\multicolumn{3}{c}{\huge{RXJ0142+2131}} \\

{A}\includegraphics[width=7.5cm,height=7.5cm,clip=,angle=0.]{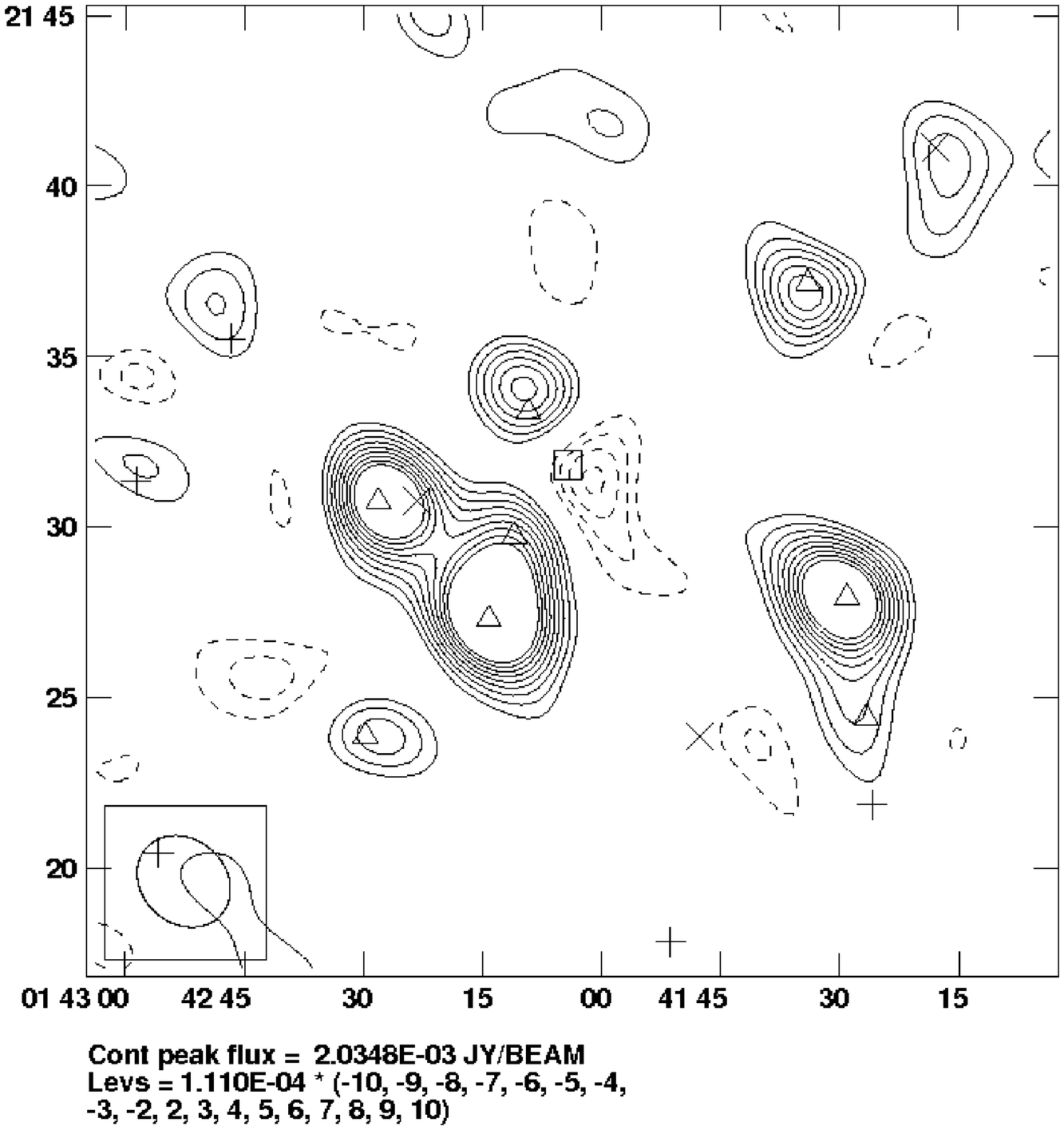}
& \quad &
{D}\includegraphics[width=7.5cm,height=7.5cm,clip=,angle=0.]{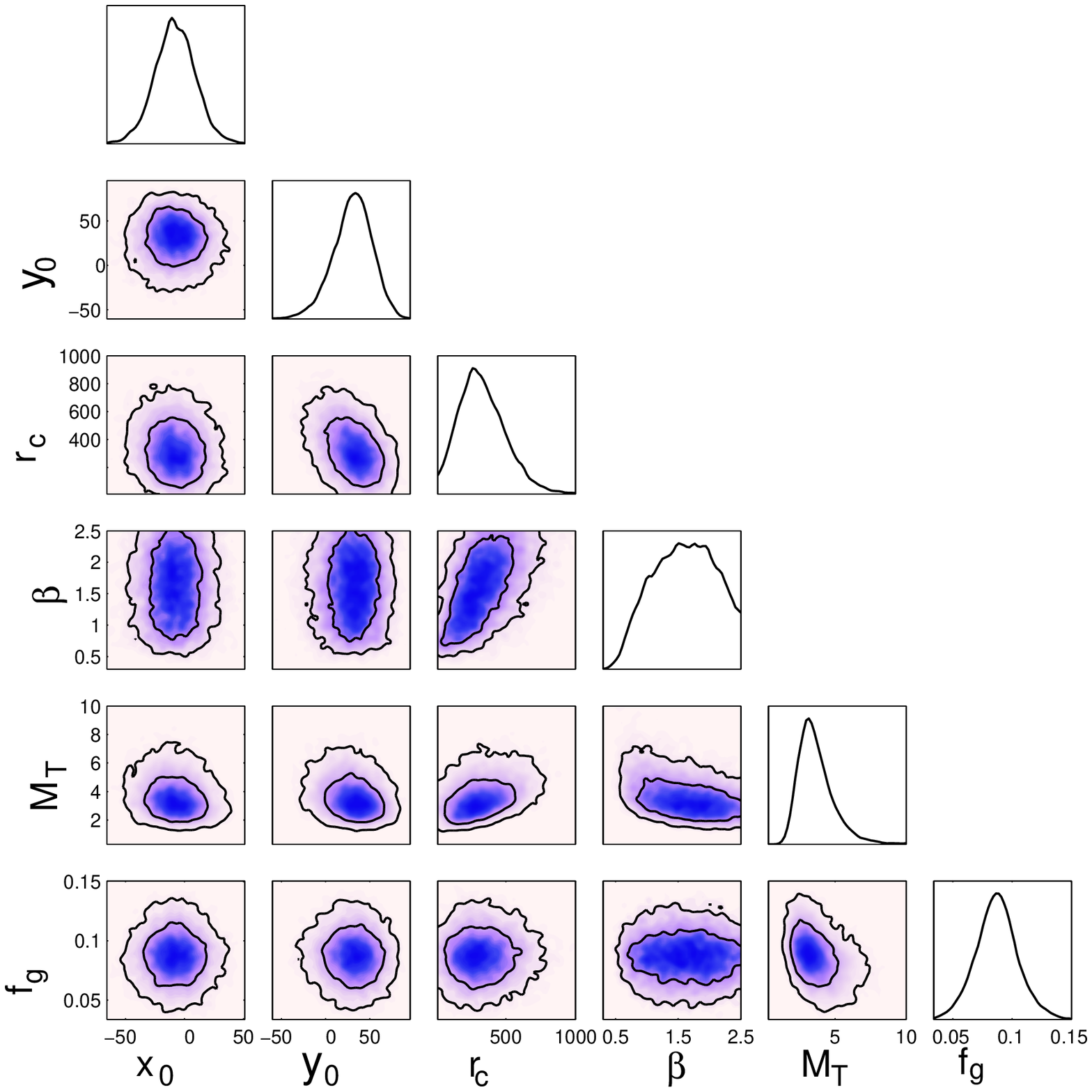} \\
{B}\includegraphics[width=7.5cm,height=7.5cm,clip=,angle=0.]{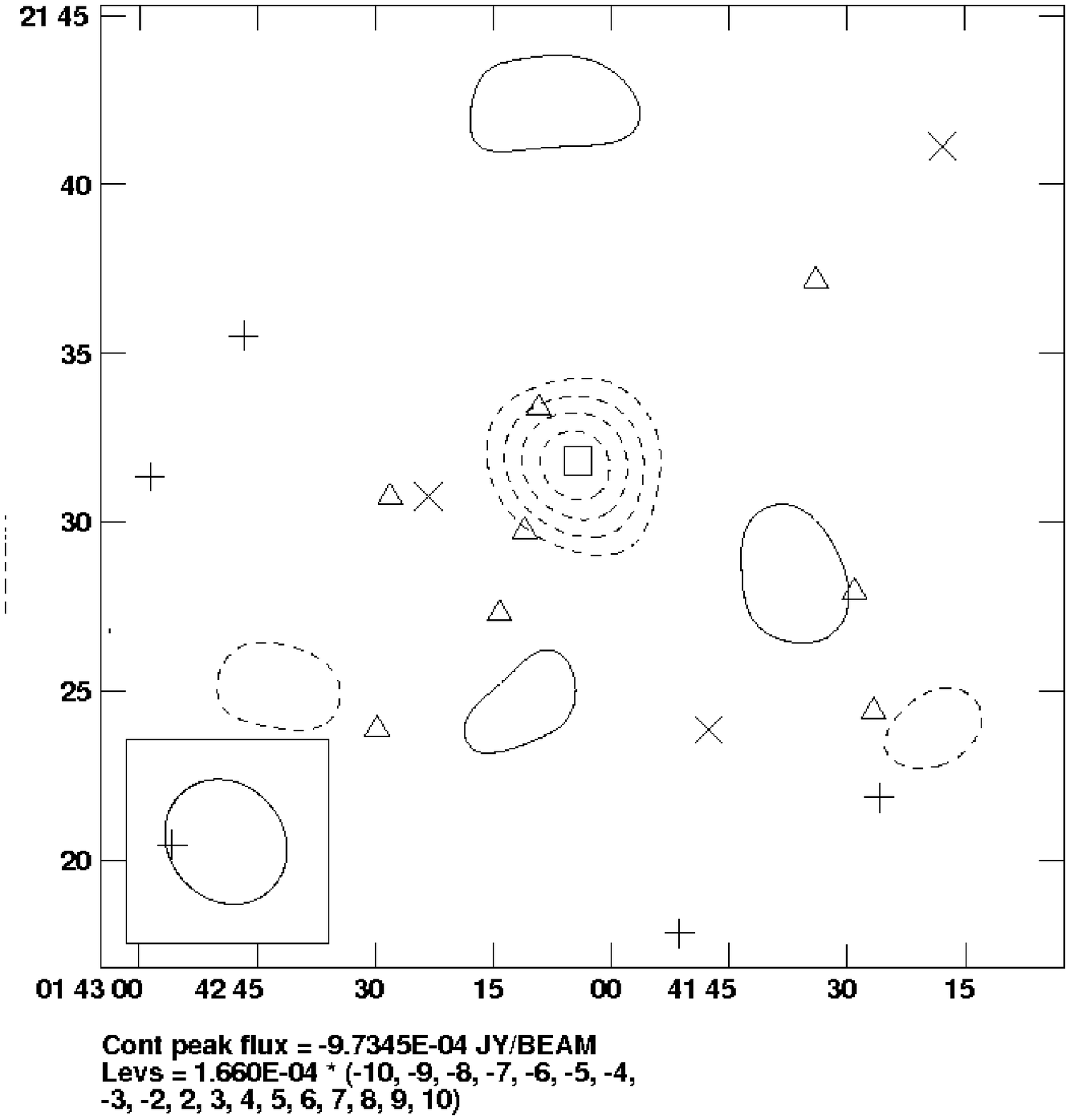}
& \quad &
{E}\includegraphics[width=7.5cm,height=7.5cm,clip=,angle=0.]{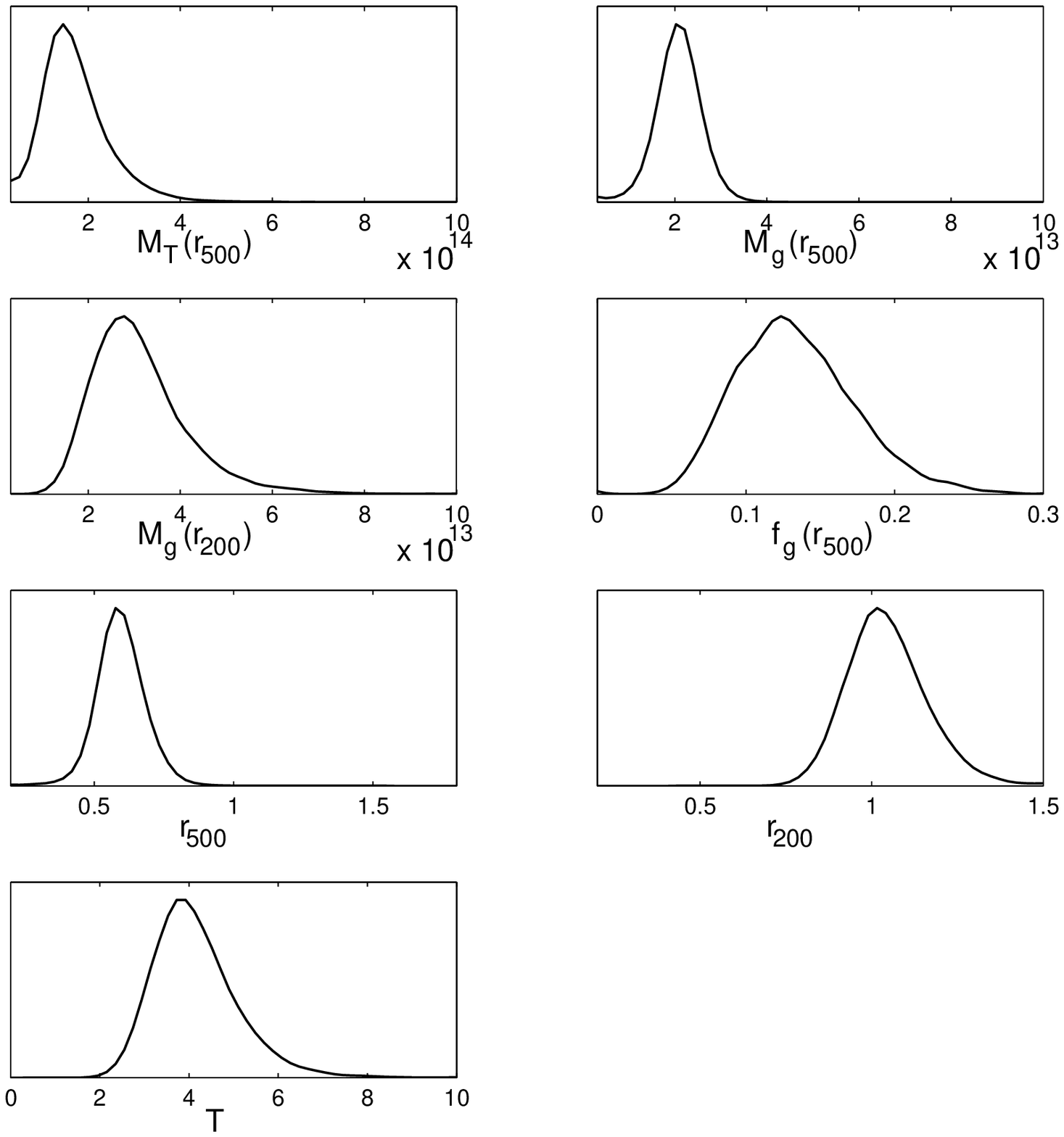} \\
%
{C}\includegraphics[width=7.0cm,height=6.5cm,clip=,angle=0.]{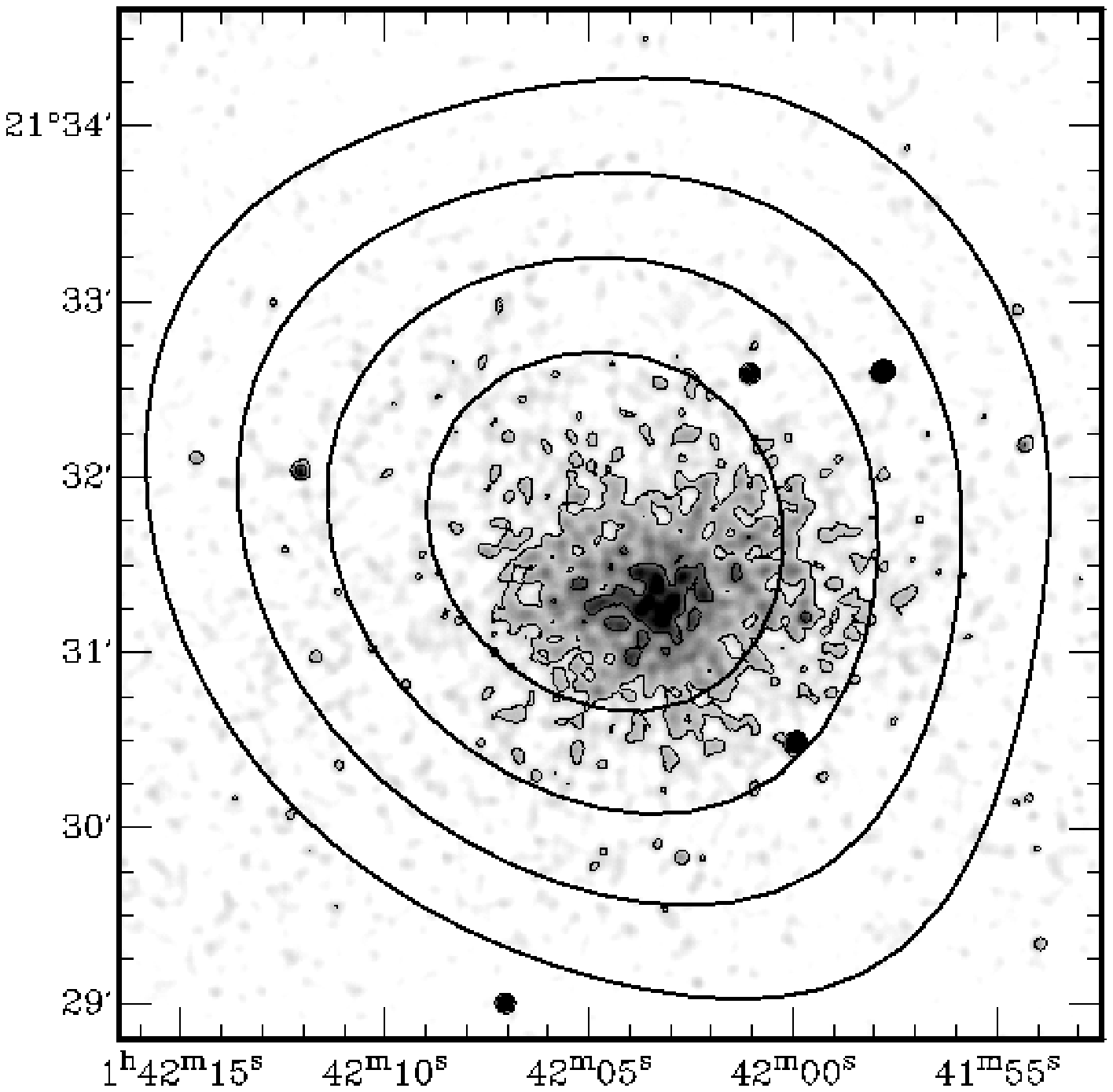}
& \quad &
{F}\includegraphics[width=7.0cm,height=6.5cm,clip=,angle=0.]{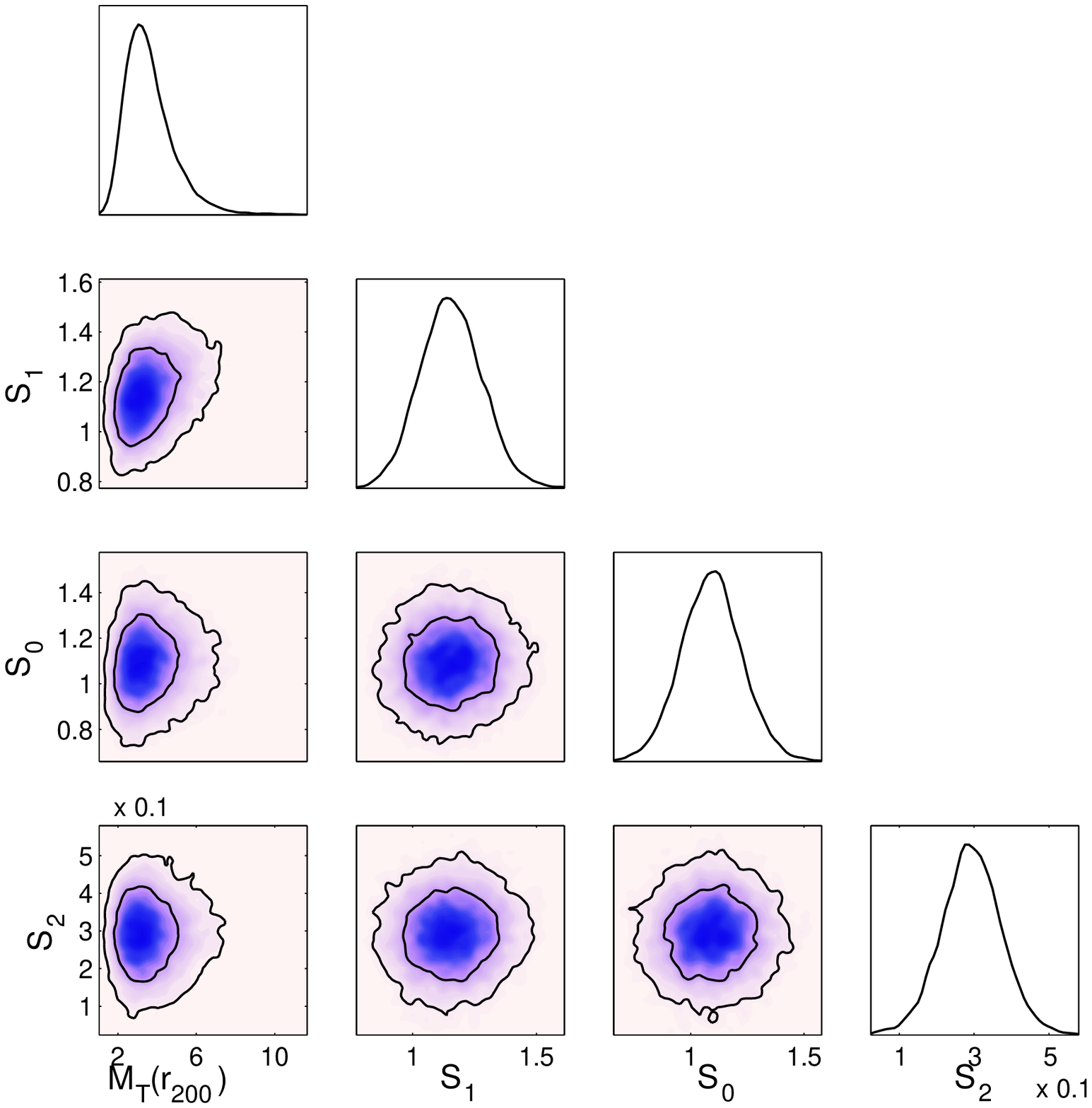}
\end{tabular}
\caption{Results for RXJ0142+2131. Panels A and B show the SA map before and after source subtraction, respectively; a $0.6$\,k$\lambda$ taper has been applied to B. The box in panels A and B indicates the cluster SZ centroid, for the other symbols see Tab. \ref{tab:sourcelabel}. The smoothed {\sc{Chandra}} X-ray map overlaid with contours from B is shown in image C. Panels D and E show the marginalized posterior distributions for the cluster sampling and derived parameters, respectively. F shows the 1 and 2-D marginalized posterior distributions for source flux densities (in Jys) within $5\arcmin$ of the cluster SZ centroid (see Tab. \ref{tab:source_info}) and $M_{\rm{T}}(r_{200})$ (in $h_{100}^{-1}\times 10^{14}M_{\odot}$). In panel D $M_{\rm{T}}$ is given in units of $h_{100}^{-1}\times10^{14}$ and $f_{\rm{g}}$ in $h_{100}^{-1}$; both parameters are estimated within $r_{200}$. In E $M_{\rm{g}}$ is in units of $h_{100}^{-2}M_{\odot}$, $r$ in $h_{100}^{-1}$Mpc and $T$ in KeV.}
\label{fig:xrayszrxj0} 
\end{center}
%
\end{figure*}


\subsection{RXJ1720.1+2638}\label{sec:R1720}

Results for RXJ1720.1+2638 are given in Fig. \ref{fig:RXJ1720.1+2638}.
At 16\,GHz the source environment around the cluster is challenging: in our LA
data we detect a 3.9\,mJy source at the same position as the
cluster, and several other
 sources with comparable flux densities within $4\arcmin$ from the cluster
centre. The difficulty of modelling this system is clear from the degeneracies between
some of the source flux densities and the cluster mass (Fig. \ref{fig:RXJ1720.1+2638} F).
However, we always recover a similarly asymmetric SZ decrement.

RXJ1720.1+2638 has been studied by \cite{RXJ1720_CHANDRA} and \cite{mazzotta2008}
through {\sc{Chandra}} observations. This cool-core cluster has two cold fronts
within $100\arcsec$ of the X-ray
 centroid and a regular morphology away from the core region; the authors
attribute the dynamics of this cluster to the sloshing scenario, in agreement
with later work by \cite{owers2011} using optical spectroscopy.
 Merger activity has also been suggested by Okabe et al. whose weak lensing
data reveal a second mass concentration to the North of the main cluster, while
the analysis of SDSS data by \cite{miller2005} finds no evidence of
substructure. Our data reveal a strong abundance of radio emission towards
this cluster
, including some extended emission, which might support the suggestion in
Mazzotta et al. (2001) that this cluster contains a low-frequency radio halo that did not
disappear after the merger event.

 Mazzotta et al. 2001 determined the mass profile for the cluster assuming
hydrostatic equilibrium to be $M_{\rm{T}}(< r=1000\rm{kpc})= 5^{+3}_{-2}\times
10^{14}h^{-1}_{50}\rm{M_{\odot}}$. We find 
 $M_{\rm{T}}(r_{500}) = 1.2\pm 0.2\times 10^{14}h_{100}^{-1}\rm{M_{\odot}}$.

 \begin{figure*}
\centerline{\huge{RXJ1720.1+2638}}
\centerline{\includegraphics[width=7.5cm,height=7.4cm,clip=,angle=0.]{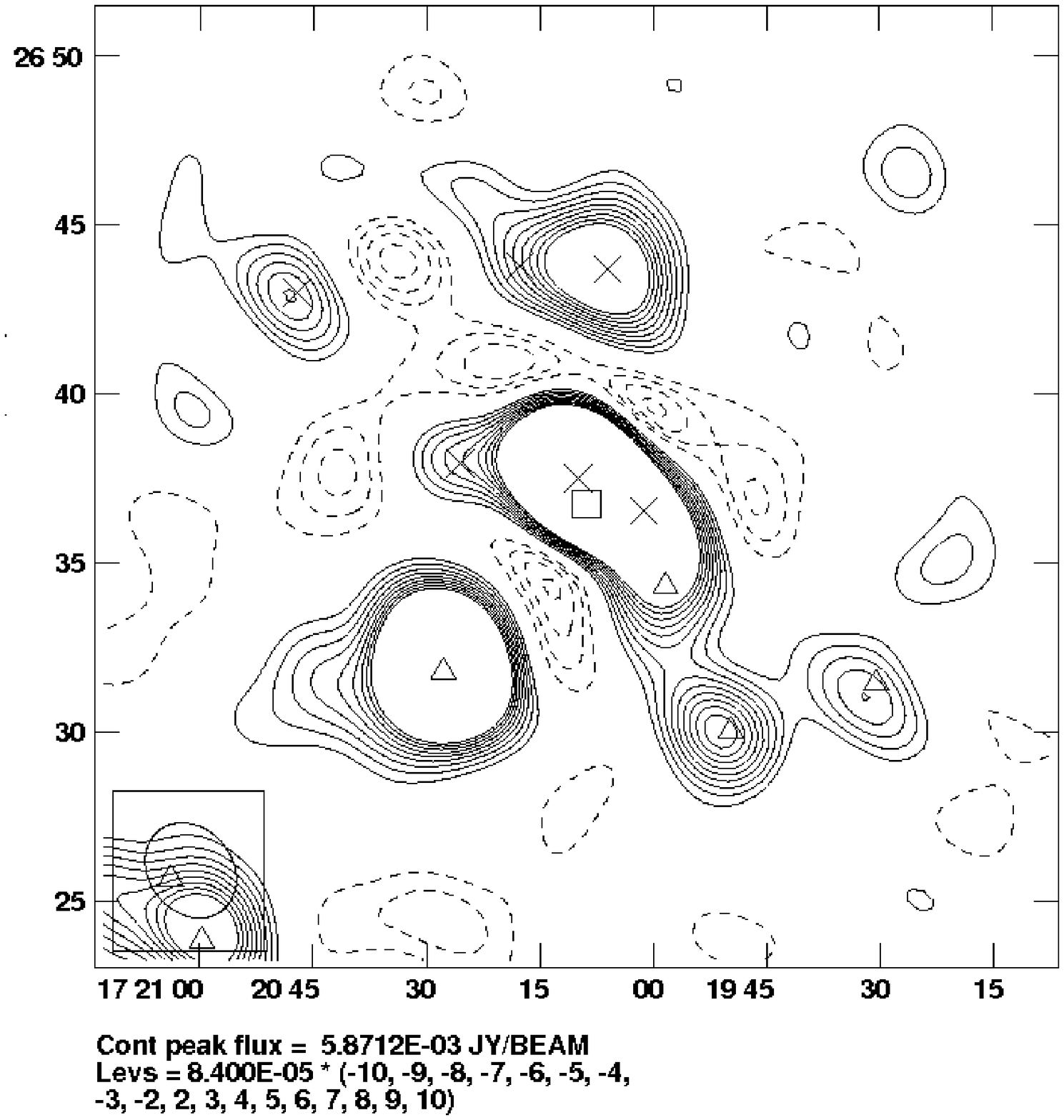}\qquad\includegraphics[width=7.5cm,height=7.4cm,clip=,angle=0.]{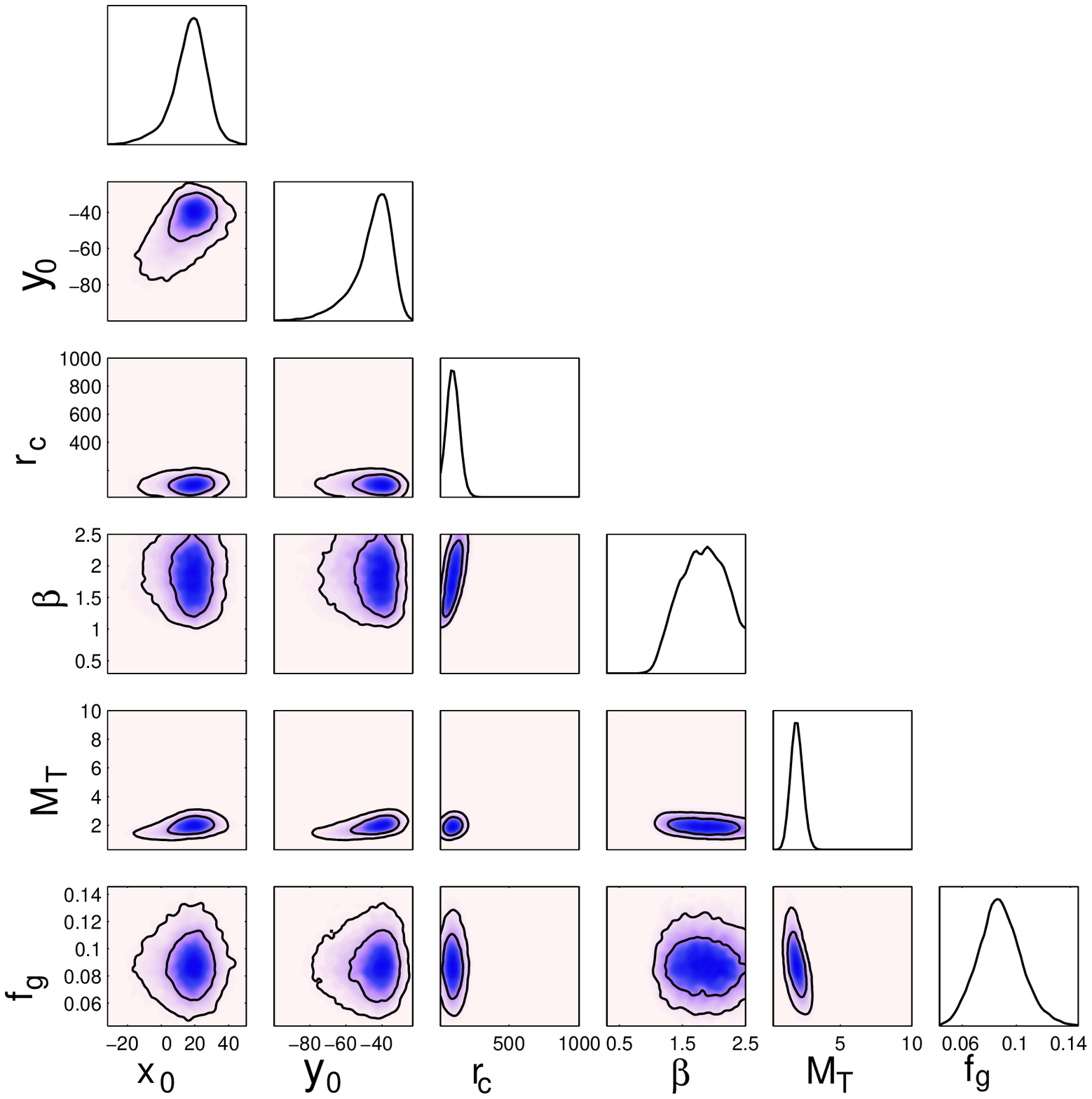}}
 \centerline{\includegraphics[width=7.3cm,height=7.3cm,clip=,angle=0.]{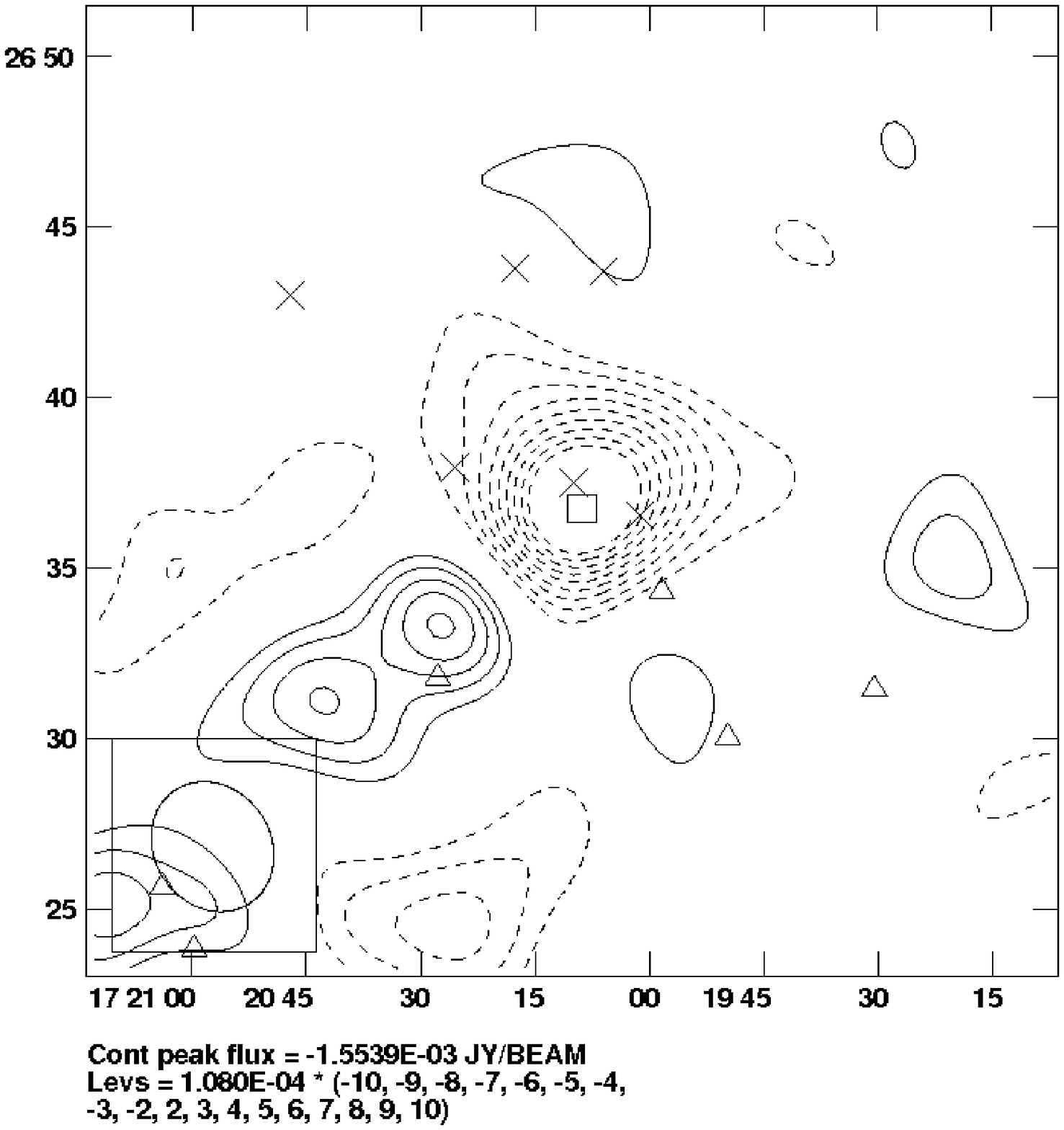}\qquad\includegraphics[width=7.3cm,height=7.3cm,clip=,angle=0.]{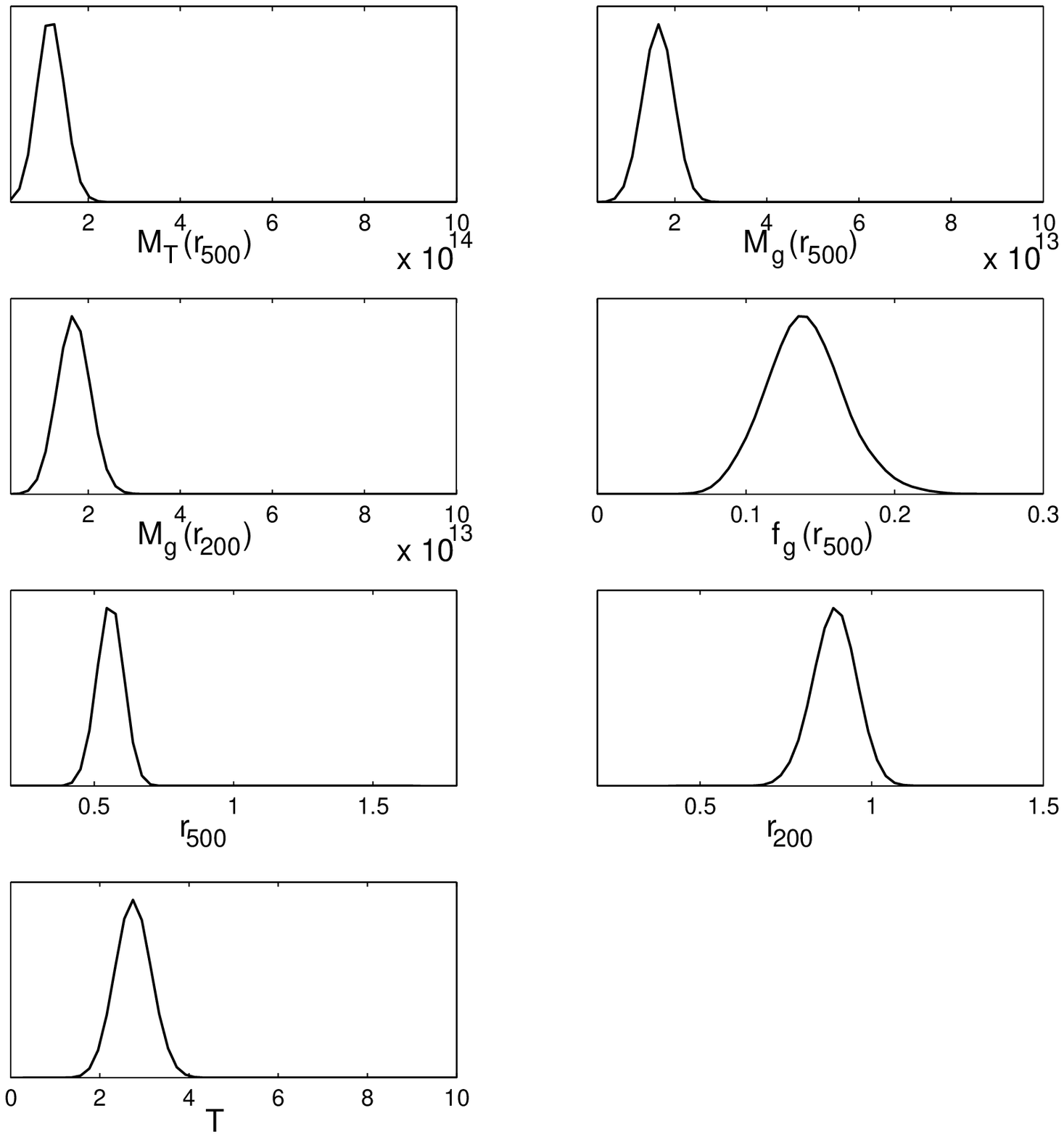}}
 \centerline{\includegraphics[width=7.1cm,height=7.1cm,clip=,angle=0.]{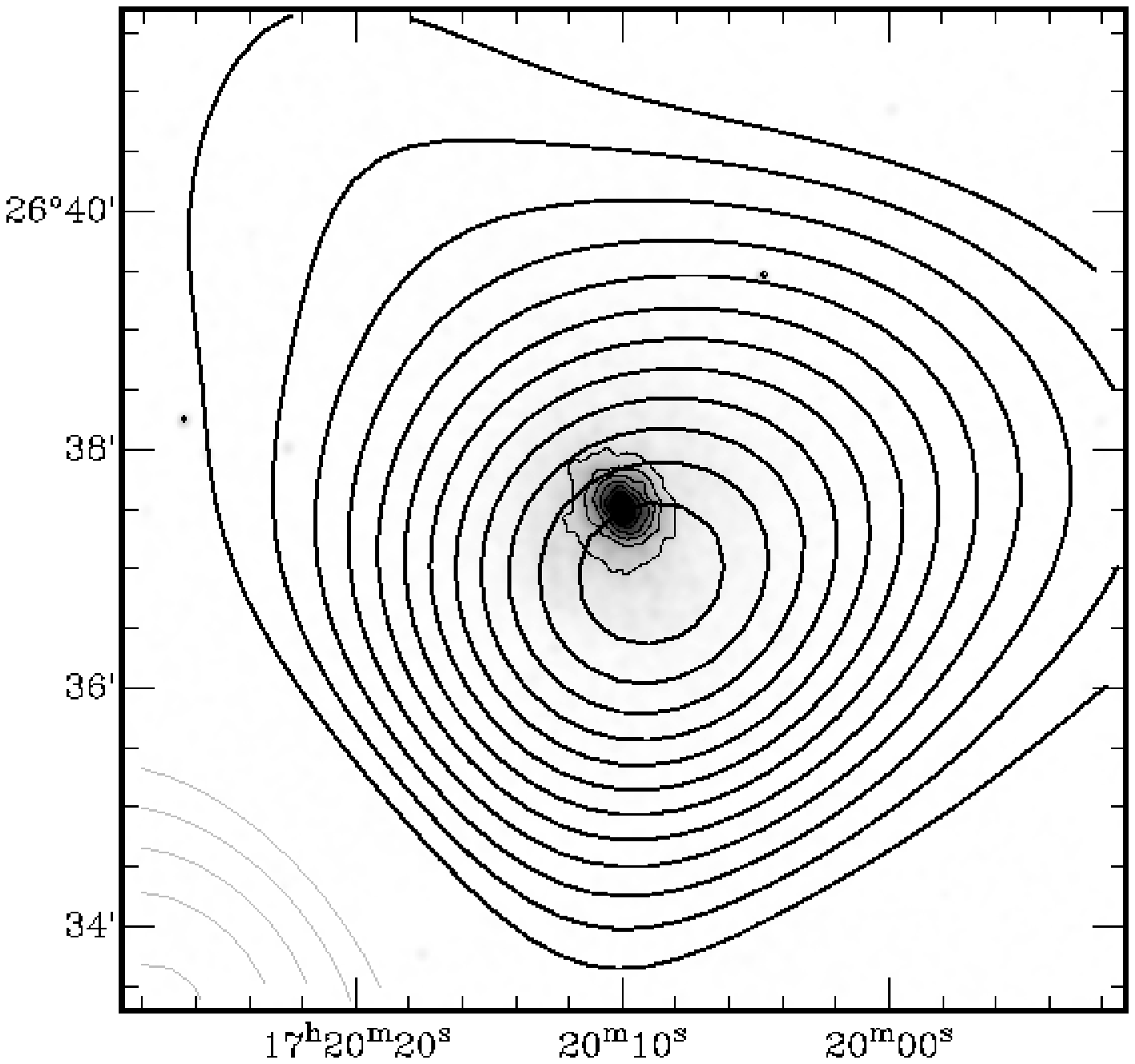}\qquad\includegraphics[width=7.1cm,height=7.1cm,clip=,angle=0.]{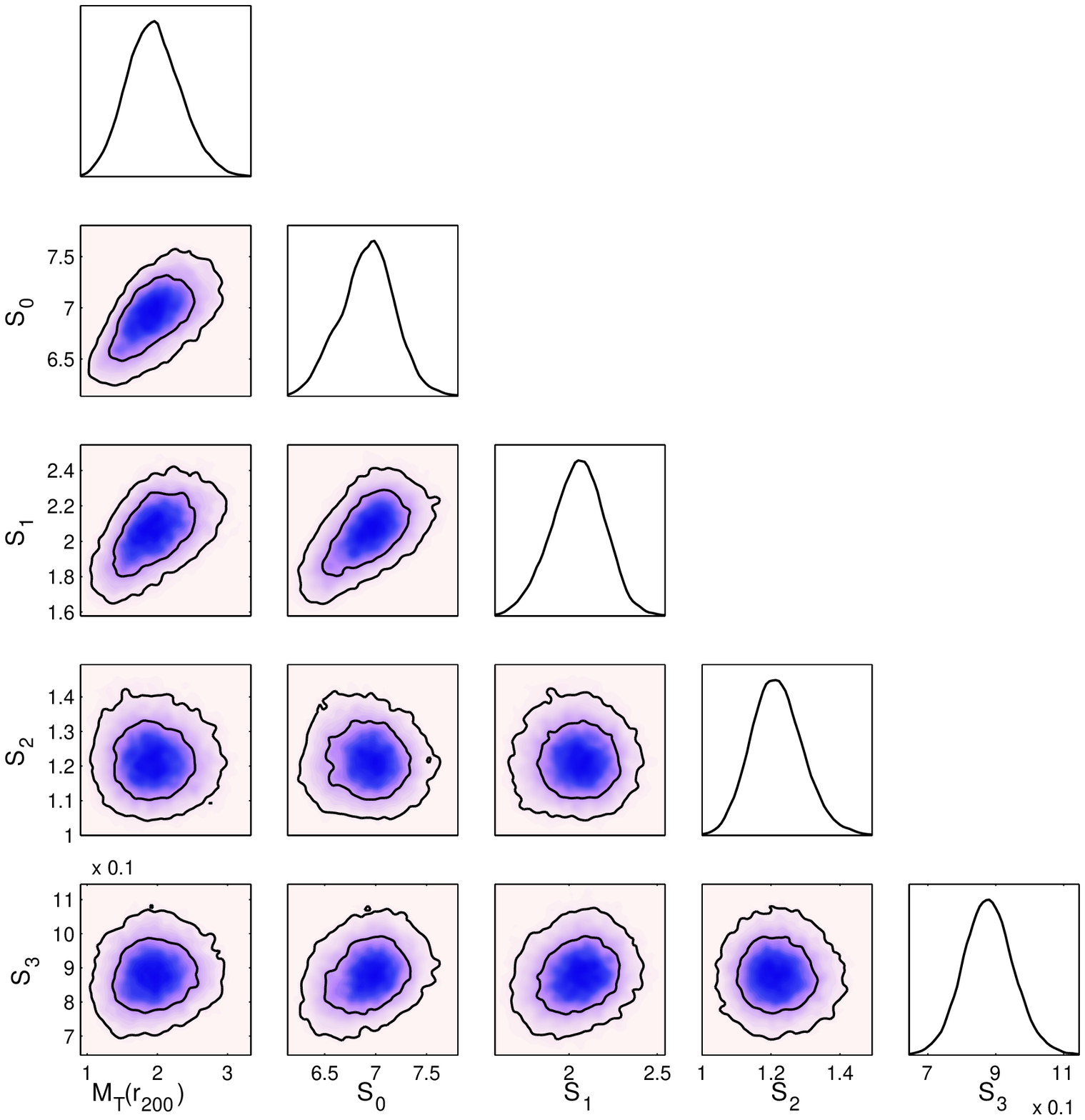}}

\caption{Results for RXJ1720.1+2638. Panels A and B show the SA map before and after source subtraction, respectively; a $0.6$\,k$\lambda$ taper has been applied to B. The box in panels A and B indicates the cluster SZ centroid, for the other symbols see Tab. \ref{tab:sourcelabel}. The smoothed {\sc{Chandra}} X-ray map overlaid with contours from B is shown in image C. Panels D and E show the marginalized posterior distributions for the cluster sampling and derived parameters, respectively. F shows the 1 and 2-D marginalized posterior distributions for source flux densities (in Jys) given in Tab. \ref{tab:source_info} and $M_{\rm{T}}(r_{200})$ (in $h_{100}^{-1}\times 10^{14}M_{\odot}$). In panel D $M_{\rm{T}}$ is given in units of $h_{100}^{-1}\times10^{14}M_{\odot}$ and $f_{\rm{g}}$ in $h_{100}^{-1}$; both parameters are estimated within $r_{200}$. In E $M_{\rm{g}}$ is in units of $h_{100}^{-2}M_{\odot}$, $r$ in $h_{100}^{-1}$Mpc and $T$ in KeV.}
\label{fig:RXJ1720.1+2638}
\end{figure*}

\subsection{  Zw0857.9+2107} \label{resultsZW}
\begin{figure}

\centerline{\huge{Zw0857.9+2107}}
\includegraphics[width=7.5cm,height=7.0cm,clip=,angle=0.]{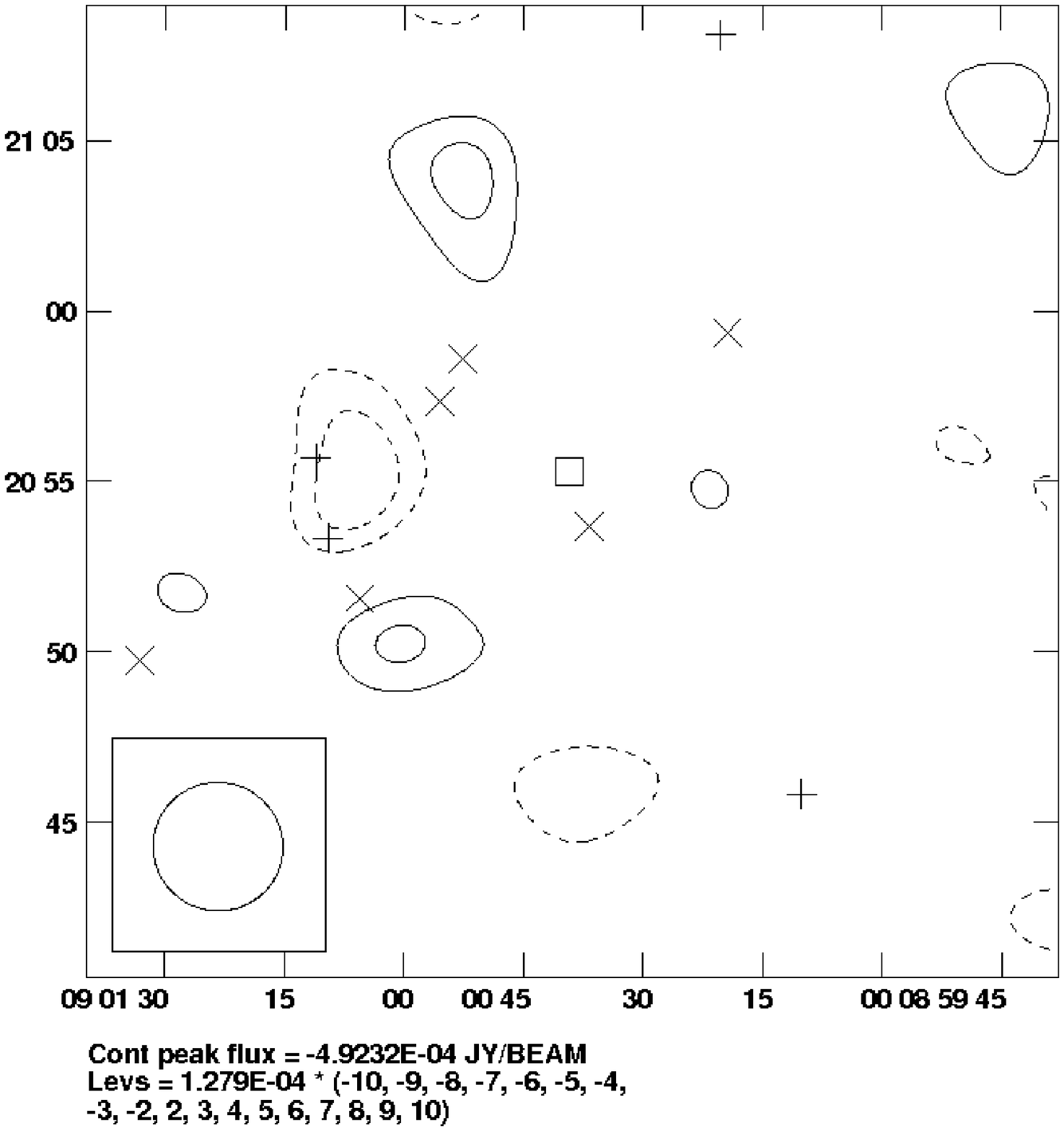}
\includegraphics[width=7.5cm,height=7.0cm,clip=,angle=0.]{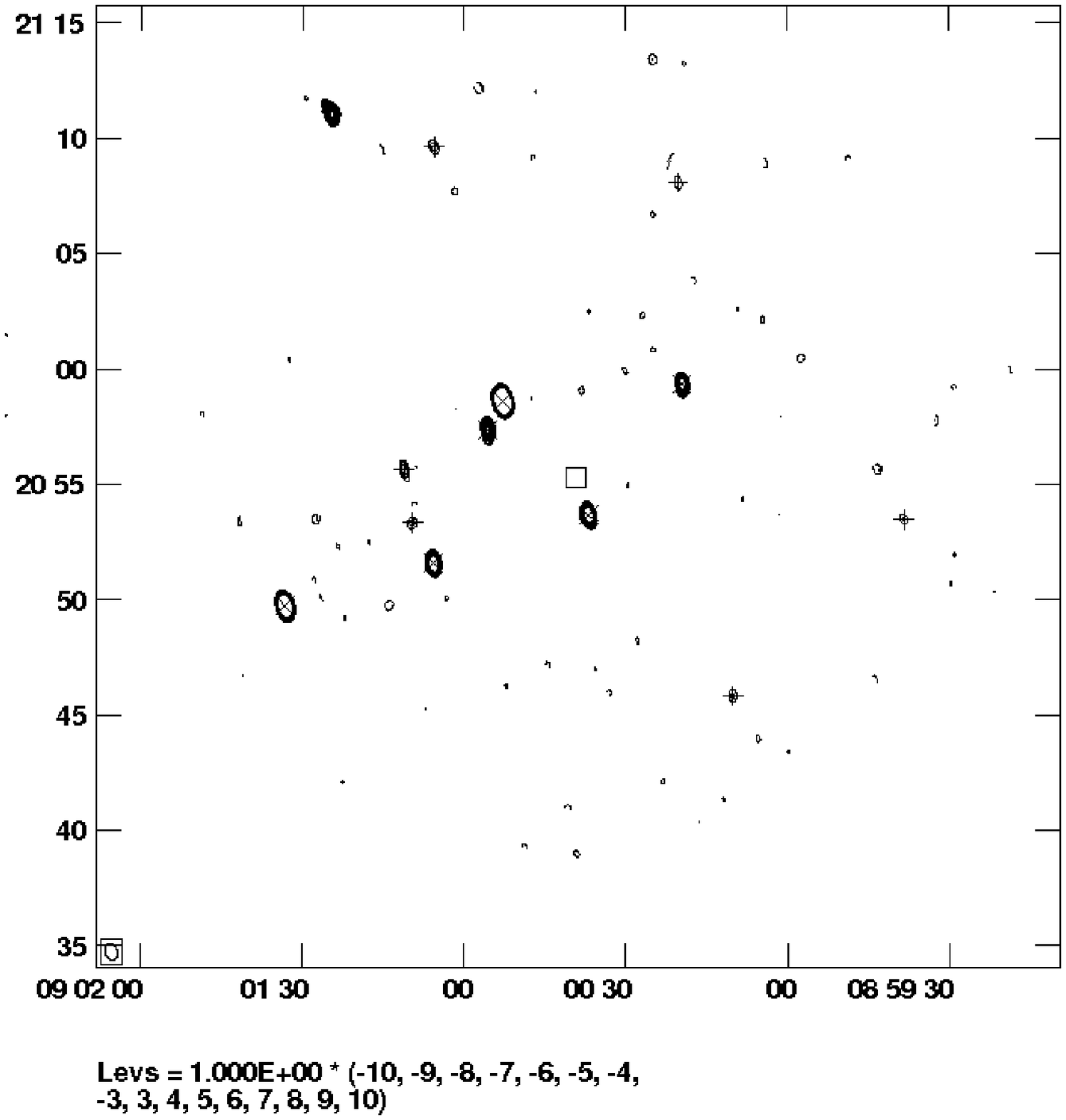}
\includegraphics[width=7.5cm,height=7.0cm,clip=,angle=0.]{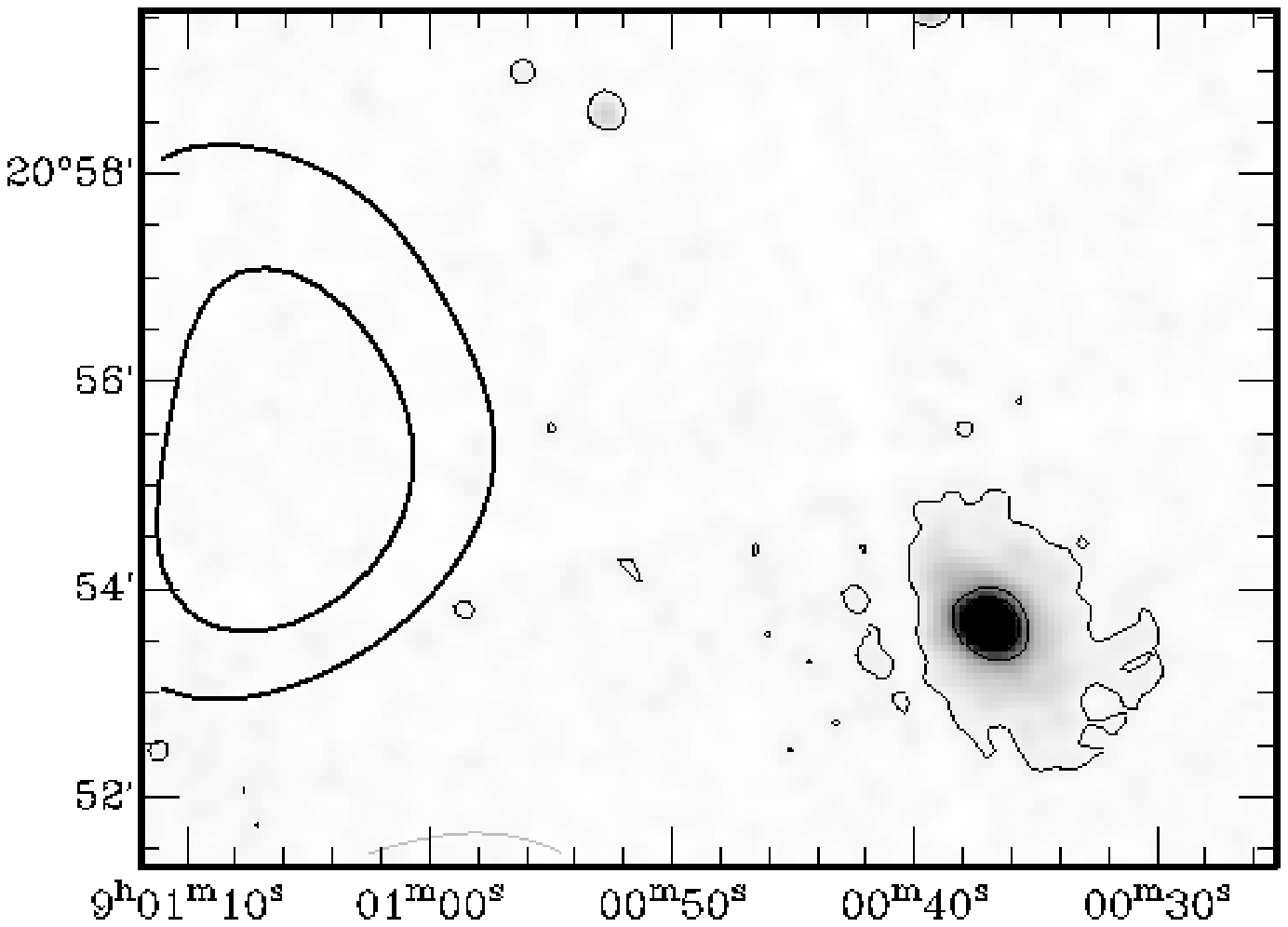}
\caption{Top  panel: Source-subtracted SA map for Zw0857.9+2107 produced using a 0.6-k$\lambda$
  taper. The contours increase linearly in units of
  $\sigma_{SA}$. Middle panel: LA signal-to-noise map. Contours start at $3\sigma$ and increase linearly to $10\sigma$, 
 where $\sigma=97 mu$Jy\,Beam$^{-1}$.
  Bottom panel: SA contours overlaid onto the Chandra
  X-ray image. The SA contours are the same as in the upper
  panel.  }
\label{fig:ZW2089}
\end{figure}

We report a null detection of an SZ
signal towards this cluster, despite the low noise levels on our SA maps and a
seemingly benign source environment. We reached a noise level ($1\sigma$) of
 97$\mu$Jy\,beam$^{-1}$ on the LA map (Fig. \ref{fig:ZW2089}, middle panel) and found no evidence
for sources below our $4\sigma_{LA}$ detection threshold.
 We detect a $1.4$\,mJy radio source at the location of the peak
X-ray
signal (see the electronic version in the ACCEPT {\sc{Chandra}}
data archive for a higher resolution X-ray image) but we seem to be able to
subtract it well from the SA maps.

Zw0857.9+2107 is not a well-studied cluster.
There are two temperature measurements for the cluster gas in
Zw0857.9+2107 from the ACCEPT {\sc{Chandra}} data archive
\citep{cavagnolo2009}: $T\approx 3\pm4$\,keV between
$\approx 10<r<100$\,kpc and
  $T\approx4.2\pm2.2$\,keV between
$\approx 100<r<600$\,kpc. One might expect the average temperature for
the cluster to be even lower at larger radii, such that $T(r_{200})<3$\,keV.
The absence of an
SZ signal could be explained by a sharp radial drop in $T$ or, perhaps, this
cluster is particularly
dense and compact such that it is X-ray bright but does not produce a strong SZ
signal on the scales
AMI is sensitive to (Alastair Edge, private communication). Fig. \ref{fig:ZW7160} illustrates what the marginalized
parameter distributions look like for non-detections such as this.

\subsection{Zw1454.8+2233}

We detect no SZ effect in the AMI data towards Zw1454.8+2233, despite the low
noise levels of our SA maps.
We detect several sources close to the cluster centre, including ones with a
flux density of 1.64\,mJy, 1.55\,mJy and
8.4\,mJy (at 13$\arcsec$, $1.8\arcmin$ and $4.3\arcmin$
 away from the pointing centre, respectively).
The SA maps and derived parameters are shown in Fig. \ref{fig:ZW7160}. The
derived parameters for this non-detection are as expected: we find that
$M_{\rm{T}}(r_{200})$ approaches our lower prior limit and that $M_g$ shows similiar behaviour (see e.g., Fig. \ref{fig:ZW7160}).

Zhang et al. found $M_{\rm{T}}(r_{500})= 2.4\pm
0.7\times 10^{14}\rm{M_{\odot}}$ using \emph{XMM-Newton}, assuming
isothermality, spherical symmetry and $h_{70}=1$. {\sc{Chandra}}
X-ray observations by \cite{Zw1454.8+2233_COOLING} suggest the cluster has a cooling flow and 
\cite{Zw1454.8+2233_HALO}
 find from 610-MHz GMRT observations that the cluster
has a core-halo radio source.

\begin{figure*}
\centerline{\huge{Zw1454.8+2233}}
\centerline{{A}\includegraphics[width=7.5cm,height=7.5cm,clip=,angle=0.]{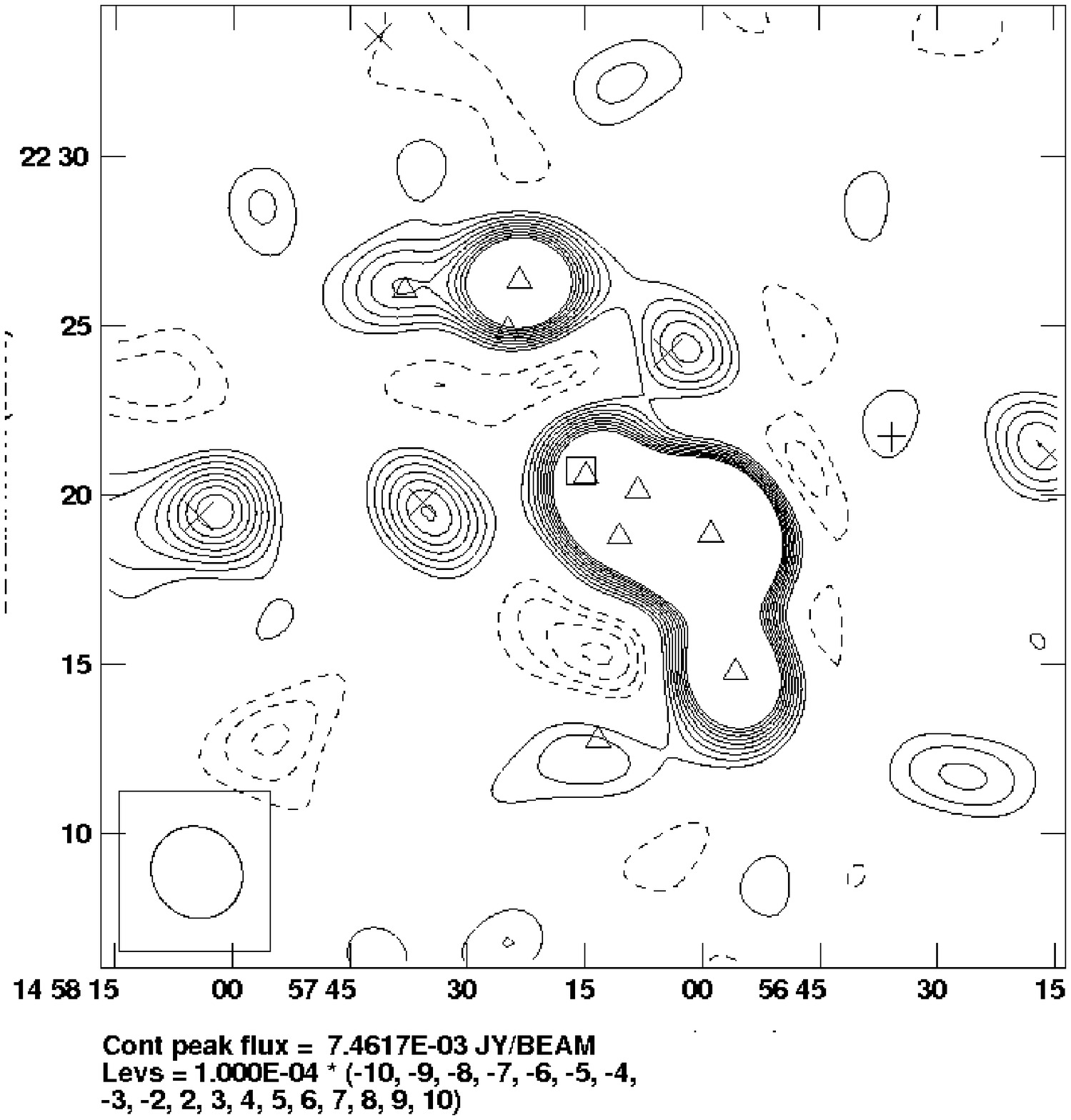}\qquad{D}\includegraphics[width=7.5cm,height=7.5cm,clip=,angle=0.]{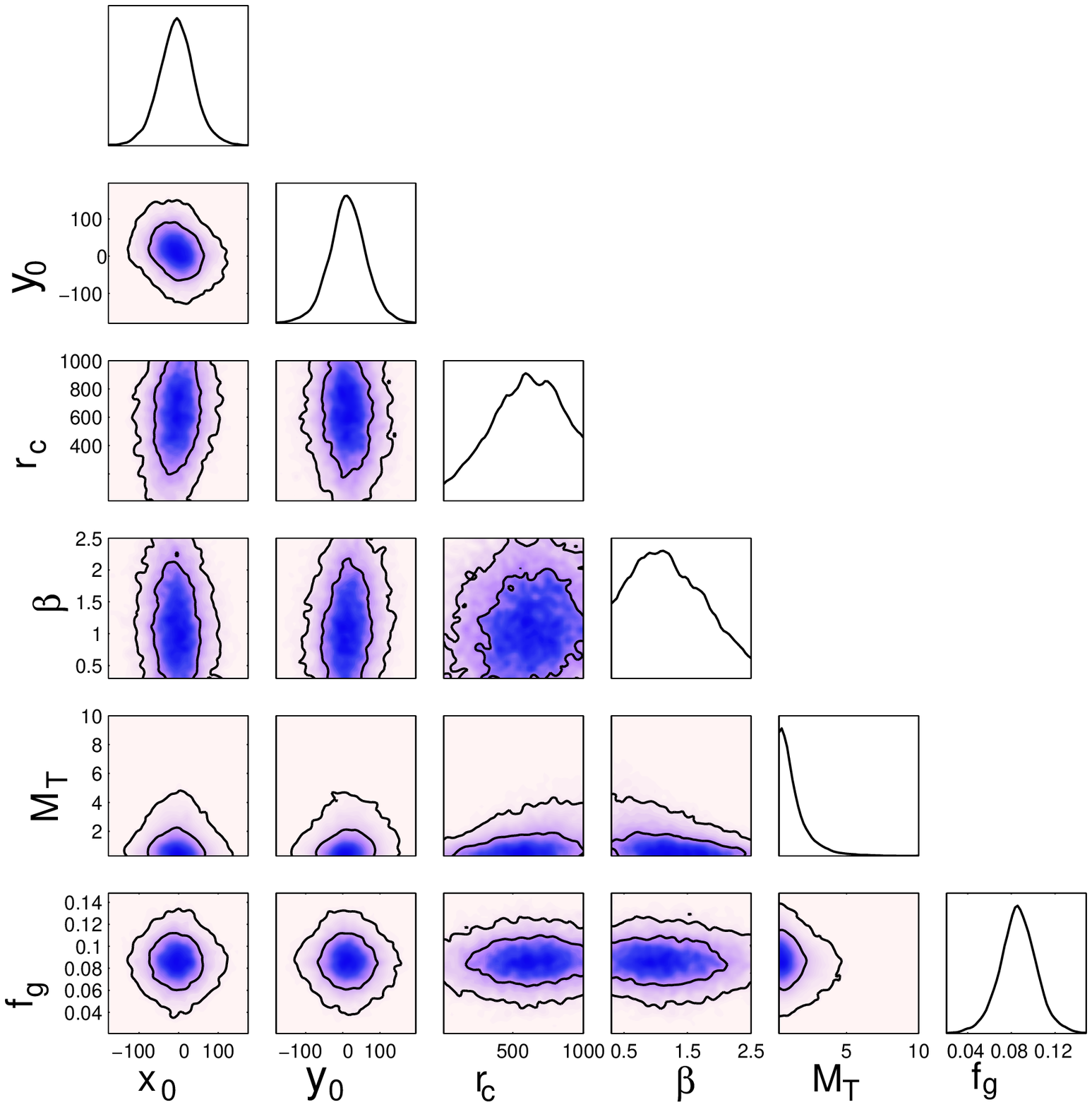}}
 \centerline{{B}\includegraphics[width=7.5cm,height=7.5cm,clip=,angle=0.]{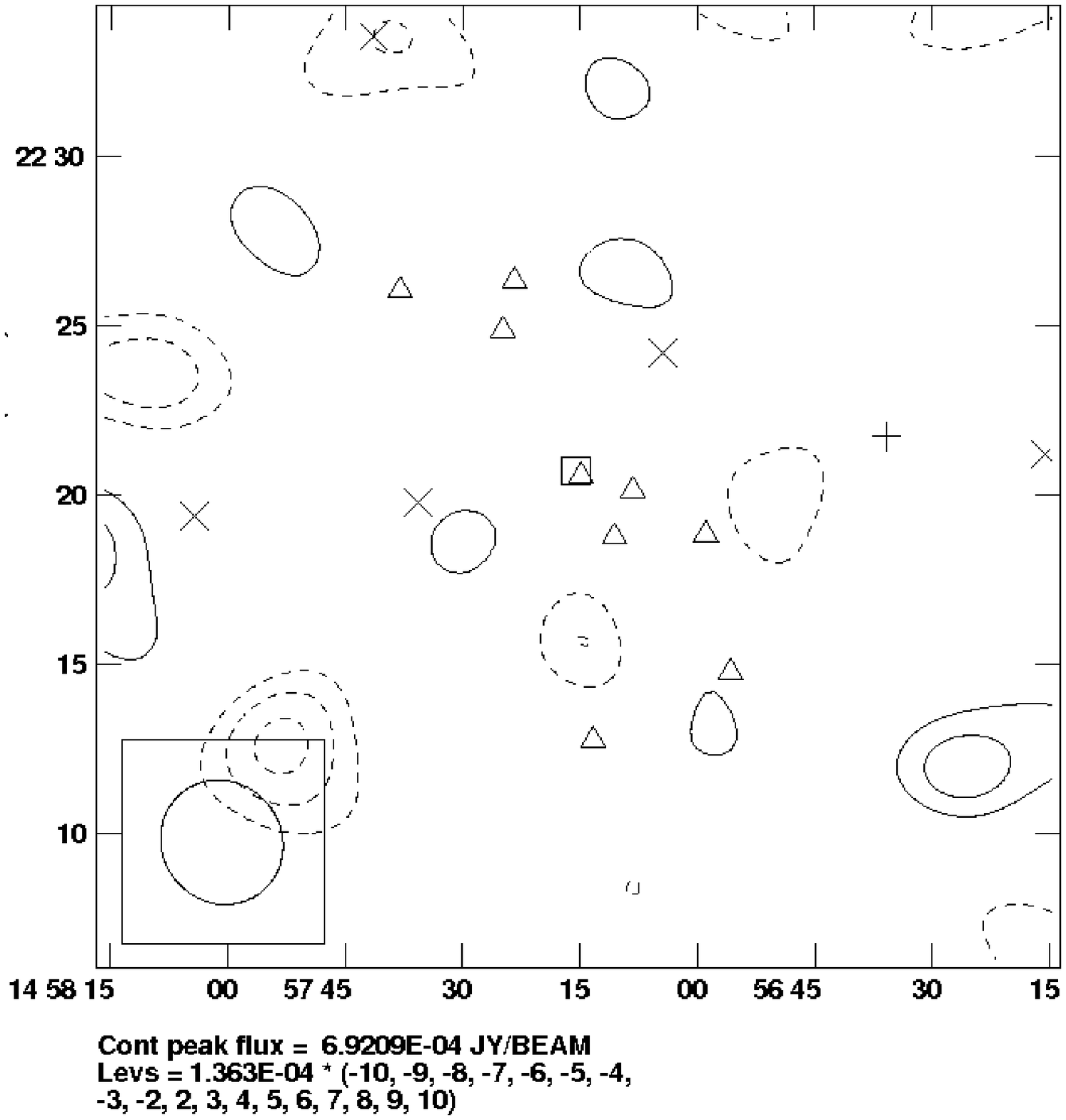}{E}\qquad\includegraphics[width=7.5cm,height= 7.5cm,clip=,angle=0.]{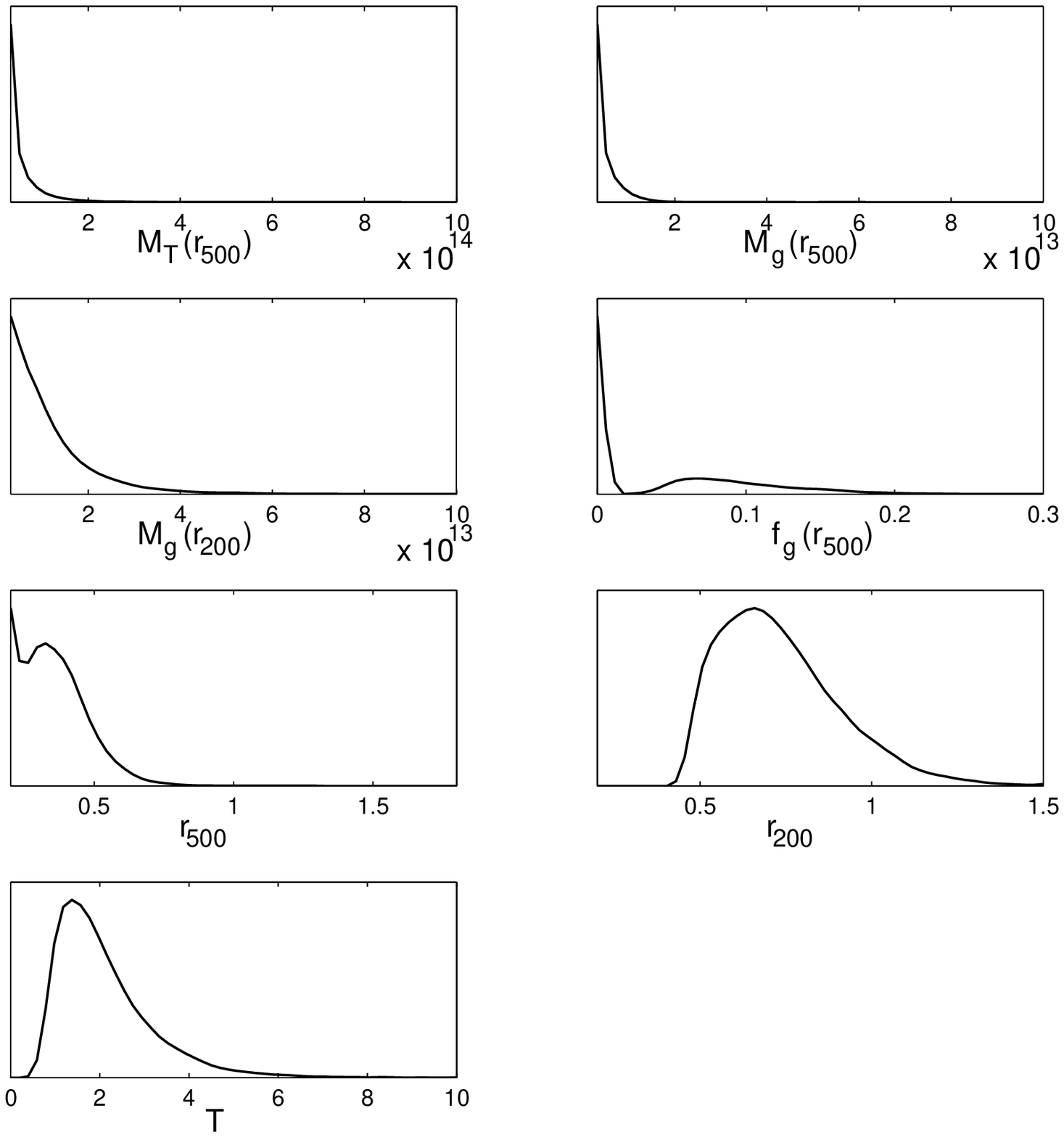}} \centerline{{C}\includegraphics[width=7.5cm,height=6.5cm,clip=,angle=0.]{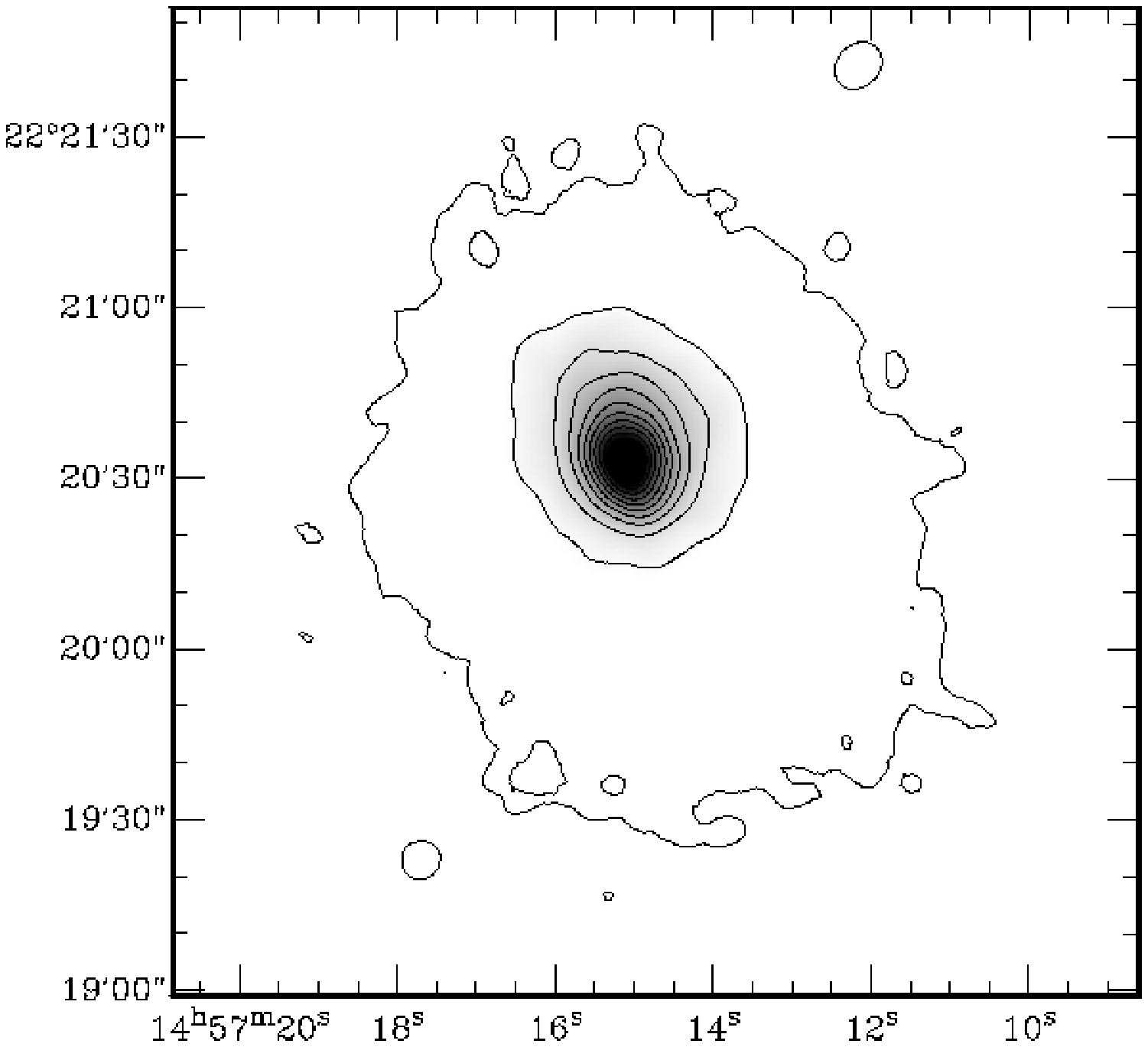}}
\caption{The null detection of Zw1454.8+2233 in SZ. Panel A 
shows the SA map before subtraction, which reveals the challenging source environment towards this cluster.
The SA map after source subtraction is shown in panel B; no convincing SZ decrement
is visible. Image C shows the {\sc{Chandra}} X-ray map overlaid with SA contours from panel B. Panels D and E show the distributions for the sampling and derived parameters respectively; 
 such distributions are consistent with a null detection. In panel D $M_{\rm{T}}$ is given in units of $h_{100}^{-1}\times10^{14}M_{\odot}$ and $f_{\rm{g}}$ in $h_{100}^{-1}$; both parameters are estimated within $r_{200}$. In E $M_{\rm{g}}$ is in units of $h_{100}^{-2}M_{\odot}$, $r$ in $h_{100}^{-1}$Mpc and $T$ in KeV.}
\label{fig:ZW7160}
\end{figure*}

\section{Source-subtraction simulation}\label{sec:sim}

Extracting robust cluster parameters for a system like Abell~2146 with
bright sources lying at or very close to the cluster is extremely
challenging. Many factors can affect the reliability of
the detection and of the recovered parameters. Aside from model
assumptions, other important factors are:
 the SNR of the decrement in our maps, the $uv$-coverage, the size of the
cluster and the distance of the sources from the cluster, their
flux-densities and their morphologies. From Sec. \ref{results},
one can appreciate
that at 16\,GHz the SZ signal is potentially strongly contaminated by
 radio sources.
We have examined some of the effects of these sources in a controlled
environment through
simulations. For this purpose, we generated mock visibilities between hour
angles -4.0 to 4.0,
with an RMS noise per channel per baseline per second of 0.54\,Jy.
 Noise contributions from a CMB realisation
and from confusion from faint sources lying below our subtraction limit were included; 
for the former we used a $\Lambda$CDM
model and for the latter we
integrated the 10C LA source counts from 10$\mu$Jy to 300$\mu$Jy. A cluster at
$z=0.23$ was simulated
using an isothermal $\beta$-profile to model the
gas distribution, with a central electron density of
$9\times 10^{3}$\,m$^{-3}$, $\beta=1.85$, $r_{\rm{c}}=440h_{70}^{-1}$\,kpc and
$T=4.8$\,keV. Integrating the density profile out to $r_{200}$ (Eq.
\ref{eq:den}) assuming a spherical cluster geometry yields
$M_{\rm{g}}(r_{200})=6.25\times
10^{13}h_{70}^{-2}$M$_{\odot}$. From the virial $M-T$ relation given in
Eq. \ref{eq:virtemp} $M_{\rm{T}}(r_{200})=5.70\times
10^{14}h_{70}^{-1}$M$_{\odot}$ and using these two estimates and
Eq. \ref{eq:fg}, we find $f_{\rm{g}}(r_{200})=0.11h^{-1}_{70}$.

Three point sources were included into the simulation. Their positions,
 flux-densities and spectral indices are given in Tab. \ref{tab:simso}.
\begin{table}
\caption{Source parameters for the three simulated sources.}
 \label{tab:simso}
\centering
 \begin{centering}
 \begin{tabular}{lccccc}\hline
{Source} & { RA (h m s)} & {Dec ($^{o}$ $^{'}$ $^{''}$)} & {$S_{16}$ (Jy)} &
{Spectral index} \\ \hline
{1}  &{15 56 04.23}  &{66 22 12.94}  &{5.92}  &{0.6}\\
{2}  &{15 56 14.30}  &{66 20 53.45}  &{1.83}  &{0.1}\\
{3}  &{15 55 57.42}  &{66 20 03.11}  &{1.65}  &{-0.2}\\ \hline
\end{tabular}
\end{centering}
\end{table}
 The map of the data is shown in Fig. \ref{fig:sim1}.
\begin{figure}
\centering
\centerline{\includegraphics[width=8.0cm,height=8.0cm,clip=,angle=0.]{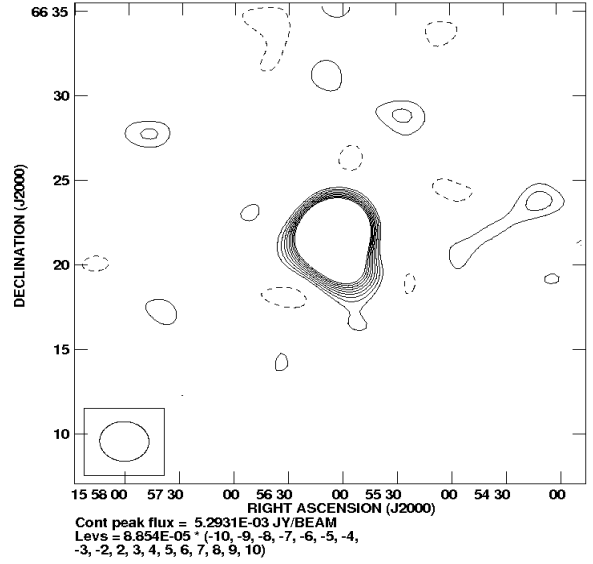}}
\caption{SA contour map of simulated data containing thermal
  noise + confusion noise + CMB + cluster + resolved point sources. Contours increase linearly in units of $\sigma_{\rm{SA}}$.}
\label{fig:sim1}
\end{figure}

The data for the simulation were run through the same analysis as described in
Sec. \ref{analysis}. In this case the source
priors were centred on the simulated values (Tab. \ref{tab:simso}) and the
cluster priors were the same of those in
 Tab. \ref{tab:Summary-of-cluster}, with the delta-prior on $z$ set to 0.23.
The  1-D and 2-D marginalized posterior distributions
for the sampling parameters are presented in  Fig. \ref{fig:sim1b}.
It can be seen that  the cluster position and gas fraction are
 recovered well by the sampler;
the core radius and $\beta$ cannot be constrained by AMI data alone, thus,
as expected, the agreement between the input and output mean
values for these paramaters is poor; the total cluster mass, on
the other hand, is very well-constrained and the
recovered value is consistent with the input value. Hence, despite the
challenging source environment, and the degeneracies between the cluster mass
and the source flux densities,
our analysis is able to provide robust cluster mass estimates.

\begin{figure*}
\centering
\centerline{\includegraphics[width=8.0cm,height=8.0cm,clip=,angle=0.]{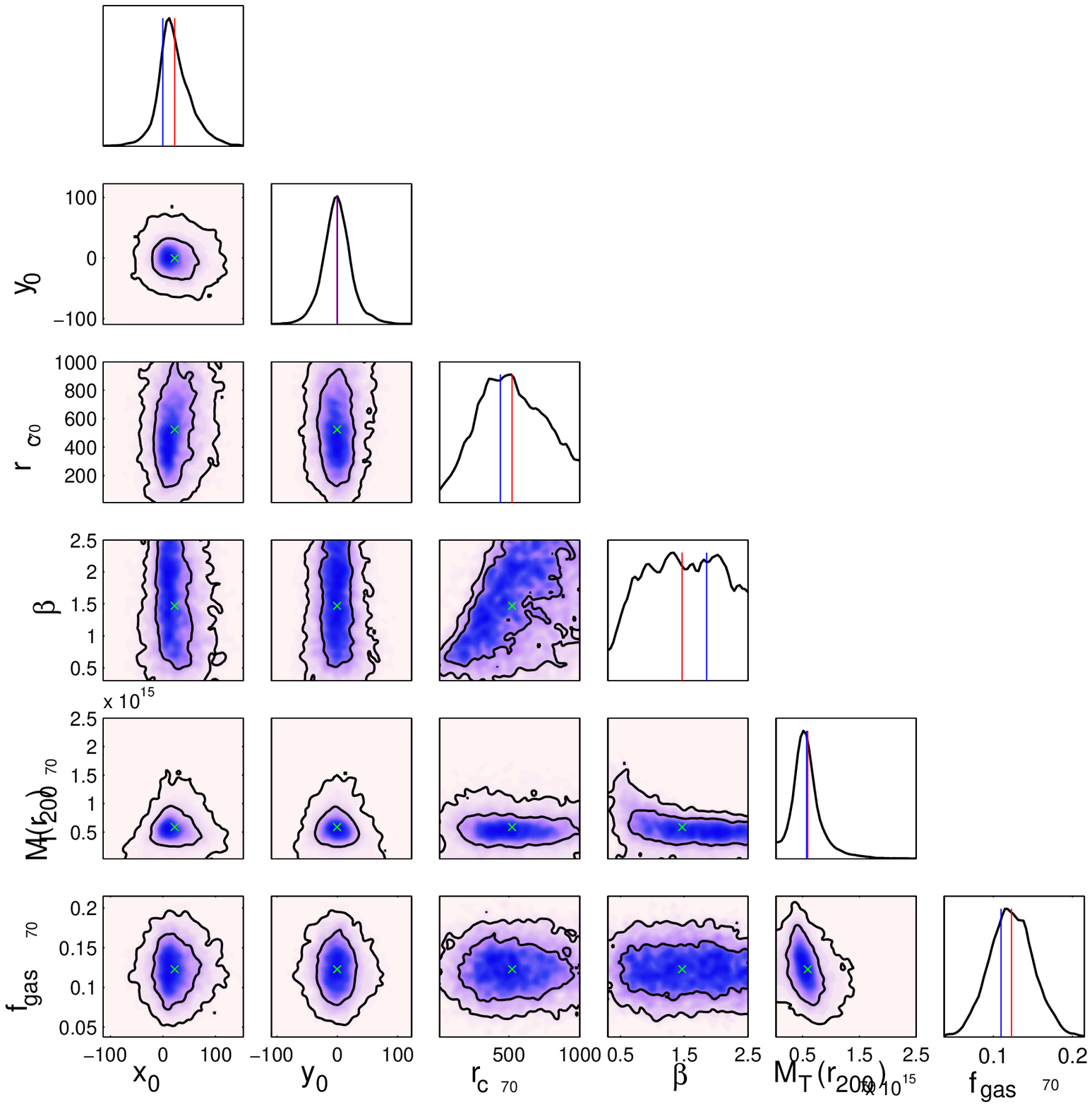}\qquad\includegraphics[width=8.0cm,height=8.0cm,clip=,angle=0.]{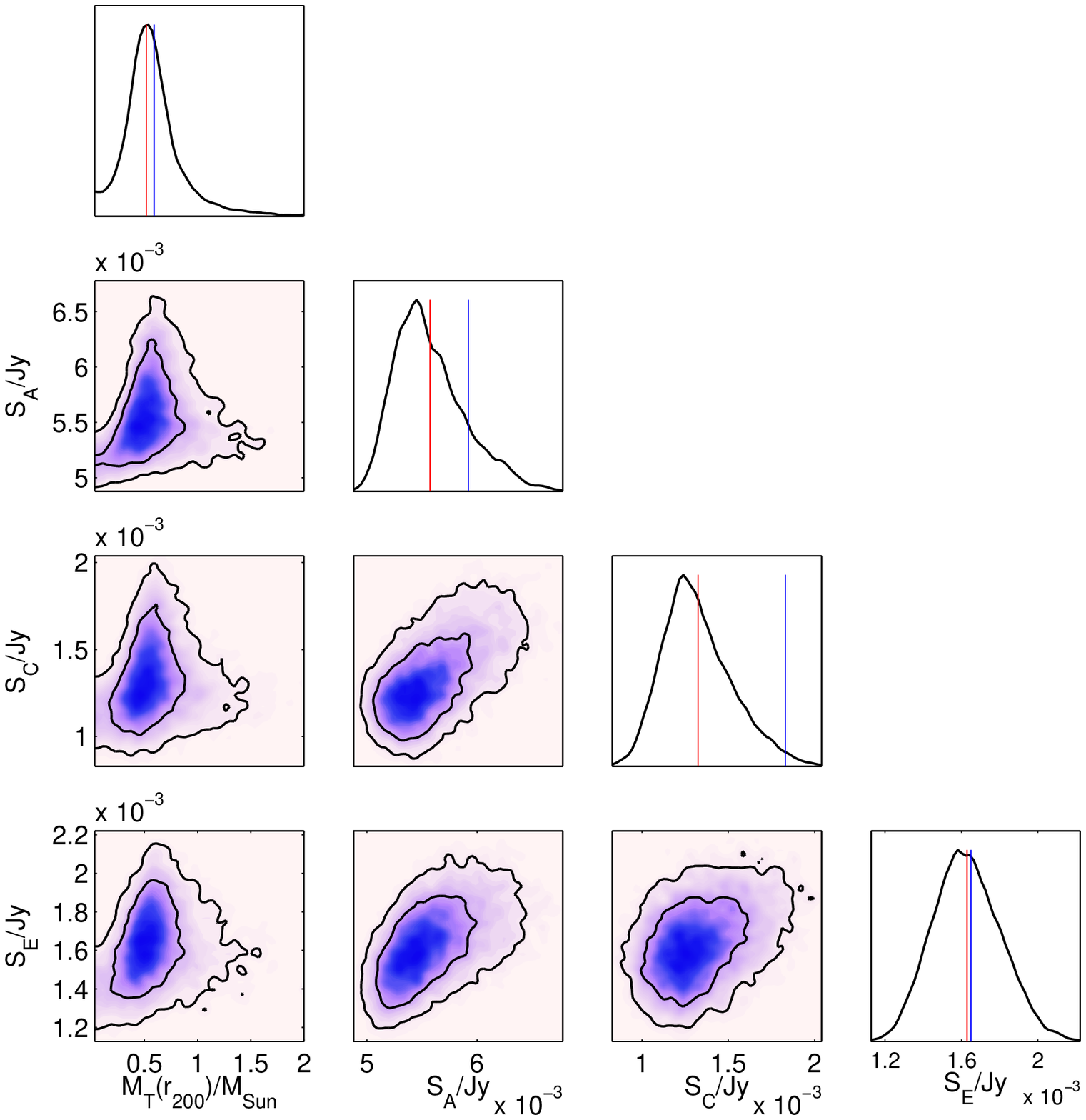}}
\caption{Left: One and two-dimensional marginalized posterior distributions for
the cluster sampling parameters from our simulation. $M_{\rm{T}}$ and $f_{\rm{g}}$ are estimated within $r_{200}$ and $M_{\rm{T}}$ is given in units of $\times 10^{14}$. The green crosses 
 in the 2-D marginals denote the mean of the distribution.
 Right: One and two-dimensional marginalized posterior distributions for the
source flux densities and $M_{\rm{T}}(r_{200})$ for our simulation. Red lines
indicate the mean of the marginalized
distribution and the blue lines represent the input value.}
\label{fig:sim1b}
\end{figure*}

\section{Discussion}\label{discussion}

Of the 20 target clusters, we have detected SZ towards 17, all of which are resolved, 
and with ``peak'' detections between 5$\sigma_{SA}$ and 23$\sigma_{SA}$.
 The analysis
has produced robust parameter extraction for 16 of the 17 -- this was not possible for Abell~2409
because of nearby extended radio emission that distorts the SZ signal and gives an unacceptable
fit for a spherical $\beta$-model. The three null detections are of Abell~1704 (difficult source environment),
Zw0857.9+2107 (it is unclear to us why we have not detected this), and Zw1458.8+2233 (difficult source environment).

\subsection{Cluster morphology and dynamics}\label{sec:morph}

The images frequently show significant differences in position of the SZ peak (and of the SZ centroid)
and the X-ray peak, indicating that the densest part of a cluster is not at the centre
of the large-scale gas distribution.
In Abell~773 and Abell~2146, both mergers, there is evidence of SZ extension 
perpendicular to the X-ray emission. Abell~1758a and Abell~A1758b are both 
major mergers and there is a hint of an SZ signal between a \& b.  
Unlike what one might naively expect, there are cases of 
SZ extensions in non-mergers and cases of near-circular SZ map structures
in mergers.

To attempt to quantify the cluster morphology from the AMI data, we ran our analysis with 
an ellipsoidal model for the cluster geometry. This model simply fits for two additional parameters: an ellipticity parameter, $\eta$,
 which is the ratio between the semi major and semi minor axes and an angle $\theta$ measured anticlockwise from the West; these values 
are given in Tab. \ref{tab:ellip}. For further details on this model see e.g., \cite{hurley2011}.

As a check that switching from spherical to ellipsoidal SZ analysis does not itself introduce significant bias in mass, we give in Tab. \ref{tab:new} the
ratios $M_{SZ, sph}/M_{SZ, ellip}$ within $r_{200}$ and $r_{500}$ and $T_{AMI, sph}/T_{AMI, ellip}$: no significant bias is evident. Of course,
elsewhere in this paper we use spherical SZ estimates because the X-ray and almost all the optical total cluster mass estimates also
 assume spherical symmetry.

\begin{table}
\caption{Median, mean and standard deviation for $M_{SZ, sph}/M_{SZ, ellip}$ within $r_{200}$ and $r_{500}$ and $T_{AMI, sph}/T_{AMI, ellip}$. Data 
 for all clusters in Tab. \ref{tab:clusmass2} were included, except for Abell~1758a and b. Ratios for each cluster at these two overdensities are given in Tab. \ref{tab:clusmass2}.}
\label{tab:new}
\begin{tabular}{lccc}
\hline 
                                                & Median & Mean & Standard deviation \\ \hline

$M_{SZ, sph}/M_{SZ, ellip}$ within $r_{500}$    & 0.96   & 0.96 & 0.16         \\ 
$M_{SZ, sph}/M_{SZ, ellip}$ within $r_{200}$    & 0.97   & 0.99 & 0.16           \\ 
$T_{AMI, sph}/T_{AMI, ellip}$                   & 0.98   & 0.98 & 0.10          \\ \hline
\end{tabular}
\end{table}


Tab. \ref{tab:ellip} also  includes other possible indicators of dynamical state. 
The presence of cooling cores (CC) is associated with relaxed clusters since it is widely accepted that merger events tend to disrupt cooling flows, e.g., \cite{fabian1984}.
We have used {\sc{Chandra}} data from the ACCEPT database, where available, to compute three CC indicators described in \cite{hudson2010}: the central entropy,
 the central cooling time and the ratio of (approximately) the central  cluster temperature to the virial temperature; Tab. \ref{tab:ellip} also includes other assessments  of dynamical
 state that we have found in the literature.

 The projected separation of the brightest cluster galaxy (BCG) and the peak of the X-ray emission has been shown to correlate
 with the dynamical equilibrium state of the the host cluster (\citealt{katayama2003} and \citealt{sanderson2009}). Similarly, the offset between the SZ centroid and the X-ray peak 
 can also be a diagnostic for cluster disturbance. For this purpose, the separation between the AMI SZ centroid, X-ray peak cluster position and the position of the BCG are
 given in Tab. \ref{tab:dyn}; in Tab. \ref{tab:dynstat} some sample statistics are provided. Large offsets between these measurements have been reported in observations 
(e.g.,  \citealt{massardi2010}, \citealt{menanteau2011}, and \citealt{korn2011}) and in simulations (e.g., \citep{molnar2012}).

 Examination of Tabs. \ref{tab:ellip}-\ref{tab:dynstat} indicates that even
 for well-studied clusters there are conflicting indications as to whether the cluster is a merger or not, e.g., Abell~773 does not 
 appear to have a CC, has high degree of ellipticity,  the X-ray and SZ signals appear to be oriented quasi-perpendicularly to 
each other and yet the relatively small position offsets in Tab. \ref{tab:dyn}
 might suggest the cluster is relaxed. 

\begin{table*}
\caption{Dynamical indicators: $\theta$, the angle measured anticlockwise from the West, $\eta$ the ratio
 between the semi major and semi minor axes (these values arise from fitting the SZ data with an elliptical geometry (see text)).  
  Cooling core information: CC denotes the presence of a cooling core and NCC the lack of ($`-'$ means this information is not clear or not known); 
  $\rm{Core}_1$ is a result from this study obtained by using three CC indicators described in Hudson et al. 2010 -- the central entropy,
 the central cooling time and the ratio of approximately the central  cluster temperature to the virial temperature, $T_0/T_{\rm{vir}}$, where all 
 the data have been taken from the {\sc{Chandra}} ACCEPT database; Core$_2$ is cooling core information on the cluster available from other studies.
 $\dag$: the core type of Abell 1423 is unclear; the ratio of $T_0/T_{\rm{vir}}$ taken from ACCEPT suggests it is not a cool-core cluster but, a CC 
 cannot be ruled out due to the large uncertainty in the X-ray temperature measurements; the central entropy and cooling time are unclear.}
\label{tab:ellip}
\begin{tabular}{lcccccc}
\hline
Cluster Name  & $\theta$              & $\eta$             & Core$_1$ & Core$_2$                                  \\\hline                  
Abell 586     & 136 $\pm$ 33    & 0.73 $\pm$ 0.13  & NCC    & CC \cite{allen2000}, NCC \cite{marrone2011}  \\
Abell 611     & 79 $\pm$ 39    & 0.80 $\pm$ 0.12  & NCC    & NCC \cite{marrone2011}                     \\
Abell 621     & 64  $\pm$ 61    & 0.73 $\pm$ 0.13  & -      & -                                         \\
Abell 773     & 41  $\pm$ 12    & 0.59 $\pm$ 0.10  & NCC    & NCC \cite{allen2000}                       \\
Abell 781     & 132 $\pm$ 32    & 0.70 $\pm$ 0.13  & -      & -                                           \\
Abell 990     & 109 $\pm$ 40    & 0.78 $\pm$ 0.13  & -      & -                                         \\
Abell 1413    & 101 $\pm$ 21    & 0.75 $\pm$ 0.12  & CC     & CC \cite{allen2000}, \cite{richard2010}                           \\
Abell 1423($\dag$)& 66 $\pm$ 28 & 0.70 $\pm$ 0.14  & NCC?   & $\&$ CC? \cite{sanderson2009}                   \\
Abell 1704    & -                   & -                & -      & CC \cite{allen2000}                           \\
Abell 1758a   & 72 $\pm$ 31     & 0.73 $\pm$ 0.14  & NCC    & -                                          \\
Abell 1758b   & 85  $\pm$ 49    & 0.77 $\pm$ 0.13  & -      & -                                          \\
Abell 2009    & 88 $\pm$ 49    & 0.78 $\pm$ 0.12  & -      & -                                           \\
Abell 2111    & 90 $\pm$ 29    & 0.77 $\pm$ 0.12  & NCC    & -                                          \\
Abell 2146    & 126 $\pm$ 4    & 0.56 $\pm$ 0.05  & -      & `Bullet-like merger' \cite{russel09}                                          \\
Abell 2218    & 107 $\pm$ 80    & 0.87 $\pm$ 0.07  & NCC    & NCC \cite{richard2010}                                       \\
Abell 2409    & -                   & -                & -      & -                                           \\
RXJ0142+2131  & 87  $\pm$ 41    & 0.77 $\pm$ 0.13  & -      & -                                         \\
RXJ1720.1+263 & 26  $\pm$ 12    & 0.58 $\pm$ 0.07  & CC     & CC  \cite{richard2010}                                          \\
Zw0857.9+2107 & -                   & -                & -      & -                                          \\
Zw1454.8+2233 & -                   & -                &        & CC \cite{Zw1454.8+2233_COOLING}                             \\ \hline 
\end{tabular}
\end{table*}

\begin{table*}
\caption{X-ray cluster position; SZ centroids from our analysis; position of the BCG from SDSS (the BCG was identified as the brightest galaxy in the central few hundred kpc from the cluster X-ray position. For clusters labelled with (*) the BCG could not be identified unambiguously.
Entries filled with a '-' indicate there is no available information.}
\label{tab:dyn}
\begin{tabular}{lcccccccccc}
\hline
Cluster Name & BCG        &              & X-ray       &              & SZ       &           &    & Position offsets ($\arcsec$)   & \\
             & RA (Deg)   & Dec (Deg)    & RA (Deg)    & Dec (Deg)    & RA (Deg) & Dec (Deg) &   SZ-X-ray  & SZ-SDSS & X-ray-SDSS\\ \hline
Abell 586    & 113.0844  & 31.6334     & 113.0833    & 31.6328     & 113.0833 & 31.6264  &   23.0    &25.7  & 4.5 \\  
Abell 611    & 120.2367  & 36.0563     & 120.2458    & 36.0503     & 120.7958 & 36.0531  &   10.1    &34.9  & 39.4\\
Abell 621    & -          & -            & 122.8000    & 70.0408     & 122.7875 & 70.0458  &   48.5    &-        & -  \\
Abell 773    & 139.4724  & 51.7270     & 139.4666    & 51.7319     & 139.4667 & 51.7331  &   4.0    &29.9  & 27.3\\
Abell 781(*)    & 140.1073  & 30.4941  & 140.1083    & 30.5147     & 140.1000 & 30.5314  &   67.0    &136.7  & 74.2\\ 
Abell 990    & 155.9161  & 49.1438     & 155.9208    & 49.1439     & 155.9125 & 49.1369  &   39.0    &27.9  & 16.9\\
Abell 1413   & 178.8250  & 23.4050     & 178.8250    & 23.4078     & 178.8250 & 23.3894  &   66.0    &55.8  & 10.2\\
Abell 1423   & 179.3222  & 33.6110     & 179.3416    & 33.6319     & 179.3375 & 33.6189  &   49.2    &62.2  & 103.0\\
Abell 1704   & 198.6025  & 64.5753     & 198.5917    & 64.5750       & -        & -         &   -         &-        & 39.0 \\
Abell 1758a(*) & 203.1189   & 50.4697  & 203.1500    & 50.4806     & 179.3375 & 50.5264  &   209.3   &209.3 & 118.6 \\ 
Abell 1758b & -           & -            & -           & -            & 203.1250 & 50.4003  &   209.3   &209.3 & -\\
Abell 2009  & 225.0833    & 21.3678     & 225.0811    & 21.3692     & 225.0875 & 21.3553  &   55.2   &47.5  & 9.5\\
Abell 2111  & 234.9333    & 34.4156     & 234.9187    & 34.4240     & 234.9125 & 34.4331  &   39.4   &97.9   & 60.8\\
Abell 2146  & -           & -            & 239.0291    & 66.3597     & 239.0250 & 66.3589  &   15.1   &-        & -\\   
Abell 2218  & -           & -            & 248.9666    & 66.2139     & 248.9375 & 66.2186  &   106.1   &-        & -\\
Abell 2409  & 330.2189   & 20.9683     & 330.2208    & 20.9606     & -        & -         &   -         &-        & 28.5\\
RXJ0142+2131& -           & -            & 25.51250    & 21.5219     & 25.51667 & 21.5303  &   33.6   &-        & - \\
RXJ1720.1+2638 & 260.0418  & 26.6256   & 260.0416    & 26.6250     & 260.0333 & 26.6125   &   54.0   &56.0  & 2.1\\
Zw0857.9+2107  & 135.1536  & 20.8943   & 135.1583    & 20.9158     & -        & -         &   -         &-        & 79.5\\
Zw1454.8+2233  & 224.3130  & 22.3428   & 224.3125    & 22.3417     & -        & -         &   -         &-        & 4.3\\ \hline 
\end{tabular}
\end{table*}

\begin{table*}
\caption{Mean, standard deviation and median for the differences in X-ray and SZ cluster centroids and the position of the BCG from SDSS maps.
 Abell 1758 (a and b) has been excluded from this analysis due to its exceptionally disturbed state.}
\label{tab:dynstat}
\begin{tabular}{lcccc}
\hline
           & Mean ($\arcsec$) & Standard Deviation ($\arcsec$) & Median ($\arcsec$) \\ \hline
SZ - Xray  & 43.9       & 26.8            & 43.6    \\
SZ - SDSS  & 51.6       & 35.3            & 43.6    \\
Xray - SDSS& 27.9       & 32.2            & 35.7    \\ \hline 
\end{tabular}
\end{table*}

\subsection{SZ temperature, large-radius X-ray temperature, and dynamics}

In Fig.  \ref{fig:tempcomp}  we compare the AMI SA observed cluster temperatures within $r_{200}$ ($T_{\rm{AMI}}$) with large-radius X-ray values ($T_{X}$) from {\sc{Chandra}} or {\sc{Suzaku}} that we have been able to find in the literature. We use \emph{large-radius} ($\approx$ 500\,kpc) X-ray temperature values to be consistent with the angular scales measured by AMI. (For Abell~611 we have plotted two X-ray values from {\sc{Chandra}} data -- one from the ACCEPT archive (\citealt{cavagnolo2009}), which is higher than our AMI SA measurement, and a second X-ray measurement from {\sc{Chandra}} (Donnarumma et al.), which is consistent with our measurement). 

There is reasonable correspondence between SZ and X-ray temperatures at lower X-ray luminosity,
 with excess (over SZ) X-ray temperatures at higher X-ray luminosity. The mean, median and standard deviation for the
 ratio of $T_{\rm{AMI}}/T_X$ were found to be 0.7, 0.8 and 0.2, respectively, when considering all the cluster in Fig. \ref{fig:tempcomp} )except for
 Abell~1758a, due to it being a complex double-merger). The numbers are obviously small, but the two
 systems that are strong mergers by clear historical consensus -- Abell~773 and Abell~1758a -- are unambiguously clear outliers with much
 higher large-radius X-ray temperatures than SZ temperatures.
 \cite{smith2004} investigate the
 scatter between lensing
 masses within $\leq 500$\,kpc with {\sc{Chandra}} X-ray temperatures
 averaged over $0.1-2$\,Mpc for ten clusters and also find that
 disturbed systems have higher temperatures. However, \cite{marrone2009} measure
 the relationship between SZ-$Y_{\rm{sph}}$ and lensing masses within
 $350$\,kpc for 14 clusters and find no segregation between disturbed
 and relaxed systems. \cite{kravtsov2006} analysed a cluster sample extracted from cosmological simulations
 and noticed that X-ray temperatures of disturbed clusters were biased high, while the X-ray analogue of SZ-$Y_{\rm{sph}}$,
 did not depend strongly on cluster structure.
Taken together, these results suggest
 that, even at small distances from the core, SZ-based mass (or temperature) is
 a less sensitive indicator of disturbance disturbance than is X-ray-based mass. 

Major mergers in our sample have large-radius X-ray temperatures (at
$\approx 500$\,kpc) \emph{higher} than the SZ temperatures (averaged over the whole cluster). 
This suggests that the mergers affect the
$n^2$-weighted X-ray temperatures more than the $n$-weighted SZ temperatures 
and do so out to large radius. This is evidence for shocking or clumping or both 
\emph{at large radius} in mergers. Indications that clumping at large $r$ might have
a significant impact on X-ray results have been found by e.g., \cite{kawaharada2010}, who find 
a flattening of the entropy profile around the virial radius, contrary to the theoretical predictions (e.g., \citealt{voit2005}).
Hydrodynamical simulations by \cite{nagai2011} have shown that gas clumping
can indeed introduce a large bias in large-$r$ X-ray measurements and could help explain the results by e.g., Kawaharada et al. 
It should be noted, however, that \cite{mazzotta2004} expect X-ray temperatures to be lower than mass-weighted temperatures
 for clusters with temperature structure since the detectors of {{\sc{Chandra}} and {{\sc{XMM-Newton}} are more efficient on the soft bands, which leads to an upweighting of the cold gas.
 However, in simulations by \cite{rasia2012} mass-weighted temperatures were shown to be larger than X-ray temperatures
 for the vast majority of their clusters, particularly for very the most  disturbed clusters in their sample.

 Examination of Fig. \ref{fig:tempcomp} given Tab. \ref{tab:ellip} is suggestive of another relation, again with obviously small numbers. Fig. \ref{fig:new}, 
 shows AMI cluster temperature versus large-scale X-ray temperature but with each cluster X-ray luminosity replaced by AMI ellipicity, $\eta$, and its error (note that
 we have removed the two Abell~611 points because of their apparently discrepant X-ray values). With one exception (Abell~2146), the clusters with
 large-radius X-ray temperature $\geq 6$\,keV have $\eta$ values $\leq 0.70$, whereas the first two outliers to the right (RXJ1720.1+2638 and Abell~773) 
 have significantly smaller values of AMI ellipticity. The rightmost outlier (Abell~1758a) itself has the ellipticity value $0.73\pm 0.14$ but this will be
 misleadingly high if we should instead be considering the ellipticity of the Abell~1758a+b taken as a merging pair. The true relationship 
 between SZ ellipticity and merger state is bound to be influenced by the collision geometry, the time since the start of the merger (Fig. 1 in \citealt{nelson2012} illustrates 
 how SZ $\eta$ and $\theta$ can vary with merger evolution), the mass ratio, and so on. Far more data, including data on clusters not selected by X-ray luminosity, are essential.

\begin{figure*}
\centering
\includegraphics[width=12.0cm,clip=,angle=270.]{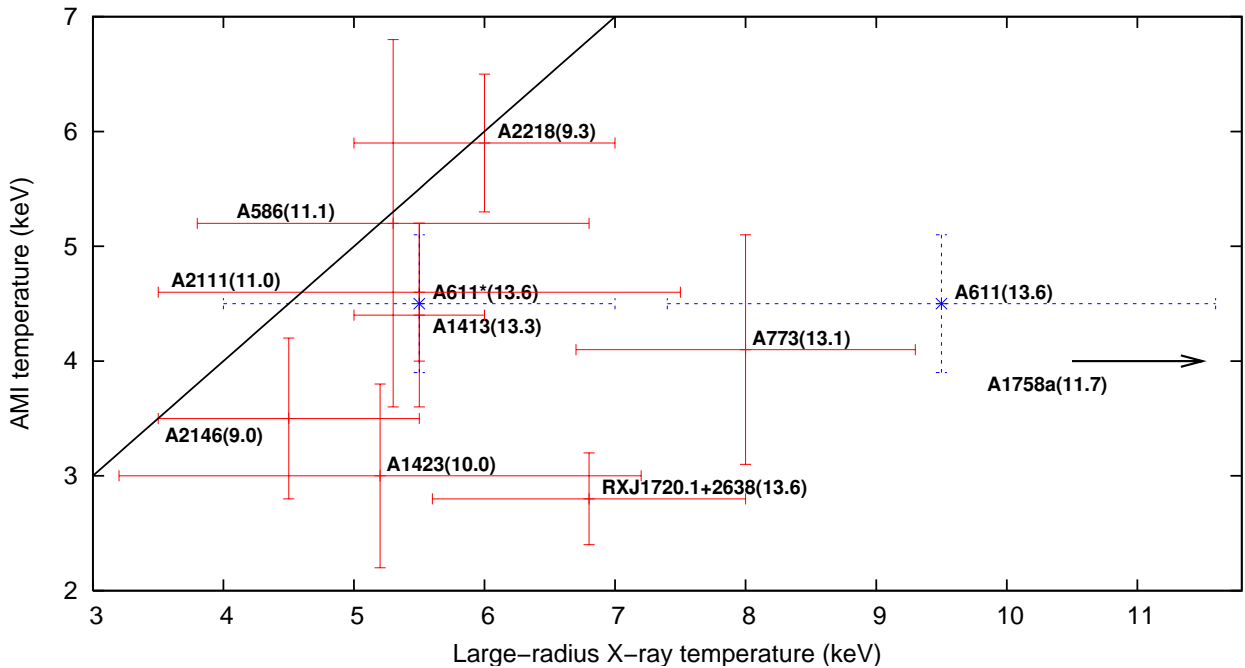}
\caption{The AMI mean temperature within $r_{200}$ versus the X-ray temperature. Each point is labelled with the cluster name and X-ray luminosity. Most of the X-ray measurements are large-radius temperatures from the ACCEPT archive (\citealt{cavagnolo2009}) with 90\% confidence bars. The radius of the measurements taken from the ACCEPT archive are 400-600\,kpc for Abell~586,  300-700\,kpc for Abell~611, 300-600\,kpc for Abell~773, 450-700\,kpc for Abell~1423, 500-1000\,kpc for Abell~2111, 450-550 for Abell~2218 and for RXJ1720.1+2638 r = 550-700\,kpc. The Abell~611* temperature is the 450-750\,kpc value with 68$\%$ confidence bars (\citealt{Xray_lense_Abell 611_2}). The Abell~2146 temperature measurement is from Russell et al. 2010 (with 68$\%$ confidence bars). The Abell~1413 X-ray temperature is estimated from the 700-1200\,kpc measurements made with the {\sc{Suzaku}} satellite (\citealt{hoshino2010}), this value is consistent with \citealt{Chandra-A1413} and \citealt{XMM_snowden}. The ACCEPT archive temperature for Abell~1758A is 16$\pm$7\,keV at r= 475-550\,kpc, and with SZ temperature 4.5$\pm$0.5, is off the right-hand edge of this plot. Abell~611 has been plotted using dashed blue lines to emphasize that this cluster has two X-ray-derived large-$r$ temperatures. The black diagonal solid line is the 1:1 line.}
\label{fig:tempcomp}
\end{figure*}

\begin{figure*}
\centering
\includegraphics[width=12.0cm,clip=,angle=270.]{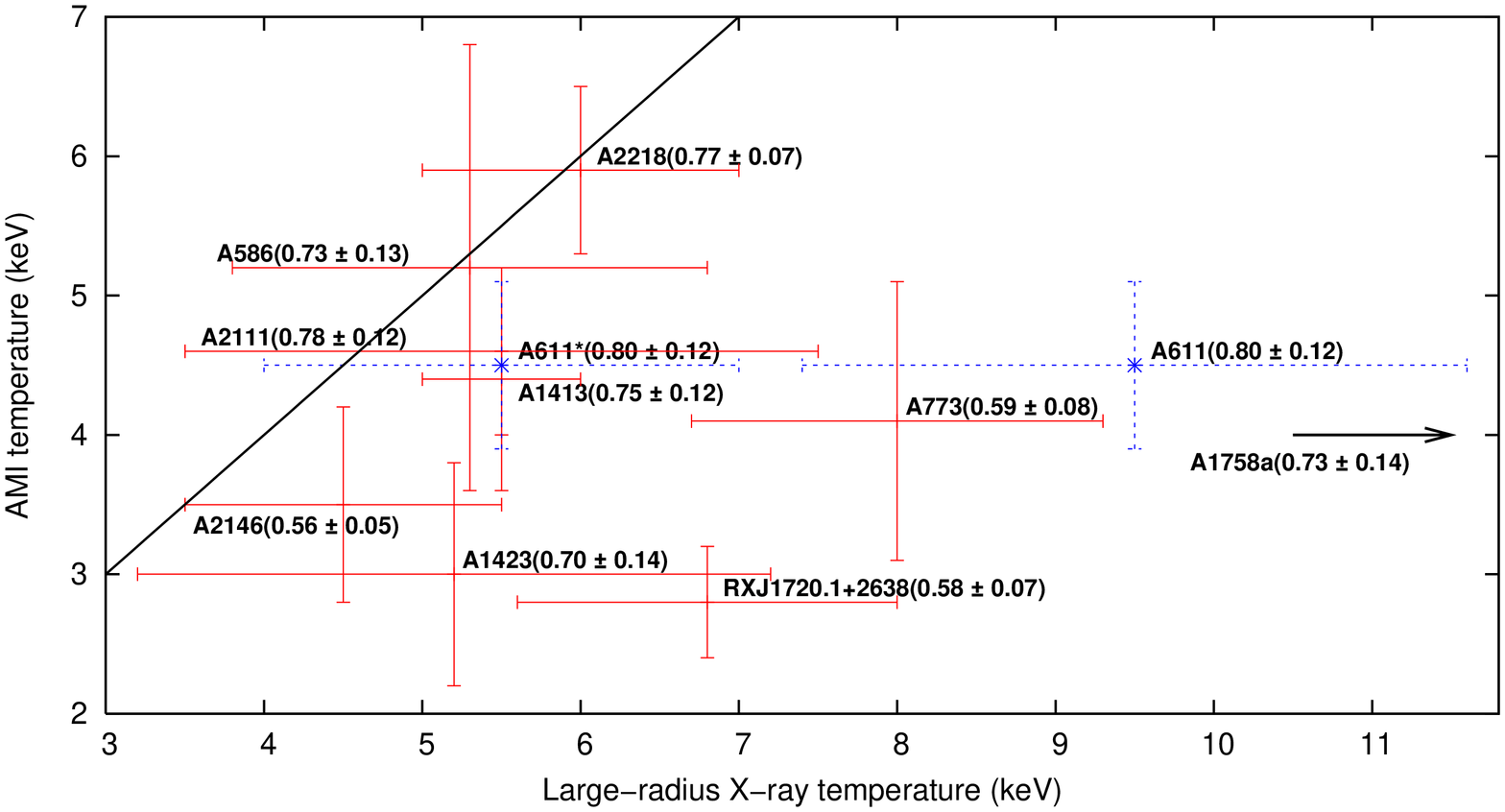}
\caption{Plot analogous to Fig. \ref{fig:tempcomp} but with the X-ray luminosity values replaced by SZ $\eta$ (ellipticity).}
\label{fig:new}
\end{figure*}

\subsection{Comparison of masses within $r_{500}$ and within the virial radius ($\approx r_{200}$)}
The classical virial radius, $\approx r_{200}$, found is typically 1.2$\pm$0.1\,Mpc. Values for $M_{\rm{T}}(r_{200})$ range from
 $2.0^{+0.4}_{-0.1} \times 10^{14}h^{-1}_{70}M_{\odot}$ to  $6.1 \pm 0.9 \times 10^{14}h^{-1}_{70}M_{\odot}$
 and are typically 2.0-2.5$\times$ larger than $M_{\rm{T}}(r_{500})$. In Fig.  \ref{fig:m500comp} and \ref{fig:m200comp} AMI
mass estimates at two overdensity radii are compared with other published mass estimates. The scarcity of mass measurements at large $r$ is apparent from these figures. 

\begin{itemize}
\item  For $M_{\rm{T}}(r_{500})$, there is good agreement between optical and AMI (HSE) mass estimates. In contrast, the X-ray (HSE) estimates tend to be higher, sometimes substantially so.
\item  For $M_{\rm{T}}(r_{200})$, there is very good agreement between optical estimates, the {\sc{Suzaku}} X-ray (HSE) estimate, and the AMI ($M-T$) estimates. 
Good agreement between AMI and optical masses has previously been reported by \cite{hurley2011}.
\item  From our sample we cannot determine whether the disagreement of masses is a function of radius.
%
\item The discrepancy between the X-ray and AMI masses for Abell~1413 is reduced at $r_{200}$, with the X-ray mass being larger than the AMI mass by $\approx 50\%$ at $r_{500}$ and smaller than the AMI mass by $\approx 10\%$ at $r_{200}$.
\item The largest discrepancies between mass measurements in SZ, optical and X-ray correspond to the strongest mergers within our sample but the 
 X-ray masses are always higher than our SZ masses, even for the few relaxed clusters in our sample. Given that the lensing masses agree well with our SZ estimates,
 this might be an indication of a stronger bias in masses estimated from X-ray data than from SZ or lensing data, especially for disturbed systems. However, most recent
 simulations and analyses indicate that X-ray HSE masses are underestimated with respect to lensing masses (e.g., \citealt{nagai2007}, \citealt{meneghetti2010}, \citealt{rasia2012}).
\end{itemize}

\subsubsection{Related results from the literature}
 To illustrate some of the issues in mass estimation, we bring together some of the other results in the literature.

\begin{itemize}
\item{{\bf{X-ray and weak-lensing masses}}\\
Observational studies by \cite{mahdavi2008}, \cite{zhang2008} and \cite{zhang2010} find systematic differences between X-ray HSE-derived and weak-lensing masses, with the lensing masses
 typically exceeding the X-ray masses. Madhavi et al. report a strong radial dependence for this difference, with weak-lensing masses being $\approx 3\%$ smaller within $r_{2500}$ but 
$\approx 20\%$ larger within $r_{500}$ than the X-ray masses, yet find no correlation between the difference level and the presence of cool cores. Zhang et al. (2010) find that X-ray masses seem 
 underestimated by $\approx 10\%$ for undisturbed systems and overestimated by $\approx 6\%$ for disturbed clusters within $r_{500}$. For relaxed clusters, they find the discrepancy is reduced at 
 larger overdensities. 
The underestimate of HSE X-ray masses with respect to lensing masses has been widely produced in simulations (e.g., \citealt{nagai2007}, \citealt{meneghetti2010})
 and \citealt{rasia2012}.

 In Tab. \ref{tab:mratios} we follow Mahdavi et al. to calculate a weighted best-fit ratio of two mass estimates at different overdensities for different data. The simulations by Rasia et al.
 and Meneghetti et al. yield significantly lower $M_X/M_{WL}$ at $r_{500}$ than the observational data. \cite{sijacki2007} suggest that a higher incidence of temperature substructure in the 
 simulations might be responsible for this effect. It is interesting to see how the mass agreement for the study by Zhang et al. seems to weaken when excluding disturbed systems.
What is very different from the literature is that we find HSE X-ray masses to be consistently higher than our HSE SZ masses within $r_{500}$.
 Modelling our clusters with an elliptical model for the cluster geometry does not substantially improve the agreement.}

\item{{\bf{SZ Y with X-ray and lensing masses}} \\
\cite{bonamente2011} find good agreement between $Y_{\rm{sph}}(r_{500})$ estimated from a joint SZ and X-ray analysis and from SZ data alone, in support of results by \cite{planck2011b}.  
For their sample of massive, relaxed clusters there appears to be no significant systematics affecting the ICM pressure measurements from X-ray or SZ data. But, of course, this result
 might not be reproduced for a sample of disturbed clusters.

\cite{marrone2009} measure the scaling between $Y_{\rm{SZ}}$ and weak lensing mass measurements within 350\,kpc ($\approx r_{4000-8000}$) for 14 LoCuSS clusters. They find it 
 behaves consistently with the self-similar predictions, has considerably less scatter than the relation between lensing mass and $T_X$
 and does not depend strongly on the dynamical state of the cluster.  They suggest SZ parameters derived from observations near the 
cluster cores may be less sensitive to the complicated physics of these regions than those in X-ray. A later study by \cite{marrone2011} 
comparing two $Y_{\rm{SZ}}-M$ scaling relations using weak-lensing masses and X-ray (HSE) masses at $r_{2500}, r_{1000}$ and $r_{500}$ indicates the latter has more scatter and is more sensitive to cluster
 morphology, with the mass estimates of undisturbed clusters exceeding those of disturbed clusters at fixed $Y_{\rm{sph}}$ by $\approx 40\%$ at large overdensities.
 However, this division is not predicted by comparing SZ and true masses from simulations and is
 could due to the use of a simple spherical lens model. Moreover, recently, \cite{rasia2012} have shown through simulations that selecting relaxed clusters for weak-lensing studies 
 based on X-ray morphology is not optimal since there can be mass from, e.g., filaments not associated to X-ray counterparts biasing the lensing mass estimates even for
 systems which appear to be regular in X-rays.}

\item{{\bf{Simulations}} \\
Simulations of cluster mergers have shown these events generate turbulence, bulk flows and complex temperature structure, all of which can 
  result in cluster mass biases (e.g., \citealt{poole2007}). Predominantly, simulations indicate that X-ray HSE masses tend to be underestimated (e.g., \citealt{krause2012})
  particularly in disturbed clusters, though the amount of the bias varies depending on the the simulation details, particularly on the physical processes taken into consideration.
Projections effects, model assumptions and the dynamical state of the cluster are some of the factors affecting how well the true cluster mass can be measured. 
 As shown by e.g., \cite{takizawa2010}, even mass estimates for spherical X-ray systems are not always recovered well. Recent simulations by
\cite{nelson2012} have investigated in detail the evolution of the non-thermal support bias as function of radius and of the merger stage.
 They reveal a very complex picture: the
 HSE bias appears to vary in amplitude and direction radially and as the merger evolves
 (and the shocks propagate through); for the most part, the HSE bias leads to an underestimate for the mass, there
 are times when it has the opposite effect.

 From simulations there appear to be two main, competing effects that can lead to a mass bias from the effects of a merger.
 Firstly, the merger event can boost the X-ray luminosity and temperature (e.g., \citealt{ricker2001}) such that if the cluster is observed during this period its X-ray mass will be overestimated.
 Secondly, the increase in non-thermal pressure support during the merger can
 lead to X-ray (HSE) cluster masses being underestimated (e.g., \citealt{lau2009}).
 The cluster sample derived from simulations studied by \cite{kravtsov2006} 
 showed that the X-ray temperatures were biased high for disturbed clusters, unlike $Y_X$, the product of the gas mass and temperature as deduced from X-ray observations 
 (the X-ray analogue of the SZ $Y$) which did not appear to depend strongly on cluster structure.}

\end{itemize}

\begin{table}
\caption{ Best-fit mass ratios calculated following Mahdavi et al. 2008. R12 are the results from simulations by Rasia et al. 2012, ME10 are the simulations from Meneghetti et al. 2010, Z10 from Zhang et al. 2010 and MA10 from Mahdavi et al 2008. For our results we have used for simplicity {\emph{sph}} to denote our SZ masses derived using a spherical geometry and {\emph{ellip}} when assuming an elliptical model. We have excluded Abell~1758 (A and B) from the analysis, given its abnormally disturbed and complex nature.}
\label{tab:mratios}
\begin{tabular}{lccc}
\hline 
                           & $r_{500}$       & $r_{200}$ \\
$M_X/M_{WL}$               &                &            \\ \hline
R12- full sample           & $0.75\pm 0.02$ & -          \\
R12- regular clusters      & $0.75\pm 0.04$ & -          \\
ME10- full sample          & $0.88\pm 0.02$ & -          \\
Z10- full sample           & $0.99\pm 0.07$ & -          \\
Z10- relaxed               & $0.91\pm 0.06$ & -          \\
MA10- all                  & $0.78\pm 0.09$ & -           \\ 
This work                  &                &           \\ \hline
$M_X/M_{SZ, sph}$          & $1.7 \pm 0.2$  & -          \\
$M_X/M_{SZ,ellip}$         & $1.6\pm 0.3$   & -          \\
$M_{SZ,sph}/M_{WL}$        & $1.2^{+0.2}_{-0.3}$   & $1.0 \pm 0.1$          \\
$M_{SZ, ellip}/M_{WL}$     & $1.2^{+0.2}_{-0.3}$   & $0.9 \pm 0.1$          \\ \hline
\end{tabular}
\end{table}

\begin{figure*}
\centering
\includegraphics[width=10.0cm,height=16.0cm,clip=,angle=270.]{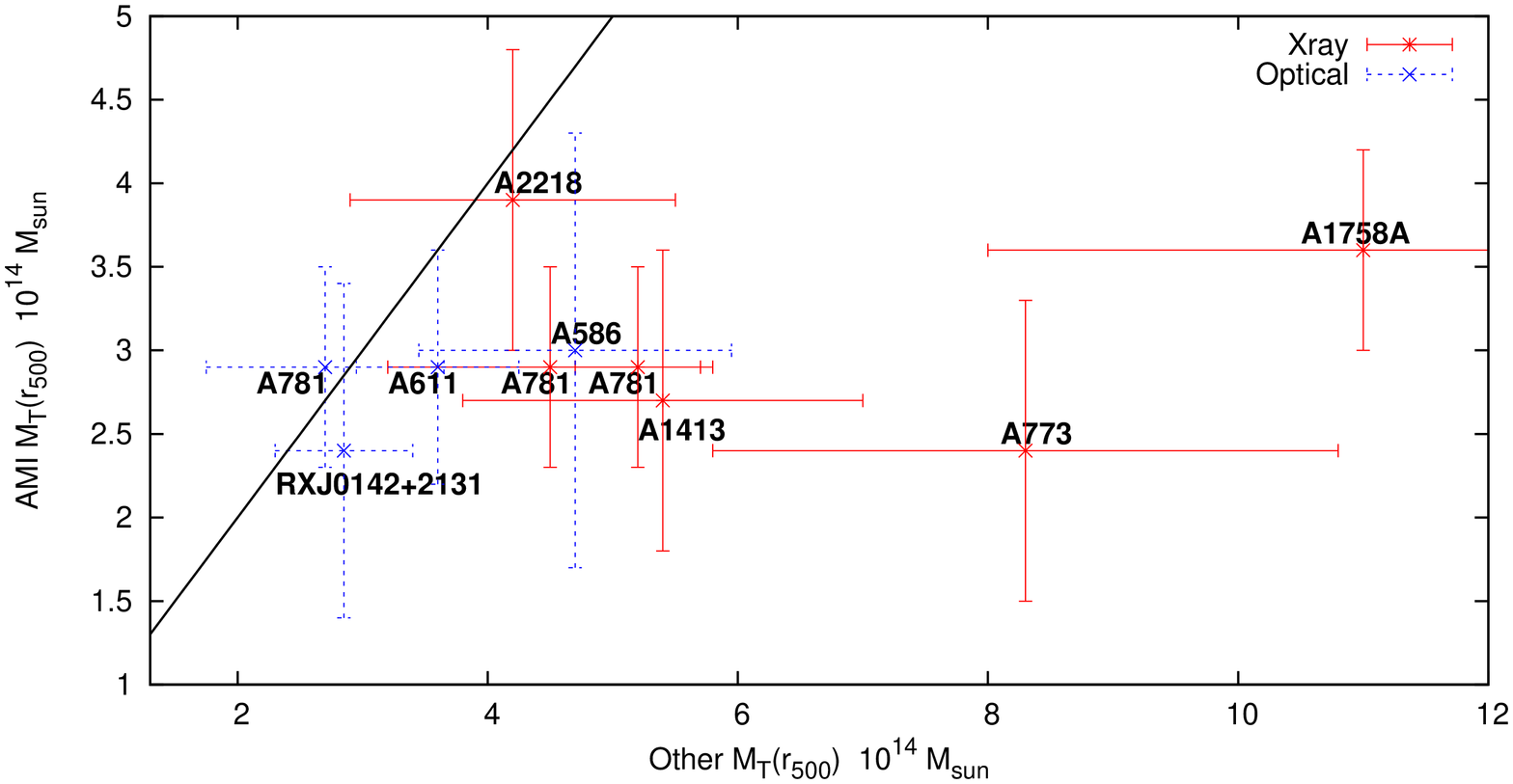}
\caption{Comparison of AMI $M_{\rm{T}}(r_{500})$ measurements with others. Methods used for estimating $M_{\rm{T}}(r_{500})$ are given in the legend.
 The line of gradient one has been included to aid the comparison. The references are as follows: Abell~586 \citep{okabe2010}; Abell~611  \citep{okabe2010}; Abell~773 \citep{zhang2010}; Abell~781 (\citealt{Abell 781_XMM} and \citealt{zhang2010}); Abell~1413 \citep{zhang2010}, Abell~1758A \citep{zhang2010}, Abell~2218 \citep{zhang2010} and RXJ0142+2131 \citep{okabe2010}. AMI values are given in Tab. \ref{tab:clusparams}. These were the $M(r_{500})$ from X-ray and weak lensing data that we found in the literature.}
\label{fig:m500comp}
%
\includegraphics[width=10.0cm,height=12.0cm,clip=,angle=270.]{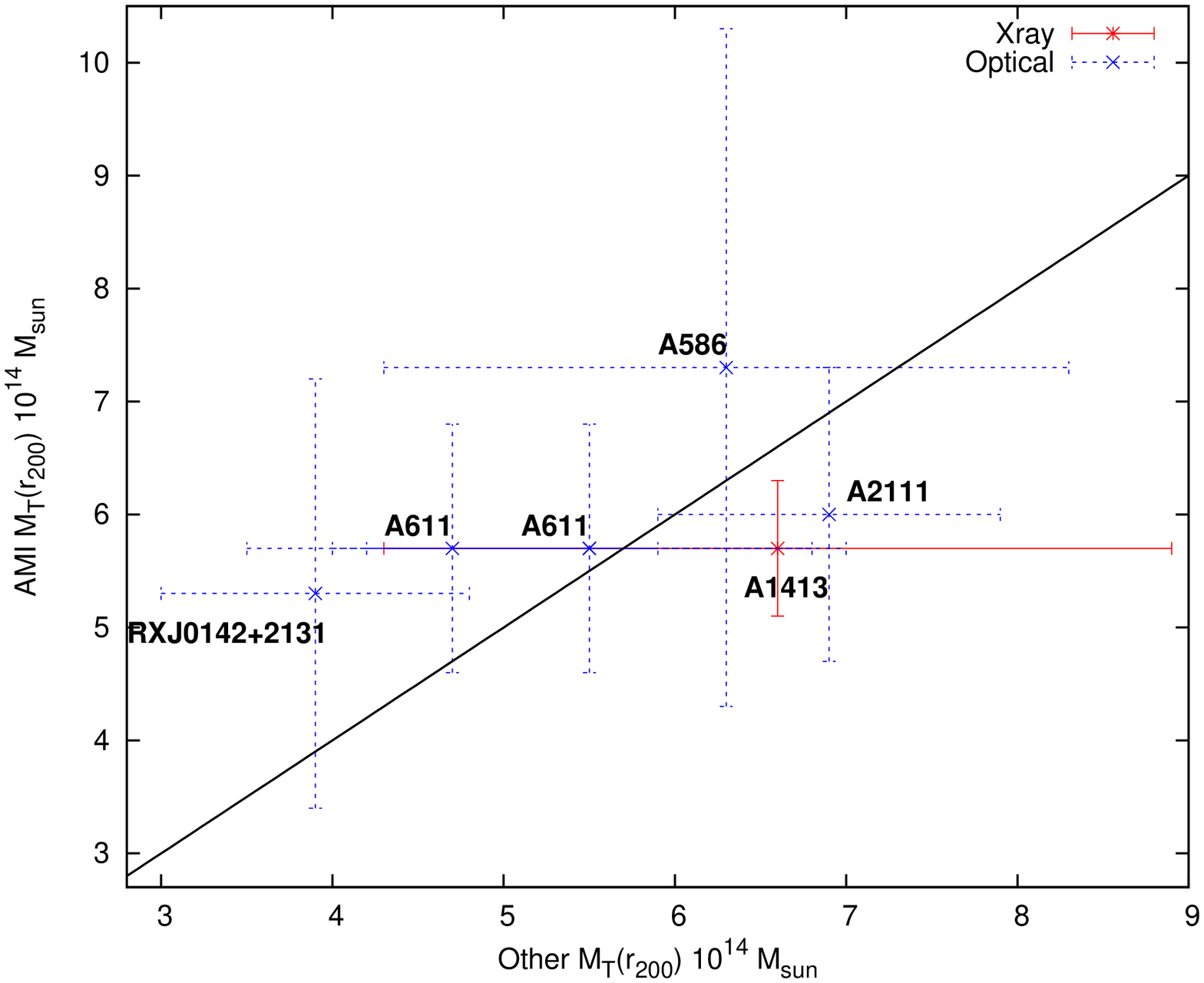}
\caption{Comparison of AMI $M_{\rm{T}}(r_{200})$ measurements with others. Methods used for estimating $M_{\rm{T}}(r_{500})$ are given in the legend. Mass is given in units of $\times 10^{14}M_{\odot}$.
 The line of gradient one has been included to aid comparison. The references are as follows: Abell~586 \citep{okabe2010}; Abell~611 (\citealt{okabe2010}, \citealt{romano_a611} and \citealt{hurley2011}); Abell~1413 \citep{hoshino2010}; Abell~2111 \citep{hurley2011} and RXJ0142+2131 \citep{okabe2010}. AMI values are given in Tab. \ref{tab:clusparams}. }
\label{fig:m200comp}
\end{figure*}

`

\begin{table*}
\begin{center}
\caption{Comparison of cluster masses at $r_{500}$ and $r_{200}$ for a spherical and an elliptical model for the cluster geometry. Ratio refers to the ratio between spherical and elliptical $M_{\rm{T}}$}
\label{tab:clusmass2}
\begin{tabular}{lccccccc}
\hline
                               &   $M_{\rm{T}}(r_{200})/\times10^{14}M_{\odot}$ & & &  $M_{\rm{T}}(r_{500}/\times10^{13}M_{\odot}$) &  &    \\
Cluster Name                   &   Spherical         & Elliptical            & Ratio                    & Spherical          & Elliptical           &  Ratio \\ \hline
A586                           & $7.3 \pm 3.0$       & $7.5^{+3.0}_{-3.1}$   & $0.97 \pm 0.59$          & $3.0 \pm 1.3$      & $3.1 \pm 1.4$        & $0.97 \pm 0.65$ \\
A611                           & $5.7 \pm 1.1$       & $5.8 \pm 1.2$         & $0.98 \pm 0.30$          & $2.9 \pm 0.7$      & $2.9 \pm 0.7$        & $1.00 \pm 0.34$ \\
A621                           & $6.8^{+2.4}_{-2.5}$ & $7.2^{+2.3}_{-2.4}$    & $0.94 \pm 0.53$          & $2.0 \pm 1.3$      & $2.2^{+1.3}_{-1.4}$  & $0.91^{+0.97}_{-1.00}$ \\ 
A773                           & $5.1 \pm 1.7$       & $7.4 \pm 2.3$         & $0.69 \pm 0.66$          & $2.4 \pm 0.9$      & $3.1^{+1.4}_{-1.3}$  & $0.77^{+0.76}_{-0.73}$ \\
A781                           & $5.9 \pm 1.1$       & $7.2 \pm 1.8$         & $0.82 \pm 0.38$          & $2.9 \pm 0.6$      & $3.2 \pm 1.0$        & $0.91 \pm 0.41$ \\
A990                           & $2.9_{-0.1}^{+0.6}$ & $2.9 \pm 0.6$         & $1.00^{+0.29}_{-0.21}$    & $1.6 \pm 0.3$      & $1.6 \pm 0.3$        & $1.00 \pm 0.27$ \\
A1413                          & $5.7 \pm 1.4$       & $5.8 \pm 1.5$         & $0.98 \pm 0.36$          & $2.7 \pm 0.9$      & $2.8 \pm 0.8$        & $0.96 \pm 0.46$ \\
A1423                          & $3.1 \pm 1.1$       & $4.3 \pm 1.8$         & $0.72 \pm 0.76$          & $1.6 \pm 0.6$      & $2.0 \pm 0.9$        & $0.80 \pm 0.73$ \\
A1758a                         & $5.9_{-1.1}^{+1.0}$ & $6.2 \pm 1.2$         & $0.95^{+0.27}_{-0.28}$   & $3.6 \pm 0.6$      & $3.7 \pm 0.7$        & $0.97 \pm 0.26$ \\
A1758b                         & $6.3 \pm 2.7$       & $5.8 \pm 2.5$         & $1.09 \pm 0.56$          & $3.1 \pm 1.4$      & $2.8 \pm 1.2$        & $1.11 \pm 0.56$ \\
A2009                          & $6.6 \pm 2.1$       & $5.7^{+2.6}_{-2.9}$   & $1.16^{+0.48}_{-0.52}$   & $2.9_{-0.9}^{+0.3}$& $2.0^{+1.2}_{-1.4}$  & $1.45^{+0.42}_{-0.53}$ \\
A2111                          & $6.0 \pm 1.3$       & $6.0 \pm 1.4$         & $1.00 \pm 0.32$          & $2.6 \pm 0.7$      & $2.7 \pm 0.9$        & $0.96 \pm 0.44$ \\
A2146                          & $7.1 \pm 1.0$       & $7.5 \pm 1.5$         & $0.95 \pm 0.26$          & $3.9 \pm 0.7$      & $3.8 \pm 0.8$        & $1.03 \pm 0.27$ \\
A2218                          & $8.7 \pm 1.3$       & $9.0^{+1.6}_{-1.5}$   & $0.97^{+0.24}_{-0.23}$   & $3.9 \pm 0.9$      & $4.0^{+1.0}_{-0.9}$  & $0.97^{+0.35}_{-0.33}$ \\
RXJ0142+2131                   & $5.3_{-1.7}^{+1.6}$ & $5.4^{+1.8}_{-1.9}$   & $0.98^{+0.46}_{-0.49} $          & $2.4 \pm 1.0$      & $2.5 \pm 1.0$        & $0.96 \pm 0.6$ \\
RXJ1720+2638                   & $2.9 \pm 0.6$       & $3.6 \pm 0.7$         & $0.81 \pm 0.35$          & $1.7 \pm 0.3$      & $2.1 \pm 0.4$        & $0.81 \pm 0.32$ \\ \hline

\end{tabular}
\end{center}
\end{table*}




\section{ Conclusions}\label{conclusion}
We observe 19 LoCuSS clusters with $L_X > 7\times 10^{37}$\,W ($h_{50}=1.0$) and present SZ images before and after source subtraction for 16 of them (and  for Abell~1758b, which was found within the FoV of Abell~1758a).
 We do not detect SZ effects towards Zw1458.8+2233 and  Abell~1704, due to difficult source environments, nor towards Zw0857.9+2107, for reasons unclear to us. We have produced marginalized posterior distributions at $r_{500}$ and $r_{200}$ for 16 clusters (since Abell~2409 can not be fitted adequately by our model).

\begin{itemize}
\item Measurements of $M_{\rm{T}}(r_{200})$ are not common in the literature but are very important for testing large-radius scaling relations and understanding the physics in the outskirts of clusters. Consequently, the 16 measurements presented here, from a sample with narrow redshift-range, represent a significant increment to what already exists.

\item For the clusters studied, we find values for $M_{\rm{T}}(r_{200})$ span $2.0-6.1\pm 0.9 \times 10^{14}h^{-1}_{70}M_{\odot}$ and are typically 2-2.5 times larger than $M_{\rm{T}}(r_{500})$; we find $r_{200}$ is typically $1.1 \pm 0.1h_{70}^{-1}$\,Mpc.

\item AMI measurements of $M_{\rm{T}}(r_{500})$ are consistent with published optical results for 3 out of 4 clusters in our sample, with the weighted best-fit ratio\footnote{With the exception of Abell~1758a+b}
 between AMI SZ masses and lensing masses being $1.2^{+0.2}_{-0.3}$ within $r_{500}$  and $1.0 \pm 0.1$ within $r_{200}$. They are systematically lower than existing X-ray measurements
 of $M_{\rm{T}}(r_{500})$ for 6 clusters with available
 X-ray estimates and are only consistent with one of these measurements. The more discrepant masses correspond to the stronger mergers of the sample. The ratio of the X-ray masses to the AMI SZ
 masses is $1.7 \pm 0.2$ for the sample. The agreement with optical measurements improves for $M_{\rm{T}}(r_{200})$, though there are few data. We have investigated the AMI vs X-ray discrepancy by comparing
 $T_{\rm{AMI}}$ estimates with $T_{X}$ estimates, when available, \emph{at $r \approx$} \emph{500\,kpc}. There tends to be good agreement in less X-ray luminous clusters and in non-mergers but large-radius $T_{X}$ can be
 substantially larger than $T_{\rm{AMI}}$ in mergers. This explains why some X-ray mass estimates are significantly higher than the AMI estimates: the use of a higher
 temperature  will give a consequently higher mass  in the hydrostatic equilibrium model used. Another implication of a higher large-radius $T_{X}$ than $T_{\rm{AMI}}$ (given the respective $n^2$ and $n$ emission weightings) is that, even at around $r_{500}$, the gas is clumped or shocked or both. There is a clear need for more large-scale measurements.

\item We have investigated the effects of our main contaminant, radio sources, by searching for degeneracies in the posterior distributions of source flux densities for sources within $5\arcmin$ of the cluster SZ centroids. We find small or negligible degeneracies between source flux densities and cluster mass for all clusters, with the exceptions of Abell~781, Abell~1758a and RXJ1720.1+2638, which have
sources with flux densities of 9, 7 and 7\,mJy at $\lesssim 2\arcmin$ from the cluster SZ centroids. By simulating a cluster with a challenging source environment, we have shown that our AMI analysis can approximately recover the true mass, even in a degenerate scenario.

\item We often find differences in the position of SZ and X-ray peaks, with an average offset of $35\arcsec$, a median of $34\arcsec$ and a sample standard deviation of $24\arcsec$ for the entire sample (excluding Abell~1758a+b), confirming what has been seen in previous observational studies
 and in simulations.
 We emphasize that our sample size is small, but we find 
 no clear relation (except for Abell~1758) between position difference and merger activity. There is however an indication of a relation between merger activity and SZ ellipticity.

\item We have analysed the AMI data for two clusters: Abell~611 and Abell~2111, with a $\beta$ parameterization and with five gNFW parameterizations,
 including the widely used  \cite{arnaud2010} ``universal'' and the \cite{nagai2007} ones. This has revealed very different degeneracies in $Y_{\rm{sph}}(r_{500})-r_{500}$
 for the two types of cluster parameterization. For both clusters, the $\beta$ parameterization, which allows the shape parameters to be fitted,
  yielded stronger constraints on $r_{500}$ than any of the gNFW
paramaterizations. The Nagai et al. and Arnaud et al. gNFW parameters produced consistent results, with the latter giving slightly better constraints. Setting the gNFW parameters to different, but
 reasonable, values altered the degeneracies significantly. 
 This illustrates the risks of using a single set of fixed, averaged profile shape parameters to model all clusters.

\end{itemize}

\section*{Acknowledgments}\label{acknowledgements}

We thank an anonymous referee for very quick and helpful comments
 and suggestions, and Alastair Edge for helpful discussion. 
We are grateful to the staff of the Cavendish Laboratory and the Mullard Radio
Astronomy
Observatory for the maintenance and operation of AMI.
 We acknowledge support from Cambridge  University and
 STFC for funding and supporting AMI.  MLD, TMOF, MO, CRG, MPS and 
 TWS are grateful for support from STFC studentships. This work was
carried out using
 the Darwin Supercomputer of Cambridge University High
 Performance Computing Service (http://www.hpc.cam.ac.uk/), provided
 by Dell Inc. using Strategic Research Infrastructure Funding from the
 Higher Education Funding Council for England and the Altix 3700
 supercomputer at DAMTP, Cambridge University, supported by HEFCE
 and STFC. We thank Stuart Rankin and Andrey Kaliazin for their
 computing support.
 This research has made use of data from the {\sc{Chandra}} Data
Archive (\emph{ACCEPT}) \citep{cavagnolo2009}.
 We acknowledge the use of NASA's SkyView facility
     (http://skyview.gsfc.nasa.gov) located at NASA Goddard
     Space Flight Center.


\begin{thebibliography}{}
\setlength{\labelwidth}{0pt} 

\bibitem[\protect\citeauthoryear{Allen \& Fabian}{1998}]{allen1998} Allen S.~W., Fabian A.~C., 1998, MNRAS, 297, L63

\bibitem[\protect\citeauthoryear{Allen}{2000}]{allen2000} Allen~S.~W., 2000, MNRAS, 315, 269

\bibitem[\protect\citeauthoryear{Allen, Evrard, \& Mantz}{2011}]{allen2011} Allen S.~W., Evrard A.~E., Mantz A.~B., 2011, ARA\&A, 49, 409 

\bibitem[\protect\citeauthoryear{Allison et al.}{2011}]{allison2011} Allison J.~R., Taylor A.~C., Jones M.~E., Rawlings S., Kay S.~T., 2011, MNRAS, 410, 341

\bibitem[\protect\citeauthoryear{AMI Consortium: Davies et al.}{2010}]{davies2010} AMI Consortium, et al., 2011, MNRAS, 415, 2708 

\bibitem[\protect\citeauthoryear{AMI Consortium: Franzen et al.}{2010}]{franzen2010} AMI Consortium, et al., 2011, MNRAS, 415, 2699 

\bibitem[\protect\citeauthoryear{AMI Consortium: Hurley-Walker et al.}{2011}]{hurley2011} AMI Consortium, et al., 2012, MNRAS, 419, 2921 

\bibitem[\protect\citeauthoryear{AMI Consortium: Olamaie et al.}{2011}]{olamaie2011} AMI Consortium: Olamaie~M. et al., 2012, MNRAS, 421, 1136

\bibitem[\protect\citeauthoryear{AMI Consortium: Rodr\'{i}guez-Gonz\'{a}lvez et al.}{2010}]{carmen2010} AMI Consortium: Rodr\'{i}guez-Gonz\'{a}lvez~C et al., 2011, MNRAS, 414, 3751 

\bibitem[\protect\citeauthoryear{AMI Consortium: Shimwell et al.}{2010}]{SHIMWELL} AMI Consortium: Shimwell. T.~W., et al., 2010, arXiv:1012.4441

\bibitem[\protect\citeauthoryear{AMI Consortium: Zwart et al.}{2008}]{zwart2008}  AMI Consortium: Zwart~J.~T.~L. et al.,  2008, MNRAS, 391, 1545

\bibitem[\protect\citeauthoryear{AMI Consortium: Zwart et al.}{2010}]{zwart2010}  AMI Consortium: Zwart~J.~T.~L., et al., 2011, MNRAS, 418, 2754


\bibitem[\protect\citeauthoryear{Arnaud et al.}{2010}]{arnaud2010} Arnaud M., Pratt G.~W., Piffaretti R., B{\"o}hringer H., Croston J.~H., Pointecouteau E., 2010, A\&A, 517, A92

\bibitem[\protect\citeauthoryear{Baars et al.}{2005}]{baars2005} Baars~J., Davies~R., Jorgensen~I., Bergmann~M, Crampton~D., AJ, 130, 445

\bibitem[\protect\citeauthoryear{Barrena et al.}{2007}]{2007A&A...467...37B} Barrena R., Boschin W., Girardi M., Spolaor M., 2007, A\&A, 467, 37

\bibitem[\protect\citeauthoryear{Bauer et al.}{2005}]{Zw1454.8+2233_COOLING} Bauer, F, E., et al., 2005, MNRAS, 359, 1481

\bibitem[\protect\citeauthoryear{Bautz et al.}{2009}]{bautz2009} Bautz M.~W., et al., 2009, PASJ, 61, 1117 

\bibitem[\protect\citeauthoryear{Birkinshaw}{1999}]{birkinshaw1999} Birkinshaw~M., 1999, Physics Reports, 310, 97

\bibitem[\protect\citeauthoryear{Birkinshaw et al.}{1981}]{birkinshaw1981} Birkinshaw~M., Gull~S.~F. \& Northover~K., 1981, MNRAS, 197, 571

\bibitem[\protect\citeauthoryear{Birkinshaw et  al.}{1984}]{birkinshaw1984} Birkinshaw~M., Gull~S.~F., 1984, MNRAS, 206, 359

\bibitem[\protect\citeauthoryear{Birkinshaw}{1994}]{birkinshaw1994} Birkinshaw~M., Hughes~J.~P., 1994, ApJ, 420, 33

\bibitem[\protect\citeauthoryear{B\"ohringer et al.}{2004}]{bohringer2004} B\"ohringer~H., Schuecker~P., Guzzo~L. et al., 2004, A\&A, 469, 363

\bibitem[\protect\citeauthoryear{Bonamente et al.}{2004}]{OCRO_Abell 611} Bonamente M., et al., 2005, ApJ, 614, 56-63

\bibitem[\protect\citeauthoryear{Bonamente et al.}{2006}]{bona_chandra} Bonamente M., Joy M.~K., LaRoque S.~J., Carlstrom J.~E., Reese E.~D., Dawson K.~S., 2006, ApJ, 647, 25

\bibitem[\protect\citeauthoryear{Bonamente et al.}{2012}]{bonamente2011} Bonamente M., et al., 2012, NJPh, 14, 025010 

\bibitem[\protect\citeauthoryear{Browne et al.}{1998}]{browne1998} Browne~I.~W.~A., Wilkinson~P.~N., Patnaik~A.~R., Wrobel~J.~M., 1998, MNRAS, 293, 257

\bibitem[\protect\citeauthoryear{Browne et al.}{2000}]{Browne_2000} Browne I.~W., Mao S., Wilkinson P.~N., Kus A.~J., Marecki A., Birkinshaw M., 2000, SPIE, 4015, 299

\bibitem[\protect\citeauthoryear{Buote}{2001}]{buote2001} Buote D.~A., 2001, ApJ, 553, L15 

\bibitem[\protect\citeauthoryear{Carlstrom et al.}{1992}]{carlstrom1996} Carlstrom~J.~E., Joy~M., Grego~L., 1996, ApJ, 456, 75-78

\bibitem[\protect\citeauthoryear{Carlstrom, Holder \& Reese}{2002}]{carlstrom2002} Carlstrom J. E., Holder G. P., \& Reese E. D. 2002, ARA\&A, 40, 643

\bibitem[\protect\citeauthoryear{Carlstrom et al.}{2011}]{carlstrom2011} Carlstrom J.~E., et al., 2011, PASP, 123, 568

\bibitem[\protect\citeauthoryear{Cavagnolo et al.}{2000}]{cavagnolo2009} Cavagnolo~K.~W., Donahue~M., Voit~G.~M. et al., 2009, ApJS, 182, 12

\bibitem[\protect\citeauthoryear{Cavaliere \& Fusco-Fermiano}{1978}]{cavaliere1978} Cavaliere~A. \& Fusco-Fermiano~R., 1978, A \& A, 70, 677

\bibitem[\protect\citeauthoryear{Corless, King, \& Clowe}{2009}]{corless2009} Corless V.~L., King L.~J., Clowe D., 2009, MNRAS, 393, 1235 

\bibitem[\protect\citeauthoryear{Cypriano}{2005}]{cypriano2004} Cypriano~E.~S., Sodr\'{e}~Jr., Kneib~L. et al, 2005, ApJ, 630, 38-49

\bibitem[\protect\citeauthoryear{Donnarumma et al.}{2010}]{Xray_lense_Abell 611_2} Donnarumma A., et al., 2010, arXiv:1002.1625

\bibitem[\protect\citeauthoryear{David \& Kempner}{2004}]{2004ApJ...613..831D} David L.~P., Kempner J., 2004, ApJ, 613, 831

\bibitem[\protect\citeauthoryear{Ebeling et al.}{1998}]{ebeling1998} Ebeling~H., Edge~A.~C., Bohringer~H. et al., 1998, MNRAS, 301, 881

\bibitem[\protect\citeauthoryear{Ebeling et al.}{2000}]{ebeling2000} Ebeling~H., Edge~A.~C., Allen~S.~W. et al.,  2000, MNRAS, 318, 333

\bibitem[\protect\citeauthoryear{Eke, Cole, \& Frenk}{1996}]{eke1996} Eke V.~R., Cole S., Frenk C.~S., 1996, MNRAS, 282, 263 

\bibitem[\protect\citeauthoryear{Evrard}{1997}]{evrard1997} Evrard A.~E., 1997, MNRAS, 292, 289 

\bibitem[\protect\citeauthoryear{Fabjan et al.}{2011}]{fabjan2011} Fabjan D., Borgani S., Rasia E., Bonafede A., Dolag K., Murante G., Tornatore L., 2011, MNRAS, 416, 801 

\bibitem[\protect\citeauthoryear{Feroz \& Hobson}{2008}]{feroz2008} Feroz~F., Hobson~M.~P., 2008, MNRAS, 384, 449

\bibitem[\protect\citeauthoryear{Feroz, Hobson \& Bridges}{2009}]{bridges2009} Feroz~F.,  Hobson~M.~P., Bridges~M., 2009, MNRAS, 398, 1601

\bibitem[\protect\citeauthoryear{Feroz et al.}{2009}]{feroz2009}  Feroz~F., Hobson~M.~P., Zwart~T.~L. et al., 2009, MNRAS, 398, 2049

\bibitem[\protect\citeauthoryear{George et al.}{2009}]{george2009} George M.~R., Fabian A.~C., Sanders J.~S., Young A.~J., Russell H.~R., 2009, MNRAS, 395, 657 

\bibitem[\protect\citeauthoryear{Giovannini, Tordi, \& Feretti}{1999}]{Gio_A773_halo} Giovannini G., Tordi M., Feretti L., 1999, NewA, 4, 141

\bibitem[\protect\citeauthoryear{Giradi \& Mezzetti}{2001}]{girardi2001} Girardi~M., Mezzetti~M., 2001, ApJ, 548, 79

\bibitem[\protect\citeauthoryear{Govoni et al.}{2001}]{govoni2001} Govoni F., Feretti L., Giovannini G., B{\"o}hringer H., Reiprich T.~H., Murgia M., 2001, A\&A, 376, 803

\bibitem[\protect\citeauthoryear{Govoni et al.}{2004}]{govoni2004} Govoni~F., Markevitch~M., Vikhlinin~A., VanSpeybroeck~L., Feretti~L., Giovannini~G., 2004, ApJ, 605, 695

\bibitem[\protect\citeauthoryear{Govoni et al.}{2009}]{govoni2009} Govoni F., Murgia M., Markevitch M., Feretti L., Giovannini G., Taylor G.~B., Carretti E., 2009, A\&A, 499, 371

\bibitem[\protect\citeauthoryear{Grainge et al.}{1993}]{1993MNRAS.265L..57G} Grainge K., Jones M., Pooley G., Saunders R., Edge A., 1993, MNRAS, 265, L57

\bibitem[\protect\citeauthoryear{Grainge et al.}{1996}]{RT_A1413} Grainge K.G.B., et al., 1996, MNRAS, 228, 17

\bibitem[\protect\citeauthoryear{Grainger et al.}{2002}]{RT_Abell 611} Grainger W.F., et al., 2002, MNRAS, 337, 1207

\bibitem[\protect\citeauthoryear{Haines et al.}{2009}]{LoCuSS_A1758} Haines C. P., Smith G. P., Egami E., Okabe N., Takada M., Ellis R. S., Moran S. M., Umetsu K., 2009, MNRAS, 396, 1297

\bibitem[\protect\citeauthoryear{Hanisch}{1980}]{hanisch1980} Hanisch~R.~J., 1980, AJ, 85, 1565

\bibitem[\protect\citeauthoryear{Henriksen et al.}{1999}]{henriksen1999} Henriksen~M., Wnag~Q.~D. \& Ulmer~M.P., 1998, AJ, 116, 1529

\bibitem[\protect\citeauthoryear{Herbig et al.}{1995}]{herbig1995}  Herbig T., Lawrence C.~R., Readhead A.~C.~S., Gulkis S., 1995, ApJ, 449, L5

\bibitem[\protect\citeauthoryear{Hincks et al.}{2010}]{Hincks_2010} Hincks A.~D., et al., 2010, ApJS, 191, 423


\bibitem[\protect\citeauthoryear{Hoshino et al.}{2010}]{hoshino2010}  Hoshino A., et al., 2010, PASJ, 62, 371

\bibitem[\protect\citeauthoryear{Hudson et al.}{2010}]{hudson2010} Hudson D.~S., Mittal R., Reiprich T.~H., Nulsen P.~E.~J., Andernach H., Sarazin C.~L., 2010, A\&A, 513, A37 

\bibitem[\protect\citeauthoryear{Jones et al.}{1993}]{jones1993} Jones~M.~E, Saunders~R., Alexander~P. et al., 1993, Nature, 365, 320

\bibitem[\protect\citeauthoryear{Jones et al.}{2005}]{jones2005} Jones~M.~E. et al., 2005, MNRAS, 357, 518

\bibitem[\protect\citeauthoryear{Joy et al.}{2001}]{joy2001} Joy M., et al., 2001, ApJ, 551, L1 

\bibitem[\protect\citeauthoryear{Kaiser}{1986}]{kaiser1986} Kaiser N., 1986, MNRAS, 222, 323 

\bibitem[\protect\citeauthoryear{Katayama et al.}{2003}]{katayama2003} Katayama H., Hayashida K., Takahara F., Fujita Y., 2003, ApJ, 585, 687 

\bibitem[\protect\citeauthoryear{Kawaharada et al.}{2010}]{kawaharada2010} Kawaharada M., et al., 2010, ApJ, 714, 423

\bibitem[\protect\citeauthoryear{Klein et al.}{1991}]{klein1991} Klein~U., Rephaeli~Y., Schlickeiser~R. et al., 1991, A \& A, 244, 43

\bibitem[\protect\citeauthoryear{Korngut et al.}{2011}]{Korngut_2011} Korngut P., Dicker S., Reese E.~D., Mason B.~S., Devlin M.~J., Mroczkowski T., Sarazin C.~L., Sun M., 2011, AAS, 43, \#227.05 

\bibitem[\protect\citeauthoryear{Korngut et al.}{2011}]{korn2011} Korngut P.~M., et al., 2011, ApJ, 734, 10 

\bibitem[\protect\citeauthoryear{Kotov \& Vikhlinin}{2005}]{kotov2005} Kotov O., Vikhlinin A., 2005, ApJ, 633, 781 

\bibitem[\protect\citeauthoryear{Krause et al.}{2012}]{krause2012} Krause E., Pierpaoli E., Dolag K., Borgani S., 2012, MNRAS, 419, 1766 

\bibitem[\protect\citeauthoryear{Kravtsov, Vikhlinin, \& Nagai}{2006}]{kravtsov2006} Kravtsov A.~V., Vikhlinin A., Nagai D., 2006, ApJ, 650, 128 

\bibitem[\protect\citeauthoryear{Lancaster}{2007}]{lancaster2008} Lancaster~K., Birkinshaw~M, Gawro\'{n}ski, Browne~I, Feiler~R. et al., 2007, MNRAS, 378, 673

\bibitem[\protect\citeauthoryear{Lancaster et al.}{2011}]{Lancaster_2011} Lancaster K., et al., 2011, MNRAS, 418, 1441

\bibitem[\protect\citeauthoryear{LaRoque et al.}{2006}]{laroque2006} LaRoque~S.~A., Bonamente~M., Carlstrom~J.~E., Marshall~K.~J. et al., 2006, ApJ, 652, 917-936

\bibitem[\protect\citeauthoryear{Larson et al.}{2011}]{larson2010} Larson~D., Dunkley~J., Hinshaw~G., Komatsu~E. et al., 2011, ApJS, 192, 16

\bibitem[\protect\citeauthoryear{Lo, Martin, \& Chiueh}{2001}]{Lo_2001} Lo K., Martin R., Chiueh T., 2001, aprs.conf, 235

\bibitem[\protect\citeauthoryear{Partridge et al.}{1987}]{partridge1987} Partridge~R.~B., Perley~R.~A., Mandolezi~N., 1987, ApJ, 317, 112

\bibitem[\protect\citeauthoryear{Pratt et al.}{2004}]{pratt2004} Pratt~G.~W., B\"{o}hringer~H. \& Finoguenov~A, 2004, A \& A,  429, 791-806


\bibitem[\protect\citeauthoryear{Machacek et al.}{2002}]{machacek2002} Machacek~M.~E., Bautz~M.~W., Canizares~C. et al., 2002, ApJ, 567, 188

\bibitem[\protect\citeauthoryear{Mahdavi et al.}{2008}]{mahdavi2008} Mahdavi A., Hoekstra H., Babul A., Henry J.~P., 2008, MNRAS, 384, 1567 


\bibitem[\protect\citeauthoryear{Mandelbaum \& Seljak}{2007}]{mandelbaum2007} Mandelbaum R., Seljak U., 2007, JCAP, 6, 24 

\bibitem[\protect\citeauthoryear{Markevitch}{1997}]{markevitch1997} Markevitch~M., 1997, ApJ, 483, L17

\bibitem[\protect\citeauthoryear{Markevitch et al.}{2002}]{markevitch2002} Markevitch~M., Gonzalez~A.~H., David~L., Vikhlinin~A., Murray~S., Forman~W., Jones~C. \& Tucker~W., 2002, ApJ, 567, L27

\bibitem[\protect\citeauthoryear{Marriage et al.}{2011}]{Marriage_2011} Marriage T.~A., et al., 2011, ApJ, 737, 61

\bibitem[\protect\citeauthoryear{Marrone et al.}{2009}]{marrone2009} Marrone D.~P., et al., 2009, ApJ, 701, L114

\bibitem[\protect\citeauthoryear{Marrone et  al.}{2011}]{marrone2011} Marrone D.~P., et al., 2011, arXiv,  arXiv:1107.5115 

\bibitem[\protect\citeauthoryear{Marshall et al.}{2003}]{marshall2003} Marshall~P.~J., Hobson~M.~P. \& Slozar~A., 2003, MNRAS, 346, 489

\bibitem[\protect\citeauthoryear{Massardi et al.}{2010}]{massardi2010} Massardi M., Ekers R.~D., Ellis S.~C., Maughan B., 2010, ApJ, 718, L23 

\bibitem[\protect\citeauthoryear{Maughan et al.}{2008}]{maughan2008} Maughan~B.~J., Jones~C., Forman~W., Van Speybroeck~L., 2008, ApJS

\bibitem[\protect\citeauthoryear{Mazzotta et al.}{2001}]{RXJ1720_CHANDRA} Mazzotta P., Markevitch M., Vikhlinin A., Forman W.~R., David L.~P., van Speybroeck L., 2001, ApJ, 555, 205

\bibitem[\protect\citeauthoryear{Mazzotta et al.}{2004}]{mazzotta2004} Mazzotta P., Rasia E., Moscardini L., Tormen G., 2004, MNRAS, 354, 10 

\bibitem[\protect\citeauthoryear{Mazzotta \& Giacintucci}{2008}]{mazzotta2008} Mazzotta P., Giacintucci S., 2008, ApJ, 675, L9

\bibitem[\protect\citeauthoryear{McGlynn \& Fabian}{1984}]{fabian1984} McGlynn T.~A., Fabian A.~C., 1984, MNRAS, 208, 709 

\bibitem[\protect\citeauthoryear{Menanteau et al.}{2011}]{menanteau2011} Menanteau F., et al., 2011, arXiv, arXiv:1109.0953

\bibitem[\protect\citeauthoryear{Meneghetti et al.}{2010}]{meneghetti2010} Meneghetti M., Rasia E., Merten J., Bellagamba F., Ettori S., Mazzotta P., Dolag K., Marri S., 2010, A\&A, 514, A93 

\bibitem[\protect\citeauthoryear{Miller et al.}{2005}]{miller2005} Miller C.~J., et al., 2005, AJ, 130, 968

\bibitem[\protect\citeauthoryear{Molnar, Hearn, \& Stadel}{2012}]{molnar2012} Molnar S.~M., Hearn N.~C., Stadel J.~G., 2012, arXiv, arXiv:1201.1533 

\bibitem[\protect\citeauthoryear{Morandi, Ettori \& Moscardini}{2007}]{MORANDI_1_Abell 611} Morandi, A., Ettori, S., Moscardini, L. 2007, MNRAS, 379, 518

\bibitem[\protect\citeauthoryear{Morandi \& Ettori}{2007}]{MORANDI_2_Abell 611} Morandi, A., Ettori, S., 2007, MNRAS, 708

\bibitem[\protect\citeauthoryear{Motl et al.}{2005}]{motl2005} Motl P.~M., Hallman E.~J., Burns J.~O., Norman M.~L., 2005, ApJ, 623, L63 

\bibitem[\protect\citeauthoryear{Mroczkowski et al.}{2009}]{mrocz2009} Mroczkowski T., et al., 2009, ApJ, 694, 1034

\bibitem[\protect\citeauthoryear{Mroczkowski et al.}{2011}]{Mroczkowski_2011} Mroczkowski T., et al., 2011, MmSAI, 82, 485

\bibitem[\protect\citeauthoryear{Mroczkowski}{2011}]{mroc2011} Mroczkowski T., 2011, ApJ, 728, L35

\bibitem[\protect\citeauthoryear{Muchovej et al.}{2011}]{Muchovej_2011} Muchovej S., et al., 2011, ApJ, 732, 28

\bibitem[\protect\citeauthoryear{Nagai et al.}{2007}]{nagai2007} Nagai~D., Vikhlin~A \& Kravtsov~A., 2007, ApJ, 655, 98

\bibitem[\protect\citeauthoryear{Nagai \& Lau}{2011}]{nagai2011} Nagai D., Lau E.~T., 2011, ApJ, 731, L10 

\bibitem[\protect\citeauthoryear{Navarro, Frenk, \& White}{1996}]{NFW}  Navarro J.~F., Frenk C.~S., White S.~D.~M., 1996, ApJ, 462, 563

\bibitem[\protect\citeauthoryear{Nelson et al.}{2011}]{nelson2012} Nelson K., Rudd D.~H., Shaw L., Nagai D., 2011, arXiv, arXiv:1112.3659 

\bibitem[\protect\citeauthoryear{Okabe \& Umetsu}{2008}]{okabe2008} Okabe N., Umetsu K., 2008, PASJ, 60, 345 

\bibitem[\protect\citeauthoryear{Okabe et al.}{2010}]{okabe2010} Okabe N., Takada M., Umetsu K., Futamase T., Smith G.~P., 2010, PASJ, 62, 811

\bibitem[\protect\citeauthoryear{Olamaie, Hobson, \& Grainge}{2011}]{malakprof} Olamaie M., Hobson M.~P., Grainge K.~J.~B., 2011, arXiv, arXiv:1109.2834

\bibitem[\protect\citeauthoryear{Owers, Nulsen, \& Couch}{2011}]{owers2011} Owers M.~S., Nulsen P.~E.~J., Couch W.~J., 2011, ApJ, 741, 122

\bibitem[\protect\citeauthoryear{Patnaik et al.}{1992}]{patnaik1992} Patnaik~A.~R., Browne~I.~W.~A., Wilkinson~P.~N., Wrobel~J.~M., 1992, MNRAS, 254, 655

\bibitem[\protect\citeauthoryear{Planck Collaboration et al.}{2011}]{planck2011} Planck Collaboration, et al., 2011, A\&A, 536, A8 

\bibitem[\protect\citeauthoryear{Planck Collaboration et al.}{2011}]{planck2011b} Planck Collaboration, et al., 2011, A\&A, 536, A10 

\bibitem[\protect\citeauthoryear{Planck and AMI Collaborations et al.}{2012}]{planckami} Planck and AMI collaborations et al., 2012, in preparation

\bibitem[\protect\citeauthoryear{Poole et al.}{2007}]{poole2007} Poole G.~B., Babul A., McCarthy I.~G., Fardal M.~A., Bildfell C.~J., Quinn T., Mahdavi A., 2007, MNRAS, 380, 437 

\bibitem[\protect\citeauthoryear{Pratt \& Arnaud}{2005}]{XMM-A1413} Pratt G.W., Arnaud M., 2005, A\&A, 429, 791-806

\bibitem[\protect\citeauthoryear{Lau, Kravtsov, \& Nagai}{2009}]{lau2009} Lau E.~T., Kravtsov A.~V., Nagai D., 2009, ApJ, 705, 1129 

\bibitem[\protect\citeauthoryear{Rasia, Tormen, \& Moscardini}{2004}]{rasia2004} Rasia E., Tormen G., Moscardini L., 2004, MNRAS, 351, 237 

\bibitem[\protect\citeauthoryear{Rasia et al.}{2012}]{rasia2012} Rasia E., et al., 2012, arXiv, arXiv:1201.1569 

\bibitem[\protect\citeauthoryear{Richard et al.}{2010}]{richard2010} Richard J., et al., 2010, MNRAS, 404, 325 

\bibitem[\protect\citeauthoryear{Ricker \& Sarazin}{2001}]{ricker2001} Ricker P.~M., Sarazin C.~L., 2001, ApJ, 561, 621 

\bibitem[\protect\citeauthoryear{Rines et al.}{2010}]{rines2010} Rines~K., Geller~M.~J., Diaferio~A., 2010, The Astrophysical Journal Letters, 715, 180

\bibitem[\protect\citeauthoryear{Rizza el al.}{1998}]{rizza1998} Rizza~E., Burns~J.~O., Ledlow~M.~J. et al., 1998, MNRAS, 301, 328

\bibitem[\protect\citeauthoryear{Romano et al.}{2010}]{romano_a611} Romano A., et al., 2010, A\&A, 514, A88

\bibitem[\protect\citeauthoryear{Rossetti et al.}{2011}]{rossetti2011} Rossetti M., Eckert D., Cavalleri B.~M., Molendi S., Gastaldello F., Ghizzardi S., 2011, A\&A, 532, A123

\bibitem[\protect\citeauthoryear{Rozo, Wu, \& Schmidt}{2011}]{rozo2011} Rozo E., Wu H.-Y., Schmidt F., 2011, ApJ, 735, 118 

\bibitem[\protect\citeauthoryear{Rudnick \& Lemmerman}{2009}]{rudnick2009} Rudnick L., Lemmerman J.~A., 2009, ApJ, 697, 1341

\bibitem[\protect\citeauthoryear{Russell et al.}{2010}]{russel09} Russell, H.R. and Sanders, J. S. and Fabian, A. C. and Baum, S. A. and Donahue, M. and Edge, A. C. and McNamara, B. R. and O'Dea, C. P.,  2010, MNRAS

\bibitem[\protect\citeauthoryear{Sanderson et al.}{2009}]{sanderson2009} Sanderson, A. J. R., Edge, A. C., Smith, G. P., 2009, MNRAS, 398, 1698

\bibitem[\protect\citeauthoryear{Saunders et al.}{2003}]{2003MNRAS.341..937S} Saunders R., et al., 2003, MNRAS, 341, 937

\bibitem[\protect\citeauthoryear{Schmidt \& Allen}{2007}]{schmidt_A611} Schmidt R.~W., Allen S.~W., 2007, MNRAS, 379, 209

\bibitem[\protect\citeauthoryear{Sehgal et al.}{2008}]{Abell 781_XMM} Sehgal N., Hughes J.~P., Wittman D., Margoniner V., Tyson J.~A., Gee P.,  dell'Antonio I., 2008, ApJ, 673, 163

\bibitem[\protect\citeauthoryear{Sifon et al.}{2012}]{sifon2012} Sifon C., et al., 2012, arXiv, arXiv:1201.0991 

\bibitem[\protect\citeauthoryear{Sijacki et al.}{2007}]{sijacki2007} Sijacki D., Springel V., Di Matteo T., Hernquist L., 2007, MNRAS, 380, 877 

\bibitem[\protect\citeauthoryear{Squires et al.}{1996}]{squires1996} Squires~G., Kaiser~N., Babul~A, Fahlman~G, Woods~D, Neumann~D.~M., B\"{o}hringer~H., 1996, ApJ, 461, 572

\bibitem[\protect\citeauthoryear{Smith et al.}{2003}]{smith2003} Smith~G.~P., Edge~A.~C., Eke~V.~R., Nichol~R.~C., Smail~I., Kneib~J.~P., 2003, ApJ, 590, L79

\bibitem[\protect\citeauthoryear{Smith et al.}{2004}]{smith2004} Smith~G.~P., Kneib~J.~P., Smail~I, Mazzotta~P., Ebeling~H., 2005, MNRAS, 359, 417

\bibitem[\protect\citeauthoryear{Snowden et al.}{2008}]{XMM_snowden} Snowden, S. L., Mushotzky, R. F., Kuntz, K. D., Davis, D. S., 2008, A\&A, 478, 615

\bibitem[\protect\citeauthoryear{Sunyaev \& Zel'dovich}{1970}]{sunyaev1970} Sunyaev~R.~A., \& Zel'dovich~Y.~B., 1970, Comments on Astrophysics and Space, 2, 66

\bibitem[\protect\citeauthoryear{Takizawa, Nagino, \& Matsushita}{2010}]{takizawa2010} Takizawa M., Nagino R., Matsushita K., 2010, arXiv, arXiv:1004.3322 

\bibitem[\protect\citeauthoryear{Tauber et al.}{2010}]{tauber2010} Tauber J.~A., et al., 2010, A\&A, 520, A1 

\bibitem[\protect\citeauthoryear{Tsuboi et al.}{1998}]{tsuboi1998} Tsuboi~M., Miyazaki~A., Kasuga~T. et al., 2004, PASJ, 56, 711

\bibitem[\protect\citeauthoryear{Venturi et al}{2008}]{Zw1454.8+2233_HALO} Venture, T., et al., 2008, A\&A 484, 327-340

\bibitem[\protect\citeauthoryear{Venturi et al.}{2008}]{GMRT_HALO} Venturi et al., 2008, A\&A, 484, 327

\bibitem[\protect\citeauthoryear{Vikhlinin et al.}{2005}]{Chandra-A1413} Vikhlinin A., et al., 2005, ApJ, 628, 655-672

\bibitem[\protect\citeauthoryear{Voit}{2005}]{2005RvMP...77..207V}  Voit G.~M.,2005, RvMP, 77, 207

\bibitem[\protect\citeauthoryear{Wang et al.}{1997}]{wang1997} Wang~Q.~D, Ulmer~M.~P. \& LAvery~R.~J., 1997, MNRAS, 288, 702

\bibitem[\protect\citeauthoryear{White et al.}{1993}]{white1993} White S.~D.~M., Navarro J.~F., Evrard A.~E., Frenk C.~S., 1993, Natur, 366, 429 

\bibitem[\protect\citeauthoryear{White}{2000}]{white2000} White~D.~A., 2000, MNRAS, 312, 663

\bibitem[\protect\citeauthoryear{Wilkinson et al.}{1998}]{wilkinson1998} Wilkinson~P.~N., Browne~I.~W.~A., Patnaik~A.~R., Wrobel~J.~M., Sorathia~B., 1998, MNRAS, 300, 790

\bibitem[\protect\citeauthoryear{Williamson et al.}{2011}]{Williamson_2011} Williamson R., et al., 2011, ApJ, 738, 139

\bibitem[\protect\citeauthoryear{Wu et al.}{2008}]{Wu_2008} Wu J.-H.~P., et al., 2008, MPLA, 23, 1675

\bibitem[\protect\citeauthoryear{Vikhlinin et al.}{2006}]{vikhlinin2006} Vikhlinin A., Kravtsov A., Forman W., Jones C., Markevitch M., Murray S.~S., Van Speybroeck L., 2006, ApJ, 640, 691

\bibitem[\protect\citeauthoryear{Voit, Kay, \& Bryan}{2005}]{voit2005} Voit G.~M., Kay S.~T., Bryan G.~L., 2005, MNRAS, 364, 909 

\bibitem[\protect\citeauthoryear{Zhang et al.}{2008}]{zhang2008} Zhang~Y.~Y., Okabe~N., Finoguenov~A. et al., 2008, A \& A, 482, 451

\bibitem[\protect\citeauthoryear{Zhang et al.}{2010}]{zhang2010} Zhang~Y.~Y., Okabe~N., Finoguenov~A. et al., 2010, ApJ, 711, 1033-1043

\end{thebibliography}
\end{document}